\documentclass[11pt,openany,amsbook]{mybook}
\usepackage[a4paper]{geometry}
\usepackage{xcolor}
 \usepackage{exscale,relsize}
  \usepackage{framed}
  \usepackage{fix-cm}

\usepackage{amsthm}
\usepackage{multirow}

 \usepackage{hhcount}
\usepackage{amssymb}
\usepackage{cancel}
\usepackage{bm}
\usepackage{latexsym}
\usepackage{amssymb}
\usepackage{amsmath}
\usepackage{amsfonts}
\usepackage{colortbl}
\usepackage{multirow}
\usepackage{array}
\usepackage{booktabs}
\usepackage{rotating}
\usepackage{graphicx}
\usepackage{wrapfig}
\usepackage{slashed}
\usepackage{subfigure}
 \usepackage{exscale,relsize}
 
\usepackage{fancyhdr}   
\usepackage[titles]{tocloft}            

\usepackage{makeidx}                    

\newcolumntype{x}[1]{%
>{\centering\hspace{0pt}}p{#1}}%

\usepackage[margin=1cm,labelfont={sf,bf,scriptsize},textfont={sl,scriptsize}]{caption}

\newcommand{\be}{\begin{equation}}
\newcommand{\ee}{\end{equation}}
\newcommand{\bea}{\begin{eqnarray}}
\newcommand{\eea}{\end{eqnarray}}
\newcommand{\bml}{\begin{subequations}}
\newcommand{\eml}{\end{subequations}}
\newcommand{\bfig}{\begin{figure}}
\newcommand{\efig}{\end{figure}}

\newcommand{\del}{\delta}

\newcommand{\Del}{\Delta}

\RequirePackage{color}
\definecolor{dullpurple}{rgb}{0.431,0.188,0.534}
\definecolor{darkgreen}{rgb}{0.075,0.302,0.047}
\definecolor{dullred}{rgb}{0.706,0.208,0.192}
\RequirePackage[colorlinks=true
,urlcolor=dullpurple
,anchorcolor=blue
,citecolor=dullred
,filecolor=blue
,linkcolor=darkgreen
,menucolor=blue
,pagecolor=blue
,linktocpage=true
,pdfproducer=medialab
]{hyperref}

\definecolor{shadecolor}{rgb}{0.706,0.208,0.192}

  \makeatletter
\newlength{\apb@width}
\newcommand{\autoparbox}[2][c]{\settowidth{\apb@width}{#2}\parbox[#1]{\apb@width}{#2}}

\makeatother

\renewcommand{\chaptermark}[1]          
  {\markboth{\sf\thechapter.\ #1}{}}    
\renewcommand{\sectionmark}[1]          
{\markright{\sf\thesection\ #1}}            

\renewcommand{\theequation}{\thesection.\arabic{equation}}

\renewcommand{\headrulewidth}{0 pt}     
\renewcommand{\footrulewidth}{0 pt}     

\fancypagestyle{plain}{                 
  \fancyhf{}                              
  \fancyfoot[C]{\bfseries\thepage}        
  \renewcommand{\headrulewidth}{0 pt}     
  \renewcommand{\footrulewidth}{0 pt}     
}

\makeatletter                           
\def\cleardoublepage{\clearpage\if@twoside \ifodd\c@page\else
   \hbox{}
   \thispagestyle{empty}                
   \newpage
   \if@twocolumn\hbox{}\newpage\fi\fi\fi}
\makeatother

\newcommand{\mychapter}[1]{
 \chapter*{#1\markboth{\sf #1}{\sf #1}}
 \addcontentsline{toc}{chapter}{#1}
}

\newcounter{secapp}[chapter]            
\newcommand{\sectionappendix}[1]        
{
  \newpage                              
  \renewcommand{\thesection}{\thechapter.\Alph{secapp}}
  \stepcounter{secapp}                  
  \refstepcounter{section}              
  \addcontentsline{toc}{section}{\protect\numberline{\thesection}#1} 
  {\flushleft\Large\sffamily\bfseries\thesection$\;\;$ #1\par}
  \sectionmark{#1}\vspace{\baselineskip}  
}

\begin{document}

%
%

\pagestyle{empty}

\begin{center}
\vspace*{15mm}
{
    {\fontsize{35}{50}\selectfont \sffamily \bfseries Field Theoretic Approaches To Early Universe} \\[45mm]

{\fontsize{18}{20}\selectfont \sffamily \bfseries Thesis submitted for the Degree of}\\[4mm]
 { \fontsize{18}{20}\selectfont \sffamily \bfseries Doctor of Philosophy (Science)}\\[4mm]
{\fontsize{18}{20}\selectfont \sffamily \bfseries in}\\[4mm]
{\fontsize{18}{20}\selectfont \sffamily \bfseries Physics (Theoretical)}\\[4mm]
{\fontsize{18}{20}\selectfont \sffamily \bfseries by}\\[4mm]
{\fontsize{20}{28}\selectfont \sffamily \bfseries Sayantan Choudhury}
 \\[80mm]

  {\fontsize{18}{20}\selectfont \sffamily \bfseries University of Calcutta, India}\\[4mm] 
{\fontsize{18}{20}\selectfont \sffamily \bfseries December, 2014}
%
%

\vspace{2cm}

\vfill
}\end{center}
\newpage
\begin{center}
\vspace*{85mm}
{
{ \textsc{ \LARGE\cal \bf I am dedicating my thesis to my beloved parents and well wishers}}
}\end{center}
\newpage
\begin{center}
{
{ \textsc{ \fontsize{20}{28}\selectfont \sffamily \bfseries \bf \underline{ACKNOWLEDGEMENT}}}
}\end{center}
I would like to express my gratitude to the Council of Scientific and Industrial Research (C.S.I.R.)
for supporting me financially throughout the tenure of my research work. I am also grateful to Indian Statistical Institute (I.S.I.), Kolkata
for giving me an opportunity to forge ahead with my research work in a decent ambiance. 

First and foremost of all, I would like to convey my heart-felt thanks to my supervisor Prof. Supratik Pal for his continuous support, constant guidance 
and never ending inspiration in course of my Ph.D. study and research. I really owe, a grate deal, to him for motivating me to carry on with my research.
Indeed, his ingenuity, insight and providence have showed me the path as to how to tread as a researcher.  

During the tenure of research work I have, besides my supervisor, come across a good number of distinguished collaborators namely-
Prof. Soumitra SenGupta, Prof. Anupam Mazumdar, Prof. Sayan Kar, Prof. Sudhakar Panda, Prof. Banasri Basu, Prof. Pratul Bandyopadhyay who excel in their 
own field of activity. I am also thankful to my young friends Mr. Arnab Dasgupta, Mr. Soumya Sadhukhan, Mrs. Trina Chakraborty, Dr. Barun Kumar Pal and Dr. Joydip Mitra for collaboration. 
I, therefore, wish to record my appreciation and gratitude towards them for their excellent cooperation.

It would be unjust for me if do not convey my gratefulness to all the faculty members of Physics and Applied Mathematics Unit, Indian Statistical Institute, Kolkata 
for helping me in various aspects. It cannot be gainsaid that I have learned a great deal from them through constant interaction. 

I shall be failing in my duty if I do not appreciate the role of fellow research scholars who have never hesitated to share their knowledge and wisdom with me 
from time to time which, in turn, have helped me to consolidate my ideas and to translate them in writing. 

I would like to acknowledge my sincere gratitude to Centre for Theoretical Physics, Jamia Millia Islamia,  Tata Institute of Fundamental Research, 
 Inter-University Centre for Astronomy and Astrophysics, Harish Chandra Research Institute, Institute of Physics, Institute of Mathematical Science,
 Indian Institute of Science, Indian Institute of Technology, Chennai and Kharagpur and many others for inviting me to participate in various research programs and 
to visit for a short period since the inception of my joining as a research scholar. I am also thankful to The Abdus Salam International Centre for Theoretical Physics, Trieste, Italy 
for giving me the opportunity to visit there for taking part in meaningful discussion on two occasions. 

My special thanks to Indian Association for the Cultivation of Science, Kolkata for allowing me to carry on my research activity under the aegis of a distinguished
Professor Soumitra SenGupta.  

Finally and most of all, I would like to thank my parents for their selfless sacrifice and unflinching support sans it would be difficult for me 
to arrive at this stage of my carrier.   
\newpage
\begin{center}
{
{ \textsc{ \fontsize{20}{28}\selectfont \sffamily \bfseries \bf \underline{ABSTRACT}}}
}\end{center}

This thesis compiles the results of six works which deal with - inflationary model building and estimation of cosmological parameters from various field theoretic setup, 
quantification of reheating temperature, studies of leptogenesis in braneworld and 
estimation of primordial non-Gaussianity from ${\cal N}=1$ supergravity using $\delta N$ formalism.

We start our discussion with exploring the possibility of 
MSSM inflation in the light of recent observed data from various D -flat directions using the saddle and inflection point techniques.
 The effective inflaton potential
 around saddle point and inflection point have been utilized in estimating 
 the observable parameters and confronting them with
WMAP7 and Planck dataset.

 Next we explore the possibility of inflation from the five dimensional ${\cal N}=2$ supergravity setup by deriving the effective potential
 in the context of Randall-Sundrum like braneworld model and Dirac Bonn Infeld Galileon. After deriving an four dimensional effective potential,
 we obtain the inflationary observables from both the scenarios and confront them with the 
WMAP7 data. Further we fit the CMB angular power spectra from TT anisotropy and other polarization data obtained from WMAP7. 

Further, we discuss the non-trivial features of reheating from supergravity inspired braneworld model,
where the results are to some extent different from that of the usual low energy General Relativistic 
counterpart, because of the modified Friedmann equations in this setup. We explicitly derive the analytical expressions for the reheating temperature and further solve the evolution equation
of the number density of thermal gravitino which results in the gravitino abundance.

Finally, we study the primordial non-Gaussian features using $\delta N$ formalism 
of unavoidable higher dimensional non-renormalizable K\"ahler operators for ${\cal N}=1$ supergravity framework.
In particular we study the nonlinear evolution of cosmological perturbations on
large scales which enable us to compute the curvature perturbation,  
without solving the exact perturbed field equations.
Hence we compute the various non-Gaussian parameters 
for local type of non-Gaussianities, 
for a generic class of sub-Planckian models dominated by the Hubble-induced corrections.
 \newpage
\begin{center}
{
{ \textsc{ \fontsize{18}{24}\selectfont \sffamily \bfseries \underline{PUBLICATIONS}}}\\
(Thesis is based on the papers marked by the symbol $*$)
}\end{center}
\begin{enumerate}
 \item {\textsc{ \large\cal \bf  Brane inflation in background supergravity$*$}}\\
{\large\cal  $~$ \textcolor{red}{{\it Sayantan Choudhury}} and Supratik Pal }\\
{\large\cal   $~$\textcolor{blue}{{ Physical Review D 85, 043529 (2012)}}. }
\item {\textsc{ \large\cal \bf  Reheating and leptogenesis in a SUGRA inspired brane inflation$*$}}\\
{\large\cal  $~$ \textcolor{red}{{\it Sayantan Choudhury}} and Supratik Pal }\\
{\large\cal  $~$ \textcolor{blue}{{ Nuclear Physics B 857 (2012) pp. 85-100}}. }
\item {\textsc{ \large\cal \bf  Fourth level MSSM inflation from new flat directions$*$}}\\
{\large\cal  $~$ \textcolor{red}{{\it Sayantan Choudhury}} and Supratik Pal }\\
{\large\cal  $~$ \textcolor{blue}{{ Journal of Cosmology and Astroparticle Physics 04 (2012) 018}}. }
\item {\textsc{ \large\cal \bf  DBI Galileon inflation in background SUGRA$*$}}\\
{\large\cal  $~$ \textcolor{red}{{\it Sayantan Choudhury}} and Supratik Pal }\\
{\large\cal  $~$ \textcolor{blue}{{ Nuclear Physics B 874 (2013) pp. 85-114}}. }
\item {\textsc{ \large\cal \bf  Primordial non-Gaussian features from DBI Galileon inflation}}\\
{\large\cal  $~$ \textcolor{red}{{\it Sayantan Choudhury}} and Supratik Pal }\\
{\large\cal  $~$ \textcolor{blue}{{ arXiv:1210.4478}}. }
\item {\textsc{ \large\cal \bf  Features of warped geometry in presence of Gauss-Bonnet coupling}}\\
{\large\cal  $~$ \textcolor{red}{{\it Sayantan Choudhury}} and Soumitra SenGupta }\\
{\large\cal  $~$ \textcolor{blue}{{ Journal of High Energy Physics 02 (2013) 136}}. }
\item {\textsc{ \large\cal \bf  Higgs inflation from new K\"ahler potential}}\\
{\large\cal  $~$ \textcolor{red}{{\it Sayantan Choudhury}}, Trina Chakraborty  and Supratik Pal }\\
{\large\cal  $~$ \textcolor{blue}{{ Nuclear Physics B 880 (2014) pp. 155-174}}. }
\item {\textsc{ \large\cal \bf  Low \& High scale MSSM inflation, gravitational waves and constraints from Planck$*$}}\\
{\large\cal  $~$ \textcolor{red}{{\it Sayantan Choudhury}}, Anupam Mazumdar and Supratik Pal }\\
{\large\cal  $~$ \textcolor{blue}{{ Journal of Cosmology and Astroparticle Physics 07 (2013) 041}}. }
\item {\textsc{ \large\cal \bf  Thermodynamics of Charged Kalb Ramond AdS black hole in presence of Gauss-Bonnet coupling}}\\
{\large\cal  $~$ \textcolor{red}{{\it Sayantan Choudhury}} and Soumitra SenGupta }\\
{\large\cal  $~$ \textcolor{blue}{{  arXiv:1306.0492}}. }
\item {\textsc{ \large\cal \bf  An accurate bound on tensor-to-scalar ratio and the scale of inflation}}\\
{\large\cal  $~$ \textcolor{red}{{\it Sayantan Choudhury}} and Anupam Mazumdar }\\
{\large\cal  $~$ \textcolor{blue}{{ Nuclear Physics B 882 (2014) pp. 386-396}}. }
\item {\textsc{ \large\cal \bf  Primordial blackholes and gravitational waves for an inflection-point model of inflation}}\\
{\large\cal  $~$ \textcolor{red}{{\it Sayantan Choudhury}} and Anupam Mazumdar }\\
{\large\cal  $~$ \textcolor{blue}{{  Phyics Letters B 733 (2014) 270-275}}.}
\item {\textsc{ \large\cal \bf  Collider constraints on Gauss-Bonnet coupling in warped geometry model}}\\
{\large\cal  $~$ \textcolor{red}{{\it Sayantan Choudhury}}, Soumya Sadhukhan and Soumitra SenGupta }\\
{\large\cal  $~$ \textcolor{blue}{{   arXiv:1308.1477}}. }
\item {\textsc{ \large\cal \bf  Galileogenesis: A new cosmophenomenological zip code for reheating through R-parity violating coupling}}\\
{\large\cal  $~$ \textcolor{red}{{\it Sayantan Choudhury}} and  Arnab Dasgupta }\\
{\large\cal  $~$ \textcolor{blue}{{ Nuclear Physics B 882 (2014) pp. 195-204}}. }
\item {\textsc{ \large\cal \bf  A step towards exploring the features of Gravidilaton sector in Randall-Sundrum scenario via lightest Kaluza-Klein graviton mass}}\\
{\large\cal  $~$ \textcolor{red}{{\it Sayantan Choudhury}} and Soumitra SenGupta }\\
{\large\cal  $~$ \textcolor{blue}{{  Europian Physical Journal C 74 (2014) 3159}}. }
\item {\textsc{ \large\cal \bf  Constraining ${\cal N}=1$ supergravity inflationary framework with non-minimal K\"ahler operators}}\\
{\large\cal  $~$ \textcolor{red}{{\it Sayantan Choudhury}}, Anupam Mazumdar and Ernestas Pukartas}\\
{\large\cal  $~$ \textcolor{blue}{{  Journal of High Energy Physics 04 (2014) 077}}. }
\item {\textsc{ \large\cal \bf  Constraining ${\cal N}=1$ supergravity inflation with non-minimal K\"ahler operators using $\delta N$ formalism$*$}}\\
{\large\cal  $~$ \textcolor{red}{{\it Sayantan Choudhury}}}\\
{\large\cal  $~$ \textcolor{blue}{{  Journal of High Energy Physics 04 (2014) 105}}. }
\item {\textsc{ \large\cal \bf  Inflamagnetogenesis redux: Unzipping inflection-point inflation via various cosmoparticle probes}}\\
{\large\cal  $~$ \textcolor{red}{{\it Sayantan Choudhury}}}\\
{\large\cal  $~$ \textcolor{blue}{{  Physics Letters B 735 (2014) 138-145}}.}
\item {\textsc{ \large\cal \bf  Reconstructing inflationary potential from BICEP2 and running of tensor modes}}\\
{\large\cal  $~$ \textcolor{red}{{\it Sayantan Choudhury}} and Anupam Mazumdar }\\
{\large\cal  $~$ \textcolor{blue}{{   arXiv:1403.5549}}. }
\item {\textsc{ \large\cal \bf  Sub-Planckian inflation \& large tensor to scalar ratio with $r\geq 0.1$}}\\ 
{\large\cal  $~$ \textcolor{red}{{\it Sayantan Choudhury}} and Anupam Mazumdar }\\
{\large\cal  $~$ \textcolor{blue}{{   arXiv:1404.3398}}. }
\item {\textsc{ \large\cal \bf  Modulus stabilization in higher curvature gravity}}\\
{\large\cal  $~$ \textcolor{red}{{\it Sayantan Choudhury}}, Joydip Mitra and Soumitra SenGupta }\\
{\large\cal  $~$ \textcolor{blue}{{   Journal of High Energy Physics 08 (2014) 004}}.}
\item {\textsc{ \large\cal \bf  Can Effective Field Theory of inflation generate large tensor-to-scalar ratio within Randall Sundrum single braneworld? }}\\
{\large\cal  $~$ \textcolor{red}{{\it Sayantan Choudhury}}}\\
{\large\cal  $~$ \textcolor{blue}{{  arXiv:1406.7618}}.}
\item {\textsc{ \large\cal \bf  Measuring CP violation within EFT of inflation from CMB }}\\
{\large\cal  $~$ \textcolor{red}{{\it Sayantan Choudhury}}, Barun Kumar Pal, Banasri Basu and Pratul Bandyopadhyay }\\
{\large\cal  $~$ \textcolor{blue}{{   arXiv:1409.6036}}.}
\end{enumerate}

\newpage 
\pagestyle{fancy}


\pagenumbering{roman}\setcounter{page}{1}

\tableofcontents
\cleardoublepage

\ifodd\value{page}
\typeout{Mainmatter starts on odd page. Fine.}
\else
\typeout{Mainmatter starts on even page. Invert odd/even}
\fancyhf{}                              
\fancyhead[LO]                          
{{\bfseries\thepage$\quad$}\leftmark}   
\fancyhead[RE]                          
{\rightmark{$\quad$\bfseries\thepage}}  
\evensidemargin=22mm
\oddsidemargin=22mm
\fi

\pagenumbering{arabic} \setcounter{page}{1}

\newpage

\mychapter{List of figures}

\textbullet Fig.~\ref{fig1}: \textcolor{blue}{Examples of different classes of slow-roll potentials: (a) hilltop inflation, (b) infection point
inflation, (c) chaotic inflation and (d) natural inflation.........................................................}\textcolor{red}{30}
\textbullet Fig.~\ref{fig2z}: \textcolor{blue}{Schematic diagram of reheating phenomena which starts just after the end of inflation
.......................................................................................................................................................}\textcolor{red}{38}
\textbullet Fig.~\ref{fvdd1}: \textcolor{blue}{\subref{fddd1} Visual representations of triangles forming the primordial bispectrum,
with various combinations of wave numbers satisfying $k_{3}\leq k_{2}\leq k_{1}$
 and \subref{fddd2} shapes of the primordial bispectra in which we show the normalized
amplitude of ${\cal S} (k_{1} , k_{2} , k_{3} )(k_{2} /k_{1} )^2 (k_{3} /k_{1})^2$ as a function of $k_{2}/k_{1}$ and $k_{3}/k_{1}$ for a given
$k_{1}$, with a condition that $k_{3}\leq k_{2}\leq k_{1}$ is satisfied..............................................}\textcolor{red}{47}
\textbullet Fig.~\ref{fvd1}: \textcolor{blue}{Visual representations of flat potential \ref{vf1} 
 near the saddle point 
 and \ref{vf2} near the inflection point..................................................................................................................................}\textcolor{red}{52}
\textbullet Fig.~\ref{figVr1181}: \textcolor{blue}{Variation of CMB angular power spectrum $~$ ($l(l+1)C^{TT}_{l}/2\pi$) for
best fit low scale MSSM model and WMAP seven years data with the multipoles $l$ for scalar mode......................................}\textcolor{red}{57}
\textbullet Fig.~\ref{figVr8167}: \textcolor{blue}{Variation of the 
 fraction of the energy density of the universe collapsing into PBHs as a function
of the PBH mass, for three different values of the threshold for MSSM..........................}\textcolor{red}{58}
\textbullet Fig.~\ref{figVr18}: \textcolor{blue}{Running of gaugino mass ($m_{i}(\mu)$)
 in one loop RGE for MSSM with the logarithmic scale $\log_{10}\left(\mu\right)$..........................................................................................................................................}\textcolor{red}{59}
\textbullet Fig.~\ref{figVr1568}: \textcolor{blue}{Running of soft mass squared ratio $\left(\frac{m^{2}_{\phi}(\mu)}{m^{2}_{\phi}(\mu_{0})}\right)$ in one loop RGE for MSSM 
with the logarithmic scale $\log_{10}\left(\mu\right)$.....................................................................................................................}\textcolor{red}{60}
\textbullet Fig.~\ref{figVr118}: \textcolor{blue}{Running of trilinear A -term ratio $\left(\frac{A_{\beta}(\mu)}{A_{\beta}(\mu_{0})}\right)$ in
 one loop RGE for MSSM with the logarithmic scale $\log_{10}\left(\mu\right)$..........................................................................................................................}\textcolor{red}{61}
\textbullet Fig.~\ref{figVr178}: \textcolor{blue}{Running of the ratio of the Yukawa coupling $\left(\frac{\lambda_{\beta}(\mu)}{\lambda_{\beta}(\mu_{0})}\right)$
 in one loop RGE for MSSM with the logarithmic scale $\log_{10}\left(\mu\right)$........................................................................................................}\textcolor{red}{62}
\textbullet Fig.~\ref{fig2}: \textcolor{blue}{For large scale MSSM inflation, $H\gg m_\phi$, we have shown the variation of $P_{S}$~vs~$n_{S}$.........}\textcolor{red}{66}
\textbullet Fig.~\ref{fig3}: \textcolor{blue}{We show the joint $1\sigma$ and $2\sigma$ CL. contours in $r-n_{S}$ plane using
 \subref{x2}~Planck+WMAP-9 data with~$\Lambda$CDM+r, and $\Lambda$CDM+$r+\alpha_{S}$,
 \subref{x3}~Planck+WMAP-9+BAO data with~$\Lambda$CDM+r and $\Lambda$CDM+$r+\alpha_{S}$ for MSSM...............................................................................................................}\textcolor{red}{68}
\textbullet Fig.~\ref{fig4}: \textcolor{blue}{We show the joint $1\sigma$ and $2\sigma$ CL. contours in $\alpha_{S}-n_{S}$ and $\kappa_{S}-\alpha_{S}$ plane using
 \subref{z1}~Planck+WMAP-9+BAO data with~$\Lambda$CDM+$\alpha_{S}$ and 
$\Lambda$CDM+$\alpha_{S}+r$,
 \subref{z2}~Planck+WMAP-9+BAO data with~$\Lambda$CDM+$\alpha_{S}+\kappa_{S}$ for MSSM..............................................................................}\textcolor{red}{69}
\textbullet Fig.~\ref{fig6}: \textcolor{blue}{TT-power spectrum within $\ell$~($2<l<2500$) for MSSM.........................................}\textcolor{red}{71}
\textbullet Fig.~\ref{vfcv1}: \textcolor{blue}{ Variation of \subref{figVr845} one loop corrected potential ($V(\phi)$) vs inflaton field ($\phi$),
  \subref{figVr9} 1-$|\eta_{V}|$ vs inflaton field $\phi$ for $C_{4}=-0.68$ and \subref{figVr241} number of e-foldings ($N$) vs inflation field ($\phi$) within the
 range $-0.70<D_{4}<-0.60$ for brane inflation......................................................................................}\textcolor{red}{78}
\textbullet Fig.~\ref{tt}: \textcolor{blue}{ We show the variation of CMB angular power spectrum for \subref{zc1}~TT,
\subref{zc2}~TE and \subref{zc3}~EE correlation with
 respect to multipole $l$ for brane inflationary model parameters...................................}\textcolor{red}{80}
\textbullet Fig.~\ref{vfcv}: \textcolor{blue}{  Variation of \subref{figgv} one loop corrected potential ($V(\phi)$) vs inflaton field ($\phi$)
 and \subref{fig:subfiggh1} number of
 e-foldings ($N$) vs inflation field ($\phi$) for best fit values of $C_{i}\forall i$ for DBI Galileon....................}\textcolor{red}{87}
\textbullet Fig.~\ref{figdbi}: \textcolor{blue}{  Variation of CMB angular power spectrum for \subref{zc1dbi}~TT,
\subref{zc2dbi}~TE and \subref{zc3dbi}~EE correlation with respect to multipole $l$ for DBI Galileon model parameters.
Also in \subref{zc4dbi} we show the variation of matter power spectrum with respect to the momentum scale.........................................................}\textcolor{red}{92}
\textbullet Fig.~\ref{figVr400}: \textcolor{blue}{Variation of the Hubble parameter
  with respect to dimensionless parameter $\frac{T}{T^{bh}_{ex}}$ in the domain $-0.70<D_{4}<-0.60 $ for brane inflationary model..................................................................}\textcolor{red}{99}
\textbullet Fig.~\ref{figVr900}: \textcolor{blue}{Variation of gravitino number density in
 a physical volume  vs scaled temperature in braneworld scenario......................................................................................................................}\textcolor{red}{101}
\textbullet Fig.~\ref{figVr600}: \textcolor{blue}{Variation of total gravitino abundance vs
 temperature in the domain $-0.70<D_{4}<-0.60$ for brane inflationary model.........................................................................................................}\textcolor{red}{104}
\textbullet Fig.~\ref{fz}: \textcolor{blue}{We show the constraints on 
  the non-renormalizable  K\"ahler operators, ``a'',``b'',``c'' and ``d'' 
with respect to the tensor-to-scalar ratio $r_{\star}$ at the pivot scale $k_{\star}=0.002~{\rm Mpc}^{-1}$...........................}\textcolor{red}{123}
\textbullet Fig.~\ref{fza}: \textcolor{blue}{Behaviour
 of the local type of non-Gaussian parameter $f_{NL}^{\mathrm{local}}$ computed from the effective theory of ${\cal N}=1$ supergravity
with respect to the sound speed $c_{s}$ in the Hubble induced inflection point inflationary regime, represented
by $H>>m_{\phi}$.............................................................}\textcolor{red}{124}
\textbullet Fig.~\ref{fzat}: \textcolor{blue}{Behaviour
 of the local type of non-Gaussian parameter $\tau_{NL}^{\mathrm{local}}$ computed from the effective theory of ${\cal N}=1$ supergravity
with respect to the sound speed $c_{s}$ in the Hubble induced inflationary regime is represented
by $H>>m_{\phi}$...........................................................................................}\textcolor{red}{125}
\textbullet Fig.~\ref{fzata}: \textcolor{blue}{Behaviour
 of the CMB dipolar asymmetry parameter $A_{CMB}$ computed from the effective theory of ${\cal N}=1$ supergravity
with respect to the tensor-to-scalar ratio $r_{*}$ at the pivot scale, $k_{*}\sim 0.002~{\rm Mpc}^{-1}$ for the Hubble induced inflation................................................................................}\textcolor{red}{126}


\mychapter{List of tables}

\textbullet Table~\ref{tabc1}: \textcolor{blue}{Present observational constraint
 for various inflationary observables.....................}\textcolor{red}{36}
\textbullet Table~\ref{tab2}: \textcolor{blue}{Input parameters in CAMB for MSSM model.........................................................}\textcolor{red}{56}
\textbullet Table~\ref{tab3}: \textcolor{blue}{Output
 obtained from CAMB for MSSM model.......................................................}\textcolor{red}{56}
\textbullet Table~\ref{tab5}: \textcolor{blue}{Entries of $f_{F}^{i}$ matrix obtained from the
 solution of RGE in MSSM.............................}\textcolor{red}{63}
\textbullet Table~\ref{tab6}: \textcolor{blue}{Entries of $(C_{\beta}^{i})^{ab}$ matrix obtained
 from the solution of RGE in MSSM..........................}\textcolor{red}{63}
\textbullet Table~\ref{tab7}: \textcolor{blue}{Entries of $K^{\beta i}$ matrix in MSSM....................................................................................}\textcolor{red}{63}
\textbullet Table~\ref{tab8}: \textcolor{blue}{MSSM parameter values obtained from RG
 flow for n=4 level flat directions................}\textcolor{red}{64}
\textbullet Table~\ref{tabv1}: \textcolor{blue}{Different observational  parameters related to the cosmological perturbation
  for brane inflation model...................................................................................................................................}\textcolor{red}{81}
\textbullet Table~\ref{tabv2}: \textcolor{blue}{Input
 in CAMB for brane inflation model.....................................................................}\textcolor{red}{82}
\textbullet Table~\ref{tabv3}: \textcolor{blue}{Output from CAMB for brane inflation........................................................................}\textcolor{red}{82}
\textbullet Table~\ref{tabg21}: \textcolor{blue}{DBI Galileon model dependent observational parameters.............................................}\textcolor{red}{92}
\textbullet Table~\ref{tabg2}: \textcolor{blue}{Input parameters
 in CAMB in DBI Galileon model.......................................................}\textcolor{red}{93}
\textbullet Table~\ref{tabg3}: \textcolor{blue}{Output parameters
 from CAMB in DBI Galileon model................................................}\textcolor{red}{93}


\chapter{Introduction}
\label{ch:FRW}

\section{Prologue}
\label{sec:geometry}
Mayhap the most important picture of the modern day science is the evolution of the
field  ``particle cosmology'' in thereotical physics. Within this subject area particle physics is one of the prime components which specifically tests the nature on
the very very smallest length scales (or equivalently in a very high energy scale),
on the other hand the subject cosmology deals with the description of the universe on the very very largest length scales. Apart from this fact that these two fields of theoretical physics are distinct in the
length scales of the specific features they analyse, it is totally impracticable to comprehend the exact theretical origin as well as the
growth of large-scale structure formation in the universe without knowing the exact ``initial conditions'' that finally led
to the structures that we observe today through various observational probes. The preferred ``initial conditions'' was set in the very early epoch of universe when all the four fundamental 
forces are active so that all of them can able to produce the cosmological perturbations in the smoothed density field \cite{kolb}. A detailed knowledge of the
presently observed large-scale structure of the universe will also be impractical without considering the specific effects of the dark component in
the cosmological density field in perturbation theory. Though things are not fully settled within this subject area, possibly this omnipresent dark component of the universe can be treated as an elementary
particle constituent from the early universe. Adiitionally it is important to note that the detailed knowledge of the structure of the presently observed universe may divulge
 intution into various facts which happened in the very
early universe. This also implies that the essence of the four fundamental forces and various particle contents at an energy scale 
afar the extend of terrestrial accelerators. Mayhap the early universe was the ideal particle
accelerator which will provide the first glance of theoretical physics at the scale of Grand Unified Theories (GUTs) \cite{Raby:2006sk,Langacker:1981},
or comparable to the Planck energy scale. The detailed study of the early universe deals with the physics of Standard Model and its beyond.
 This involves exploring ideas with Supersymmetry (SUSY) and Supergravity \cite{Nilles:1983ge} (SUGRA), String Theory \cite{Green1,Green2,Polchinski1,Polchinski2},
 Braneworld gravity \cite{Maartens:2010ar} and analyzing fundamental
 aspects like- reheating \cite{Bassett:2005xm,Allahverdi:2010xz,Mazumdar:2010sa}, leptogenesis \cite{Fong:2013wr,Davidson:2008bu}, baryogenesis 
\cite{Cline:2006ts,Morrissey:2012db} etc.
 In fact, the exciting opportunities presented by
 links to the early Universe and fundamental physics provided essential motivation for
 collecting the new data in the first place. 
Since the information about the early universe is encoded in the spectrum of cosmological fluctuations
 about the expanding background, it is these fluctuations which provide a link between the
physics of the very early universe and cosmological observations till date.
At the present stage observational cosmology is one of the strongest probes by which one can, at least attempt to check
 the fundamental components of the early universe. A wealth of new observational
results are being uncovered. The cosmic microwave background (CMB) has been measured to
high precision, the distribution of visible and dark matter is being mapped out to greater and
greater depths. New observational windows to probe the structure of the universe are opening up.
To explain these observational results it is necessary to consider processes which happened in the
very early universe.

The most natural proposal to explore the physics of early universe is Inflationary Cosmology. Originally {\it inflation}
was introduced by Guth \cite{guth} in order to explain the initial conditions
forced on the standard big bang model, but later, it has been found to have
 more important role: a favored candidate for the origin of structure
in our universe. Thirty years later {\it inflation} is still alive in a much stronger position than
ever based on highly precise observational data available of late. Inflation is said to complement
 the older standard big bang cosmology (SBBC) in a much sophisticated way.
 Although the idea of Standard Big Bang Cosmology (SSBC) very successfully explains expanding universe, nucleosynthesis,
formation of Cosmic Microwave Background Radiation (CMBR)\cite{wilson}, the
 temperature anisotropies \cite{silk} in the CMB challenged SBBC
as it can not satisfactorily explain those observations. The CMB which
gives us a snapshot of the very early universe shows that at last
scattering surface the universe was almost perfectly homogeneous
and isotropic on all scales. Then what initial conditions could lead to such homogeneity and
isotropy? Inflation not only provides natural answers to those but
also solves an additional puzzle of the SBBC, i. e., of the  generation of
cosmological perturbations.

 Cosmic $inflation$ was proposed to solve those pathological problems of SBBC and to get rid of
unwanted relics generally predicted by high energy theories. The simplest definition of
inflation is that it corresponds to a phase of super-fast acceleration in the
very early (time scale $t\sim10^{-34}~sec$) universe. 
During inflation, the potential energy of slowly rolling scalar field(s), called inflaton, is supposed to dominate the
energy density of our universe by a false vacuum. Also at that time the quantum fluctuations imprinted
on space-time are stretched outside the Hubble patch. These primordial fluctuations eventually
 re-enter the Hubble patch, whence their form can be extracted by observing the
perturbations in the CMBR.
In the past three decades numerous inflationary models have
been proposed, some of them \cite{lyth} are more or less consistent with the observations but the nature of
the field driving inflation is still unknown. The prime input of Inflationary Cosmology is inflaton potential which is
originated from the various background field theoretic prescriptions \cite{Mazumdar:2010sa,riotto}.  
The end of inflation can be considered as a paradigm for the origin of matter, since all matter arises from the vacuum energy stored in the inflaton field.
Slow-roll inflationary framework generically predicts almost Gaussian adiabatic perturbations with a nearly flat
spectrum, which
has met with an unprecedented success with the observational results obtained from CMB experiments like WMAP \cite{WMAP}, PLANCK \cite{Ade:2013zuv,Ade:2013uln} etc.
 Apart from the cosmological issues, one of the theoretical challenges is to understand the proper origin of
the scale of inflation which is in principle below the UV cut-off scale of the gravity in the effective theory prescription.
But if the inflation is embedded on the SM framework then in such scenario the masses are not protected from the
quantum (loop) corrections and in theoretical physics this which is known as the {\it hierarchy } or {\it fine-tuning } problem \cite{ArkaniHamed:1998rs,ArkaniHamed:1998nn}.
The most popular and successful remedy
is the extension of SM by including new symmetry, commonly this known as supersymmetry (SUSY) \cite{Nilles:1983ge,Martin:1997ns,Gates:1983nr}, which is presumably broken first at high scales in some
hidden sector and then transmitted to the global SUSY extension of the SM, by gravitational or gauge interactions \cite{Mazumdar:2010sa}. 
When SUSY is broken locally just similarly like a gauge symmetry, an intimate connection with
gravity emerges, commonly known as the supergravity (SUGRA) \cite{Nilles:1983ge,Martin:1997ns,Gates:1983nr}, which is valid in the sub-Planckian
scale of the effective low energy field theory. In a much broader sense, the unification of gravity along with the other gauge interactions, treated 
all the fundamental particles as the excitations of extended objects within the
framework of string theory \cite{Green1,Green2,Polchinski1,Polchinski2}. Therefore, it is a significant question to ask whether beyond the
SM physics prescription can provide all the right ingredients for inflation or not?
As there is still no unique model of inflation consistent with all
the theoretical and observational requirements, there are
 lots of open spaces to work on in this field.

Additionally it is important to note that, observational cosmology has indeed made very rapid progress in recent years.
The ability to quantify the universe has largely improved due to various available observational constraints
coming from large scale structure formation of the universe. The transition to precision cosmology has been spear-headed by 
measurements of the anisotropy in the cosmic microwave background (CMB)
over the past decade. Observations of the large scale structure formation of the universe in the distribution of galaxies,
 high red-shift supernova, have provided the required complementary information in this context. 
But there are also other open as well as challenging issues are appearing,
 which strongly motivates us to do the work on inflationary cosmology and on the particle physics and
 cosmology overlap. These are-
 \begin{itemize}
  \item breaking the degeneracy among the cosmological
 parameters using the CMB polarization data,
 \item dealing with sophisticated and precise data obtained from various probes with cosmic variances etc.  
 \end{itemize}
 In a nutshell, the aim of our work presented in the thesis is two fold:- 
\begin{itemize}
 \item To derive inflationary models from beyond the SM prescription, specifically from
SUSY and SUGRA, DBI Galileon framework and explore their pros and cons in the light of recent observational data,
\item To propose new observational tools for inflationary cosmology 
i. e. non-Gaussianity, precise constraint on scale of inflation, primordial gravitational waves (PGW) via tensor-to-scalar ratio and primordial black hole (PBH) formation 
etc. We have addressed some of these issues in this thesis. 
\end{itemize}

\section{Field theoretic tools for early universe}
\label{a1}

\subsection{Supersymmetry}
Supersymmetry (SUSY) is a generalization of the space time
symmetries in quantum field theory that transforms fermions into
bosons and vice versa \cite{Nilles:1983ge,Martin:1997ns,Gates:1983nr}. 
SUSY is appealing for various reasons both for phenomenological and cosmological points of view:
\begin{itemize}
\item It provides
a framework for the unification of the particle physics and
gravity, which is governed by the Planck scale where the gravitational
interactions of elementary particles is supposed to be comparable
to the gauge interactions. 

\item It is possible to explain the large
hierarchy of the energy scales from W and Z boson masses to the Planck
scale using SUSY.

\item  Stability of the hierarchy in presence
of radiative corrections is not possible in SM, but it
is possible for supersymmetric theories. SUSY eliminates the
quadratic divergence in the mass of fundamental light
scalar fields, \be \delta m^{2}\sim \Lambda^{2}_{UV},\ee where $\Lambda_{UV}$
is the scale where low energy theories no longer apply.

\item  SUSY is the
only framework in which we seem to be able to understand the light
fundamental scalars in the context of model building in particle physics and cosmology interface.

\item Although initially investigated for other reasons, SUSY turns out to have a
 significant impact on the cosmological constant problem, and may even be said to solve it halfway.

\end{itemize}

 Only to mention a few drawbacks, the theory
has too many parameters, it does not estimate the fermion masses
and why the number of generations is three. However these problems
can be addressed in the modern framework of SUSY. If SUSY is an
exact symmetry of nature, then particles and their superpartners
would have degenerated in mass. Since this is usually not observed in
experimental data from particle accelerators, SUSY is not an exact
symmetry and must be broken. The stability of hierarchy of scales
mentioned above can still be maintained if the symmetry breaking
is soft (corresponding symmetry breaking mass terms are no
longer more than a few TeV). The most interesting theories of this
type are theories of low energy or weak-scale SUSY where
the effective scale of SUSY breaking is tied to the scale of
electroweak symmetry breaking. At present, there are no experimental
support for the existence of low energy SUSY. However the 
unification of the three gauge couplings at an energy scale close
to the Planck scale may serve as the indirect evidence. The
unification of gauge coupling is not possible in Standard model
but it is achievable in minimal supersymmetric extension of the
Standard model (MSSM) \cite{Nilles:1983ge,Mazumdar:2010sa,Martin:1997ns,Csaki:1996ks,Aitchison:2005cf} and provides an additional motivation for low
energy SUSY. If Large Hadron Collider (LHC) \cite{lhc} uncovers the
evidence of SUSY, this would have a profound effect on the study of
TeV scale physics and development of more fundamental theory of
mass and symmetry breaking in particle physics. During the early
stages of the evolution of the universe, more precisely during
inflation SUSY plays an important role.

 The generator (Q) of the supersymmetric transformation is characterized by the
spin-$\frac{1}{2}$ degrees of freedom. The SUSY generator
$Q_{\alpha}(\alpha=1,2)$ can be chosen in such a way that it
belongs to the family of a left handed Weyl spinor which transforms
as $(\frac{1}{2},0)$
 under Lorentz transformations. Its Hermitian conjugate is
designated by $\overline {Q}_{\dot{{\beta}}}$, which belongs
to the family of right handed Weyl spinor. Since the
anti-commutator of any operator with its Hermitian adjoint is
nonzero, therefore, it  does not vanish in this
context. It transform as ($\frac{1}{2},\frac{1}{2}$) under
Lorentz transformations. Similarly it also transforms
as ($\frac{1}{2},\frac{1}{2}$) under Lorentz transformation
like the anti-commutator stated above. The superalgebra is
defined by the following expression \cite{Nilles:1983ge,riotto,Martin:1997ns,Gates:1983nr}:
\be \label{comm}
[Q_{\alpha},P_{\mu}]=0 \ee 
\be \label{anticomm}
\{Q_{\alpha},\overline{Q}_{\dot{\beta}}\}=2\sigma^{\mu}_{\alpha\dot{\beta}}P_{\mu}
\ee
 where $\sigma^{\mu}=(-1,\vec{\sigma})$ and $\vec{\sigma}$
denotes Pauli spin matrices and $P_{\mu}$ are conserved by Coleman Mandula theorem \cite{Coleman}. 
Fermionic behavior of the SUSY generator indicates that the
probability of appearing bosonic (spin-0) and fermionic
(spin-$\frac{1}{2}$) states are equal in
number. Here we introduce the anti-commutating
parameters as $\theta_{\alpha} (\alpha=1,2)$ and
$\overline{\theta}^{\dot{\beta}}(\dot{\beta}=1,2)$, which are
the elements of Grassmann super-algebra
\be \label{grass}
\{\theta^{\alpha},\theta^{\beta}\}=\{\overline{\theta}^{\dot{\alpha}},\overline{\theta}^{\dot{\beta}}\}=
\{\theta^{\alpha},\overline{\theta}^{\dot{\beta}}\}=0 .\ee
Including the anti-commutating parameters the Grassmann
super-algebra can be modified to \cite{Nilles:1983ge,riotto,Martin:1997ns,Gates:1983nr}:\bea \label{theta} [\theta Q
,\overline{Q}
 ~\overline{\theta}]&=& 2\theta\sigma_{\mu}\overline{\theta}P^{\mu}, \\
\label{beta} [\theta Q,\theta Q] &=& [\overline{Q}~
\overline{\theta},\overline{Q}~ \overline{\theta}]=0 \eea where
\be \theta
Q=\theta^{\alpha}Q_{\alpha}=\theta^{\alpha}Q^{\beta}\epsilon_{\alpha\beta}\ee
and $\epsilon_{\alpha\beta}$ is an antisymmetric tensor with $\epsilon_{12}=1$.
Using this input further one can decompose the superfield into left and right
handed representation by the
following relation\be
\label{lrfield}\phi(x_{\mu},\theta,\overline{\theta})=\phi_{L}(x_{\mu}+i\theta\sigma_{\mu}\overline{\theta},\theta,\overline{\theta})=\phi_{R}(x_{\mu}
-i\theta\sigma_{\mu}\overline{\theta},\theta,\overline{\theta}).\ee
L and R superfield transform in the same way as the superfield
$\phi(x_{\mu},\theta,\overline{\theta})$ transform under SUSY
transformation. Let us now expand
$\phi(x,\theta)$ in power series keeping in mind 
that $\theta$ is anti-commuting parameter. As a consequence the
power series terminates after the third term of the left
chiral superfield. So finally we may write \be \label{taylor}
\phi(x,\theta)=
\varphi(x)+\theta^{\alpha}\psi_{\alpha}(x)+\theta^{\alpha}\theta^{\beta}\epsilon_{\alpha\beta}F(x),\ee
where $\varphi$ and $F$(auxiliary field) are complex scalar
fields and $\psi$ is a left/right-handed Weyl spinor. Taking the
infinitesimal transformation of equation(\ref{taylor}) we obtain \cite{Nilles:1983ge,riotto,Martin:1997ns,Gates:1983nr}:\be \label{intaylor}
\delta\phi(x,\theta)=\delta\varphi+\theta\delta\psi+\theta\theta\delta
F \ee Under a SUSY transformation the component fields can be
shown to transform as \cite{Nilles:1983ge,riotto,Martin:1997ns,Gates:1983nr}: \bea \label{co1}
\delta\varphi&=&\sqrt{2}\eta\psi, \\ \label{co2}
\delta\psi &=& \sqrt{2}\eta
F+\sqrt{2}i\sigma_{\mu}\overline{\eta}\partial^{\mu}\varphi, \\ \label{co3} \delta
F &=& -\sqrt{2}i\partial^{\mu}\psi\sigma_{\mu}\overline{\eta}. \eea

 A scalar superfield contains spin 0 bosons and
spin $\frac{1}{2}$ fermions. For a gauge theory one has to introduce spin 1 vector superfields, which can be expressed in the reducible 
representation as:
 \cite{Nilles:1983ge,fayet}:
\be \begin{array}{llll}\label{vec}
\displaystyle V(x,\theta,\overline{\theta})=\left(1+\frac{1}{4}\theta~\theta~\overline{\theta}~\overline{\theta}\Box\right)C+\left(i\theta+\frac{1}{2}\theta\theta\sigma^{\mu}
\overline{\theta}\partial_{\mu}\right)\chi+\left(-i\overline{\theta}+\frac{1}{2}\overline{\theta}~\overline{\theta}~\theta\sigma_{\mu}
\partial^{\mu}\right)\overline{\chi}\\\displaystyle ~~~~~~~~~~~~~~~~+\frac{1}{2}i\theta\theta
(M+iN)-\frac{i}{2}\overline{\theta}~\overline{\theta} (M-i
N)-\theta\sigma_{\mu}\overline{\theta}V^{\mu}+i\theta\theta\overline{\theta}~\overline{\lambda}-i\overline{\theta}~\overline{\theta}
\theta\lambda+\frac{1}{2}\theta\theta\overline{\theta}~\overline{\theta}
D \end{array}\ee 
where $\Box=\partial_{\mu}\partial^{\mu}$; C, M, N, D
are all real scalar fields; $\chi$ and $~\lambda$ are Weyl
spinors and $V^{\mu}$ is a vector field. In an Abelian gauge theory (e. g.
QED) the gauge field $V^{\mu}$ transforms under gauge
transformation as, \be V_{\mu}\rightarrow
V_{\mu}+\partial_{\mu}\varphi,\ee where $\varphi$ is real
scalar field. Here the gauge field  $V^{\mu}$ is a partner of
the supermultiplet $ V(x,\theta,\overline{\theta})$. In the
context of Abelian gauge theory we need the supersymmetric
extension of the gauge theory. So $\varphi$ is also the part of
the supermultiplet. For supersymmetric generalization Wess and
Zumino \cite{wz} showed that the
 vector superfield transforms as, \be V
\rightarrow V+i(\Lambda-\Lambda^{\dag}),\ee where $\Lambda$
represents  a chiral superfield. Further, using Eq~(\ref{vec}) one can show that the
fields C, $\chi$, M and N are gauge artifacts which can be
gauged away using the Wess-Zumino gauge while $\lambda$ and D
are gauge invariant quantities. Consequently, the vector
superfield reduces to \cite{Nilles:1983ge,wz}:
\be \label{wzumino}
V_{WZ}(x,\theta,\overline{\theta})=-\theta\sigma^{\mu}\overline{\theta}V_{\mu}+i\theta\theta\overline{\theta}~\overline{\lambda}-i\overline{\theta}~
\overline{\theta}\theta\lambda+\frac{1}{2}\theta\theta\overline{\theta}~\overline{\theta}D
\ee 
where
$\lambda_{\alpha}$ is the gaugino, D is the auxiliary field
and $V_{\mu}$ is the gauge
field. However, in strict sense the Wess-Zumino gauge is
incomplete in the context of gauge fixation. Here our goal is to obtain the supersymmetric
generalization of the electromagnetic field strength. To
fulfill our need we will start with a spinor chiral superfield \cite{Nilles:1983ge,Martin:1997ns,Csaki:1996ks}:\be \label{comp}
W_{\alpha}=4i\lambda_{\alpha}-4\theta_{\alpha}D+4i\theta^{\beta}\sigma_{\nu\alpha\dot{\beta}}\sigma^{\dot{\beta}}_{\mu\beta}\partial^{[\mu}V^{\nu]}
-4(\theta\theta)\sigma_{\mu\alpha\dot{\beta}}\partial^{\mu}\overline{\lambda}^{\dot{\beta}}
\ee 
where 
the field strength tensor is defined as
\be F_{\mu\nu}=V_{\mu\nu}=\partial_{[\mu}V_{\nu]}.\ee Under the infinitesimal SUSY transformation the chiral field
transform as \cite{Nilles:1983ge,riotto,Csaki:1996ks}::
 \be \label{ch} \delta\lambda=i\eta D
+\eta\sigma^{\mu}\overline{\sigma}^{\nu}F_{\mu\nu}. \ee

  The generic
     Lagrangian for global supersymmetry (GSUSY) can be decomposed into two parts \cite{Nilles:1983ge,riotto,Martin:1997ns,Csaki:1996ks}: \be
     \label{lag} \cal{L}=\cal{L_{F}}+\cal{L_{D}} ,\ee where
     $\cal{L_{F}}$ is the sum of scalar superfields and
     $\cal{L_{D}}$ is a sum of vector superfields. These are
     called F and D terms respectively. F and D terms are
     defined through the following integral over $\theta$ and
     $\overline{\theta}$  as: \be \label{theta} \int d\theta=0
     , \int d\theta^{\alpha}~\theta_{\alpha}=1 .\ee 
      Using equation (\ref{comp}) one
     can have \cite{Nilles:1983ge,Martin:1997ns,Csaki:1996ks}:\be \label{wcomp}
     {\cal
L}=\frac{1}{32}(W^{\alpha}W_{\alpha})=-\frac{1}{4}V_{\mu\nu}V^{\mu\nu}-\frac{i}{2}\lambda^{\alpha}\sigma_{\mu\alpha\dot{\gamma}}
\partial^{\mu}\overline{\lambda}^{\dot{\gamma}}-\frac{i}{2}\sigma^{\alpha}_{\mu\dot{\beta}}(\partial^{\mu}\overline{\lambda}^{\dot{\beta}})\lambda_{\alpha}
+\frac{D^{2}}{2}.\ee which is the Lagrangian
density of pure SUSY extension of Abelian
Yang-Mills theory. To make the theory more
general we introduce here non-Abelian gauge theory
\cite{ferrera}. Here we redefine the vector and chiral field
as \be V_{\mu}=V^{a}_{\mu}T_{a}\ee and \be \Lambda^{a}=\Lambda T^{a}\ee
where $T^{a}$ are the generators of the non-Abelian gauge
group and satisfy the basic propositions of Lie
algebra. Thus in the context of pure SUSY
Yang-Mills theory we get \cite{ferrera}:

\be\label{gg} \begin{array}{lcl} \displaystyle {\cal
L}=\frac{1}{32g^{2}}W_{\alpha}W^{\alpha}
=-\frac{1}{4}G^{\mu\nu}G_{\mu\nu}+\frac{D^{2}}{2}-\frac{i}{2}[\lambda^{\alpha}\sigma_{\mu\alpha\dot{\beta}}
(\partial^{\mu}\overline{\lambda}^{\dot{\beta}}+ig[V^{\mu},\overline{\lambda}^{\dot{\beta}}])\\ \displaystyle ~~~~~~~~~~~~~~~~~~~~~~~~
~~~~~~~~~~~~~~~~~~~~~~~~~~~~~~~~~~~~~~~~~~~~~~~-(\partial^{\mu}\overline{\lambda}^{\dot{\beta}}+
ig[V^{\mu},\overline{\lambda}^{\dot{\beta}}])\sigma_{\mu\alpha\dot{\beta}}\lambda^{\alpha}]
\end{array}
\ee

where
\be G^{a}_{\mu\nu}=\partial_{[\mu}V^{a}_{\nu]}+ig[V_{\mu},V_{\nu}]^{a}=\partial_{[\mu}V^{a}_{\nu]}
-gf^{abc}V_{b\mu}V_{c\nu}\ee and \be \lambda=\lambda^{a}T_{a}\ee in
the adjoint representation of the gauge group. Gauge group generator
satisfies the commutation relation
\be [T^{a},T^{b}]=if^{abc}T_{c},\ee where $T^{a}$'s are the
generator of the gauge group and $f^{abc}$
 are real constants, called the structure constants of the
group which satisfy the famous Jacobi
identity. So the most general renormalizable Lagrangian in
superspace can be
written as \cite{Nilles:1983ge,riotto}: \be \label{totallag} {\cal L}=\sum_{n}\int
d^{2}\theta~d^{2}\overline{\theta}\phi^{\dag}_{n}e^{gV}\phi_{n}+\frac{1}{64g^{2}}\int
d^{2}\theta W_{\alpha}W^{\alpha}+\int d^{2}\theta W(\phi_{n})
+h.c. \ee where {\it h.c.} signifies the hermitian conjugate and in the adjoint representation
\be Tr(T^{a}T_{b})=\delta^{a}_{b}.\ee Here  $W(\phi_{n})$
is the holomorphic superpotential which will play the most crucial role
in deriving the F and D term potentials, as revealed
 subsequently. The single $U(1)$
gauge group with coupling $g$ is simple. But in the case of
several $U(1)$'s, there are no cross-terms in the potential from the D-terms, i. e.
\be V_{D}=\sum_{n}(V_{D})_{n}.\ee In terms of component fields the
Lagrangian for Abelian theory can be written as \cite{Nilles:1983ge,riotto}:\be\begin{array}{llll}
\label{compla} \displaystyle {\cal
L}=\sum_{n}[(D_{\mu}\phi_{n}^{\star})(D^{\mu}\phi_{n})+i\overline{\psi}_{n}
D_{\mu}\overline{\sigma}^{\mu}\psi_{n}
+|F_{n}|^{2}] \displaystyle -\left[\frac{1}{4}F_{\mu\nu}F^{\mu\nu}+i\lambda\sigma^{\mu}\partial_{\mu}\overline{\lambda}-\frac{1}{2}D^{2}-\frac{gD}{2}\sum_{n}q_{n}|\phi_{n}|^{2}
\right]
\\ \displaystyle~~~~~~~~~~~~~~~~~~~~~~~~~~~~~~~~~~~ -\left[i\sum_{n}\frac{g}{\sqrt{2}}\overline{\psi_{n}}~\overline{\lambda}\phi_{n}-\sum_{n,m}\frac{1}{2}\frac{\partial^{2}W}{\partial\phi_{n}\partial\phi_{m}}
\psi_{n}\psi_{m}+\sum_{n}F_{n}\left(\frac{\partial W}{\partial
\phi_{n}}\right)\right]+h.c. \end{array}\ee where the covariant
derivative is defined as $ D_{\mu}=\partial\mu+igV_{\mu}$ and
its non-Abelian generalization can be written as \cite{ferrera}:\be\begin{array}{llll}
\label{noncompla} \displaystyle {\cal
L}=\sum_{n}[(D_{\mu}\phi_{n}^{\star})(D^{\mu}\phi_{n})+i\overline{\psi}_{n}
D_{\mu}\overline{\sigma}^{\mu}\psi_{n}
+|F_{n}|^{2}]\displaystyle -\frac{1}{4} G_{\mu\nu}G^{\mu\nu}
+\frac{1}{2}D^{2}-\frac{i}{2}[\lambda^{\alpha}\sigma_{\mu\alpha\dot{\beta}}
(\partial^{\mu}\overline{\lambda}^{\dot{\beta}}
+ig[V^{\mu},\overline{\lambda}^{\dot{\beta}}])
\\ \displaystyle ~~~~~~~~~~~~~~~~~~~~~~~~~~~~~~~~-(\partial^{\mu}\overline{\lambda}^{\dot{\beta}}+
ig[V^{\mu},\overline{\lambda}^{\dot{\beta}}])\sigma_{\mu\alpha\dot{\beta}}\lambda^{\alpha}]+\frac{gD}{2}\sum_{n}q_{n}|\phi_{n}|^{2}
\displaystyle -\left[i\sum_{n}\frac{g}{\sqrt{2}}\overline{\psi_{n}}~\overline{\lambda}\phi_{n}\right.\\ \left. \displaystyle 
~~~~~~~~~~~~~~~~~~~~~~~~~~~~~~~~~~~~~~~~~~~~~~~~~~~~~~~~~~~-\sum_{n,m}\frac{1}{2}\frac{\partial^{2}W}{\partial\phi_{n}\partial\phi_{m}}
\psi_{n}\psi_{m}+\sum_{n}F_{n}\left(\frac{\partial W}{\partial
\phi_{n}}\right)\right]+h.c.\end{array}\ee where  the covariant
derivative is defined as:
\be D_{\mu}=\partial\mu+igT_{a}V_{\mu}^{a}\ee and $q_{n}$ are the
$U(1)$ charges (the response of interaction) of
the fields $\phi_{n}$. Here in this expression for Lagrangian
both for Abelian and non-Abelian gauge singlet kinetic term
appears with overall family of fields and it reflects the gauge interaction with coupling strength $g$. The
constraints of $F_{n}$ and D are given by \cite{Nilles:1983ge,riotto,Martin:1997ns}: 
\be \label{co1}
F_{n}~=-\left(\frac{\partial
W}{\partial\phi_{n}}\right)^{\star}\ee \be \label{co2}
D~=-\frac{g}{2}\sum_{n}q_{n}|\phi_{n}|^{2}. \ee
These
equations basically give the equation of motion of auxiliary
fields.Using equation(\ref{co1}) and (\ref{co2}) the scalar
field potential can be split into two parts \cite{Nilles:1983ge,riotto,Martin:1997ns}:
\be \label{pot1}
V~=~V_{F}+V_{D}\ee
where 
\be \label{pot2}
V_{F}~\equiv~\sum_{n}|F_{n}|^{2} \ee \be \label{pot3}
V_{D}~\equiv~\frac{1}{2}D^{2}.\ee 
This splitting of potential
into F and D term is very much significant in the context for
 model building for inflation. 

In order to get a
renormalizable theory, one can investigate that the
superpotential $W$ is at most cubic in the fields, which
corresponds to the potential is at most quartic. From
(\ref{pot1}) we observe that the overall phase factor appearing
in $W$ is not physically significant at all. Due to R-symmetry
a sign flip occurs in $W$. Internal symmetries respect
holomorphicity of $W$ and thus restrict its form much more
with respect to the case of potential $V$. For completeness In the
context of $U(1)$ gauge symmetry it is instructive to add an extra
contribution to the Lagrangian, which is called
Fayet-Iliopoulos term and the new form of the generalized
Lagrangian is \cite{riotto,fayetlip},
\be \label{fi}{\cal L}=\sum_{n}\int
d^{2}\theta~d^{2}\overline{\theta}\phi^{\dag}_{n}e^{gV}\phi_{n}+\frac{1}{64g^{2}}\int
d^{2}\theta W_{\alpha}W^{\alpha}+\int d^{2}\theta
W(\phi_{n})-{\underbrace{2\xi\int
d^{2}\theta~d^{2}\overline{\theta} ~V}}+h.c. \ee 
where the
underbrace term is the Fayet-Iliopoulos term 
and the corresponding modified $D$ field can be expressed as \cite{riotto,fayetlip}:
\be \label{dm}
D~=-\frac{g_{c}}{2}\sum_{n}q_{n}|\phi_{n}|^{2}-\Lambda
.\ee 
where $\Lambda$ appears as a effective coupling strength in
the Fayet-Iliopoulos term which plays similar role as in the
first term in $D$. Here one can identify $\Lambda$ as effective
$U(1)$ Fayet-Iliopoulos charge. The modified D term of the
potential can ten be written as \cite{riotto,fayetlip}:
 \be \label{modd}
V_{D}~=\frac{1}{2}\left(\frac{g_{c}}{2}\sum_{n}q_{n}|\phi_{n}|^{2}+\Lambda\right)^{2}.\ee
For convenience again we can redefine the D term in such a way
that it can identify the effective Fayet-Iliopoulos $U(1)$
charge as $g_{c}\Lambda$ and the corresponding modified potential by
redefining the charges and $\Lambda$ as \cite{riotto,fayetlip}:
\be \label{red}
D~=-g_{c}\left(\sum_{n}q_{n}|\phi_{n}|^{2}+\Lambda\right)\ee 
and the D-term potential can be written as:
\be
\label{redv}
V_{D}~=\frac{g^{2}_{c}}{2}\left(\sum_{n}q_{n}|\phi_{n}|^{2}+\Lambda\right).
\ee 
The Fayet-Iliopoulos term appears as a heavy degrees of
freedom. However, at
the end of the calculation, the heavy degrees of freedom have
been integrated out.

\subsection{Minimal Supersymmetric Standard Model (MSSM)}

The Minimal Supersymmetric Standard Model (MSSM) \cite{Nilles:1983ge,Mazumdar:2010sa,Martin:1997ns,Csaki:1996ks,Aitchison:2005cf,Randall:1998uk} is a minimal extended version
 to the Standard Model which realizes basic principles of ${\cal N}=1$ SUSY as introduced in the earlier subsection.
The prime incentive for introducing MSSM was to stabilize the physics of weak scale via solving the well known hierarchy or naturalness problem.
 This is a very well known fact that the Higgs boson mass of the Standard Model (SM) is very much unstable in presence of quantum corrections
 and the corresponding theory also predicts that weak energy scale should be much weaker than collider experiment bound.
 In the context of MSSM, the Higgs boson has a fermionic superpartner within ${\cal N}=1$ SUSY. This is known as the Higgsino which has the
 exactly same mass as it would if also ${\cal N}=1$ SUSY were an exact symmetry of the present setup.
 In this physical scenario all the fermion masses and Higgs mass are stable under radiative correction in the matter sector of the theory.
 However, within the framework of ${\cal N}=1$ SUSY based MSSM there is a need for more than one Higgs field and this is explicitly described later in detail.
There are three principal incentives for the ${\cal N}=1$ SUSY based MSSM over the other theoretical extensions
 of the Standard Model, namely \cite{Nilles:1983ge,Mazumdar:2010sa,Martin:1997ns,Haber:1984rc,Sohnius:1985qm}:
\begin{enumerate}
 \item To solve the naturalness or hierarchy problem in the context of ${\cal N}=1$ SUSY,

\item To unify all gauge couplings and

\item To identify the proper dark matter candidate in the context of ${\cal N}=1$ SUSY.

\end{enumerate}
These motivations are the primary
 reasons for considering the MSSM as the leading candidate for a new theory to be discovered
 at collider experiments such as the LHC \cite{lhc}, ILC \cite{Ilc} etc in future.

In addition to the usual quark and lepton superfields as appearing in the context of ${\cal N}=1$ SUSY,
MSSM has two Higgs fields, $H_u$ and $H_d$. In this context two Higgses are are the building block to construct the mathematical structure of the superpotential 
and for such construction  $H^\dagger$ and any other combination of this term is completely redundant to maintain the perfect holomorphicity. 
The superpotential in the context of MSSM for ${\cal N}=1$ SUSY is described by the following expression \cite{Nilles:1983ge,Martin:1997ns,Haber:1984rc,Sohnius:1985qm}:
\be\label{mssmxc}
W_{MSSM}=\lambda_uQH_u u+\lambda_dQH_d d+\lambda_eLH_d e+\mu H_uH_d,
\ee
where $H_{u}, H_{d}, Q, L, u, d, e$ in
Eq.~(\ref{mssmxc}) characterizing the chiral superfields, and the dimensionless Yukawa couplings
$\lambda_{u}, \lambda_{d}, \lambda_{e}$ are $3\times 3$ matrices in the SUSY family
space. We have suppressed the gauge and family indices for clarity in the present context. Here $H_{u}, H_{d}, Q, L$ fields are actually
$SU(2)$ doublets and $u, d, e$ fields are  the $SU(2)$ singlets. Additionally, the last term is called the $\mu$
term in MSSM, which is a ${\cal N}=1$ SUSY version of the SM Higgs boson mass in the context of MSSM.
 The prime reason for the requirement of two higgs field $H_{u}$ and $H_{d}$ are two fold in the present context of discussion-
\begin{enumerate}
 \item to give
masses to all the quarks and leptons via spontaneous SUSY breaking (SSB) in the context of MSSM and
\item for the cancellation of
gauge anomalies in the context of MSSM.
\end{enumerate}
It is important to note that as the top quark, bottom quark and tau lepton are the heaviest fermionic contents
in the SM family, only the third family element of the matrices
$\lambda_{u}, \lambda_{d}, \lambda_{e}$ plays significant role in the context of MSSM.

In this context the $\mu$ term provides masses to the fermionic superpartner of Higgs field, which is known as the Higgsino and can be represented by the following Lagrangian density:
\be
{\cal L} \supset -\mu(\tilde H_{u}^{+}\tilde H_{d}^{-}-\tilde H^{0}_{u}\tilde H^{0}_{d})+{c.c},
\ee
and actively contributes to the Higgs mass terms in the scalar potential via the following expression:
\be\label{higgsmasscvc}
-{\cal L} \supset V \supset |\mu|^2(|H^{0}_{u}|^2+|H^{+}_{u}|^2+|H^{0}_{d}|^2+|H^{-}_{d}|^2)> 0.
\ee
 Consequently, here $|\mu|^2$ should approximately cancel the
negative soft mass term in order to accommodate for a Higgs VEV
of order $\sim 246$~GeV. Additionally it is important to note that, a general gauge invariant as well as renormalizable superpotential
would also include baryon number $B$ or lepton number $L$ violating terms in the present context. 
There exists a discrete $Z_2$ symmetry in the present context, which can easily forbid baryon and lepton number violating terms. This is 
commonly known as $R$-parity \cite{Martin:1997ns,Fayet:1975ki}. For each particle one can write the expression for $R$-parity as:
\be P_{R}=(-1)^{3(B-L)+2s},\ee
with \be P_{R}=+1\ee for the SM particles and the Higgs bosons, while \be P_{R}=-1\ee
for all the sleptons, squarks, gauginos, and Higgsinos.
Additionally, $s$ signifies the spin of the particle.

Also it is important to note that the general soft ${\cal N}=1$ SUSY breaking terms in the total MSSM
Lagrangian can be expressed as \cite{Nilles:1983ge,Mazumdar:2010sa,Martin:1997ns,Haber:1984rc,Sohnius:1985qm}:
\be{\cal L}_{soft}=-\frac{1}{2}\left(M_{\lambda}\lambda^{a}\lambda^{a}+{\rm c.c.}\right)-(m^2)^{i}_{j}\phi^{j\ast}\phi_{i}-\left(\frac{1}{2}
b_{ij}\phi_{i}\phi_{j}+\frac{1}{6}a^{ijk}\phi_{i}\phi_{j}\phi_{k}+{\rm c.c.}
\right),
\ee
where $M_{\lambda}$ signifies the common gaugino mass as given by: \be (m^2)^{j}_{i}\sim m^2_{0}\sim ({\cal O}(100){\rm GeV})^2\ee
which are $3\times 3$ matrices determining the masses for
squarks and sleptons given by:
\be m^2_{Q},m^2_{u},m^2_{d},m^2_{L},m^2_{ e}, m^2_{H_u},m^2_{H_d}, b\sim m_{0}^2\sim ({\cal O}(100))^2~{\rm GeV^2}.\ee
Additionally, $b_{ij}$ is the mass term for the specific combination $H_{u}H_{d}$ and also 
$a^{ijk}$ are the complex $3\times 3$ matrices in the family space which yield the trilinear 
$A$-terms as given by:
\be a_{u},~a_{d},~a_{e}\sim m_{0}\sim {\cal O}(100)~{\rm GeV}.\ee 
It is important to note that, there are a total of $105$ new entries in the
MSSM Lagrangian which have not any counterpart in the context of SM. The only mechanism of SUSY breaking where the breaking
scale is not introduced either at the level of superpotential or in
the gauge sector is through dynamical SUSY breaking and this appears to be useful in MSSM.

Also it is important to mention here that the ${\cal N}=1$ SUSY field configurations satisfying simultaneously the following sets of constraint equation of motions in the context of MSSM \cite{Gherghetta:1995dv,Dine:1995kz,Dine:1995uk}:
\bea\label{fflatdflat}
D^{a}&\equiv& X^{\dagger}T^{a}X=0, \\
F_{X_{i}}&\equiv& \frac{\partial W}{\partial X_{i}}=0.
\eea
where these are called the 
$D$-flat and $F$-flat respectively for $N$ chiral superfields $X_{i}$ defined in MSSM supermultiplet. Also here $D$-flat directions are parameterized by gauge invariant monomials of
the chiral superfields in the context of MSSM.

In many cosmological applications of flat directions, it is important to
consider the running of mass term below the GUT scale,
\be M_{\rm GUT}\sim {\cal O}(10^{16}~{\rm GeV})\ee 
to study the effective field theory at low and high 
 energy scale. For simplicity we can also assume that it is the scale where
SUSY breaking is transmitted to the visible sector. 

A most generalized structural form of one loop Renormalization Group (RG) equations can be written in terms of $\beta$ function as \cite{Nilles:1983ge,Mazumdar:2010sa,Martin:1997ns}:
\be\label{rge}
\beta_{m^{2}_{i}}:={\partial m_i^2\over \partial t}=\sum_{g}a_{ig} m_g^2+\sum_a h^2_a(\sum_j b_{ij}m_j^2+A^2),
\ee
where $a_{ig}$ and $b_{ij}$ are two constants, $m_g$ characterizes  the gaugino mass,
$h_a$ is the Yukawa coupling, $A$ is the supersymmrtric trilinear $A$-term, and \be t=\ln M_X/q\ee characterizes the background scale of RG.

 In this context the one-loop RG equations for the three gaugino mass parameters are
determined by the following expression \cite{Nilles:1983ge,Mazumdar:2010sa,Martin:1997ns}:
\be
\frac{d}{dt}m_i=\frac{1}{8\pi^2}b_i g_i^2m_{i},~~~~(b_{i}=33/5,~1,~-3)
\ee
where $i=1,2,3$ correspond to $U(1),~SU(2),~SU(3)$. An interesting property
is that the three ratios $m_i/g_i^2$ are RG scale independent. Therefore at the
GUT scale, it is assumed that gauginos masses also unify with a value $m_{1/2}$. Then at any scale:
\be m_{i}/g_{i}^2=m_{1/2}/g^2_{GUT},\ee where $g_{GUT}$ is the unified gauge coupling at the GUT scale. The RG evolution
due to Yukawa interactions are small except for top.

\subsection{Next to Minimal Supersymmetric Standard Model (NMSSM)}

NMSSM is an acronym for Next-to-Minimal Supersymmetric Standard Model.
 It is a ${\cal N}=1$ supersymmetric extended version to the Standard Model that incorporates the effect of an additional
 singlet chiral superfield to filed content of the MSSM as already introduced in the previous section. 
This simple extended version of MSSM can be 
obtained just by including a new gauge-singlet chiral supermultiplet with even 
matter parity in ${\cal N}=1$ SUSY sector. In this context the superpotential can be expressed as \cite{King:1995vk,Ellwanger:2009dp,Maniatis:2009re}:
\bea\label{NMSSMwwwc}
W_{\rm NMSSM} &=& W_{\rm MSSM}+ \underbrace{\lambda S H_u H_d + \frac{1}{3} \kappa S^3}_{\mathbb{Z}_3~{\rm invariant~part}}+ \frac{1}{2} \mu_S S^2,
\eea
where $S$ is the new gauge-singlet chiral supermultiplet added in the filed content of MSSM. 
Additionally it is important to note that, NMSSM also introduces extra coefficients, which play crucial role for the successful realization of the electroweak symmetry breaking 
and can be easily done by choosing appropriately all such coefficients.
One of the advantage of the NMSSM framework is that it can provide an answer to solve the
$\mu$ problem~\footnote{In the context of SUSY a very important question one can ask regarding the issue that how SUSY mass parameter
$\mu$ can assume a value of the order of the SUSY breaking scale $M_\mathrm{SUSY}$. In theoretical particle physics this commonly known as  
the ``$\mu$-problem'' of the MSSM \cite{Kim:1983dt} theory. Here we mention a possible graceful way to resolve this problem.
Within this study one needs to generate an SUSY effective 
 mass term $\mu$ by exactly following the generation mechanism of quark and lepton masses in the context of SM. In this case the mass term $\mu$ is
exactly replaced by a Yukawa coupling of $H_u$ and $H_d$ to a scalar field and
specifically the  soft SUSY breaking induced VEV of the scalar field 
is consistent with the requirement.}. Additionally, it is important to mention here that an effective $\mu$-term:
\be \mu_{\rm eff}= \lambda S\ee
for the contribution $H_u H_d$ can be easily able to generate from
Eq.~(\ref{NMSSMwwwc}) in the present context. For the determination of such term 
dimensionless coupling parameters as well as the soft SUSY breaking mass term $m_{\rm soft}$ are required.
In the most generalized physical prescription, it is important to note that the natural framework of NMSSM also provides an extra sources for the large amount of CP 
violation. Apart from this fact NMSSM also provides the conditions for electroweak baryogenesis which is a very important fact for the study of physics of early universe. 
In eq.~(\ref{NMSSMwwwc}) the symbol $\underbrace{\cdots}$ characterizes the scale invariant cubic
superpotential for the $\mathbb{Z}_3$-invariant NMSSM, where accidental
$\mathbb{Z}_3$ symmetry is maintained due to a overall multiplication of a phase factor $e^{2\pi i/3}$ to all the components of
all chiral superfields as appearing in ${\cal N}=1$ extended supermultiplet. Here it is important to mention that any of the
dimensionful contributions in the general ${\cal N}=1$ superpotential (which is not cubic in general) breaks
the $\mathbb{Z}_3$ symmetry in a explicit fashion within the framework of NMSSM. As a consequence the extended version corresponding
to the general full modified superpotential will be designated as the generalzed NMSSM.


\subsection{Supergravity}

Till now in the introduction of the thesis we have discussed the various field theoretic features of global supersymmetry (GSUSY).
Now it is important to note that, in the context of the physics of early universe, especially during the epoch of inflation, one needs to consider 
effect of local version of supersymmetry, which is commonly known as supergravity (SUGRA). Within the framework of SUGRA  one can describe the most general 
non-renormalizable structure of GSUSY \cite{Nilles:1983ge,Mazumdar:2010sa,riotto,Martin:1997ns}. Sometimes this can also be treated as the low energy version of string theory.
The mathematical structure of all such SUGRA operators mimics the role of effective field theory operators which we will discuss in the context of inflation.
Within the effective description such a non-renormalizable field theory is valid upto the
ultra-violet cutoff scale $\Lambda_{UV}\sim M_{p}$, beyond which it is invalid.

\subsubsection{Prime components of SUGRA}

\label{specsug}

In the context of GSUSY we start our discussion with the definition of various crucial components-chiral and
gauge supermultiplets and also the characteristic superalgebra for GSUSY.
In the context of SUGRA~\footnote{In the rest of the thesis we will
concentrate on the ${\cal N}=1$ and ${\cal N}=2$ SUGRA sector which is relevant for both cosmological
 and particle physics model building purpose for early universe \cite{Mazumdar:2010sa,riotto}.}
 one need the following components to construct the background field theoretic setup:
 \begin{itemize}
  \item The 
superpotential for ${\cal N}=1$ SUGRA is characterized by holomorphic function $W$ which is in principle function 
of complex scalar fields.
\item The K\"{a}hler potential is characterized by the real function $K$ which is not at all holomprphic in nature.
\item The holomorphic gauge kinetic function $f$~\footnote{In the more generalized physical prescription
one can think of other class of holomorphic functions 
which incorporates the ffect of higher order simple derivatives, higher order non-minimal derivatives and various other complicated functions
in principle. But the general framework of effective field theory indicates
that those type of non-trivial complicated contributions are highly suppressed by the UV cut-off scale (i.e. $\Lambda_{UV}\sim M_{p}$) of the 
theory and consequently all such contributions are negligibly small in the present context and at the
leading order of SUGRA theories one can easily neglect such complicated non-trivial functions from the computation. 
See \cite{Choudhury:2014sxa,Choudhury:2014uxa,Assassi:2013gxa,Baumann:2011nm} for the details on these aspects.}.
 \end{itemize}
Within the framework of SUGRA one can start discussion with two crucial components- the superpotential $W$ and K\"{a}hler potential $K$. In terms of these crucial 
components one can further define a generating functional using which one can easily able to generate the effective potential required for the study of the physics of early universe, specifically 
the physics of inflation. Such generating functional for ${\cal N}=1$ SUGRA can be expressed as \cite{Nilles:1983ge,riotto}:
\be
 G\equiv\frac{ K}{M^{2}_{p}}+ \ln\frac{|W|^2}{M_{p}^6}
\ee
where the K\"{a}hler potential $K$ and superpotential $W$
transform under the K\"{a}hler transformation defined within the framework of ${\cal N}=1$ SUGRA as \cite{Nilles:1983ge,riotto}:
\bea \frac{ K}{M^{2}_{p}}&\rightarrow&\frac{ K}{M^{2}_{p}}-X-\bar X,
\\ W&\rightarrow& e^X W 
\eea
 where $X$ and $\bar X$ characterize any arbitrary holomorphic and anti-holomorphic function
of the superfield contents of ${\cal N}=1$ SUGRA multiplet. 
Throughout the thesis henceforth we will follow the following list of crucial conventions within the framework of ${\cal N}=1$ SUGRA:
\begin{itemize}
 \item All the scalar components $\phi^n$ and also the auxiliary components 
$F^n$ in the SUGRA multiplet are 
identified by a superscript.

\item In this context a subscript $n$ characterizes 
$\partial/\partial\phi^n$, and a subscript $n_{\dagger}$ 
characterizes $\partial/\partial\phi^{n{\dagger}}$.

\item Lowering operation of the indices can be performed within the present framework
 as, \be \phi_n\equiv K_{nm^{\dagger}}\phi^{m{\dagger}}\ee
and \be F_n\equiv K_{nm^{\dagger}} F^{m^{\dagger}}.\ee

\end{itemize}
Let us briefly discuss about the technical details of superpotential $W$,  K\"{a}hler potential $K$ and gauge kinetic function $f$ within the framework of ${\cal N}=1$ SUGRA.
To serve this purpose first of all in this context we consider a most generalized expansion about a preferred choice of
origin in ${\cal N}=1$ superfield space within the framework of ${\cal N}=1$ SUGRA
\cite{Nilles:1983ge,Mazumdar:2010sa,riotto,Martin:1997ns}.
\paragraph{\bf \# The superpotential:}
In the most general physical prescription the holomorphic superpotential can be expanded
as a power series expansion \cite{Nilles:1983ge,Mazumdar:2010sa,riotto} within the framework of ${\cal N}=1$ SUGRA as:
\be
W =  \sum_{d=0}^\infty \frac{W_d(\phi^n)}{M^{d-3}_{p}}\,
\label{wexp}
\ee
where $W_d$ is the contribution in the ${\cal N}=1$ SUGRA superpotential from the superfield sector in $d$ mass dimension.
Additionally it is important to note that, for $d\geq 4$ the above mentioned ${\cal N}=1$ SUGRA operators are non-renormalizable
in nature and also suppressed by the UV cut-off scale i.e. Planck scale, $M_{p}$ in the present context.
Now here one can also think of a physical situation where the computation starts with an expression in which the ${\cal N}=1$ SUGRA field contents only appears
at lower order. In that case one can easily forbid additional 
terms appearing in the above mentioned generic series expansion and truncate the series up to a finite order by imposing an additional discrete $Z_N$
symmetry in the present context. Also one can truncate the series up to a finite order by
imposing a continuous symmetry in the context of ${\cal N}=1$ SUGRA.

\paragraph{\bf \# The K\"{a}hler potential:}

Within the framework of ${\cal N}=1$ SUGRA the K\"{a}hler potential determines the structure of the
 kinetic terms of the superfields as given by \cite{Nilles:1983ge,Mazumdar:2010sa,riotto}:
\be
{\cal L}_{\rm kin} = (\partial_\mu\phi^{n^{\dagger}})K_{n^{\dagger} m}(\partial^\mu\phi^m).
\ee
In the ${\cal N}=1$ superfield space one can use the following series expansion ansatz of K\"{a}hler potential as given by \cite{Nilles:1983ge,riotto}:
\be
K=K_{nm^{\dagger}}\phi^n\phi^{m{\dagger}}
+ \underbrace{\sum_{d=3}^{\infty} 
\frac{K_{d}(\phi^n,\phi^{n{\dagger}})}{M_{p}^{d-2}}}_{Planck~scale~suppressed~operators} \,,
\label{Kalerexp}
\ee
where $K_{nm^{\dagger}}$ characterizes the ${\cal N}=1$ K\"{a}hler metric evaluated at the specified preferred choice of origin~\footnote{It is important to note that, in the present context any constant term 
or linear contribution has been absorbed into the ${\cal N}=1$ SUGRA superpotential by a K\"ahler transformation in this context.}.
In the most simplest physical situation the scalar fields are chosen to be canonically normalized at the 
preferred choice of origin, which directly implies that in such a case ${\cal N}=1$ K\"{a}hler metric is diagonal and can be expressed as:
\be K_{nm^{\dagger}}=\delta_{nm^{\dagger}}.\ee In Eq~(\ref{Kalerexp}) the higher mass dimensional ${\cal N}=1$ SUGRA  K\"{a}hler operators as denoted by $\underbrace{\cdots}$ symbol and most importantly
are highly suppressed by the Planck scale.
Additionally, it is important to note that ${\cal N}=1$ K\"{a}hler metric $K$ is not a holomorphic function and consequently using the physical symmetries of the setup it is not possible to constrain the 
mathematical structural form of the function in a very strong fashion.

\paragraph{\bf \# The gauge kinetic function:}

Finally, within the framework of ${\cal N}=1$ SUGRA the gauge kinetic function determines the kinetic terms of the gauge and 
gaugino field contents. Here one can easily choose all of them to be canonically normalized
when the scalar fields are specified at a preferred choice of origin and in that case one can use the following series expansion ansatz of gauge kinetic function as given by 
\cite{Nilles:1983ge,riotto}:
\be
f= 1+ \underbrace{\sum_{d=1}^\infty \frac{f_d(\phi^n)}{M_{p}^{d}}}_{Planck~scale~suppressed~operators} \,.
\label{gkin}
\ee
where it is important to note that the mathematical structural form of $f$ is constrained via various physical symmetries of the ${\cal N}=1$ SUGRA setup due to its holomorphicity.

\subsubsection{Scalar potential for SUGRA}

Here we will explicitly write down the total structure of ${\cal N}=1$ SUGRA effective potential which is very relevant for the discussion of the physics of early universe. 
It is very well know fact that ${\cal N}=1$ SUGRA can be broken only using the mechanism of spontaneously SUSY breaking but not explicitly. But in the case of
GSUSY it is possible to break SUSY explicitly. Within the framework of ${\cal N}=1$ SUGRA the auxiliary fields which are determined from the equations of motion are are given by \cite{Nilles:1983ge,Mazumdar:2010sa,riotto}:
\bea\label{fdef1}
D &=& - g( q_n K_n \phi^n +\xi ), \\
F^n &=& -e^{K/2} K^{n m^{\dagger}}\left( W_{m}+M_{p}^{-2}WK_{m}\right)^{\dagger}. 
\eea
Using Eq~(\ref{fdef1}) the D-term and F-term potentials~\footnote{Within the framework of ${\cal N}=1$ SUGRA, if we choose canonical SUGRA, $K_{nm^{\dagger}}=\delta_{nm}$ then one can explicitly show that 
$V_D$ is proportional to $D^2$ while $F^2$ is equal to $\sum|F_n|^2$.} are given by \cite{Nilles:1983ge,Mazumdar:2010sa,riotto}:
\bea\label{ass}
V_D &\equiv& \frac12({\rm Re},f)^{-1} g^2\left(q_n K_n \phi^n+\xi \right)^2,\\
\label{ass2}
V_F &=& F^2-3e^{K/M_{p}^2} \frac{|W|^2}{M_{p}^{2}}  \nonumber \\
&=& e^{K/M_{p}^2} \left[ \left( W_n+\frac{W}{M_{p}^{2}}K_n \right)
K^{m^{\dagger} n}\left( W_m+\frac{W}{M_{p}^{2}}K_m\right)^{\dagger}- 3\frac{|W|^2}{M_{p}^{2}} \right].
\eea
Finally adding the contribution from Eq~(\ref{ass}) and Eq~(\ref{ass2}) the total tree level potential within ${\cal N}=1$ SUGRA
is given by the following expression \cite{Nilles:1983ge,Mazumdar:2010sa,riotto}:
\be
V = V_D + V_F .
\label{sugpotxxc}
\ee

\subsection{String theory and its low energy realizations}

In this subsection we will highlight specific aspects of the ultra-violet (UV) completion
 of quantum field theoretic prescription and gravity sector. Here the field content and the corresponding interactions
are described in the vicinity of Planck scale, which can be treated as the UV cut-off scale in the present context. This type of crucial issues are most naturally studied in the vicinity of 
Planck-scale and for this purpose string theory is the conceivably the best developed
field theoretic prescription \cite{Baumann:2009ds,Silverstein:2013wua,McAllister:2007bg,
Linde:2014nna,Kallosh:2007ig,Baumann:2014nda,Burgess:2011fa}. 
This motivates one to understand the various aspects of the physics of early universe within the framework of string theory.
String theory is a very very successful theoretical framework within which all the point like particles as
 appearing in the context of particle physics are replaced by one dimensional objects called ``strings''.
 Within this prescribed field theoretic setup different types of observed elementary particles in various experiments usually appear from the
 different quantum mechanical states of these one dimensional strings. Additionally it is importnat to note that, apart from  hypothesizing various types of particles within SM of
 particle physics, string theory normally includes the effect of gravity. Consequently this can be treated as a most sucessful candidate
 for a self-contained and well established mathematical framework that unifies all the four fundamental
 forces and various forms of matter. Additionally, string field theoretic framework
 is now extensively used as a theoretical tool in various areas in theoretical physics and it has shed light on many unknown aspects of
 quantum field theory as well as quantum gravity.
 It is very well known fact that initially string theory is described by the bosonic sector only which incorporated the effcet of 
 bosonic degrees of freedom. Later a more sophisticated and upgraded version of the string theory was proposed, known as ``superstring theory'' which
 incorporates SUSY within this present field theoretic framework. Also it is important to mention here that, String theory requires the existence of extra spatial dimensions
 for its mathematical consistency. But all such extra dimensions are compactified to extremely small scales in realistic physical models constructed from the basic priciples from string theory.
In our discussion we will mainly focus on the four-dimensional effective actions
which can be used in the context of inflation.

\subsubsection{D-branes}
In this subsection let us briefly discuss about the very crucial component of the string theory which is made up of various solitonic degrees of freedom. 
It is important to note that, in this context of discussion $D$ branes are the most successful candidate. In case of D{\it p}-brane one can describe this as an object which has 
$p$ spatial dimensions and also this is electrically charged under $C_{p+1}$. Additionally this is described by the Chern Simons (CS) action
\cite{Bachas:1998rg,Giveon:1998sr,Polchinski:1996na,Johnson} as given by:
\be\begin{array}{lll}\label{cs}
    \displaystyle S_{CS}=\mu_{p}\int_{\Sigma_{p+1}}C_{p+1}
   \end{array}\ee
where $\Sigma_{p+1}$ physically represents the D{\it p}-brane worldvolume and also $\mu_{p}$ characterizes the brane charge.

It is very well know fact that D-branes can be treated as surfaces on which open strings can end. Usually
D-brane stands for Dirichlet brane which deals with the open strings and satisfies
Dirichlet boundary conditions in the transverse directions D-brane. Also it is important to note that the open strings satisfies the Neumann boundary conditions as well 
in the directions along the spatial extension of a D{\it p}-brane with the additional restriction $p>0$.
 For the sake of simplicity let us concentrate on the bosonic sector for the discussion in the present context.
In the context of our discussion a charge less p-dimensional membrane moving in a curved space-time with the curved background metric
$G_{MN}$ can be directly explained by the well known Dirac action and within the framework of string theory this can be described by a extra dimensional generalization of
the Polyakov action \cite{Bachas:1998rg,Giveon:1998sr,Polchinski:1996na,Johnson} given by:
\be\begin{array}{llll}\label{ew1}
    \displaystyle S_{D}=-T_{p}\int d^{p+1}\sigma\sqrt{-\det(G_{ab})}~~~~~~~~~~~~~~{\rm where}~~~~ G_{ab}\equiv \partial_{\sigma^{a}}X^{M}\partial_{\sigma^{b}}X^{N} G_{MN}.
   \end{array}\ee
Here in Eq~(\ref{ew1}), $G_{ab}$ characterizes the usual pullback of the metric of the target space-time and also $T_{p}$ describes the membrane tension in the present context.
To construct a most generalized version of D-brane within the framework of string theory one can firstly consider the effect of non-linear electromagnetism and this can be 
successfully described by Born-Infeld theory in the present context and described by the following action in 
$p + 1$ flat space-time dimensions \cite{Savvidy:1999yf,Tseytlin:1999dj,CaminoMartinez:2002tj}:
\be\begin{array}{llll}\label{ew2}
    \displaystyle S_{BI}=-Q_{p}\int d^{p+1}\sigma\sqrt{-\det(\eta_{ab}+2\pi \alpha^{'}F_{ab})}=-Q_{p}\int d^{p+1}\sigma
\left[1+\frac{(2\pi\alpha^{'})^{2}}{4}F_{ab}F^{ab}\right],
   \end{array}\ee
where $F_{ab}$ characterizes the field
strength for Abelian gauge field $A_{a}$ and also $Q_{p}$ serves the role of a constant in the present discussion with the dimensions of brane tension.
By computing string field theoretic amplitudes one can explicitly show that the effective action for
a D{\it p}-brane in a most generalized string theoretic prescription can be written as a combination of the Dirac actions and Born-Infeld
actions. This combined version is commonly known as the Dirac-Born-Infeld action \cite{Johnson,Cederwall:1996uu,Tong:2009np} given by:
\be\begin{array}{llll}\label{ew3}
    \displaystyle S_{DBI}=-g_{s}T_{p}\int d^{p+1}\sigma e^{-\Phi}\sqrt{-\det(G_{ab}+{\cal F}_{ab})}.
   \end{array}\ee
where \be {\cal F}_{ab}\equiv B_{ab}+2\pi\alpha^{'}F_{ab}\ee is a gauge invariant object. Here $\alpha^{'}$ is the Regge slope parameter.
Also the D{\it p}-brane tension is given by \be T_{p}\equiv (2\pi)^{-p}g^{-1}_{s}(\alpha^{'})^{-(p+1)/2}.\ee
It is important to note that when $N$ number of D{\it p}-branes coincide in the present context, then the world volume gauge theory transformed to a non-Abelian gauge theory and in that case the representative action
takes the following form \cite{Johnson,Tong:2009np}:
\be\begin{array}{llll}\label{ew4}
    \displaystyle S_{DBI}=-g_{s}T_{p}\int d^{p+1}\sigma~ {\rm Tr}\left[e^{-\Phi}\sqrt{-\det(G_{ab}+{\cal F}_{ab})}\right]
   \end{array}\ee

where the mathematical operation for the trace is performed over the gauge field indices. In that case the generalized version of the Chern-Simons action for $N$ number of
D{\it p}-branes in the presence
of various background fields can be written as \cite{Bachas:1998rg,Giveon:1998sr,Polchinski:1996na,Johnson}:
\be\begin{array}{lll}\label{ew5}
    \displaystyle S_{CS}=i\mu_{p}\int _{\Sigma_{p+1}}{\rm Tr}\left[\sum_{n}C_{n}\wedge e^{\cal F}_{2}\right]
   \end{array}\ee
Finally, it is importnat to mention here that the complete bosonic action 
for D-branes in a low energy supergravity background can be expressed as the sum of the Dirac-Born-Infeld action and the
Chern-Simons action as \be S_{D{\it p}}=S_{DBI}+S_{CS}.\ee

\subsubsection{Dimensional Reduction}
Here our prime objective is to compute the effective action in $D=4$ effective action using the basic principles of string compactification. To serve this purpose in this context
one starts with the $D=10$ string theory action and then use Kaluza-Klein (KK) reduction technique to derive an effective action in $D=4$. To give clear picture about this technique here
we will discuss a very specific example in the present context.

to implement this technique let us start with a $D=10$ geometry described by the following background metric:
\be\begin{array}{llll}\label{warmet}
    \displaystyle G_{MN}dX^{M}dX^{N}=e^{-6U(x)}g_{\mu\nu}dx^{\mu}dx^{\nu}+e^{2U(x)}\hat{g}_{mn}dy^{m}dy^{n}
   \end{array}
\ee
where $e^{U(x)}$ characterizes a ``breathing mode'' and  $\hat{g}_{mn}$ signifies a reference metric with fixed six volume as given by \be \int_{X_{6}}d^{6}y \sqrt{\hat{g}}\equiv {\cal V}.\ee 
Now it is important to mention here that, using the background metric mentioned in Eq~(\ref{warmet}) one can explicitly study the 
dimensional reduction technique of $D=10$ Einstein Hilbert action to its $D=4$ counterpart. To show this explicitly here we start with $D=10$ Einstein Hilbert action
which can be written as:
\be\begin{array}{lll}\label{ac10}
    \displaystyle S^{(10)} = \frac{1}{2k^{2}_{10}}\int d^{10}X\sqrt{-G}e^{-\Phi}R_{10} 
\end{array}
\ee
where $k_{10}$ be the ten dimensional gravitational coupling strength and $R_{10}$ be the Ricci scalar at $D=10$.
After performing dimensional reduction Eq~(\ref{ac10}) can be recast as:
\be\begin{array}{lll}\label{ac10new}
    \displaystyle S^{(10)} = \frac{1}{2k^{2}_{10}}\int d^{4}x\sqrt{-g}\int_{X_{6}}d^{6}y \sqrt{\hat{g}}e^{-2\Phi}
\left[R_{4}+e^{-8U(x)}\hat{R}_{6}+\frac{12}{U^{2}(x)}\partial_{\mu}U(x)\partial^{\mu}U(x)\right] 
\end{array}
\ee
Here we define $R_{4}$ to be the $D=4$ counterpart of the Ricci scalar constructed from the metric $g_{\mu\nu}$ and also $R_{6}$ characterizes $D=6$ counterpart of the Ricci scalar
constructed from the metric $\hat{g}_{mn}$. Here $U(x)$ are the scalar moduli field appearing from string theory. Most importantly the second term in the above mentioned dimensionally reduced version of the effective action
is non-canonical in nature and interacted with the dilaton modes represented by $e^{-2\Phi}$. Also it is important to note that $D=6$ counterpart of the Ricci scalar $R_{6}$ interacted in a non-minimal fashion with both scalar field $U(x)$ and 
the dilaton modes. For the sake of simplicity here one can also consider a physical situation where the string theoretic coupling strength is given by \be g_{s}\equiv e^{\Phi}\ee
and in that case the effective action corresponding to the 4D counterpart of the Einstein-
Hilbert action term can be written in the following form as:
\be\begin{array}{lll}\label{ac4}
    \displaystyle S^{(4)} = \frac{M^{2}_{p}}{2}\int d^{4}x\sqrt{-g}e^{-\Phi}R_{4} 
\end{array}
\ee
where it is important to note that the four-dimensional effective (reduced) Planck mass scale appearing through dimensional reduction is defined as:
\be\begin{array}{llll}\label{plk}
    \displaystyle M^{2}_{p}\equiv \frac{{\cal V}}{g^{2}_{s}\kappa^{2}_{10}}.
   \end{array}\ee

\subsection{Braneworld gravity}

A very well known low energy version of String Theoretic prescription is Braneworld Gravity within which the observable universe is usually described by 1+3-surface. In technical language this is known as
``brane'' or ``membrane''. According to construction of braneworld gravity such membranes are 
  embedded in a 1+3+\textit{d}-dimensional space-time within the ``bulk'', with
  Standard Model degrees of freedom and fields trapped on the brane. On the other hand in this context graviton 
  and other field contents (example: dilaton, scalar field, antisymmetric higher rank tensor fields etc.) are freely propagate within
 the bulk \cite{Maartens:2010ar}. In this discussion \textit{d} characterizes the extra spatial
  dimensions. Brane-world gravity models includes a purely phenomenological way to study the features of modifications
  to general theory of relativity via warp geometry and extra dimensions. 
  The Randall Sundrum braneworld model (RS) \cite{Randall:1999ee,Randall:1999vf},
one of the pioneering warped geometry model, was proposed to resolve the long 
standing problem in connection with the fine tuning of the mass of
Higgs (also known as gauge hierarchy or naturalness problem) in an otherwise successful Standard Model of elementary particles.
In particle phenomenology, one of the important experimental signatures of such extra dimensional models is the search of the  Kaluza-Klien (KK) gravitons
in pp collision leading to dilepton decays in the Large Hadron Collider (LHC) \cite{lhc}. The couplings of the zero mode as well as the higher KK modes are
determined by assuming the standard model fields to be confined on a 3-brane located at an orbifold fixed point. Such a picture is rooted in a string-inspired model
where the standard model fields being open string-excitations are localized on a 3-brane. This led to the braneworld description of extra dimensional 
models with gravity only propagating in the bulk as a closed string excitation. But apart from graviton, string theory admits of various 
higher rank antisymmetric tensor excitations as 
closed string modes which can also propagate as a bulk field. It was found that remarkably such fields are heavily suppressed on the brane in such warped
geometry model and thus offers a possible explanation of invisibility of these fields in
 current experiments \cite{Mukhopadhyaya:2002jn,Mukhopadhyaya:2001fc,Mukhopadhyaya:2007jn,Mukhopadhyaya:2009gp}.
Subsequently going beyond the stringy description, the implications of the presence of standard model fields in the bulk were also studied
in different variants of warped braneworld models. All these models in general assumed the 3-brane hypersurface to be flat.   
These models were subsequently generalized to include non-flat branes \cite{Koley:2008hs,Mitra:2009jw,Das:2010xx} and also
 braneworld with larger number of extra dimensions \cite{Choudhury:2006nj,Mukhopadhyaya:2011gn,Koley:2008dh}.
  
From a theoretical standpoint, warped geometry model has its underlying motivation
in the backdrop of string theory where the Klevanov-Strassler (KS) throat geometry 
solution exhibits warping character. While the Randall-Sundrum model starts with Einstein's gravity in ${\bf AdS_{5}}$ manifold in five dimensional
space-time, there have been efforts to include the higher curvature effects in the nature of the warped geometry.
Such corrections originate naturally in string theory where power expansion in terms of inverse string tension 
yields the higher order corrections to pure Einstein's gravity.
Supergravity, as the low energy limit of heterotic string theory yields the Gauss-Bonnet (GB) term as the leading order correction and therefore 
became an active area of interest as a modified theory of gravity.
 In addition
 to this string theory admits of higher loop corrections which is
a further modification on Einstein-Gauss-Bonnet correction on Einstein's gravity \cite{Choudhury:2013aqa,Choudhury:2013eoa,Choudhury:2011jt}.

\subsubsection{Higher dimensional gravity}
 For an Einstein--Hilbert gravitational action in (4+d) dimension
we have \cite{Maartens:2010ar}:
\bea
  S^{(4+d)}_\mathrm{EH} &=&{1\over 2\kappa_{4+d}^2}\int d^4x\, d^dy\,\sqrt{-^{(4+d)\!}g}\left[ {}^{(4+d)\!}R- 2\Lambda_{4+d} \right].
  \eea
After varying this action with respect to the (4+d) dimensional metric ${}^{(4+d)\!}g_{AB}$ one can derive the equation of motion as \cite{Maartens:2010ar}:
\bea
  {}^{(4+d)\!}G_{AB} & \equiv & \;{}^{(4+d)\!}R_{AB}-{1\over2}{}^{(4+d)\!} R \;{}^{(4+d)\!}g_{AB}\nonumber \\&=& -\Lambda_{4+d}{}^{(4+d)\!}g_{AB}+ \kappa_{4+d}^2 \;{}^{(4+d)\!}T_{AB},
\eea
where the $4+d$ coordinates can be identified as $X^A=(x^\mu,y^1, \dots, y^d)$, and in the present context $\kappa_{4+d}^2$ signifies the (4+d) dimensional
gravitational coupling constant and defined as:
\be
  \kappa_{4+d}^2=8\pi G_{4+d}={1\over M_{4+d}^{2+d}}\sim \frac{L^d}{M^2_p}.
\ee
 where the length scale of the extra dimensions is $L$ and we have used the fact that the fundamental scale via the
volume of the extra dimensions can be written as:
\be
  M_{p}^2 \sim M_{4+d}^{2+d}\,L^d.
\ee
If we incorporate the further modification on Einstein-Gauss-Bonnet dilaton correction on Einstein's gravity then the (4+d) dimensional higher dimensional 
gravity is described by the following action \cite{Choudhury:2013aqa,Choudhury:2011jt}:
\be\begin{array}{llllll}\label{actionEHGB}
   \displaystyle S^{(4+d)}_\mathrm{EHGB} =
  {1\over 2\kappa_{4+d}^2}\int d^4x\, d^dy\,\sqrt{-^{(4+d)\!}g}
  \left[ {}^{(4+d)\!}R- 2\Lambda_{4+d}\right.\\ \left.~~~~~~~~~~~~~
~~~~~~~~~~~~~~~~~~~~~~~~~~\displaystyle +\alpha_{4+d}(1-A^{loop}_{4+d}\exp[\Theta_{4+d}\chi(y)])\left\{{}^{(4+d)\!}R^{ABCD}{}^{(4+d)\!}R_{ABCD}
\right.\right.\\ \left.\left.~~~~~~~~~~~~~
~~~~~~~~~~~~~~~~~~~~~~~~~~~~~~~~~~~~~~~~~~~~~~~~~~~~~~~~~~~\displaystyle -4{}^{(4+d)\!}R^{AB}{}^{(4+d)\!}R_{AB}+{}^{(4+d)\!}R^{2}\right\}\right],
   \end{array}\ee
where $\alpha_{4+d}$, $A^{loop}_{4+d}$ and $\Theta_{4+d}$ represent the {\it Gauss-Bonnet}, {\it string two-loop} and {\it dilaton} coupling 
respectively coming from the interaction with
dilatonic degrees of freedom coming from the computation of Conformal Field Theory (CFT) disk amplitude in the bulk geometry. 
After varying Eq~(\ref{actionEHGB}) with respect to the (4+d) dimensional metric ${}^{(4+d)\!}g_{AB}$ one can derive the equation of motion as \cite{Choudhury:2013aqa,Choudhury:2011jt}:
\bea\label{defeEHGB}
  {}^{(4+d)\!}G_{AB}+\alpha_{4+d}(1-A^{loop}_{4+d}\exp[\Theta_{4+d}\chi(y)]){}^{(4+d)\!}H_{AB} & = & -\Lambda_{4+d}
  \;{}^{(4+d)\!}g_{AB}\nonumber\\ &&+ \kappa_{4+d}^2 \;{}^{(4+d)\!}T_{AB},
  \eea

where the Gauss-Bonnet tensor ${}^{(4+d)\!}H_{AB}$ in (4+d) dimension is defined as \cite{Choudhury:2013aqa,Choudhury:2011jt}:

\be\begin{array}{llll}\label{gbp}
\displaystyle {}^{(4+d)\!} H_{AB}=2{}^{(4+d)\!}R_{ACDE}{}^{(4+d)\!}R_{B}^{CDE}-4{}^{(4+d)\!}R_{ACBD}{}^{(4+d)\!}R^{CD}
-4{}^{(4+d)\!}R_{AC}{}^{(4+d)\!}R_{B}^{C}\\ ~~~~~~~~~~~~~~~~~~~~~~~~~~~~~~~~~~~~+2{}^{(4+d)\!}R{}^{(4+d)\!}R_{AB}-\frac{1}{2}{}^{(4+d)\!}g_{AB}
\left({}^{(4+d)\!}R^{ABCD}{}^{(4+d)\!}R_{ABCD}\right.\\ \left. \displaystyle~~~~~~~~~~~~~~~~~~~~~~~~~~~~~~~~~~~~~~~~~~~~~~~~~~~~~~~~~~~~~~~~
-4{}^{(4+d)\!}R^{AB}{}^{(4+d)\!}R_{AB}+{}^{(4+d)\!}R^{2}_{(5)}\right).
   \end{array}\ee


Let us now discuss about three well known braneworld gravity models elaborately:

\subsubsection{A. Randall--Sundrum braneworlds}
\label{section_2}

Let us start our discussion with Randall--Sundrum (RS) braneworld gravity model. Before going to details of 
RS brane-worlds it is importnat to mention here that such type of frameworks do not rely on the compactification technique to localize the 
gravity at the orbifold fixed points where the brane is fixed. But within this framework the curvature of the bulk plays a significant role as 
in warped geometry background one can easily implement the method of compactification very easily in the present context. In the slice of $AdS_{\bf 5}$ to make the curvature of the space-time real and positive 
one puts necessary restriction on the signature of the five dimensional bulk cosmological 
constant and in the present context it is given by:
\be
  \Lambda_5=-\frac{6}{f^2}=-6q^2<0,
\ee
where $f$ signifies the radius of curvature ${AdS}_{\bf 5}$ and $q$ characterizes the
corresponding energy scale associated with this setup. Now introducing Gaussian normal coordinates $X^A=(x^\mu,y)$ localized on the brane
at the orbifold point $y=0$ one can re-express the ${AdS}_{\bf 5}$ metric in the following form as \cite{Randall:1999ee,Randall:1999vf}:
\be
  {}^{(5)\!}ds^2=e^{-\frac{2|y|}{f}} \eta_{\mu\nu}dx^\mu dx^\nu + dy^2,
  \label{mwth}
\ee
with $\eta_{\mu\nu}$ being the usual flat Minkowski metric. It is important to mention here that here $Z_2$-symmetry is imposed at the orbifold fixed point $y=0$.
In the bulk the 5D Einstein equations \cite{Randall:1999ee,Randall:1999vf} can be written as:
\be
  {}^{(5)\!}G_{AB}=- \Lambda_5 \;{}^{(5)\!}g_{AB}.
\ee
Here it is additionally important to note that to avoid back reaction effect we set \be {}^{(5)}T_{AB}=0\ee in the 5D bulk Einstein equations.

Similarly neglecting the contribution from back reaction effect within the bulk the Einstein Gauss Bonnet in presence of dilaton coupling turns out to be \cite{Choudhury:2013aqa,Choudhury:2011jt}:
\bea\label{defeEHGB1}
  {}^{(5)\!}G_{AB}+\alpha_{5}(1-A^{loop}_{5}\exp[\Theta_{5}\chi(y)]){}^{(5)\!}H_{AB} & = & -\Lambda_{5}
  \;{}^{(5)\!}g_{AB},
  \eea

where the Gauss-Bonnet tensor ${}^{(5)\!}H_{AB}$ is defined in (4+1) dimension and the dilaton field $\chi(y)$ is given by \cite{Choudhury:2013aqa,Choudhury:2011jt}:
\be\begin{array}{llll}\label{dil}
    \displaystyle \chi(y)=c_{1}|y|+c_{2}
   \end{array}\ee
where $c_{1}$ and $c_{2}$ are dimensionful arbitrary integration constants which can be precisely determined by imposing the boundary condition at the 
orbifold fixed points. Now using these solutions in the higher curvature gravity setup, the warped metric in ${AdS}_{\bf 5}$ slice takes the form \cite{Choudhury:2013aqa,Choudhury:2011jt}:
\be
  {}^{(5)\!}ds^2=e^{-\frac{|y|}{2}\sqrt{\frac{3}{\alpha_{5}(1-A^{loop}_{5}\exp[\Theta_{5}\chi(y)])}\left[1\pm \sqrt{1-\frac{8\alpha_{5}}{3l^{2}}(1-A^{loop}_{5}\exp[\Theta_{5}\chi(y)])}\right]}}
 \eta_{\mu\nu}dx^\mu dx^\nu + dy^2
\ee
where we get two branch of warping solutions. In the limit $\alpha_{5}\rightarrow 0$, $A^{loop}_{5}\rightarrow 0$ and $\Theta_{5}\rightarrow 0$ the -ve branch of solution 
exactly reduces to RS metric as stated in Eq~(\ref{mwth}). On the other hand, the +ve branch diverges in such weak coupling limit and incorporates ghost fields.

The two RS models are distinguished as follows:
\begin{enumerate}
\item \underline{\bf RS 2-brane:} In this prescription there are two branes are localized at two orbifold fixed points 
  $y=0$ and $y=L\sim \pi$.
  Most importantly in this context the branes separated by a distance $r_{c}$ have equal and opposite membrane tensions $\pm\lambda_{b}$ and it is given by the following expression:
  \be
    \lambda_{b}={3M_\mathrm{p}^2 \over 4\pi f^2}.
    \label{rst}
  \ee
  The positive-tension brane has fundamental scale $M_5$ called 
  hidden and negative-tension brane is called visible brane where all the SM fields are embedded. In this context the effective 4D mass scale can be expressed in terms of 5D quantum gravity cut-off scale as: 
  \be
    M_\mathrm{p}^2=M_5^3\left[e^{\frac{2\pi}{f}}-1\right]f.
  \ee
  Additionally it is important to note that using this set of braneworld construction it is possible to address the well known hierarchy
  problem \cite{Choudhury:2013aqa,Choudhury:2011jt}. 

\item \underline{\bf  RS 1-brane:} In this prescription there is a single brane which is localized at the orbifold fized point $y=0$ with 
  positive membrane tension. Here one can think that the single braneworld prescription is arises from the two braneworld prescription, provided the negative brane tension brane is infinitely far from the other brane which 
  is fixed at $y=0$. In this context the effective 4D mass scale can be expressed in terms of 5D quantum gravity cut-off scale as: 
  \be
    M_5^3=\frac{M_{p}^2}{f}.
  \ee
  Surprisingly within the single brane framework we get a finite contribution to the 5D world
  volume because of the warp factor $f$ and :
  \be
    V_{5}=\int d^5X\sqrt{-{}^{(5)\!}g} = \frac{f}{2}\int d^4x.
  \ee
  This directly implies that the effective size of the extra dimension realized by the 5D
  graviton degrees of freedom is of the order of $f$.
\noindent

\end{enumerate}
\subsubsection{B. Other contemporary models}

In this section we will give a short introduction about other well known contemporary techniques applicable in the context of braneworld (modified) gravity.

\begin{enumerate}

\item \underline{\bf  DGP model:}

The Dvali-Gabadadze-Porrati model is a model of modified gravity theory which
consists of a 4D Minkowski brane embedded in a 5D Minkowski bulk like Randall Sundrum (RS) 2 model as discussed earlier. In the present context
the 5th extra dimension is flat as well as infinitely large. Most importantly in this context one can easily recover the 
Newton's law by just adding a 4D Einstein-Hilbert term generated by the braneworld curvature contribution to the 5D action.
Additionally it is important to note that, in the present context at very small length scales the standard 4D gravity theory can be retrieved. On the other hand  
the effect from the 5D gravity plays significant role in large length scales.

The DGP model is described by the following 5D action \cite{Dvali:2000hr} as given by:
\be\label{DGPaction}
S=\frac{1}{2\kappa_{(5)}^{2}}\int d^{5}X\sqrt{-\tilde{g}}\,
\tilde{R}+\frac{1}{2\kappa^{2}}\int d^{4}x\sqrt{-g}\,R
+\int d^{4}x\sqrt{-g}\,{\cal L}_{M}^{\mathrm{brane}}\,,
\ee
where the Lagrangian ${\cal L}_{M}^{\mathrm{brane}}$ describes 
matter localized on the 3-brane. Additionally $\tilde{g}_{AB}$ characterizes the metric in the 5D bulk and the corresponding 4D counterpart of the metric can be written in terms of the 5D metric as:
\be g_{\mu\nu}=\partial_{\mu}X^{A}\partial_{\nu}X^{B}\tilde{g}_{AB}.\ee
This can be physically interpreted as the induced metric on the brane with bulk coordinates $X^{A}(x^{c})$ and the brane coordinates are labeled by $x^{c}$.
Also it is important to note that, the first and second terms as appearing in Eq.~(\ref{DGPaction}) correspond to
Einstein-Hilbert actions in the 5D bulk and on the brane respectively.
In the present context $\kappa_{(5)}^{2}$ and 
$\kappa^{2}$ are the 5D and 4D gravitational coupling constants, 
respectively, which are related with 5D and 4D Planck masses,
$M_{5}$ and $M_{4}$, via  the following relationship:
\bea \kappa_{(5)}^{2}&=&\frac{1}{M_{5}^3},\\ 
\kappa^{2}&=&\frac{1}{M_{4}^2}=\frac{1}{M^{2}_{p}}= 8\pi G.\eea
The bulk equations of motion in 5D can be expressed as:
\be
G^{(5)}_{AB}=0,
\ee
where $G^{(5)}_{AB}$ represents the 5D counterpart of the Einstein tensor.  
Further applying the well known Israel junction conditions for non compact extra dimension on the brane the effective equation of motion can be expressed as:
\be
\label{bcxcx}
G_{\mu\nu}-\frac{1}{r_c}(K_{\mu\nu}-g_{\mu\nu} K)=\kappa_{(4)}^2T_{\mu\nu},
\ee
where $K_{\mu\nu}$ characterizes the extrinsic curvature computed on the brane,
 $T_{\mu \nu}$ signifies the energy-momentum tensor for localized matter on the brane. Additionally, here $r_{c}$ characterizes the crossover or critical length scale defined by:
 \be r_{c} = \frac{M_{4}^{2}}{2M_{5}^{3}}.\ee 
 Such length scale sets an upper limit above which the effect of extra dimension becomes important in the present discussion and below such scale it is also possible to 
 retrieve the GR limiting results.

The modified Friedmann equations in this DGP model are given by \cite{Dvali:2000hr}~\footnote{In the next section we have introduced the Friedmann equation in the context of General Relativity (GR) for flat FLRW spacetime.}:
\bea
H^2 &=&
 \Bigg(\sqrt{ \frac{\kappa^2 \rho}{3} + \frac{1}{4r_{c}^{2}} }  + \frac{1}{2r_{c}}
 \Bigg)^2,\\  
 \label{DGPfr1}
\dot{H} + H^2 &=& -\frac{\kappa^2}{6}(\rho + p) \left[ 1+
\left(\kappa^2 \frac{\rho}{3}   + \frac{1}{4r_{c}^2}\right)^{-1/2}
\frac{1}{2r_{c}} \right] + \left[ \sqrt{ \kappa^2
\frac{\rho}{3} + \frac{1}{4r_{c}^2}  } + \frac{1}{2r_{c}}
\right]^2 .  
 \label{DGPfr2}
\eea
In the present analysis, we have set the extra dimension as time-like in DGP model. But in generalized prescription in principle one can consider both space and time like extra dimensions in the present context. 

However apart from the various success of the DGP model in the context of modification in the gravity sector it has crucial problems as well. These are:
\begin{itemize}
 \item The model deals with ghost degrees of freedom, which cannot be possible to get read off from the present setup.
 Additionally, applying a quasi-static approximation to linear cosmological perturbations on the length scales required for the large scale structure formation directly  
shows that the DGP model contains a ghost mode.
 \item In the quantum regime of gravity the model has strong coupling problem for 
the length scale smaller than ${\cal O}(1000)$ km. 
\item Additionally this model contains super-luminal modes. 
\end{itemize}

\item \underline{\bf  ADD model:}

This model proposes that the scale of gravity
is $M_{ew}$ in the bulk space-time. To make this proposal consistent with observations we
need to change the nature of gravity so that this scale is related to 4D Planck mass in a
natural way. This is achieved by assuming ordinary matter and gauge fields are localized
on a 3-brane which are embedded in a higher-dimensional bulk. Only gravity propagates in
higher dimensions and becomes strongly coupled at the TeV scale. This proposal ignores
brane tension (mass per unit volume) and consequently back-reaction of it on the geometry
transverse to the brane. In the simplest form it considers compact extra dimensions of
radius R. Only gravity becomes higher dimensional at length scales smaller than $R$. Since
Newton's law in 4D is experimentally tested up to 0.2 mm, to be consistent with it, extra
dimensions should be smaller than 0.2 mm but can be as large as, say, 0.1 mm. In other
words, fundamental scale of higher-dimensional gravity is $M\sim 10^{3}$~GeV. One starts with
a (4+d) dimensional Einstein action,
\bea\label{eqhzx}
  S^{(4+d)} &=& \frac{1}{2\kappa_{4+d}^2}\int d^4x\, d^dy\,\sqrt{-^{(4+d)\!}g}{}^{(4+d)\!}R.
  \eea
4D gravity is obtained by simple dimension reduction and by
ignoring massive KK modes. Since d-dimensional space is compactified on a flat torus,
wave function of the zero mode is homogeneous in extra dimensions. 4D effective action
for gravity is as:
\bea\label{eff}
  S_\mathrm{eff} &=&\frac{V_d}{2\kappa_{4+d}^2}\int d^4x\sqrt{-^{(4)\!}g}{}^{(4)\!}R
  \eea
where $V_d\sim R^{d}$ is the volume of extra dimensions. We can now read off the 4D Newton's
constant \be G_{N}=G^{4+d}/V_d=M^{-(2+d)}R^{-d}\ee or equivalently, \be M_{p}= M(MR)^{d/2}.\ee Planck mass
$M_{p}$ is a derived quantity and can be made larger by having large extra dimensions. 
While this proposal removes $M_{ew}/M_{p}$ hierarchy, it resurfaces as a hierarchy of scales between
the compactification scale $\sim 10^{-2}$ cm and the electroweak scale $\sim 10^{-16}$ cm.
Alternatively, we have very light KK mass spectrum $m\sim$ MeV. In ADD model \cite{ArkaniHamed:1998rs,ArkaniHamed:1998nn}, 
KK modes are uniformly distributed in the internal dimensions. We
can get weak gravity coupling if we make them spend most of their time away from the
visible brane thereby reducing their overlap with the physical brane. 

\end{enumerate}

\section{Applications of Field Theory to Early Universe}
\label{a2}

\subsection{Inflation}

Inflation is a well accepted proposal to explain the physics of early universe based on the prescription of quantum field theory minimally or non-minimally coupled to the gravity sector.
It is a very well known fact that for a given a field theoretic prescription, it is generally simple to compute the time evolution of the homogeneous cosmological background,
and many many quantum field theoretic prescriptions that support the existence of cosmological inflationary phases have been proposed. However, deriving the effective action in four dimension for the 
the inflationary paradigm from a more fundamental physical principle, or specifically in the context of a well-motivated original
background theory, notwithstanding an important problem.

There are two parallel and well known approaches which can be implemented to finally obtain an ultimate quantum field theoretic prescription
well acceptable in the context of inflation \cite{Baumann:2009ds,Weinberg:2008hq,Cheung:2007st,Senatore:2010wk,Noumi:2012vr,Gwyn:2012mw}.
 Specifically in the `top-down' approach, the general notion is to start the computation with a UV-complete field theoretic prescription,
 and using this approach one finally attempts to derive the theory of inflation as one of the low energy effective theory. On the other hand in the `bottom-up' approach
one generally start the computation from the low-energy (IR) degrees of freedom and try to predict the bacground UV complete field theory.
Most importantly both the parallel approaches appear at an {\it effective field theory} prescription which is justifiable at inflationary energy scales. 

The original
incentive for implementing the idea of Inflation was to explain the initial conditions for the
Big-Bang theory or more precisely the initial condition for large scale structure formation observed at present epoch, by hypothesizing a much earlier epoch during which our Universe expanded
and the scale factor varies quasi exponentially. But, at present inflation is considered as the most favored theoretical prescription in physics
to describe the cosmological evolution of Early Universe.
Throughout the thesis we will focus only on the single field models of inflation which are embedded within the framework of effective field theory
prescription \cite{Baumann:2009ds,Weinberg:2008hq,Cheung:2007st,Senatore:2010wk,Noumi:2012vr,Gwyn:2012mw}.
 \subsubsection{Conditions for inflation}
The conditions for inflation are described by the following conditions \cite{Mazumdar:2010sa,riotto,Baumann:2009ds}:
\begin{itemize}
 \item \underline{\bf I. Decreasing comoving horizon:}\\ During the epoch of inflation the
comoving Hubble sphere shrinks and after inflation it expands. So one can explain inflation as
a physical mechanism to ``zoom-in'' on a smooth sub-horizon patch which finally generates required amount of 
cosmological fluctuations consistent with various observational probes. Using the Friedmann Equations a shrinking comoving Hubble radius can be expressed in terms of the acceleration
and the pressure of the universe:
\be\begin{array}{llll}\label{condition1}
    \displaystyle \frac{d}{dt}(aH)^{-1}<0.
   \end{array}\ee
\item \underline{\bf II. Accelerated expansion:} \\Using the relation \be \frac{d}{dt}(aH)^{-1}=-\frac{\ddot{a}}{(\dot{a})^{2}}\ee it can be easily shown that
shrinking comoving Hubble radius directly implies accelerated expansion of universe given by:
\be\begin{array}{llll}\label{condition2}
    \displaystyle \frac{d^{2}a}{dt^{2}}>0.
   \end{array}\ee

\item \underline{\bf III. Slowly-varying Hubble parameter:}\\ Similarly using the relation \be \frac{d}{dt}(aH)^{-1}=-\frac{1}{a}(1-\epsilon),\ee
 where \be \epsilon=-\frac{\dot{H}}{H^{2}}>0\ee it can be easily shown that
shrinking comoving Hubble radius implies the following constraint:
\be\begin{array}{llll}\label{condition3}
    \displaystyle \epsilon=-\frac{\dot{H}}{H^{2}}=-\frac{d\ln H}{dN}<1.
   \end{array}\ee
where we have defined $dN=Hdt=d\ln a$, which measures the number of e-folds $N$ of inflationary
expansion. Also to achieve sufficient number of e-foldings during inflationary epoch the necessary condition is the slow roll parameter $\epsilon$ must be small compared to unity for a sufficient large number 
of Hubble times. This condition is explicitly taken care of by the second slow roll parameter defined as: 
\be\begin{array}{llll}\label{condition4}
    \displaystyle \eta=\frac{\dot{\epsilon}}{H\epsilon}=\frac{d\ln \epsilon}{dN}.
   \end{array}\ee
For $|\eta|< 1$ the fractional change of $\epsilon$ per Hubble time is small compared to unity and inflation continues.

\item \underline{\bf IV. Negative pressure:}\\ Although this condition essentially follows from the above discussion, assuming a
perfect fluid with pressure $p$ and density $\rho$, we can write the Friedmann
Equations in the following form as:
\be\begin{array}{llll}\label{condition5}
    \displaystyle \dot{H}+H^{2}=-\frac{1}{6M^{2}_{p}}(\rho +3p)=-\frac{H^{2}}{2}\left(1+\frac{3p}{\rho}\right).
   \end{array}\ee
Now here it is importnat to note that:

\be\begin{array}{llll}\label{condition6}
    \displaystyle \epsilon=-\frac{\dot{H}}{H^{2}}=\frac{3}{2}(1+w)~~~~~\Leftrightarrow w=\frac{p}{\rho}<-\frac{1}{3}.
   \end{array}\ee
which directly implies that inflation requires negative pressure or more precisely a violation of the strong energy condition (SEC).

\end{itemize}

\subsubsection{Effective field theory of inflation}

It is very well known fact that for single scalar field models, cosmological dynamics for inflation is directly explaied by the scalar inflaton field $\phi$ which is  
minimally coupled to Einstein gravity sector~\footnote{In principle, a non-minimal coupling between the inflaton and the graviton sector can be considered in the present context. However, 
in most of the realistic physical situations, such non-minimally coupled theories can be transformed to minimally coupled form by a field redefinition or via special type of conformal 
transformation in the present context. Example: Higgs inflation in presence of non-minimal coupling \cite{Choudhury:2013zna}. 
Similarly, one can think of the possibility that the leading order the UV incomplete Einstein-Hilbert part of the action is modified at high
energies in presence of perturbative UV corrections in the gravity sector. The simplest
 examples are: $f(R)$ gravity \cite{Sotiriou:2008rp,DeFelice:2010aj,Tsujikawa:2010zza}, Gauss-Bonnet gravity \cite{Sotiriou:2008rp,DeFelice:2010aj,Tsujikawa:2010zza,Lovelock}, which can 
be transformed into a minimally coupled scalar field.
}
\be\begin{array}{llll}\label{mod1}
    \displaystyle S_{EFT1}=\int d^{4}x \sqrt{-^{(4)\!}g}
  \left[\frac{M^{2}_{p}}{2}{}^{(4)\!}R+\frac{1}{2}g^{\mu\nu}\partial_{\mu}\phi\partial_{\nu}\phi-V(\phi)\right].
   \end{array}\ee
where ${}^{(4)\!}R$ is the four-dimensional Ricci scalar derived from the metric $g_{\mu\nu}$ in FLRW background and $V(\phi)$ be the inflationary potential.
In principle $V(\phi)$ can be expressed in terms of any arbitrary mathematical function of the inflation filed $\phi$ in the present context of discussion.
However, the background field theoretic prescriptions always restrict the 4D field 
theory to take such an arbitrary mathematical form. In a generic physical prescription the inflaton potential $V(\phi)$ can be expressed as:
\be\begin{array}{llll}\label{potemod}
    \displaystyle V(\phi)=V_{ren}(\phi)+\sum^{\infty}_{\alpha=5}c_{\alpha}\frac{\phi^{\alpha}}{\Lambda^{\alpha-4}_{UV}}
   \end{array}\ee
where $\Lambda_{UV}$ is the UV cut-off of gravity. In our discussion we take $\Lambda_{UV}\sim M_{p}$ and $c_{\alpha}$ be the expansion co-efficient of the non-renormalizable contribution in the potential.
In the effective field theory prescription $c_{\alpha}$ can be treated as the Wilson coeffeints as appearing in the context of RG flow.
Also the contribution from the renormalizable part of the inflationary potential can be expressed as:
\be\begin{array}{llll}\label{repot}
    \displaystyle V_{ren}(\phi)=\sum^{4}_{\delta=0}a_{\delta}\frac{\phi^{\delta}}{\Lambda^{\delta-4}_{UV}}
   \end{array}\ee
where $a_{\delta}$ is the expansion co-efficient in the renormalizable sector in EFT and can also be treated as Wilson coefficients as appearing in the context of RG flow.
However in this most generalized prescription various types of interactions between the scalar inflaton field and other 
matter degree of freedom are allowed via higher mass dimensional UV scale suppressed non-renormalizable operators in the kinetic as well the potential sector of the field theory arising from
effective field theory (EFT) \cite{Baumann:2009ds,Weinberg:2008hq,Cheung:2007st,Senatore:2010wk,Noumi:2012vr,Gwyn:2012mw} prescription. Let us briefly discuss two parallel theoretical approaches
 of EFT through which one can be able to study the details of inflationary paradigm. These approaches are-
\begin{itemize}
\item \underline{\bf Top down approach:}\\
To construct an effective field theory setup following the top down approach one needs to first identify the physical degrees of freedom that are
significant to explain the observational results. More precisely in this context we introduce a cutoff scale $\Lambda_{UV}$ of gravity sector to specify
the regime of validity of the EFT prescription in the context of inflation. In the present discussion light particles $\phi$, with very small masses $m < \Lambda_{UV}$, are included within the framework of effective field 
theory of inflation. On the other hand usually the heavy particles $\Psi$, with large masses $M > \Lambda_{UV}$, are integrated out form the framework of effective field 
theory. Within this physical prescription the most generalized action for EFT of inflation can be recast
from Eq~(\ref{mod1}) in the following form as:

\be\begin{array}{llll}\label{mod2}
    \displaystyle S_{EFT2}=\int d^{4}x \sqrt{-^{(4)\!}g}
  \left[\frac{\Lambda^{2}_{UV}}{2}{}^{(4)\!}R+{\cal L}_{inf}[\phi]+{\cal L}_{heavy}[\Psi]+{\cal L}_{int}[\phi,\Psi]\right].
   \end{array}\ee
where ${\cal L}_{inf}[\phi]$ and ${\cal L}_{heavy}[\Psi]$ represent the part of total Lagrangian density ${\cal L}$ involving only the light and heavy fields, 
and ${\cal L}_{int}[\phi,\Psi]$ includes all types of
possible interactions involving the light and heavy fields within EFT prescription for inflation. Now to integrate out the effects of all such heavy field from EFT of inflation picture here one need to preform
the path integral over the heavy degrees of freedom and over the all possible high frequency contributions of the light degrees of freedom within the framework of EFT of inflation 
and it can be written in terms of the following effective action
for inflation as:
\be\begin{array}{lll}\label{mod3}
    \displaystyle e^{i S_{EFT2}[\phi]}=\int [{\cal D}\Psi] e^{iS_{EFT2}[\phi,\Psi]}.
   \end{array}
 \ee
 In EFT prescription this serves the purpose for Wilsonian effective action from which one can easily study RG flow.
The effective
action for inflation in 4D admits a systematic series expansion in terms of the light degrees of freedom for which the total action within EFT action for inflation as stated in Eq~(\ref{mod2}) using Eq~(\ref{mod3}) 
can be re-expressed as:
\be\begin{array}{llll}\label{mod2a}
    \displaystyle S_{EFT2}=\int d^{4}x \sqrt{-^{(4)\!}g}
  \left[\frac{\Lambda^{2}_{UV}}{2}{}^{(4)\!}R+{\cal L}_{inf}[\phi]+\sum_{\alpha}J_{\alpha}(g)\frac{{\cal O}_{\alpha}[\phi]}{M^{\Delta_{\alpha}-4}}\right].
   \end{array}\ee
where $J_{\alpha}(g)\forall \alpha$ are the dimensionless coupling constants that depend on the coupling parameter $g$ of the background UV complete higher dimensional field theoretic prescription.
Also ${\cal O}_{\alpha}[\phi]$ characterizes the
local EFT renormalizable or non-renormalizable higher mass dimensional EFT operators for inflation of mass dimension $\Delta_{\alpha}$. 
Following this prescription one can easily able to generate all such relevant local EFT renormalizable or non-renormalizable higher mass dimensional EFT operators for inflation ${\cal O}_{\alpha}[\phi]$ physically 
consistent with the symmetries of the background UV complete higher dimensional field theory. If we switch off all such relevant contribution from the EFT of inflation 
which are physically allowed
by the symmetries of the the background UV complete higher dimensional field theoretic prescription is generally 
described as a ``fine-tuning''\cite{Baumann:2009ds,Weinberg:2008hq,Cheung:2007st,Senatore:2010wk,Noumi:2012vr,Gwyn:2012mw} mechanism or ``naturalness'' in EFT picture. 

\item \underline{\bf Bottom up approach:} \\ In the bottom up approach the situation is completely opposite as discussed above.
Here we do not know the structural form of the complete UV complete higher dimensional field theoretic prescription and due to this reason one cannot at all construct the EFT explicitly by
integrating out the heavy degrees of freedom always from the background field theoretic picture. In the present context we usually parameterize the various choices of UV field theoretic setup and then
we explicitly use various assumptions about the symmetries of the UV field theoretic setup and finally we write down the most general
EFt action for inflation which is consistent with all of these symmetries appearing in the context of UV theory. In this case the 4D effective action can be expressed as:
\be\begin{array}{llll}\label{mod2b}
    \displaystyle S_{EFT3}=\int d^{4}x \sqrt{-^{(4)\!}g}
  \left[\frac{\Lambda^{2}_{UV}}{2}{}^{(4)\!}R+{\cal L}_{inf}[\phi]+\sum_{\alpha}z_{\alpha}(g)\frac{{\cal O}_{\alpha}[\phi]}{M^{\Delta_{\alpha}-4}}\right].
   \end{array}\ee
where it is importnat to note that the sum runs over all local EFT renormalizable or non-renormalizable higher mass dimensional EFT operators for inflation
${\cal O}_{\alpha}[\phi]$ of mass dimension $\Delta_{\alpha}$, favored by the symmetries appearing in the context of UV field theoretic setup. 
The size of the local EFT renormalizable or non-renormalizable higher mass dimensional EFT operators for inflation is estimated in terms of the EFT UV cutoff scale $\Lambda_{UV}$,
while the pre-factors appearing in Eq~(\ref{mod2b}) $z_{\alpha}$ serves the purpose of dimensionless Wilson coefficients in the context of EFT of inflation.

\end{itemize}

\subsubsection{Slow-roll technique}
To illustrate the features of slow-roll, we start with the Klein Gordon equation and the Friedmann equations 
computed from the simplest action ~(\ref{mod1})~\footnote{By assuming the isotropy and homogeneity in the background geometry here we use the Friedmann-Robertson
-Walker (FRW) metric with spatially flat hypersurface.}. The equations governing the cosmological dynamics are:
\begin{eqnarray}
 \ddot{\phi}+3H\dot{\phi}+V^{'}(\phi)&=&0,\\
  H^{2}&=&\frac{\rho}{3M^{2}_{p}}=\frac{1}{6M^{2}_{p}}\left[\dot{\phi}^{2}+2V(\phi)\right],\\
\dot{H}+H^{2}&=&-\frac{1}{6M^{2}_{p}}(\rho +3p)=-\frac{1}{3M^{2}_{p}}\left[\dot{\phi}^{2}-V(\phi)\right].
\end{eqnarray}
The slow-roll approximation in EFT of inflation states that-
\begin{itemize}
 \item The contribution from the kinetic term is negligibly small in the effective pressure ($p$) and density ($\rho$).
       Consequently the Friedmann equations can be recast as:
        \begin{eqnarray}
  H^{2}&\approx&\frac{V(\phi)}{3M^{2}_{p}},\\
\dot{H}+H^{2}&\approx&\frac{V(\phi)}{3M^{2}_{p}}.
\end{eqnarray}

\item The contribution from the acceleration term, $\ddot{\phi}$, is negligibly small compared to the damping term $3H\dot{\phi}$ and $V^{'}(\phi)$. Consequently 
the Klein-Gordon equation is modified to:
\begin{eqnarray}
3H\dot{\phi}+V^{'}(\phi)&\approx&0.
\end{eqnarray}
\end{itemize}
Furthermore, the slow-roll technique 
is characterized by a set of flatness parameters which can be expressed in terms of the
inflationary potential and its higher powers of the derivatives as~\footnote{The Hubble slow roll parameter $(\epsilon,\eta)$ and the potential dependent slow-roll
parameters $(\epsilon_{V},\eta_{V})$ are connected via the relations, $ \epsilon\approx\epsilon_{V}$, and $\eta\approx\eta_{V}-\epsilon_{V}$. }:
\begin{eqnarray}\label{ra1}
    \epsilon_{V}=\frac{M^{2}_{p}}{2}\left(\frac{V^{\prime}}{V}\right)^{2}\,,~~
    \label{ra2} \eta_{V}={M^{2}_{P}}\left(\frac{V^{\prime\prime}}{V}\right)\,~~ \nonumber\\
\label{ja1}\xi^{2}_{V}=M^{4}_{p}\left(\frac{V^{\prime}V^{\prime\prime\prime}}{V^{2}}\right)\,,~~
    \label{ja2} \sigma^{3}_{V}=M^{6}_{p}\left(\frac{V^{\prime 2}V^{\prime\prime\prime\prime}}{V^{3}}\right)\,
   \end{eqnarray}
where within the slow-roll regime, \be \epsilon_{V},|\eta_{V}|,|\xi^{2}_{V}|,|\sigma^{3}_{V}|<<1.\ee Inflation ends when the
slow-roll approximation is violated and the corresponding field value of the inflaton is computed
from the following equation:
\begin{eqnarray}
 \max_{\phi}\left\{\epsilon_{V},|\eta_{V}|,|\xi^{2}_{V}|,|\sigma^{3}_{V}|\right\}=1
\end{eqnarray}

The amount of inflaton is generally expressed as the logarithmic difference between the
final and initial values of the scale factor, and is called number of e-foldings, $N$:
\begin{eqnarray}
 N\equiv \ln\left(\frac{a_{f}}{a_{i}}\right)=\int^{t_{f}}_{t_{i}} Hdt\approx-\frac{1}{M_{p}}\int^{\phi_{f}}_{\phi_{i}}\frac{d\phi}{\sqrt{2\epsilon_{V}}}
\end{eqnarray}

where $a_{i}$ and $a_{f}$ and $\phi_{i}$ and $\phi_{f}$ are the values of the scale factor and inflaton field at the start
and at the end of inflation respectively. The largest scales
observed in the CMB are produced some 50 to 70 e-folds before the end of inflation
\begin{eqnarray}
 N_{cmb}=-\frac{1}{M_{p}}\int^{\phi_{f}}_{\phi_{cmb}}\frac{d\phi}{\sqrt{2\epsilon_{V}}}\approx 50 - 70.
\end{eqnarray}
A successful solution to the horizon problem requires at least $N_{cmb}$ e-folds of inflation.

\begin{figure}[t]
\centering
\includegraphics[width=14.3cm,height=9cm]{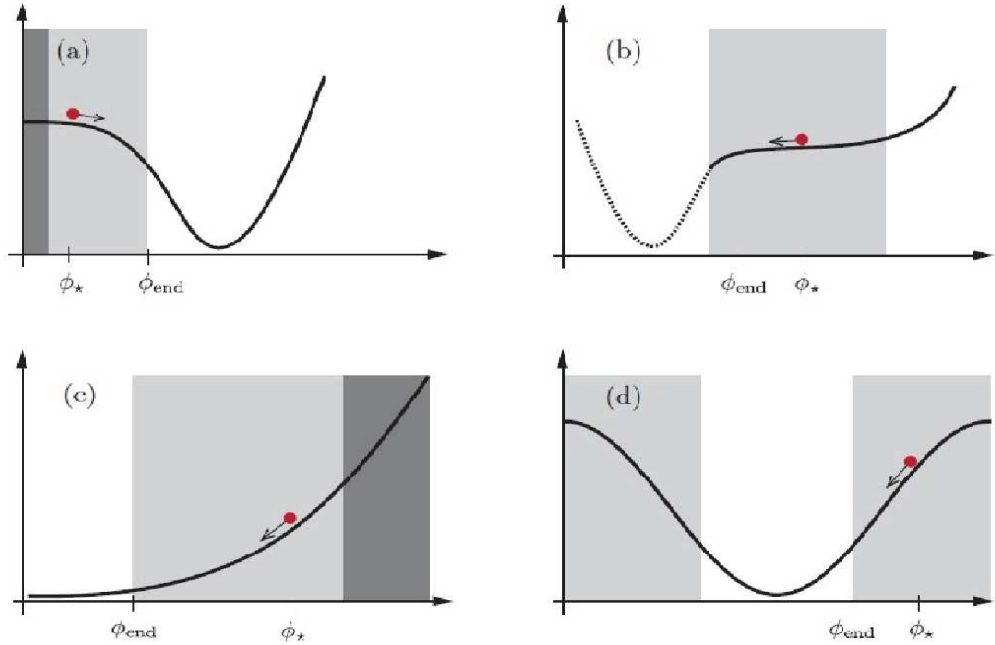}
\caption{\footnotesize Examples of different classes of slow-roll potentials: (a) hilltop inflation, (b) infection point
inflation, (c) chaotic inflation and (d) natural inflation. The light gray regions indication the parts of
the potential where slow-roll inflation occurs. The dark gray regions denote regions of eternal inflation.
Here (a) and (b) correspond to small-field models ($\Delta\phi < M_{p}$), while (c) and (d)
are large-field models ($\Delta\phi > M_{p}$) \cite{Baumann:2014nda}.
}
\label{fig1}
\end{figure}

The dynamics of the inflaton field, from the time when CMB fluctuations were created
 at $\phi_{cmb}$ to the end of inflation at $\phi_{f}$, is determined by the shape of the inflationary potential
$V(\phi)$. The different possibilities for $V(\phi)$ can be classified in a useful way by determining whether
they allow the inflaton field to move over a large or small distance via field excursion, \be \Delta\phi \equiv \phi_{cmb} - \phi_{f},\ee  as measured
in Planck units. Depending on the numerical values of the field excursion the single field inflationary
 models are classified into two categories~\footnote{In figure(\ref{fig1}) we have shown the behaviour of various types of single field inflationary
 models obtained from sub-Planckian and super-Planckian field excursion.}:
\begin{itemize}
 \item \underline{\bf A. Small field model:} Within this category the field excursion
 becomes sub-Planckian, $\Delta\phi\leq M_{p}$ and the VEV of the inflaton
 field is also sub-Planckian~\footnote{This is consistent with the EFT prescription for inflation where
the particle theory is embedded in EFT setup which is trustable below the UV cut-off of the scale of gravity \cite{Choudhury:2013iaa,Choudhury:2014kma,Choudhury:2014wsa}.},
\be \langle \phi \rangle \equiv \phi_{0}\leq M_{p}.\ee 
 Small field models are characterized by the concave potentials i.e., \be V^{''}(\phi)\leq 0.\ee Example: MSSM inflation from saddle point ($V^{'}(\phi_0)=0=V^{'''}(\phi_{0})$)
\cite{Allahverdi:2006iq,Allahverdi:2006we} and inflection point ($V^{''}(\phi_0)=0$) \cite{Enqvist:2010vd,Choudhury:2013jya}, Assisted inflation \cite{Liddle:1998jc,Copeland:1999cs} satisfy this condition.

\item \underline{\bf B. Large field model:} Within this category the field excursion becomes super-Planckian, $\Delta\phi>M_{p}$ and the VEV of the inflaton
 field is super-Planckian~\footnote{This is not consistent with the EFT prescription for inflation. However string theory
models can be embedded within this prescription within which it is possible to go beyond the UV cut-off of the scale of gravity \cite{Lyth:1996im}.},
\be \langle \phi \rangle \equiv \phi_{0}>M_{p}.\ee 
 Small field models are characterized by the concave potentials i.e., \be V^{''}(\phi)>0.\ee Example: Chaotic inflation \cite{Linde:1986fc,Linde:1983gd}, Natural inflation
 \cite{Freese:1990rb} satisfy this condition.
\end{itemize}


\subsubsection{Possible extensions of the paradigm}
In this subsection we mention the various physical possibilities for getting inflationary paradigm in extended picture. It is very well known fact that inflation is
 a theoretical framework for the early universe, but it is not at all a unique prescription to describe the physics of early universe. From phenomenological point of view various
models of inflation have been proposed with distinctive theoretical incentives as well to describe various  
observational predictions. In this context the most straightforward inflationary paradigm described by single field effective actions in 4D can be modified the following ways:
\begin{itemize}
 \item \underline{\bf A. Non-minimal coupling to gravity:}\\ In principle, we could imagine a non-minimal
coupling between the inflaton and the graviton, however, in practical purpose, non-minimally coupled gravity 
theories can be easily expressed as minimally coupled gravity theories by just a field re-definition via conformal transformations on the background FLRW metric. Higgs inflation \cite{Choudhury:2013zna,Bezrukov:2007ep} is one of the examples
of the inflation in presence of non-minimal coupling with the Einstein-Hilbert term in the representative action of EFT.

\item \underline{\bf B. Modified gravity:}\\ In this specific case the Einstein-Hilbert part of the effective action is
modified at high energy scales in presence of higher curvature perturbative corrections motivated from quantum gravity sector. Nevertheless, the simplest possibilities for this UV extension of gravity sector,
so-called $f(R)$ theory of gravity and Gauss-Bonnet gravity non-minimally coupled with a scalar field in 4D. Here both of the extended theories
of gravity can again easily be transformed into a gravity theory minimally coupled scalar field in 4D. Starobinsky inflation \cite{Starobinsky:1980te,Starobinsky:1983zz} is one
of the famous example of the inflationary model driven by modified gravity prescription, where the usual Einstein-Hilbert term, $R$, is modified by $R+\alpha R^{2}$.

\item \underline{\bf C. Non-canonical kinetic term:}\\ Within this prescription the inflaton action in presence of non-canonical term is modified as \cite{Baumann:2009ds}:
\begin{eqnarray}
 S_{\phi}=\frac{M^{2}_{p}}{2}\int d^{4}x \sqrt{-^{(4)\!}g}\left[P(\phi, X) - V(\phi)\right]
\end{eqnarray}
where $P(\phi, X)$ is arbitrary function of the inflaton field and its derivative, 
\be X=\frac{1}{2}g^{\mu\nu}\partial_{\mu}\phi\partial_{\nu}\phi.\ee
In this specific situation one can also consider a possibility where it is easily possible that inflation is driven by the kinetic term as appearing in the effective action in 4D 
and occurs even in the presence
of a steep inflationary potential, which is different from usual slow-roll inflationary models explained through canonical kinetic terms in the matter sector. Single field DBI inflation \cite{Shandera:2006ax,Firouzjahi:2005dh} and Galileon
 inflation \cite{Burrage:2010cu,Ohashi:2012wf} are the famous examples of the inflation driven by $P(X,\phi)$ theory.

\item \underline{\bf D. Multifield model:}\\ In this context we incorporate more than one scalar field in the background field theoretic picture and all of them are 
dynamically relevant during the epoch of inflation. Consequently the possibilities
to explain the inflationary paradigm and the dynamical mechanisms for the production of quantum fluctuations expand in very dramatic fashion
and finally the effective theory loses various important predictions in the present context. Multifield DBI \cite{Langlois:2009ej} and DBI Galileon \cite{deRham:2010eu} inflation are the
 famous examples in this area.

\end{itemize}

\subsubsection{Quantum fluctuations within EFT}
The quantum fluctuations of inflaton
produce a spectrum that matches the CMB observations to quite a good extent.
The inflaton evolution $\phi(t)$ governs the energy density of the early universe $\rho(t)$
and hence controls the end of inflation. Essentially, $\phi$ plays the role of a local clock reading
off the amount of inflationary expansion remaining. Because microscopic clocks are quantum mechanical objects
 with necessarily some variance by the uncertainty principle, the inflaton
 will have spatially varying fluctuations \be \delta\phi(t, {\bf x}) \equiv \phi(t, {\bf x}) -\bar{\phi}(t).\ee These fluctuations imply that
different regions of space inflate by different amounts. In other words, there will be local differences
 in the time when inflation end $\delta t({\bf x})$. Moreover, these differences in the local expansion
 lead to differences in the local densities after inflation.

Herein we will briefly study both of the scalar and tensor fluctuations in the present context.
For the scalar modes
we have to be careful to identify the true physical degrees of freedom.
A priori, we have 5
scalar modes which come from 4 metric perturbations given by-$\delta g_{00}$, $\delta g_{ii}$, $\delta g_{0i}$ and $\delta g_{ij}$
and 1 scalar
field perturbation $\delta\phi$. Gauge invariance associated with the invariance of the effective action of the EFT as stated in Eq~(\ref{mod1}) under scalar
coordinate transformations: \be t \rightarrow t + \epsilon_{0}\ee and \be x_{i} \rightarrow x_{i} + \partial_{i}\theta\ee will further remove two modes.
The Einstein equations remove two more modes from this picture, so that we are left with only 1 physical scalar mode.

Here we will work
in comoving gauge, defined by the vanishing of the momentum density, \be \delta T_{0i}\equiv 0.\ee For slow-roll inflation this
 condition can be translated into \be \delta\phi=0.\ee It is important to note that the additional non-dynamical metric perturbations $\delta g_{00}$ and $\delta g_{0i}$
in terms of the comoving curvature perturbation $\zeta$~\footnote{In some literature the notation ${\cal R}$ is used for the comoving curvature perturbation, to distinguish it from the curvature
perturbation on uniform density hypersurfaces, which sometimes is also denoted by $\zeta$.} is replaced by the Einstein equations in the present context~\footnote{The constraint
 equations are solved most conveniently in the ADM formalism,
where the metric fluctuations become non-dynamical Lagrange multipliers.
See \cite{Baumann:2009ds} for more details on technical aspects.}.
In this gauge, perturbations are characterized purely by fluctuations in the metric \cite{Baumann:2009ds},
\begin{eqnarray}
 \delta g_{ij}&=&a^{2}\left[e^{-2\zeta}\delta_{ij}+h_{ij}\right]\approx a^{2}\left[(1-2\zeta)\delta_{ij}+h_{ij}\right]
\end{eqnarray}
Here, $h_{ij}$ is a spin-2 graviton degrees of freedom, which is transverse ($\nabla_{i}h_{ij}=0$), traceless ($h_{i}^{i}=0$) tensor and $\zeta$ is a scalar.
It is important to note that
in the comoving spatial slices $\phi = const$ have extrinsic three-curvature \cite{Baumann:2009ds}:
\be R_{(3)}=\frac{4}{a^{2}}\nabla^{2}\zeta. \ee
In the present context the comoving curvature perturbation $\zeta$ has the crucial property for adiabatic matter fluctuations which is
time-independent on superhorizon scales after Fourier decomposition i.e. \be \lim_{k<<aH}\dot{\zeta_{\bf k}}=0.\ee
The constancy of $\zeta$ on superhorizon scales
allows us to relate CMB observations directly to the inflationary dynamics at the time when
a given fluctuation crosses the horizon.

Further substituting the metric fluctuations $\delta g_{00}$ and $\delta g_{0i}$ into the effective action of EFT as stated in 
Eq~(\ref{mod1}) and expanding in the powers of $\zeta$ and $h_{ij}$ up to the second order, one can write the free field action for scalar and tensor modes respectively as:
\bea
\label{sd1} S^{(2)}_{\zeta}
&=&\frac{M^{2}_{p}}{2}\int dt~ d^{3}{\bf x}~a^{3}~\frac{\dot{\phi}^{2}}{H^{2}}\left[\dot{\zeta}^{2}-\frac{1}{a^{2}}\left(\partial_{i}\zeta\right)^{2}\right],\\
\label{sd1t} S^{(2)}_{h}&=&\frac{M^{2}_{p}}{8}\int dt~ d^{3}{\bf x}~a^{3}~\left[\dot{h_{ij}}^{2}-\left(\partial_{q}h_{ij}\right)^{2}\right]
\eea
 Now we define the canonically-
normalized Mukhanov variable as, \be v=z\zeta M_p \ee and consequently one can write~\footnote{Here $\gamma$ stands for the helicity index for the transverse and traceless spin-2 graviton degrees of freedom.
In general tensor modes can be written in terms of the two orthogonal polarization basis vectors.}:
 \be u_{\gamma}=\frac{a}{\sqrt{2}}h_{\gamma}M_p,\ee where \be z=\frac{a\dot{\phi}}{H}=\sqrt{2\epsilon}a\ee  and transitioning to conformal time $\eta$
 leads to the action for a canonically normalized scalar:
\bea
\label{sd2} S^{(2)}_{\zeta}&=&\frac{1}{2}\int d\eta~ d^{3}{\bf x}~\left[(v^{'})^{2}-\left(\partial_{i}v\right)^{2}-m^{2}_{eff;\zeta}(\eta)v^{2}\right],\\
\label{sd2t} S^{(2)}_{h}&=&\frac{1}{2}\sum_{\gamma=+,\times}\int d\eta~ d^{3}{\bf x}~\left[(u^{'}_{\gamma})^{2}-\left(\partial_{i}v_{\gamma}\right)^{2}-m^{2}_{eff;h}(\eta)u^{2}_{\gamma}\right].
\eea
where $'$ represents the differentiation with respect to conformal time $\eta$. We recognize this as the action of an harmonic oscillator with a time-dependent effective 
mass for scalar and tensor modes as:
\bea
 m^{2}_{eff;\zeta}(\eta)&\equiv&-\frac{z^{''}}{z}=-\frac{H}{a\dot{\phi}}\partial^{2}_{\eta}\left(\frac{a\dot{\phi}}{H}\right),\\
m^{2}_{eff;h}(\eta)&\equiv&-\frac{a^{''}}{a}.
\eea
The time-dependence of the effective mass accounts for the interaction of the scalar field $\zeta$ and graviton field $h_{ij}$ with
the gravitational background respectively. 

After decomposing $v$ and $u_{\gamma}$ in terms of the Fourier modes and varying the action (\ref{sd2}) and (\ref{sd2t}), 
we get the following equation of motions for the scalar and tensor modes respectively:
\bea
\label{msa}v^{''}_{\bf k}+w^{2}_{k;\zeta}(\eta)v_{\bf k}&=&0,\\
\label{msat}u^{''}_{\bf k}+w^{2}_{k;h}(\eta)u_{\bf k}&=&0.
\eea
which is commonly known as the Mukhanov-Sasaki equation~\footnote{In general the Mukhanov-Sasaki equation is hard to solve
 analytically since in general $z(\eta)$ depends on the
background dynamics. For a given inflationary background, one may, of course, solve this equation
numerically. However, for the simplicity, we will discuss approximate
analytical solutions: in the pure de Sitter limit, as well as in the slow-roll expansion of quasi-de Sitter space.
}~\footnote{In Eq~(\ref{msat}) the helicity index $\gamma$ is summed over in the Fourier modes for the tensor contribution $u_{\bf k}$.}.
 Here also the effective frequency of the harmonic oscillator $w_{k}(\eta)$ can be expressed as:
\bea
 \label{eqeff1} w^{2}_{k;\zeta}&=&\left(k^{2}-\frac{z^{''}}{z}\right),\\
\label{eqeff2} w^{2}_{k;h}&=&\left(k^{2}-\frac{a^{''}}{a}\right),
\eea

which depends only on the magnitude of the momentum vector, $k\equiv |{\bf k}|$. 
The general solution of
Eq~(\ref{msa}) and Eq~(\ref{msat}) after quantization can be written as:
\bea
\label{msa1}v_{\bf k}&\equiv&a^{-}_{\bf k}v_{k}(\eta)+a^{+}_{-\bf k}v^{*}_{k}(\eta),\\
\label{msa1t}u_{\bf k}&\equiv&a^{-}_{\bf k}u_{k}(\eta)+a^{+}_{-\bf k}u^{*}_{k}(\eta).
\eea
Here, $v_{k}(\eta)$ and its complex conjugate $v^{*}_{k}(\eta)$ are two linearly independent solutions of Eq~(\ref{msa}).
Also the creation and annihilation operators $a^{+}_{-\bf k}$ and $a^{-}_{\bf k}$ satisfy the canonical commutation relations.
Now the Wronskian of the mode functions for scalar and tensor modes are given by:
\bea
\label{msa2}W_{\zeta}[v_{k},v^{*}_{k}]&\equiv & v^{'}_{k}v^{*}_{k}-v_{k}v^{*'}_{k}=2i~{\rm Im}(v^{'}_{k}v^{*}_{k}),\\
\label{msa2t}W_{h}[u_{k},u^{*}_{k}]&\equiv & u^{'}_{k}u^{*}_{k}-u_{k}u^{*'}_{k}=2i~{\rm Im}(u^{'}_{k}u^{*}_{k}),
\eea
which is time independent and by rescaling the mode functions via the scale transformation $v_{k} \rightarrow \lambda v_{k}$, $u_{k}\rightarrow \lambda u_{k}$
(giving $W_{\zeta}[v_{k},v^{*}_{k}] \rightarrow |\lambda|^{2}W_{\zeta}[v_{k},v^{*}_{k}]$ and $W_{h}[u_{k},u^{*}_{k}] \rightarrow |\lambda|^{2}W_{h}[u_{k},u^{*}_{k}]$) one can always
normalize $v_{k}$ and $u_{k}$ such that:
\bea
W_{\zeta}[v_{k},v^{*}_{k}]&\equiv&-i,\\
W_{h}[u_{k},u^{*}_{k}]&\equiv&-i.
\eea
We must choose a vacuum state for the fluctuations, \be a_{\bf k}|0\rangle =0\ee where $|0\rangle $ is the time dependent vacuum, commonly
 known as {\it Bunch-Davies} vacuum and quite different from the usual {\it Minkowski} vacuum. This corresponds to specifying an additional
 boundary conditions for $v_{k}$. Now when all comoving scales were far inside the Hubble horizon at the far past, $\eta\rightarrow -\infty$
or $|k\eta|>>1$ or $k>>aH$ fixes the solution as:
\bea
\label{msa3}\lim_{\eta\rightarrow -\infty}(v_{k},u_{k})=\frac{e^{-ik\eta}}{\sqrt{2k}}.
\eea
Further we will concentrate on two limiting situation to get the closed form of the analytical solution of $v_{k}(\eta)$ and $u_{k}(\eta)$:
\begin{itemize}
 \item \underline{\bf A. Pure de Sitter limit:} In this case using \be a = −(H\eta)^{-1},\ee the effective frequency reduces to
 \be w^{2}_{ k;\zeta}(\eta)=w^{2}_{ k;h}(\eta)=\left(k^{2}-\frac{2}{\eta^{2}}\right).\ee Consequently the exact solution of the Eq~(\ref{msa}) and Eq~(\ref{msat}) can be written as:
\bea
\label{sol1a}v_{k}(\eta)&=&\alpha\frac{e^{-ik\eta}}{\sqrt{2k}}\left(1-\frac{i}{k\eta}\right)+\beta\frac{e^{ik\eta}}{\sqrt{2k}}\left(1+\frac{i}{k\eta}\right),\\
\label{sol1at}u_{k}(\eta)&=&\mu\frac{e^{-ik\eta}}{\sqrt{2k}}\left(1-\frac{i}{k\eta}\right)+\sigma\frac{e^{ik\eta}}{\sqrt{2k}}\left(1+\frac{i}{k\eta}\right).
\eea
The above mentioned boundary condition stated in Eq~(\ref{msa3}) fixes the co-efficients $\alpha = 1, \beta = 0$, $\mu=1, \sigma=0$ and leads to the unique
solution of the Bunch-Davies mode functions at the subhorizon scale as:
\bea
\label{sol1a}v_{k}(\eta)&=&u_{k}(\eta)=\frac{e^{-ik\eta}}{\sqrt{2k}}\left(1-\frac{i}{k\eta}\right).
\eea
This determines the future evolution of the mode including its superhorizon dynamics at $k<<aH$ or $|k\eta|<<1$ or $\eta\rightarrow 0$ as:
\bea
\label{sol1anw}v_{k}(\eta)&=&u_{k}(\eta)=\frac{1}{i\sqrt{2}}\frac{1}{k^{3/2}\eta}.
\eea
\item \underline{\bf B. Quasi de Sitter limit:} In this case the effective frequency reduces to
 \be w^{2}_{ k;\zeta}(\eta)=w^{2}_{ k;h}(\eta)=\left(k^{2}-\frac{z^{''}}{z}\right),\ee where \be \frac{z^{''}}{z}\equiv\frac{ a^{''}}{a}=\frac{\left(\nu^{2}-\frac{1}{4}\right)}{\eta^{2}}\ee
 with \be \nu=\frac{3}{2}+\epsilon+\frac{\eta}{2},\ee in the first
 order of slow-roll approximation. Consequently the exact solution of the Eq~(\ref{msa}) and Eq~(\ref{msat}) can be written as:
\bea
\label{sol1b}v_{k}(\eta)&=&\sqrt{-k\eta}\left[\alpha H^{(1)}_{\nu}(-k\eta)+\beta H^{(2)}_{\nu}(-k\eta)\right],\\
\label{sol1bt}u_{k}(\eta)&=&\sqrt{-k\eta}\left[\mu H^{(1)}_{\nu}(-k\eta)+\sigma H^{(2)}_{\nu}(-k\eta)\right],
\eea
To impose the Bunch-Davies boundary condition at early times we have used:
\bea
\label{sol1c}\lim_{k\eta\rightarrow -\infty}H^{(1,2)}_{\nu}(-k\eta)&=&\sqrt{\frac{2}{\pi}}\frac{1}{\sqrt{-k\eta}}e^{\pm ik\eta}
e^{\pm\frac{i\pi}{2}\left(\nu+\frac{1}{2}\right)}.
\eea
As a result we get \be \alpha=\mu=\sqrt{\frac{\pi}{2}}\ee and \be \beta=\sigma=0\ee and this leads to the unique
solution of the Bunch-Davies mode functions  at the subhorizon scale as:
\bea
\label{sol1a}v_{k}(\eta)&=&u_{k}(\eta)=\sqrt{\frac{\pi}{2}}e^{\pm\frac{i\pi}{2}\left(\nu+\frac{1}{2}\right)}\sqrt{-k\eta}H^{(1)}_{\nu}(-k\eta).
\eea
This determines the future evolution of the mode including its superhorizon dynamics at $k<<aH$ or $|k\eta|<<1$ or $\eta\rightarrow 0$ as:
\bea
\label{sol1anw}v_{k}(\eta)&=&u_{k}(\eta)=\frac{i}{\sqrt{\pi}}e^{\pm\frac{i\pi}{2}\left(\nu+\frac{1}{2}\right)}\left(-\frac{k\eta}{2}\right)^{\frac{1}{2}-\nu}\Gamma(\nu).
\eea
\end{itemize}

\subsubsection{Inflationary observables and EFT}

 Here we will start with the two-point correlation functions computed from scalar and tensor contribution of inflationary EFT using {\it Bunch-Davies} vacuum.
For both the cases the final results are proportional to the power spectrum at any arbitrary momentum scale $k$. 
In the in-in picture (discussed in details in section \ref{ng}) the two-point correlator
can be written as~\footnote{Here to compute the two-point correlator for slow-roll limit we use expression for the comoving curvature perturbation and tensor perturbation 
in terms of time delay during inflation as,
 \be \zeta=H\frac{\delta\phi}{\dot{\phi}}\equiv-H\delta t\ee and \be h=\frac{2}{M_{p}}\psi\ee in a spatially flat gauge \cite{Baumann:2009ds}.}~
\footnote{In Eq~(\ref{pt1}) an extra 2 factor appears due to the appearance of two orthogonal polarization basis for the spin-2 helicity degrees of freedom \cite{Baumann:2009ds}.}:
\bea
 \label{sc2}\langle \zeta_{\bf k}\zeta_{\bf k^{'}}\rangle =\left(\frac{H}{\dot{\phi}}\right)^{2}\langle \delta\phi_{\bf k}
 \delta\phi_{\bf k^{'}}\rangle=\frac{1}{z^{2}M^{2}_{p}}\langle v_{\bf k}
 v_{\bf k^{'}}\rangle &=&(2\pi)^{3}\delta^{3}({\bf k}+{\bf k^{'}})\frac{1}{z^{2}M^{2}_{p}}P_{v}(k) \nonumber \\ &=&(2\pi)^{3}\delta^{3}({\bf k}+{\bf k^{'}})\frac{2\pi^{2}}{k^{3}}\left(\frac{H}{\dot{\phi}}\right)^{2}\left(\frac{H}{2\pi}\right)^{2}\nonumber,\\
&=&(2\pi)^{3}\delta^{3}({\bf k}+{\bf k^{'}})\frac{2\pi^{2}}{k^{3}}P_{s}(k),\\
 \label{sc2x}\langle h_{\bf k}h_{\bf k^{'}}\rangle =\frac{4}{M^{2}_{p}}\langle \delta\psi_{\bf k}
 \delta\psi_{\bf k^{'}}\rangle =\frac{2}{a^{2}M^{2}_{p}}\langle u_{\bf k}
 u_{\bf k^{'}}\rangle 
&=&(2\pi)^{3}\delta^{3}({\bf k}+{\bf k^{'}})\frac{2}{a^{2}M^{2}_{p}}P_{u}(k)\nonumber\\ &=&(2\pi)^{3}\delta^{3}({\bf k}+{\bf k^{'}})\frac{2\pi^{2}}{k^{3}}\frac{4}{M^{2}_{p}}\left(\frac{H}{2\pi}\right)^{2}\nonumber,\\
&=&(2\pi)^{3}\delta^{3}({\bf k}+{\bf k^{'}})\frac{2\pi^{2}}{k^{3}}P_{t}(k)
\eea
where the primordial power spectrum for scalar and tensor modes and tensor-to-scalar ratio at any arbitrary momentum scale $k$ 
can be written within the slow-roll regime as: 
\bea
\label{ps1} P_{s}(k)\equiv \frac{k^{3}P_{\zeta}(k)}{2\pi^2}&=&\frac{k^{3}|v_{k}|^{2}}{2\pi^{2}z^{2}M^{2}_{p}}=\left(\frac{H}{\dot{\phi}}\right)^{2}
\left(\frac{H}{2\pi}\right)^{2}\sim \frac{V}{24\pi^2 \epsilon_{V}(k) M^{4}_p},\nonumber\\
&=& P_{s}(k_{*})\left(\frac{k}{k_{*}}\right)^{n_{s}(k_{*})-1+\frac{\alpha_{s}(k_{*})}{2!}
\ln \left(\frac{k}{k_{*}}\right)+\frac{\kappa_{s}(k_{*})}{3!}\ln^{2}\left(\frac{k}{k_{*}}\right)+\cdots},\\
\label{pt1} P_{t}(k)\equiv 2\times\frac{k^{3}P_{h}(k)}{2\pi^2}&=&\frac{2k^{3}|u_{k}|^{2}}{\pi^{2}a^{2}M^{2}_{p}}=\frac{8}{M^{2}_{p}}\left(\frac{H}{2\pi}\right)^{2}
\sim \frac{2V}{3\pi^2 M^{4}_p},\nonumber\\
&=& P_{t}(k_{*})\left(\frac{k}{k_{*}}\right)^{n_{t}(k_{*})+\frac{\alpha_{t}(k_{*})}{2!}
\ln \left(\frac{k}{k_{*}}\right)+\frac{\kappa_{t}(k_{*})}{3!}\ln^{2}\left(\frac{k}{k_{*}}\right)+\cdots},\\
\label{rrt1} r(k)&\equiv& \frac{P_{t}(k)}{P_{s}(k)}=16\epsilon(k)\sim 16\epsilon_{V}(k)\nonumber\\
&=& r(k_{*})\left(\frac{k}{k_{*}}\right)^{n_{t}(k_{*})-n_{s}(k_{*})+1+\frac{\alpha_{t}(k_{*})-\alpha_{s}(k_{*})}{2!}\ln \left(\frac{k}{k_{*}}\right)+\frac{\kappa_{t}(k_{*})-\kappa_{s}(k_{*})}{3!}\ln^{2}\left(\frac{k}{k_{*}}\right)+\cdots}
\eea
where in CMB experiments, $k_{*}$ is chosen at the scale at which the modes
cross the horizon, $k_{*}=aH$. power spectrum for scalar and tensor modes and tensor-to-scalar ratio are given by \cite{infla,liddle}:
\bea
 P_{s}(k_{*})&=&\frac{V_{*}}{24\pi^2 \epsilon_{V}(k_{*}) M^{2}_p},\\
P_{t}(k_{*})&=&\frac{2V_{*}}{3\pi^2 M^{4}_p},\\
r(k_{*})&=&\frac{P_{t}(k_{*})}{P_{s}(k_{*})}\sim 16 \epsilon_{V}(k_{*})
\eea

and $k_{*}$ is the momentum pivot or scale of normalization of the power spectrum around which
the logarithm of the power spectrum as well as the slow-roll parameters are expanded in Taylor series. In both of the Eq~(\ref{ps1}) and Eq~(\ref{pt1}) $n_{s},n_{t}$ stand for spectral tilt, $\alpha_{s},\alpha_{t}$ represent running of the tilt
and $\kappa_{s},\kappa_{t}$ signifies the running of the running of the spectral tilt for the scalar and tensor modes respectively. 

Further the spectral tilt, running of the tilt and running of the running of spectral tilt can be expressed
 at any arbitrary momentum scale within the slow-roll regime as \cite{Choudhury:2013woa}: 
\bea
\label{ns1} n_{s}(k)-1&\equiv& \frac{d\ln P_{s}(k)}{d\ln k}\nonumber \\
&=& n_{s}(k_{*})-1+\alpha_{s}(k_{*})\ln\left(\frac{k}{k_{*}}\right)
 +\frac{\kappa_{s}(k_{*})}{2}\ln^{2}\left(\frac{k}{k_{*}}\right)+\cdots\,\\
\label{nt1} n_{t}(k)&\equiv& \frac{d\ln P_{t}(k)}{d\ln k}=n_{t}(k_{*})+\alpha_{t}(k_{*})\ln\left(\frac{k}{k_{*}}\right)
 +\frac{\kappa_{t}(k_{*})}{2}\ln^{2}\left(\frac{k}{k_{*}}\right)+\cdots\,\\
\label{asat1} \alpha_{s,t}(k)&\equiv&\frac{dn_{s,t}(k)}{d\ln k}=\alpha_{s,t}(k_{*})
+\kappa_{s,t}(k_{*})\ln\left(\frac{k}{k_{*}}\right)+\cdots\,\\
\label{ksat1} \kappa_{s,t}(k)&\equiv&\frac{d\alpha_{s,t}(k)}{d\ln k}\approx\kappa_{s,t}(k_{*})+\cdots\,
\eea

where at the scale $k_{*}$ the spectral tilt, running of the tilt, running of the running of spectral tilt for scalar and tensor modes and consistency relation for tensor-to scalar ratio 
are given by \cite{infla,liddle}:
\bea
 n_{s}(k_{*})-1&=&\left[\frac{d\ln P_{s}(k)}{d\ln k}\right]_{*}=2\eta_{V}(k_{*})-6\epsilon_{V}(k_{*}),\\
 n_{t}(k_{*})&=&\left[\frac{d\ln P_{t}(k)}{d\ln k}\right]_{*}=-2\epsilon_{V}(k_{*}),\\
 \alpha_{s}(k_{*})&=&\left[\frac{dn_{s}}{d\ln k}\right]_{*}=16\eta_{V}(k_{*})\epsilon_{V}(k_{*})-24\epsilon^{2}_{V}(k_{*})-2\xi^{2}_{V}(k_{*}),\\
 \alpha_{t}(k_{*})&=&\left[\frac{dn_{t}}{d\ln k}\right]_{*}=4\eta_{V}(k_{*})\epsilon_{V}(k_{*})-8\epsilon^{2}_{V}(k_{*}),\\
 \kappa_{s}(k_{*})&=&\left[\frac{d^{2}n_{s}}{d\ln k^{2}}\right]_{*}=192\epsilon^{2}_{V}(k_{*})\eta_{V}(k_{*})-192\epsilon^{3}_{V}(k_{*})+2\sigma^{3}_{V}(k_{*})
-24\epsilon_{V}(k_{*})\xi^{2}_{V}(k_{*})\nonumber \\ &&~~~~~~~~~~~~~~~~~~~~~~~~~~~~~~+2\eta_{V}(k_{*})\xi^{2}_{V}(k_{*})-32\eta^{2}_{V}(k_{*})\epsilon_{V}(k_{*}),\\
\kappa_{t}(k_{*})&=&\left[\frac{d^{2}n_{t}}{d\ln k^{2}}\right]_{*}=56\eta_{V}(k_{*})\epsilon^{2}_{V}(k_{*})-64\epsilon^{3}_{V}(k_{*})\nonumber\\
&&~~~~~~~~~~~~~~~~~~~~~~~~~~-8\eta^{2}_{V}(k_{*})\epsilon_{V}(k_{*})-4\epsilon_{V}(k_{*})\xi^{2}_{V}(k_{*}),\\
r(k_{*})&=&-8n_{t}(k_{*}).
\eea
\begin{table*}
\centering
\small\begin{tabular}{|c|c|c|c|c|}
\hline
\hline
Sr. & {\bf  Inflationary}& {\bf WP} &  {\bf  PLANCK+WP} & {\bf  PLANCK+WP+BICEP2}\\
No. &{\bf  observables} &  \cite{WMAP}& \cite{Ade:2013zuv,Ade:2013uln} &  \cite{Ade:2014xna}\\
\hline\hline\hline
{\bf  1} & $\ln(10^{10}P_{S})$ & $3.204^{+0.328}_{-0.328}$ &  $3.089^{+0.024}_{-0.027}$ & $3.089^{+0.024}_{-0.027}$
\\
{\bf 2} & $n_{S}$ & $0.9608\pm 0.008$ & $0.9603 \pm 0.0073$  & $0.9600 \pm 0.0071$
\\
{\bf  3} &$\alpha_{S}$ & $-0.023\pm0.011$ & $-0.013\pm 0.009$& $-0.022\pm 0.010$
\\
{\bf  4} & $\kappa_{S}$ & ? & $0.020^{+0.016}_{-0.015}$ & $0.020^{+0.016}_{-0.015}$
\\
 {\bf  5} &  $r$ & $<0.36$ & $<0.12$ & $0.2^{+0.07}_{-0.05}$\\
 &   &  &  &${\bf (r=0~ruled~out~at~7\sigma)}$\\
 {\bf  6} & $n_{T}$ & ? & ? & $1.36\pm 0.83~{\bf (Blue)}$
\\
 &  & $>-0.048~{\bf (Red)}$ & ? & $>-0.76~{\bf (Red)}$
\\
&  &  &  &${\bf  (n_{T}=0~ruled~out~at~3\sigma)}$
\\
\hline
\hline
\end{tabular}
\caption{Present observational constraint for various inflationary observables.}\label{tabc1}
\vspace{.4cm}
\end{table*}

The present observational status for the inflationary paradigm is shown in Table(\ref{tabc1}) where the constraints are quoted from 
{\bf WMAP9 (WP)} \cite{WMAP}, {\bf Planck+ WP} \cite{Ade:2013zuv} and {\bf Planck+WP}\\{\bf+BICEP2} \cite{Ade:2014xna} dataset. 
\subsection{Reheating after inflation}

\begin{figure}[t]
\centering
\includegraphics[width=12.3cm,height=6cm]{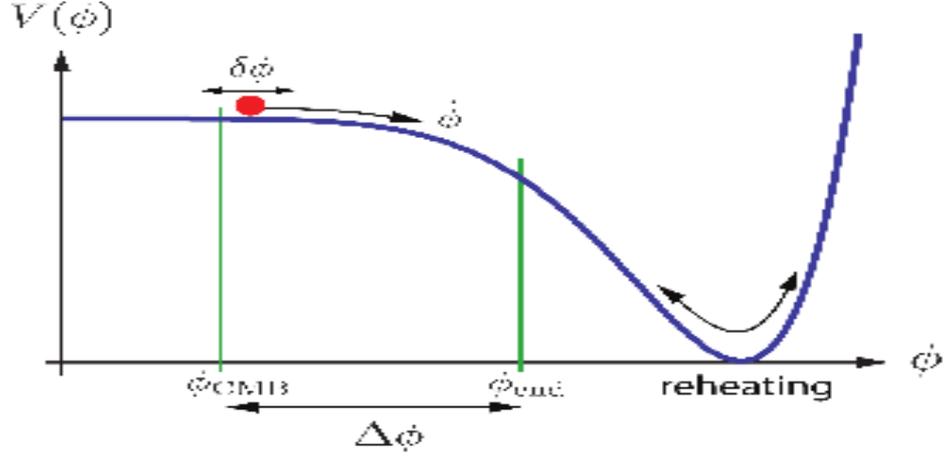}
\caption{\footnotesize Schematic diagram of reheating phenomena which starts just after the end of inflation \cite{Baumann:2009ds}.
}
\label{fig2z}
\end{figure}

\begin{figure*}[htb]
\centering
\subfigure[Visual representations of triangles forming the primordial bispectrum.]{
    \includegraphics[width=15cm,height=5cm] {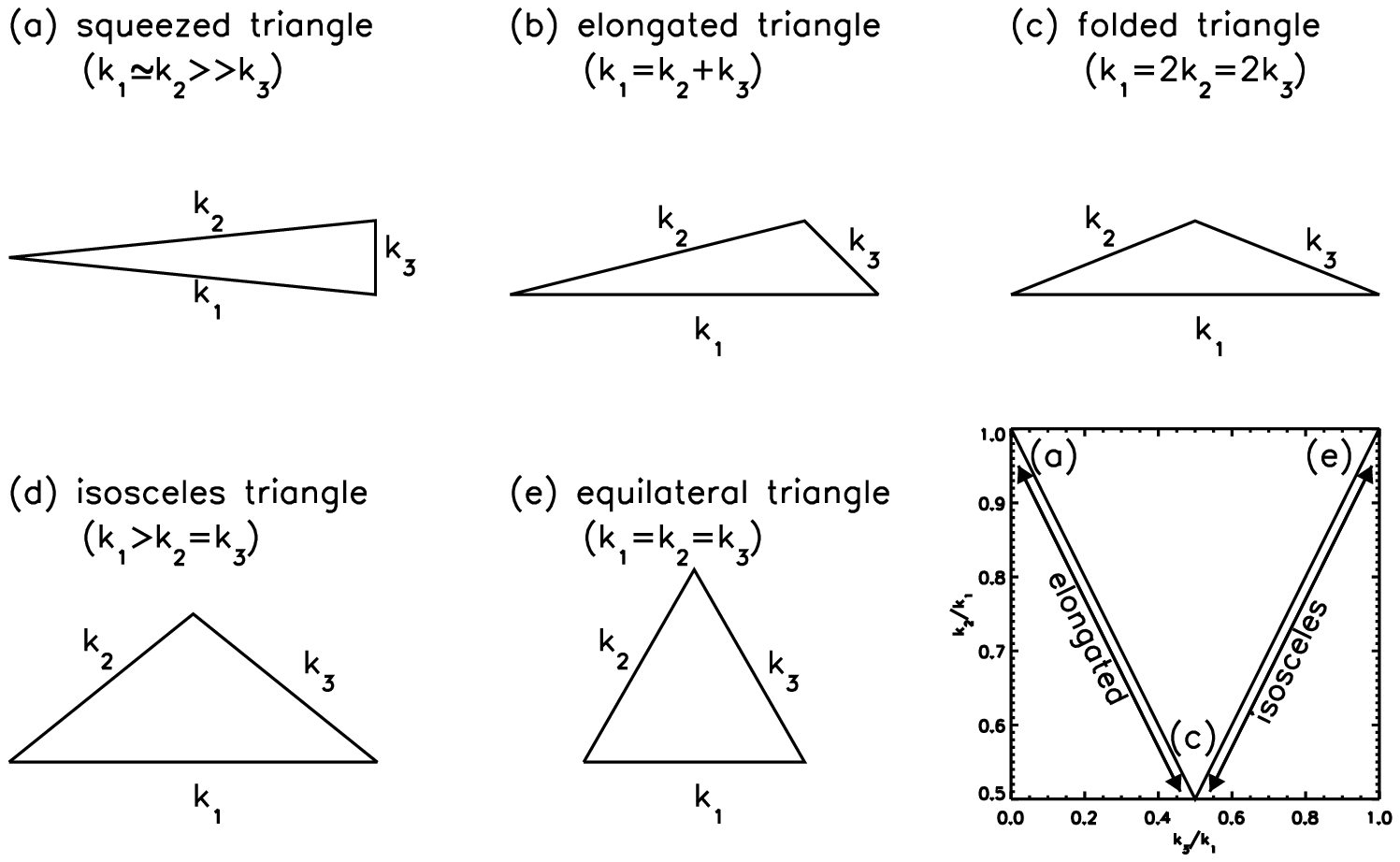}
    \label{fddd1}
}
\subfigure[Shapes of the primordial bispectra.]{
    \includegraphics[width=15cm,height=5cm] {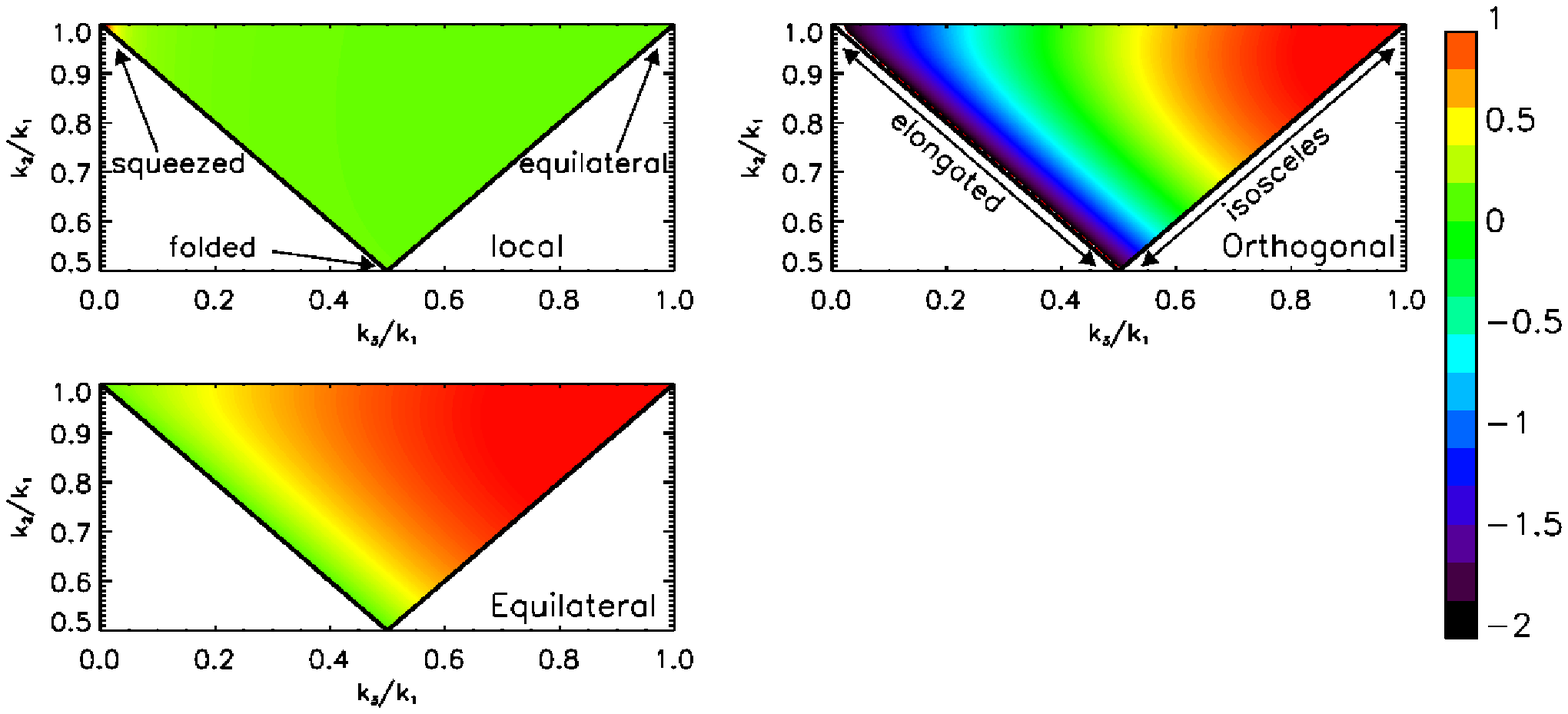}
    \label{fddd2}
}
\caption[Optional caption for list of figures]{\subref{fddd1} 
 Visual representations of triangles forming the primordial bispectrum,
with various combinations of wave numbers satisfying $k_{3}\leq k_{2}\leq k_{1}$ \cite{Komatsu:2010hc}
 and \subref{fddd2} shapes of the primordial bispectra in which we show the normalized
amplitude of ${\cal S} (k_{1} , k_{2} , k_{3} )(k_{2} /k_{1} )^2 (k_{3} /k_{1})^2$ as a function of $k_{2}/k_{1}$ and $k_{3}/k_{1}$ for a given
$k_{1}$, with a condition that $k_{3}\leq k_{2}\leq k_{1}$ is satisfied \cite{Komatsu:2010hc}. 

}
\label{fvdd1}
\end{figure*}

\begin{figure*}[htb]
\centering
\subfigure[Flat potential near the saddle point.
]{
    \includegraphics[width=7.2cm,height=5.9cm] {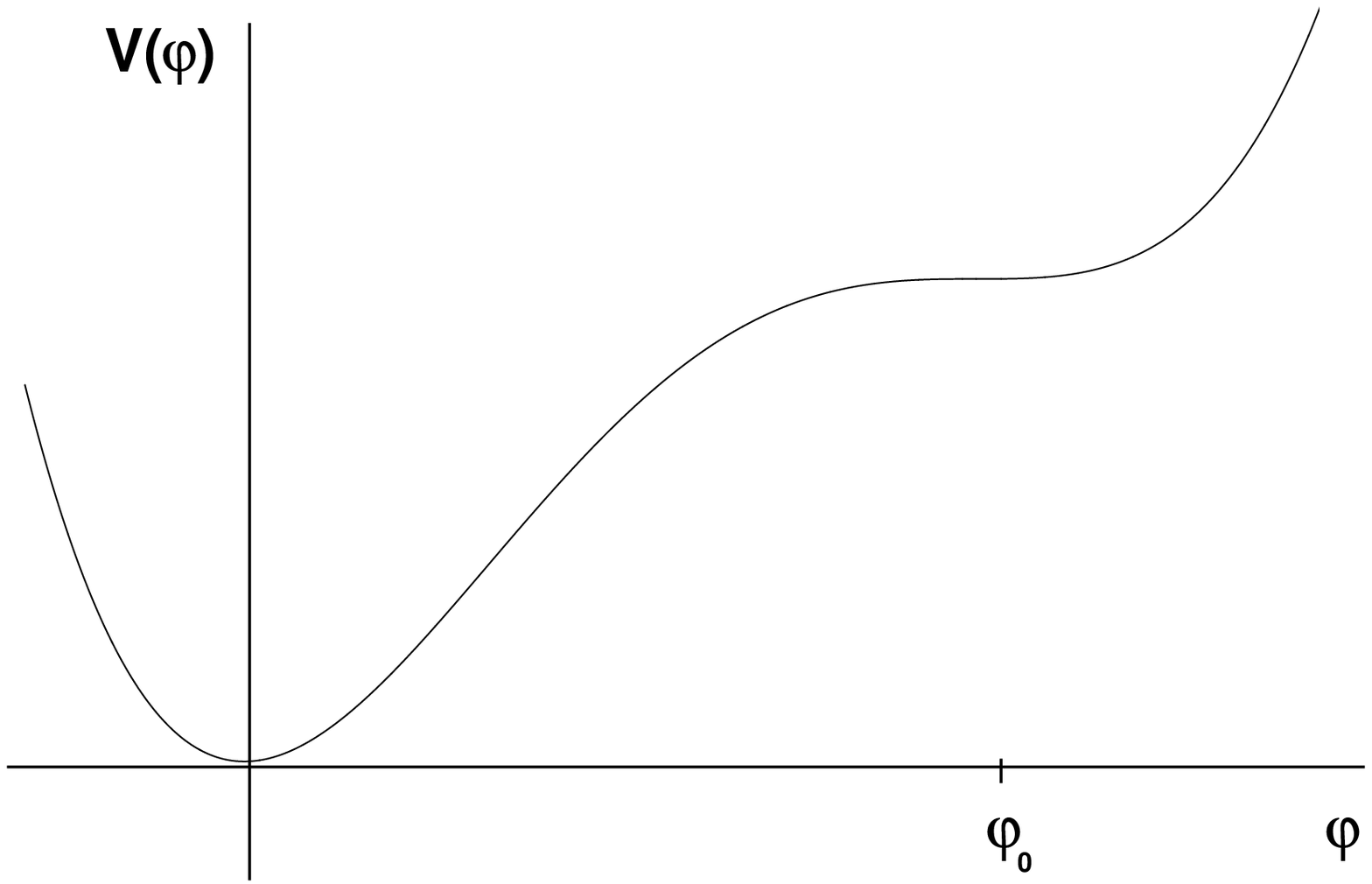}
    \label{vf1}
}
\subfigure[Flat potential near the inflection point.]{
    \includegraphics[width=7.2cm,height=5.9cm] {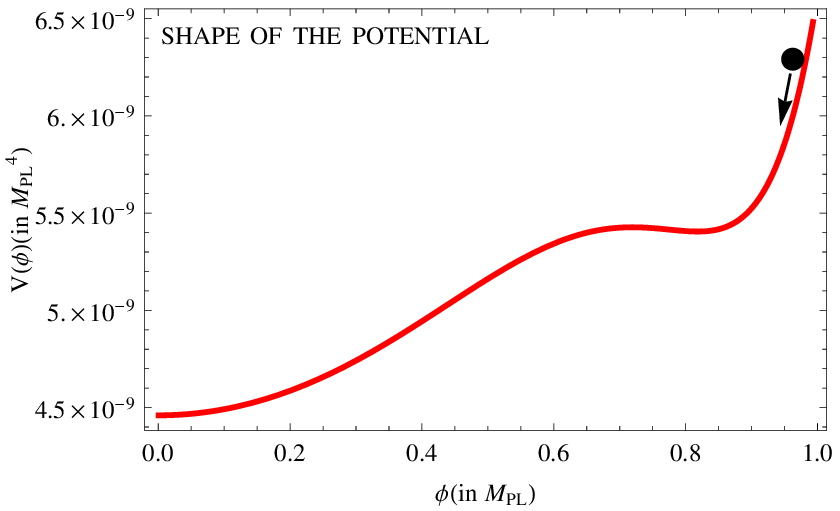}
    \label{vf2}
}
\caption[Optional caption for list of figures]{ Visual representations of flat potential \ref{vf1} 
 near the saddle point \cite{Allahverdi:2006cx}
 and \ref{vf2} near the inflection point \cite{Choudhury:2013jya}. 

}
\label{fvd1}
\end{figure*}

Reheating at the end of the period of accelerated expansion is a significant
component of inflationary paradigm~\footnote{See figure~(\ref{fig2z}) for the details.}. Without reheating phenomena, inflation cannot able to generate matter in
universe. Reheating appears through coupling of the inflaton
degrees of freedom, $\phi$, the scalar field generating the accelerated expansion,
to Standard Model (SM) matter. Such couplings must be present at least in
gravitational interactions. However, in many inflationary models, instead of such gravitational interactions direct
couplings through the matter sector play crucial role.
Reheating mechanism was initially proposed using first order perturbation
theory within EFT prescription and further analyzed in terms of the decay of an inflaton degrees of freedom into
SM particle constituents \cite{Bassett:2005xm,Allahverdi:2010xz,Mazumdar:2010sa}.

\subsubsection{Initial condition for reheating}

As discussed, inflationary paradigm generically requires a scalar field.
A scalar field is postulated to exist in the SM too, where
the Higgs field used to give elementary fermions their masses via spontaneous symmetry breaking mechanism. To serve
the scalar degrees of freedom as a Higgs field, its potential energy must have a minimum at a non-trivial
field value, known as vacuum expectation value (VEV). To demonstrate this let us start with standard Higgs potential:
\be \label{potv1}
V(\phi)=\frac{\lambda}{4}  (\phi^2 - v^2)^2
\ee
where $v$ represents the VEV of $\phi$. To fix the initial condition for reheating it is assumed that
at high temperatures the symmetry is restored by twofold components- 
\begin{enumerate}
 \item finite temperature
effects,
\item  the Higgs field value at $\phi = 0$.
\end{enumerate}
During reheating one can note down the following characteristic features:
\begin{itemize}
 \item When
the temperature falls below a critical value $T_c$, the scalar field $\phi$ ceases
to be trapped and starts to roll towards one of the lowest energy
states $\phi = \pm v$.
\item The SM Higgs
must have a self coupling constant $\lambda$ and its numerical value cannot be sufficiently small to achieve 
slow rolling of $\phi$, which is required to obtain enough
inflation.
\item Inflation takes place during the period
when $\phi$ is undergoing the symmetry-breaking phase transition
and slowly rolling towards $\phi = \pm v$.
\item Here the model of reheating was based
on scalar field dynamics obtained by replacing the potential (\ref{potv1})
by a symmetry breaking potential
of Coleman-Weinberg form, where the mass term at the
field origin is set to zero and symmetry breaking is obtained through
quantum corrections to the effective potential.
\end{itemize}
 Large-field inflation is an alternative
prescription where inflation is triggered by a period of slow-rolling of $\phi$. This situation can be described using the 
 monomial potential:
\be \label{pot2}
V(\phi)=\frac{1}{2} m^2 \phi^2,
\ee
where $m$ is the mass of $\phi$. One can avoid this situation by adding a second scalar field $\chi$ to the potential sector of the theory
responsible for inflation and to invoke a hybrid potential of the form
\be \label{hybrid}
V(\phi, \chi)=\frac{1}{2} m^2 \phi^2 + \frac{g^2}{2} \chi^2 \phi^2
+ \frac{\lambda}{4}  \bigl( \chi^2 - \theta^2 \bigr)^2,
\ee
where $g$ and $\lambda$ are dimensionless coupling constants
and $\theta$ is the VEV of $\chi$.
For large values of $|\phi|$, the potential in $\chi$
direction has a minimum
at $\chi = 0$, whereas for small values of $|\phi|<M_p$, $\chi = 0$
becomes an unstable point.
 

\subsubsection{Perturbative decay of inflaton}\label{section_reheating}

Reheating takes place at the end of inflation when the energy density is stored
overwhelmingly in the oscillations of $\phi$. We assume that the inflaton $\phi$
is coupled to another scalar field $\chi$ or fermionic field $\psi$. 
Taking the interaction Lagrangian to be
\bea \label{toy}
{\cal L}^{(1)}_{\rm int} &=& - g \sigma \phi \chi^2 ,\\
{\cal L}^{(2)}_{\rm int} &=& - h  \phi \bar{\psi}\psi ,
\eea
where $g$ and $h$ are dimensionless coupling constants and $\sigma$
is a mass scale. When the mass of the inflaton is much larger
than those of $\chi$ and $\psi$ ($m>>m_{\chi}, m_{\psi}$), the decay rate are given by:
\bea
\Gamma (\phi\rightarrow \chi \chi)&=&\frac{g^2 \sigma^2}{8 \pi m},\\
\Gamma (\phi\rightarrow \psi \bar{\psi}) &=& \frac{h^2 m}{8 \pi}.
\eea
The energy loss of the inflaton due
to the production of $\chi$ and $\psi$ particles can be taken into account by
adding a damping term to the inflaton equation of motion
which in the case of a homogeneous inflaton field is \cite{Baumann:2009ds}:
\be \label{effeom}
{\ddot \phi} + 3 H {\dot \phi} + \Gamma_{total} {\dot \phi} =  - V^{'}(\phi).
\ee
For small coupling constant, the total interaction rate 
\be \Gamma_{total}=\Gamma \,(\phi\rightarrow \chi \chi)+\Gamma \,(\phi\rightarrow \psi \bar{\psi})\ee
is typically much smaller than the Hubble parameter at the end
of inflation. Thus, at the beginning of the 
oscillations, the energy loss into particles is initially negligible
compared to the energy loss due to the expansion of space.
Once the Hubble expansion rate decreases to a
value comparable to $\Gamma_{total}$, $\chi$ and $\psi$ particle production
becomes effective. In the context of Einstein GR it is the energy density at the time when 
\be H = \Gamma_{total}=\Gamma (\phi\rightarrow \chi \chi)+\Gamma (\phi\rightarrow \psi \bar{\psi})=\sqrt{\frac{\rho}{3M^2_{PL}}}.\ee 
 Assuming all the energy density $\rho$ of
the universe is in the form of relativistic matter with
\be \rho= N^* \pi^2 T^4 /30,\ee where $N^*$ is the effective number of mass-less degrees of freedom ($N^* = 10^2 -10^3$), we obtain the
reheat temperature:
\be
T_R \sim 0.2 \left(\frac{100}{N^*}\right)^{1/4}\left( \Gamma_{total} M_{PL} \right)^{1/2}.
\ee
Since 
\be \Gamma_{total}\propto \sqrt{g},\ee
which is generally very small, perturbative
reheating is slow and produces a reheating temperature, \be T_R <<\sqrt[4]{V_*}.\ee
There are two prime problems with the perturbative approach as described above. First of all, even if the inflaton
decay were perturbative, it is not justified to use the heuristic
equation (\ref{effeom}) since it violates the fluctuation-dissipation
theorem: in systems with dissipation, there are always fluctuations,
and these are missing in (\ref{effeom}). Also the prescribed analysis does
not take into account the coherent nature of the inflaton field. However, the 
matter fields can be assumed to start off in their vacuum state. Thus,
matter fields $\chi$ and $\psi$ must be treated quantum mechanically.

\subsubsection{Preheating}

In this case the energy is rapidly transferred from the inflaton degrees of freedom to the other
bosonic fields interacting with it in the present context. In principle this process occurs very far away from thermal equilibrium and commonly known as preheating. In the context of physics of early universe 
this plays a very significant role. Preheating is in general follows the non-perturbative nature 
 of the particle creation and consequently the theory of preheating is complicated in nature and different from the reheating phenomena as introduced earlier.
We will present the preheating mechanism \cite{Kofman:1997pt,Brax:2010ai} for the simple
toy model with interaction Lagrangian:
\be \label{intLag}
{\cal L}_{\rm int} = - \frac{1}{2} g^2 \chi^2  \phi^2,
\ee
where, the parameter $g$ plays the role of a dimensionless coupling in the present context.
Here the time period of preheating is small compared
to the Hubble expansion time $H^{-1}$. 
The quantum theory of $\chi$ particle production in the external classical
inflaton background begins by expanding the quantum field $\hat{\chi}$ as:
\be
\hat{\chi}(t,\mathbf{x})=\frac{1}{(2\pi)^{3/2}}\int d^3k 
\left(\chi_k^*(t)\hat{a}_{k} e^{i\mathbf{k}.\mathbf{x}}+\chi_k(t)\hat{a}^{\dagger}_{k} e^{-i\mathbf{k}.\mathbf{x}}\right),
\ee
In the present context the mode functions $\chi_k \forall k$ satisfy
the Mathieu equation in Fourier space as:
\be\label{mat}
\chi''_k+\frac{\omega^2 _k}{m^2}\chi_k  =  0,
\ee
where
\be \omega_k =  \sqrt{k^2 + m_{\chi}^2 + g^2 \Phi(t)^2 sin^2 {z}},\ee
where we introduce a dimensionless variable $z = m t$.
Here $\Phi$ is the amplitude of oscillation of $\phi$.
The growth of the mode function
corresponds to particle production, analogous to the situation in case of an external gravitational field. 
Eq~(\ref{mat}) leads to exponential growth, 
\be \chi_k \propto {\rm exp}(\mu_k z),\ee
where $\mu_k$ is called the Floquet exponent. For \be \frac{g^2\Phi^2}{4m^2} \ll 1,\ee
resonance occurs in a narrow instability band around $k = m$.
The resonance is much more efficient if \be \frac{g^2\Phi^2}{4m^2} \gg 1\ee where it occurs in broad bands. In particular, the bands include all
long wavelength modes $k \rightarrow 0$. A condition for particle production is that the WKB approximation for
the evolution of $\chi$ is violated. Here we write,
\be \chi_k  \propto  e^{\pm i\int\omega_kdt},\ee
which is valid as long as the adiabaticity condition
\be \frac{d\omega_k^2}{dt}\leq 2 \omega^3_k\ee
is satisfied. The adiabaticity condition is violated for momenta satisfying,
\be
k^2 \leq \frac{2}{3\sqrt{3}}gm\Phi-m_{\chi}^2.
\ee
For modes with these values of $k$, the adiabaticity condition breaks
down in each oscillation period when $\phi$ is close to zero.

\subsubsection{Phenomenological consequences}

Reheating or preheating lead to non-thermal particle production, as we
have mentioned earlier. In cosmology it is usually assumed that
all particles start out in thermal equilibrium at the beginning of the
Standard Cosmology phase. However, reheating begins with out-of-equilibrium
decay of the inflaton oscillations and decay products may
 not reach full thermal equilibrium immediately \cite{Allahverdi:2011aj,Allahverdi:2007zz,Enqvist:2003qc}. During the transition from
 inflation to the Standard Cosmology various non-thermal processes take place
 and the assumption of thermal equilibrium of all particles clearly breaks down.
 In the following, we briefly mention a few applications of non-thermal particle production.

\begin{itemize}
 \item {\bf 1. Baryogenesis and Leptogenesis:}

The first application is leptogenesis \cite{Fong:2013wr,Davidson:2008bu} and baryogenesis \cite{Cline:2006ts,Morrissey:2012db}. One of the
several possible mechanisms to explain the observed asymmetry
between baryons and antibaryons is to make use of out-of-equilibrium
decay of superheavy Higgs and gauge particles. If
reheating were purely perturbative, particles as heavy as the inflaton could have been
created either in inflaton decay or from scatterings of inflaton decay products. 
Preheating, however, provides a mechanism to produce a large population
of superheavy scalar particles much heavier than the inflaton. 
Another way to generate observed baryon to entropy ratio is
via leptogenesis, a scenario in which initially an asymmetry
in the lepton number is produced which is then partially converted 
into baryon asymmetry via SM sphalerons. Preheating after inflation is a way to generate the
initial lepton asymmetry. For example, preheating can produce
a large number density of super-massive right handed neutrinos
in a model in which the inflaton couples to these neutrinos $\psi$
via the standard fermionic preheating interaction term:
\be\label{fermionic-coupling}
{\cal L}_{int} \, = -\, h \phi {\bar \psi}{\psi} \, .
\ee
If hybrid inflation occurs at a scale close to the electroweak
scale, then the non-thermal production of particles may
provide the out-of-equilibrium condition that is necessary
in order to achieve electroweak baryogenesis \cite{Morrissey:2012db,Trodden:1998ym}.
\\
\item {\bf 2. Dark matter:}

Another application of non-thermal particle creation during
reheating is to excite dark matter \cite{Olive:2003iq,Garrett:2010hd}. It is usually assumed that
the dark matter particles are thermally
distributed. This assumption is implicit in most
current analyses of the prospects for dark matter
detection in direct and indirect experiments. However,
if the dark matter particles couple to the inflaton,
then non-thermal production of dark matter during
reheating is to be expected. If the dark matter
particles have sufficiently strong interactions which
allows them to
thermalize during reheating, then the signatures
of the initial non-thermal distribution will be washed
out. However, if the interactions do not permit
thermalization after inflation, then the predictions
concerning the dark matter distribution will be
quite different. The dark matter abundance which can
be obtained by the preheating channel is very
model-dependent, whereas, direct gravitational
particle production produces dark matter of the
required abundance for particle masses of
\be M_X \sim g^{1/2} 10^{15}~{\rm GeV}.\ee
\\
\item{\bf 3. Moduli and Gravitino Production}

Preheating could also produce 
unwanted particles. An example are particles with gravitationally 
suppressed couplings and weak scale masses that
arise in many theories beyond the SM. Overproduction of these particles 
could overclose the universe, if they are stable, or ruin the success of Big Bang 
nucleosynthesis (BBN) in the case of unstable relics. Here we consider moduli 
and gravitino production during preheating. Moduli  (bosonic modulus, $\chi$, and fermionic modulus, 
$\psi$) are typically coupled to the inflaton via
non-renormalizable UV scale suppressed such as:
\be
{\cal{ L}}_{int} \, \sim \, \phi^4 \frac{\chi}{\Lambda_{UV}}~~~~
({\rm bosonic}),~~~~~~~~~
{\cal{L}}_{int} \sim  \frac{\phi^2}{\Lambda_{UV}} {\bar{\psi}}\psi ~~~~({\rm fermionic})\,.
\ee
where $\Lambda_{UV}\sim M_{PL}$ is the UV cut-off scale of the effective theory.
 It was shown that moduli
field can be parametrically amplified.
Another important example is the gravitino, the spin 3/2 partner of the graviton.
Gravitinos are produced thermally from scatterings of light particles in the thermal bath.
The number density of gravitinos thus produced 
can be obtained by solving the Boltzmann equation:
\be
\dot n_X + 3 H n_X \, \simeq \, \langle \sigma v \rangle n_l^2 \, ,
\ee
where $n_X$ is the number density of the gravitinos,
$\sigma$ is the production cross section which scales
as $M_{PL}^{-2}$, and $v \sim c$ is the relative velocity of scatterers $l$ 
whose number density is $n_l$. The
resulting abundance is found to be:
\be
\Omega_{X}=\frac{n_X}{s} \, \sim \, 10^{-2} \frac{T_R}{M_{PL}} \,\sim \, 2\times10^{-3} \left(\frac{100}{N^*}\right)^{1/4}\left( \frac{\Gamma_{total}}{ M_{PL}} \right)^{1/2},
\ee
where $s$ is the entropy density and $T_R$ is the reheat temperature of the universe.  
BBN gives rise to an absolute upper bound \be (n_X/s)< 10^{-12},\ee which in turn leads to an upper 
bound \be T_R < 10^9~{\rm GeV}.\ee
The presence of the oscillating
inflaton field leads to a periodically varying correction
to the effective gravitino mass that results in an
instability in the same way that there is an instability for spin 0 and 1/2 particle
modes. The exact strength of the instability depends
sensitively on the precise SUSY inflationary
model one is considering. 
Gravitino with helicity $\pm 1/2$ component mainly contain the Goldstino 
component- the inflatino (superpartner of the inflaton),
whose interactions are not suppressed by $\Lambda_{UV}$. One would naturally 
expect them to be created in large abundance. However,
in realistic scenarios, where the scale of inflation is much higher than the 
scale of SUSY breaking, e.g.  \be H_{inf}\gg 
{\cal O}(100~{\rm GeV})\ee and the helicity $\pm 1/2$ states that are produced during 
preheating mainly decay in the form of inflatinos along with the inflaton.

\end{itemize}

\subsection{Primordial non-Gaussianity}\label{ng}

The CMB power spectrum analysis reduces the WMAP data from about $10^{6}$ pixels to $10^{3}$
multipole moments. In principle, there
can be a wealth of information that is contained in deviations from the perfectly Gaussian
distribution. So far CMB experiments haven't had the sensitivity to extract this information from
the data. However, searches are going on with improved precision which will presumably provide accurate measurements of higher-order CMB correlations.
 So, it is instructive to study non-Gaussian features which will also constrain various classes of inflationary models.
The precise measurements of primordial non-Gaussianity are also a powerful way to exploring the
various hidden aspects of particle physics, which is to determine the action (i.e. the fields, symmetries
and couplings) as a function of energy scale. The prime sources for the non-Gaussianity in CMB are~\footnote{Here we mostly
 concerned about primordial non-Gaussianity and second order non-Gaussianity.}:
\begin{itemize}
\item {\bf 1. Primordial non-Gaussianity:}
Non-Gaussianity in the primordial curvature perturbation $\zeta$ produced in the very early
universe by inflation or in an alternative prescription.
\item {\bf 2. Second-order non-Gaussianity:}
Non-Gaussianity arising from non-linearities in the transfer function relating $\zeta$ to the
CMB temperature anisotropy $\Delta T$ at recombination.
\item {\bf 3. Secondary non-Gaussianity:}
Non-Gaussianity generated by late time effects after recombination e.g. gravitational lensing.
\item {\bf 4. Foreground non-Gaussianity:}
Non-Gaussianity created by Galactic and extra-Galactic sources.
\end{itemize}

We briefly discuss here about the
two essential tools - the in-in formalism and the $\delta N$ formalism - for computing non-Gaussianities in 
the models of the early universe.

\subsubsection*{A.~~In-In formalism}

The specific computation of $n$-point correlation functions in the field of inflationary cosmology is significantly different in various manner from
the analysis used in the context of flat space quantum field theory applicable to particle physics theory. In the framework of flat space quantum field theory or in particle
physics the prime motivation is to study various effective interactions via the S-matrix which physically describes the transition probability for a state vector $|in\rangle$
in the very far past to some state vector $|out\rangle$ realized in the very far future. This can be technically expressed by the following expression:
\be\label{out}
\langle out | S | in\rangle = \langle out~(+\infty)| in~(-\infty) \rangle
\ee
Here one needs to impose the asymptotic boundary conditions at very early and very late time scales. This is because of the fact that
in Minkowski space all the quantum states are assumed to be non-interacting in the far past and the
far future when the scattering particles are far from the interaction region. 

In the present context the quantum fluctuations by the field content is collectively denoted by the following way: \be \psi=\{\zeta,h^{+},h^{\times}\}\ee
and our here prime objective is to compute the exact $n$-point correlator or more precisely the expectation value of the filed theory operator defined as:
\be A=\prod^{n}_{i=1}\psi_{k_{i}},\ee
and in the present context this can be expressed as, \be \langle A\rangle=\langle in| A | in\rangle.\ee Here the state vector $| in\rangle$ characterizes the structure of the quantum field theoretic 
vacuum of the interacting theory at 
the moment in the far past in FLRW curved space-time. Here in this computation we introduce the interaction picture in which
the leading order time-dependence of the fields is exactly determined by the quadratic Hamiltonian or equivalently by the 
linear equations of motion. Consequently corrections arising from the various effective interactions in the quantum field theory can be treated as a
power series in the interaction Hamiltonian, $H_{int}$ in interaction picture. Consequently the expectation value of the quantum field theoretic operator can be recast in the following form~
\footnote{It is important to note that, in Eq~(\ref{expectation}) the integration
open contour exactly goes from $-\infty(1 - i\epsilon)$ to $t$ in time scale where all the correlation function is evaluated for quantum field theory in curved FLRW background and back
to $-\infty(1 + i\epsilon)$. This will directly imply that the Wick contraction
of quantum field theoretic operators finally lead to real-valued Wightman Green's functions, rather than complex-valued
Feynman Green's functions in the present context of discussion.
}:
\be\begin{array}{llll}\label{expectation}
 \displaystyle \langle A \rangle=\langle 0| \bar{T} \exp\left[i\int^{t}_{-\infty(1-i\epsilon)}
H^{I}_{int}(t^{'})dt^{'}\right]~A^{I}(t)~T \exp\left[-i\int^{t}_{-\infty(1+i\epsilon)}H^{I}_{int}(t^{''})dt^{''}\right]| 0 \rangle
\end{array}\ee
where $T$ and $\bar{T}$ signify the time ordering and anti-time ordering symbolic operations respectively. The superscript $I$ is specifically used for interaction picture in quantum field theory.
The standard $i\epsilon$ prescription in quantum field theory has been used in the present context to effectively turn off the interaction in the very far past time scale 
and finally project the interacting quantum $|in\rangle$ state vector onto the
free vacuum state vector represented by $|0\rangle$. Here we also introduce time evolution operator $U$~\footnote{The interaction Hamiltonian in interaction picture of quantum field theory 
defines the evolution of quantum
states via the well known unitary time evolution operator: \be U(\tau_{2},\tau_{1})=T~\exp\left[-i\int^{\tau_{2}}_{\tau_{1}}H^{I}_{int}(t^{''})dt^{''}\right].\ee
By directly solving the mode function from the {\it Mukhanov-Sasaki equation} one can directly choose the {\it Bunch-Davies} initial condition in the interaction Hamiltonian. In principle one can also 
study the consequences from arbitrary vacuum as well, but {\it Bunch-Davies} vacuum is more consistent with the various observed data sets. } may be used to relate
the interacting vacuum state at arbitrary any time $|\Omega(\tau)\rangle$ to the free Bunch-Davies vacuum $|0\rangle$.
We first
expand $|\Omega(\tau)\rangle$ in its eigenstates of the free Hamiltonian in the following fashion: \be |\Omega\rangle=\sum_{n}|n\rangle \langle n|\Omega(\tau)\rangle\ee and
after time evolution this can be 
expressed as:
\begin{eqnarray}
 |\Omega(\tau_{2})\rangle &=& U(\tau_{2},\tau_{1})|\Omega(\tau_{1})\rangle=|0\rangle \langle 0|\Omega\rangle +\sum_{n\geq 1}e^{iE_{n}(\tau_{2}-\tau_{1})}|n\rangle \langle n|\Omega(\tau_{1})\rangle
\end{eqnarray}
Further in the present context we choose $\tau_{2} = -\infty(1 - i\epsilon)$ which finally projects out all excited quantum states in curved FLRW space-time. Consequently we get
the following relation between the interacting vacuum at the time scale $\tau=\tau_{2}= -\infty(1-i\epsilon)$ and the free Bunch-Davies vacuum state $|0\rangle$ as:
 \be |\Omega(-\infty(1-i\epsilon))\rangle = |0\rangle\langle 0|\Omega\rangle.\ee Using this finally the interacting vacuum at an arbitrary time scale $t$ can 
be expressed as:
\begin{eqnarray}
 |in\rangle\equiv|\Omega(t)\rangle &=&U(t,-\infty(1-i\epsilon))
|\Omega(-\infty(1-i\epsilon))\rangle\nonumber\\&=&T~\exp\left[-i\int^{t}_{-\infty(1-i\epsilon)}H^{I}_{int}(t^{''})dt^{''}\right]|0\rangle \langle 0|\Omega\rangle.
\end{eqnarray}

Using this useful computational tool it can be easily shown that the three and four point correlator of curvature perturbation in momentum space can be written as
~\footnote{In this definition of bispectrum and trispectrum all the contributions of diagrams-connected and disconnected are taken care of.}: 
\begin{eqnarray}
 \label{eq3pt}\langle \zeta_{\bf k_{1}}\zeta_{\bf k_{2}}\zeta_{\bf k_{3}}\rangle &=& (2\pi)^{3} \delta^{3}({\bf k_{1}}+{\bf k_{2}}+{\bf k_{3}})B_{\zeta}(k_{1},k_{2},k_{3}),\\
\label{eq4pt}\langle \zeta_{\bf k_{1}}\zeta_{\bf k_{2}}\zeta_{\bf k_{3}}\zeta_{\bf k_{4}}\rangle &=& (2\pi)^{4} \delta^{3}({\bf k_{1}}+{\bf k_{2}}+{\bf k_{3}}+{\bf k_{4}})
T_{\zeta}(k_{1},k_{2},k_{3},k_{4}). 
\end{eqnarray}
where $B_{\zeta}(k_{1},k_{2},k_{3})$ and $T_{\zeta}(k_{1},k_{2},k_{3},k_{4})$ represent the bispectrum and trispectrum which only depend on the magnitude of the momentum vectors due to isotropy or rotational invariance. 
In case of scale invariant quantum fluctuations, the bispectrum and trispectrum are homogeneous functions of degree $-6$ and $-8$ respectively. Further,
the bispectrum and trispectrum can be expressed in terms of the non-Gaussian observable parameter $f_{NL}, \tau_{NL}, g_{NL}$ and the power spectrum $P_{\zeta}$ as
~\footnote{In Eq~(\ref{eqt}) the symbol $k_{ij}$ is defined as, $k_{ij}:=|{\bf k_{i}}-{\bf k_{j}}|=\sqrt{k^{2}_{i}+k^{2}_{j}-2k_{i}k_{j}\cos \theta_{ij}}$, where $\theta_{ij}$
be the angle between the momentum vectors ${\bf k_{i}}$ and ${\bf k_{j}}$.}:
\begin{eqnarray}
\label{eqb} B_{\zeta}(k_{1},k_{2},k_{3})&\propto& f_{NL}\sum^{3}_{i<j =1}P_{\zeta}(k_{i})P_{\zeta}(k_{j}),\\
\label{eqt} T_{\zeta}(k_{1},k_{2},k_{3},k_{4}) &\propto& 
g_{NL}\sum^{3}_{i<j<p =1}P_{\zeta}(k_{i})P_{\zeta}(k_{j})P_{\zeta}(k_{p})\nonumber\\
&&~~~~~~~~+\tau_{NL}\sum^{11}_{j<p,i\neq j,p=1}P_{\zeta}(k_{ij})P_{\zeta}(k_{j})P_{\zeta}(k_{p}).
\end{eqnarray}
Here the proportionality constants for both the cases are determined by various configurations and types of the non-Gaussianity to be discussed later.
It will be convenient to define the shape function in terms of the bispectrum to determine the configurations of non-Gaussianity by the following expression:
\begin{eqnarray}
 {\cal S}(k_{1}, k_{2}, k_{3}) \equiv N_{norm}(k_{1}k_{2}k_{3})^{2}B_{\zeta}(k_{1},k_{2},k_{3})
\end{eqnarray}
where $N_{norm}$ is the normalization factor which are different for various non-Gaussian configurations.

\subsubsection*{B.~~$\delta N$ formalism}

In the previous section we computed the non-Gaussianity generated at horizon crossing. In
this section we discuss a second source of non-Gaussianity arising from non-linearities after
horizon crossing when all modes have become classical. A convenient way to describe these
non-Gaussianities is the $\delta N$ formalism ($N$ being the number of e-foldings) \cite{Sasaki:1995aw,Wands:2000dp,Lyth:2004gb,Lyth:2005fi,Mazumdar:2012jj,Sugiyama:2012tj}
 on large scales ($k << aH$). It provides a fruitful technique to compute the expression for the curvature perturbation $\zeta$
without explicitly solving the perturbed field equations. This formalism has twofold advantages-(1) it perfectly holds good
at the super-horizon scales and (2) is also independent of any kind of intrinsic non-Gaussianities
generated at the scale of horizon crossing. Let us mention the algorithm for $\delta N$ formalism point wise:
\begin{enumerate}
 \item One can define $\delta N ({\bf x}, t)$ as the number of e-folds from a fixed flat slice~\footnote{In the earlier discussions 
we have introduced $\zeta$ as the curvature perturbation in `comoving' gauge. Now in the context of $\delta N$ formalism on the superhorizon scales this is exactly equal
to the curvature perturbation in `uniform density' gauge.
}($\psi = 0$) to a uniform density
slice ($\psi = \zeta$) at time $t$. 

\item Then, one can write, $\zeta({\bf x}, t) = \delta N ({\bf x}, t)$. This leads to a simple algorithm to compute the superhorizon evolution of the
 primordial curvature perturbation $\zeta$ to illustrate the procedure consider a set of scalars $\phi_{i}$. A linear combination
of these fields can be identified to be the inflaton degrees of freedom. The remaining background fields are known as `isocurvatons'. We also assume that all
fields have become superhorizon at some initial time at which we choose a spatially flat
time-slice, on which there are no scalar metric fluctuations, but only fluctuations in the matter
fields, $\bar{\phi}_{i}+\delta \phi_{i}({\bf x})$. 

\item Next we choose the final time-slice to have uniform density, i.e. the inflaton field is
unperturbed and all fluctuations are in the metric and the isocurvatons.

\item Further we evolve the unperturbed 
fields $\bar{\phi}_{i}$ in the initial slice `classically' to the unperturbed final slice, and denote the corresponding
 number of e-folds as $\bar{N}(\bar{\phi}_{i})$. 

\item   Next, we evolve the perturbed initial field configuration $\bar{\phi}_{i}+\delta \phi_{i}({\bf x})$
classically to the perturbed final slice. In this prescription the $\delta N$ is defined as, 
$\delta N=N(\bar{\phi}_{i}+\delta \phi_{i}({\bf x}))-\bar{N}(\bar{\phi}_{i})$. Further Taylor expanding $\delta N$, we obtain an expression for
$\zeta$ in terms of the scalar field fluctuations $\delta \phi_{i}$ and derivatives of $N$ is defined on the initial slice as,

\begin{eqnarray}
 \zeta&=&N_{i}\delta\phi_{i}+\frac{1}{2!}N_{ij}\delta\phi_{i}\delta\phi_{j}+\frac{1}{3!}N_{ijk}\delta\phi_{i}\delta\phi_{j}\delta\phi_{k}+\cdots
\end{eqnarray}
where $N_{i}\equiv \partial_{i}N$, $N_{ij}\equiv \partial_{i}\partial_{j}N$ and $N_{ijk}\equiv\partial_{i}\partial_{j}\partial_{k}N$
are derivatives evaluated on the initial slice. 
\end{enumerate}

 Consequently the three and four point correlation function can be recast as the analytical expressions written
in Eq~(\ref{eq3pt}) and Eq~(\ref{eq4pt}). In this context the non-Gaussian parameter $f_{NL}$, $\tau_{NL}$ and $g_{NL}$
 can be expressed in terms of the derivatives of $N$ as:
\begin{eqnarray}
f_{NL}&\propto& \frac{N_{i}N_{j}N_{ij}}{(N^{2}_{l})^{2}},\\
\tau_{NL}&\propto& \frac{N_{i}N_{j}N_{ik}N_{jk}}{(N^{2}_{l})^{3}},\\
g_{NL}&\propto& \frac{N_{i}N_{j}N_{k}N_{ijk}}{(N^{2}_{l})^{3}}.
\end{eqnarray}
where all the repeated indices are summed over and the proportionality constants are determined by the shapes and configurations of the non-Gaussianities.

\subsubsection*{C.~~Shapes of non-Gaussianity}

\begin{itemize}
 \item \underline{\bf Local non-Gaussianity:}
One of the first ways to parameterize non-Gaussianity phenomenologically was via a non-linear
correction to a Gaussian perturbation $\zeta_{g}$,
\bea
 \zeta({\bf x})=\zeta_{g}({\bf x})+\frac{3}{5}f^{loc}_{NL}\left[\zeta^{2}_{g}({\bf x})-\langle\zeta^{2}_{g}({\bf x})\rangle \right]
+\frac{9}{25}g^{loc}_{NL}\zeta^{3}_{g}({\bf x})+\cdots
\eea
where $\cdots$ are higher order non-Gaussian contributions. This definition is local in real space and therefore called local non-Gaussianity. The bispectrum,
 trispectrum and the shape function of local non-Gaussianity are~\footnote{For the sake of simplicity here, in this context we use the following symbols,
\bea K_p &=& \sum_i (k_i)^p \quad {\rm with} \quad K = K_1 ,\\
 K_{pq}&=& \frac{1}{\Delta_{pq}} \sum_{i, j} (k_i)^p (k_j)^q,\\
K_{pqr}&=& \frac{1}{\Delta_{pqr}} \sum_{i, j, l} (k_i)^p (k_j)^q (k_l)^q,\\
\tilde k_{ip} &=& K_p - 2 (k_i)^p \quad {\rm with} \quad \tilde k_i = \tilde k_{i1},\eea
where \be \Delta_{pq} = 1 + \delta_{pq}\ee and \be \Delta_{pqr} = \Delta_{pq}(\Delta_{qr} + \delta_{pr})\ee where no summation as introduced in \cite{Fergusson:2008ra}.}:
\bea
\label{eqb} B^{loc}_{\zeta}(k_{1},k_{2},k_{3})&=& \frac{6}{5}f^{loc}_{NL}\sum^{3}_{i<j =1}P_{\zeta}(k_{i})P_{\zeta}(k_{j}),\\
\label{eqt} T^{loc}_{\zeta}(k_{1},k_{2},k_{3},k_{4}) &=& 
\frac{54}{25}g^{loc}_{NL}\sum^{3}_{i<j<p =1}P_{\zeta}(k_{i})P_{\zeta}(k_{j})P_{\zeta}(k_{p})\nonumber\\
&&~~~~~~~~+\tau^{loc}_{NL}\sum^{11}_{j<p,i\neq j,p=1}P_{\zeta}(k_{ij})P_{\zeta}(k_{j})P_{\zeta}(k_{p}),\\
 {\cal S}^{loc}(k_{1}, k_{2}, k_{3}) &=&\frac{K_{3}}{3K_{111}}.
\eea
The bispectrum for
local non-Gaussianity is then largest when the smallest $k$ (i.e. $k_{1}$) is very small, $k_{1}<<k_{2} \sim k_{3}$
~\footnote{This is the dominant mode of models with multiple light fields during inflation \cite{Kobayashi:2010fm}, the
curvaton scenario \cite{Lyth:2002my}, inhomogeneous reheating \cite{Dvali:2003ar}, and New Ekpyrotic models \cite{Buchbinder:2007ad}.
}. By momentum conservation, the other two momenta are then nearly equal. This is known as squeezed
limit. The bispectrum, shape function and the non-Gaussian parameter $f^{loc}_{NL}$ for local non-Gaussianity become
~\footnote{In case of local non-Gaussianity a consistency relation between $f^{loc}_{NL}$ and $\tau^{loc}_{NL}$ is satisfied within usual Einstein's GR motivated EFT setup.
It is known as {\it Suyama-Yamaguchi} relation \cite{Smith:2011if} given by, \be \tau^{loc}_{NL}\geq \left(\frac{6}{5}f^{loc}_{NL}\right)^{2},\ee where the equality is valid for single field slow-roll
inflation and the inequality appears in the context of multi-field inflation. Later we have shown that in the context of DBI Galileon inflation induced by 
the non-Einsteinian framework, such consistency condition is violated.}:
\bea
\lim_{k_{1}<<k_{2}\sim k_{3}} B^{loc}_{\zeta}(k_{1},k_{2},k_{3})&=& \frac{12}{5}f^{loc,sq}_{NL}P_{\zeta}(k_{1})P_{\zeta}(k_{2}),\\
 \lim_{k_{1}<<k_{2}\sim k_{3}}{\cal S}^{loc}(k_{1}, k_{2}, k_{3}) &=&\frac{2k_{2}}{3k_{1}},\\
f^{loc,sq}_{NL}=\lim_{k_{1}<<k_{2}\sim k_{3}} f^{loc}_{NL}&=&\frac{5}{12}(1-n_{s}).
\eea

\item \underline{\bf Equilateral non-Gaussianity:}
Higher-derivative corrections during inflation can lead to large non-
Gaussianities in presence of sound speed $c_{s}<1$. A key characteristic of derivative interactions is that they are suppressed when
any individual mode is far outside the horizon. This suggests that the bispectrum is maximal
when all three moves have wavelengths equal to the horizon size. The bispectrum therefore
has a shape that peaks in the equilateral configuration, $k_{1} = k_{2} = k_{3}=k$~\footnote{Additionally the other
 limits-folded ($k_{1}=2k_{2}=2k_{3}$), elongated ($k_{1}=k_{2}+k_{3}$) and isosceles ($k_{1}>k_{2}=k_{3}$)
configurations can be studied from the non-Gaussian template.}. Consequently the bispectrum and the shape function
can be expressed as:
\bea
B^{equil}_{\zeta}(k,k,k)&=& \frac{18}{5}f^{equil}_{NL}P^{2}_{\zeta}(k),\\
 {\cal S}^{equil}(k, k, k) &=& \lim_{k_{1}=k_{2}=k_{3}=k}\frac{\tilde k_1 \tilde k_2 \tilde k_3}{K_{111}}\,=1.
\eea

\item \underline{\bf Orthogonal non-Gaussianity:} Just like in the previous case each higher-derivative interaction of the
inflaton field generically gives rise to a bispectrum with a
shape which is similar – but not identical to – the equilateral
form~\footnote{An example is provided by the two interaction terms for an inflaton with a non-standard kinetic term.}.
Therefore it has been shown, using an effective field theory approach to inflationary perturbations, that it is possible to build a
combination of the corresponding similar equilateral shapes to
generate a bispectrum that is orthogonal to the equilateral one,
the so-called orthogonal shape. This can be approximated by
the template:

\bea
 \displaystyle B^{orth}_{\zeta}(k_1,k_2,k_3) &=& 6f^{orth}_{NL}P_{s}(k_{*})
\left(-\frac{3}{k_1^3k_2^3}-\frac{3}{k_1^3k_3^3}-\frac{3}{k_2^3k_3^3}\nonumber\right.\\ &&\left.~~~~~~~~~~~~~~~~-\frac{8}{k_1^2k_2^2k_3^2}+\frac{3}{k_1 k_2^2k_3^3}+(5\ perm.)\right)\\
 \displaystyle S^{orth}_{\zeta}(k_1,k_2,k_3)&=&\frac{(k_{1}k_{2}k_{3})^{2}}{P_{s}(k_{*})}B^{orth}_{\zeta}(k_1,k_2,k_3)
\eea
where $k_{*}$ is the pivot momentum scale. 
\end{itemize}

\section{Plan of the thesis}
\label{a3}

The plan of the thesis is as follows: In chapter \ref{ch:FRWxc} we have 
studied the features of MSSM inflation from various supersymmetric D -flat directions using the saddle and inflection point techniques.
In the case of saddle point inflation we derive the one loop effective potential by introducing non-renormalizable higher mass dimensional operators 
in the context of low scale MSSM. We have explored the possibility of PBH formation from this proposed model. Also we have solved the one-loop 
Renormalization Group (RG) flow equation to extract the behaviour of one-loop couplings and then the inflationary observable parameters as
 estimated from the model are then confronted
with observational data from WMAP. 
Next introducing an inflection point feature we have derived the expression for the effective potential in the context of high scale MSSM inflation within ${\cal N}=1$
SUGRA which can able to generate large tensor-to-scalar ratio. Hence the inflationary observable parameters as
 estimated from the model are then confronted 
with observational data from Planck data and fit the observed CMB TT spectra within low and high multipole region.

 Further, in chapter \ref{ch:FRWxc1} we have explored the possibility of inflation from the five dimensional background supergravity setup,
in the context of Randall-Sundrum (RS) like two braneworld model and Dirac Bonn Infeld (DBI) Galileon embedded in
 Klebanov-Strassler (KS) throat geometry. In both the cases we have derived the model from the background five dimensional ${\cal N}=2$ supergravity setup
by implementing dimensional reduction technique. Hence we have derived the inflationary observables from both of the scenarios and confront them with the 
WMAP data. Further using a cosmological code CAMB we fit the CMB angular power spectra from TT anisotropy and other polarization data obtained from WMAP and 
also estimate various cosmological parameters from these models. 

Hence in chapter \ref{ch:FRWxc2} we have discussed the various features of reheating phenomenology in modified gravity framework embedded in the SUGRA inspired RS like two braneworld model
 where the results are 
subsequently different from that of the usual low energy General Relativity (GR) prescribed phenomenological counterpart as the Friedmann equations are 
modified in RS braneworld. 
In this chapter we have explicitly derived the analytical expressions for the reheating temperature and using this we have further solved the evolution equation
of the number density of thermal gravitino in perturbative regime which results in the gravitino abundance in RS two braneworld model.
Next we have compared the results with that of its low energy GR limiting results.

Further, we have studied the primordial non-Gaussian features using $\delta N$ formalism 
of unavoidable higher dimensional non-renormalizable K\"ahler operators for ${\cal N}=1$ SUGRA framework in chapter \ref{ch:FRWxc3}.
In particular we have studied the nonlinear evolution of cosmological perturbations on
large scales which enables us to compute the curvature perturbation,  
without solving the exact perturbed field equations.
Hence we compute the various non-Gaussian parameters 
for local type of non-Gaussianities and CMB dipolar asymmetry parameter, 
for a generic class of sub-Planckian models induced by the Hubble-induced corrections for a 
MSSM D-flat direction where inflation occurs at the point of inflection within the visible sector of effective theory. 
Next using the constraint on sound speed from Planck data we determine the stringent bound on the non-Gaussian parameters and CMB dipolar asymmetry parameter.  
 
Finally we summarize our works in chapter \ref{ch:FRWxc4}. In this we also mention the future prospects of our works and overall comments 
made from our study.

\chapter{MSSM inflation from various flat directions}
\label{ch:FRWxc}

\section{Introduction}
\label{c1}
The observational success of primordial inflation arising from the Cosmic Microwave Background (CMB) 
radiation~\cite{Ade:2013uln,WMAP} has led to an outstanding question how to embed the inflationary paradigm
within a particle theory~\cite{infl-rev}. Since inflation dilutes all matter except for the quantum vacuum fluctuations 
of the inflaton, it is pertinent that the end of inflation creates all the relevant Standard Model degrees of freedom for 
the success of Big Bang Nucleosynthesis~\cite{BBN}, without any extra relativistic degrees of freedom, 
i.e. dark radiation~\cite{Ade:2013zuv}~\footnote{Embedding the last 50 -70 e-foldings of inflation within string theory has a major disadvantage.
Due to large number of hidden sectors arising from any string compactifications, it is likely that the inflaton energy density will get dumped 
into the hidden sectors instead of the visible sector~\cite{Cicoli:2010ha,Cicoli:2010yj}. The branching ratio for the inflaton decay into the visible sector is 
very tiny, therefore, reheating the Standard Model degrees of freedom is one of the biggest challenges for any string motivated models of inflation.
Furthermore, many of the compactifications generically lead to extra dark radiation (massless axions) which are already at the verge of being 
ruled out by the present data~\cite{Angus:2013zfa}.}.

This immediately suggests that the inflationary vacuum cannot be arbitrary and the inflaton must decay {\it solely} into the 
Standard Model degrees of freedom. Furthermore, the recent WMAP \cite{WMAP} and Planck data~\cite{Ade:2013zuv} separately indicate that the perturbations in the
 baryons and the cold dark matter are adiabatic in nature, it is, therefore, evident that there must be a single source of 
perturbations, which is responsible for seeding the fluctuations in all forms of matter~\footnote{In principle more than one fields can still
participate during inflation, but they must do so in such a way that here exists an attractor solution which would yield solely adiabatic perturbations 
and no isocurvature perturbations, such as, in the case of assisted inflation~\cite{Assist}.}. 
This can be realized conveniently if the 
inflaton itself carries the Standard Model charges as in the case of Minimal Supersymmetric Standard Model (MSSM) flat-directions~\cite{Enqvist:2003gh},
 where the lightest supersymmetric particle could be the dark matter candidate and can be created from thermal annihilation of the MSSM
 degrees of freedom ~\cite{Allahverdi:2006iq,Allahverdi:2006we,Choudhury:2011jt,Chatterjee:2011qr}.
  The inflaton candidates are made up of
{\it gauge invariant } combinations of squarks (supersymmetric partners of quarks) and sleptons (supersymmetric partners of leptons).

One of the {\it key} ingredients for  embedding inflation within MSSM is that the inflaton VEV must be below the Planck scale. This justifies the application of an effective field theory treatment at low energies.
It is well-known that the potential for the MSSM flat-direction inflaton has high degree of  {\it flexibility} -- the potential can accommodate both the
{\it saddle-point} and {\it inflection-point} below the Planck VEV, which allows a rich class of flat potentials,
which has been studied analytically and numerically~\cite{Allahverdi:2006we,Bueno Sanchez:2006xk}. The application of 
saddle point and inflection point inflation is not just limited to particle theory,
 but such potentials have also found their applications in string theory~\cite{Jain:2008dw}.
 It is conceivable that, at high energies the universe is dominated by a large cosmological constant
 arising from a string theory landscape. Our own patch of the universe could
 be locked in a false vacuum within an MSSM landscape, or there could be
 hidden sector contributions, or there could also be a 
combination of these effects. For the purpose of illustration, we will model earlier phases of
 inflation driven by the superpotential of type \cite{Choudhury:2014sxa,Choudhury:2014uxa}:
\begin{eqnarray}
 W&=&W(\Phi)+W(S)\,=\frac{\lambda \Phi^n}{n M_{p}^{n-3}}+\frac{M_s}{2}S^2\,,
\end{eqnarray}
where $n\geq 3$ and $\lambda\sim {\cal O}(1)$.
Here $M_{s}$ governs the scale of heavy physics which dictates the initial vacuum energy density, and $S(=s~e^{i\theta})$ is the hidden sector superfield. 
The total K\"ahler potential can be of the form \cite{Choudhury:2014sxa,Choudhury:2014uxa} of:
\begin{equation}
 K=S^{\dagger}S+\Phi^{\dagger}\Phi +\delta K, 
 \end{equation}
where the non-minimal term $\delta K$ can be any one of these functional forms:
\begin{equation}
\delta K=f(\Phi^{\dagger}\Phi,S^{\dagger}S)\,,~f(S^{\dagger}\Phi\Phi)\,,~f(S^\dagger S^\dagger\Phi\Phi)\,,~f(S\Phi^{\dagger}\Phi).
 \end{equation}
 We will always endeavour to treat the fields $s,~\phi\ll M_{p}$. The higher order Planck scale suppressed corrections to the K\"ahler potential
 are extremely hard to compute. Within the regime of effective field theory it is possible to constrain the co-efficients of all of these effective operators
and one can easily envisage various cosmological consequences from such contributions. See the ref. \cite{Choudhury:2014sxa,Choudhury:2014uxa} for details. 
The scalar potential in ${\cal N}=1$ supergravity can be written in terms of superpotential, 
$W$, and K\"ahler potential, $K$, as
\be\begin{array}{llll}\label{weq}
\displaystyle  V=e^{K(\Phi, \Phi^{\dagger})/M_{p}^2}\Big[ \big( D_{\Phi_i} W(\Phi) \big)
K^{\Phi_i \bar{\Phi}_j} \big(D_{\bar{\Phi}_j} W^*(\Phi^{\dagger}) \big) - \frac{3}{M_{p}^2} \left| W(\Phi) \right|^2 \Big] + ({\rm D-terms}),
\end{array}\ee
where $D_{{\Phi}_{i}} W = W_{{\Phi}_{i}} + K_{{\Phi}_{i}} W/M_{p}^2$, and 
$K^{\Phi_i \bar{\Phi}_j}$ is the inverse matrix of $K_{\Phi_i \bar{\Phi}_j}$, and the subscript (${\Phi}_{i}$) denotes derivative with respect to the field. Hereafter, 
we neglect the contribution from the D-term, since the MSSM inflatons are D-flat directions. 
After minimizing the potential along the angular direction, $\theta$ ($\Phi$ = $\phi e^{i \theta}$), we get \cite{Choudhury:2013jya}:
\be\begin{array}{llll}\label{hc1}
\displaystyle V(\phi,\theta) = V_{0}+\frac{(m^2_\phi+c_{H}H^{2})}{2}|\phi|^{2}+(a_{H} H 
+a_\lambda  m_{\phi})\frac{\lambda\phi^{n}} {nM^{n-3}_{p}}\cos(n\theta+\theta_{a_{H}}+\theta_{a_\lambda})
+  \frac{\lambda^{2}|\phi|^{2(n-1)}}{M^{2(n-3)}_{p}}
   \end{array}\ee
where the cosmological constant will be determined by the overall inflationary potential,
 \be V_0=\langle V(s)\rangle= M^4_{s}\approx 3H^2M_{p}^2.\ee
 Usually this bare cosmological term can be set to zero from the beginning by tuning
the graviton mass. We will consider 
scenarios, where we will have $V_0\neq 0$ (for $n=6$) and $V_0=0$ (for $n=4$).
Note that $m_\phi$ ($\sim{\cal O}(1)$~TeV) and 
$a_\lambda$ are soft-breaking mass and the non-renormalizable $A$-term respectively ($A$ is a positive quantity 
since its phase is absorbed by a redefinition of $\theta$ during the process)~\footnote{The masses of the various flat directions are given by \cite{Allahverdi:2006we}:
$$m^2_{\phi}=\frac{\sum_{\widetilde i}m^2_{\widetilde i}}{n}$$ where
$3\leq n \leq 9$ and the symbol ${\widetilde i}$ is used for flat direction contents. In this chapter we will concentrate only on $n=4$ and $n=6$ level flat directions.
 Typically these
masses are set by the scale of SUSY, in the low scale case the masses will be typically of order ${\cal O}(1)$~TeV.}.
 The potential also obtains Hubble-induced corrections, with coefficients $c_{H},~a_{H}\sim {\cal O}(1)$. Their exact numerical values
 will depend on the nature of K\"ahler corrections and compactification, which are hard to compute for a generic scenario~\cite{Berg}, 
 but the corrections typically yield $\sim {\cal O}(1)$ coefficients. The non-renormalizable terms have a  
 periodicity of $2\pi$ in $(\phi,~\theta)$ 2D plane, $\theta_{a_{H}},~\theta_{a_\lambda}$ are the extra phase factors.

Further using the potential stated in Eq~(\ref{hc1}) we have studied the inflationary paradigm from various supersymmetric D-flat direction 
in section~(\ref{c2}) and section~(\ref{c3}) 
from the low and high scale limiting situations by implementing
saddle and inflection point techniques respectively. Hence we have estimated various cosmological parameters and studied the features of CMB TT angular power
spectra from these two limiting situations. We have also analyzed other crucial astro-particle features- Primordial Black Hole (PBH) formation, one loop Renormalization Group (RG) flow
and reheating temperature from the proposed models.

\section{Saddle point and inflection point in MSSM}
\label{tech}

Let us briefly discuss saddle point and inflection point and there importance in the context of inflationary model building within MSSM:  
\begin{enumerate}
\item {\bf \underline{Saddle point:}} It is characterized by a point on a curve which is a stationary point but not an extremum. In terms of 
potential the saddle point is defined as, $V^{'}(\phi_0)=0$ and $V^{''}(\phi_0)\neq 0$. 

\item  {\bf \underline{Inflection point:}} Point of inflection is characterized by a point on a curve at which the curve changes 
from being concave to convex, or vice versa. Equivalently, an inflection point is defined by a point
 where the tangent meets the curve to order at least three~\footnote{Additionally in the present context if at a point the curvature vanishes
 but does not change sign is sometimes called a undulation point.}. Points of inflection can also be categorized according to whether the first derivative of the 
potential with respect to inflaton field i. e. $V^{'}(\phi)$ is zero or not zero. If $V^{'}(\phi_0)=0$ then the inflection point $\phi_0$ is a stationary point. On the other hand, 
if $V^{'}(\phi_0)\neq 0$, then the inflection point $\phi_0$ is identified with a non-stationary point. A necessary condition for $\phi_0$
 to be an inflection point is $V^{''}(\phi_0)=0$. A sufficient condition requires $V^{''}(\phi_0\pm \epsilon)$ have opposite signs in the neighborhood of $\phi_0$.

\end{enumerate}
It is important to note that in this chapter we discuss the cosmological consequences from non-stationary inflection point ($V^{'}(\phi_0)\neq 0$, $V^{''}(\phi_0)=0$) 
and stationary saddle point ($V^{'}(\phi_0)=0$, $V^{''}(\phi_0)\neq 0$) within MSSM.

\section{MSSM inflation using saddle point technique}
\label{c2}
Let us start with $n=4$ level, D-flat directions given by, ${\bf QQQL, QuQd, QuLe, uude}$, where we have considered vacuum energy correction, $V_{0}=0$
and the contribution from the soft SUSY breaking mass term is larger than the Hubble induced corrections i. e. $m_{\phi}\gg H$.
\subsection{Flat potential around saddle point}
For $n=4$ level the superpotential can be written as \cite{Gherghetta:1995dv}:
\be\begin{array}{ll}\label{mssm}
\displaystyle  W^{nr}_{4}  = \frac{1}{M_{p}}\left[\sum_{I=1}^{24}\alpha_I  ({\bf QQQL})_I+
\sum_{I=1}^{81}\beta_I ({\bf QuQd})_I+
\sum_{I=1}^{81}\gamma_I ({\bf QuLe})_I
+\sum_{I=1}^{27}\delta_I ({\bf uude})_I\right]
;\end{array}\ee

The renormalizable flat directions of the MSSM at n=4 level correspond to the gauge invariant
monomials subject to the four additional complex constraints \cite{Gherghetta:1995dv} two each from
\be F^{\alpha}_{H_{u}}=\mu H^{\alpha}_{d}+\lambda^{ab}_{U}Q^{\alpha}_{a}u_{b}=0,\ee
\be F^{\alpha}_{H_{d}}=-\mu H^{\alpha}_{u}+\lambda^{ab}_{D}Q^{\alpha}_{a}d_{b}+\lambda^{ab}_{E}L^{\alpha}_{a}e_{b}=0,\ee
which can lift the flat directions which do not contain a Higgs field. Here $\lambda_{U}, \lambda_{D}$ and $\lambda_{E}$ are the Yukawa 
couplings, $H_{u}, H_{d}$ are the Higgs superfield and the $\mu$- term appears in the renormalizable part of the superpotential of MSSM . Consequently the equation(\ref{mssm}) 
breaks into four parts, each one of them now being flat:
\bea\label{xd1}W^{(1)}_{4}&=&
\frac{1}{M_{p}}\sum_{I=1}^{24}\alpha_I ({\bf QQQL})_I,
\\
\label{xd2} W^{(2)}_{4}&=&
\frac{1}{M_{p}}\sum_{I=1}^{81}\beta_I ({\bf QuQd})_I,
\\
\label{xd3} W^{(3)}_{4}&=&
\frac{1}{M_{p}}\sum_{I=1}^{81}\gamma_I ({\bf QuLe})_I,
\\
\label{xd4} W^{(4)}_{4}&=&
\frac{1}{M_{p}}\sum_{I=1}^{27}\delta_I ({\bf uude})_I,\eea

resulting in $W^{(i)}_{4}\approx \frac{\lambda_{4}}{4M_{p}}{\bf\Phi}^{4}  ~\forall i(=(1,2,3,4))$.
Considering any one of the above flat directions leading to the one loop corrected effective potential:
\bea\label{vg}
V(\phi,\theta)&=&\frac{1}{2}m^{2}_{\phi}|\phi|^{2}
+\frac{\lambda_{4}A}{4M_{p}}\phi^{4}Cos(4\theta+\theta_{A})
+\frac{\lambda^{2}_{4}}{M^{2}_{p}}|\phi|^{6},\eea

for all i. Here we define: \begin{eqnarray}\lambda_{4}&=&\lambda_{4,0}\left[1+D_{3}\log\left(\frac{\phi^{2}}{\mu^{2}_{0}}\right)\right],\\
 A&=&\frac{A_{0}\left[1+D_{2}\log\left(\frac{\phi^{2}}{\mu^{2}_{0}
}\right)\right]}{\left[1+D_{3}\log\left(\frac{\phi^{2}}{\mu^{2}_{0}}\right)\right]},\\
 m^{2}_{\phi}&=& m^{2}_{0}\left[1+D_{1}\log\left(\frac{\phi^{2}}{\mu^{2}_{0}}\right)\right]\end{eqnarray}
 and in $\bf G_{MSSM}=SU(3)_{C}\otimes SU(2)_{L}\otimes U(1)_{Y}$ the representative flat direction field content is given by
\be\begin{array}{llll}\label{uip}
 \displaystyle {\bf Q^{I_{1}}_{a}}=\frac{1}{\sqrt{2}}({\bf\Phi},0)^{T},~~~~{\bf Q^{I_{2}}_{b}}=\frac{1}{\sqrt{2}}({\bf\Phi},0)^{T},~~~~{\bf Q^{I_{3}}_{c}}=\frac{1}{\sqrt{2}}({\bf\Phi},0)^{T},\\
\displaystyle ~~~~~~~~ {\bf L^{I_{4}}_{3}}=\frac{1}{\sqrt{2}}P_{d}(0,{\bf\Phi})^{T},~~~~{\bf d^{B_{1}}_{a}}=\frac{{\bf\Phi}}{\sqrt{2}},~~~~
{\bf u^{B_{2}}_{b}}=\frac{{\bf\Phi}}{\sqrt{2}},\\ \displaystyle~~~~~~~~~~~~~~~~~~~~~~ {\bf u^{B_{3}}_{c}}=\frac{{\bf\Phi}}{\sqrt{2}},~~~~{\bf e_{3}}=\frac{{\bf\Phi}}{\sqrt{2}}.
       \end{array}\ee

Here $m_{0},A_{0}$ and $\lambda_{4,0}$ are the values of the respective parameters at the scale $\mu_{0}$ and $D_{1},D_{2}$
and $D_{3}$ ($|D_{i}|\ll 1 \forall i$) are the fine tuning parameters. Additionally in the field contents 
${\bf 1 \leq B_{1},B_{2}}$ $,{\bf B_{3} \leq 3}$ are color indices, ${\bf 1 \leq a,b,c\leq 3 }$ denote the indices for quark and lepton families and 
${\bf 1\leq I_{1},I_{2},I_{3}}$ $,{\bf I_{4}\leq 2}$ are the weak isospin indices. The
flatness constraints require that ${\bf B_{1} \neq B_{2} \neq B_{3}}$ for quarks, ${\bf I_{1}\neq I_{2}\neq I_{3}\neq I_{4}}$, 
${\bf\sum^{3}_{d=1}P^{2}_{d}=1}$ ${\bf\forall P_{d}\in \mathbb{R}}$ for leptons and ${\bf a\neq b\neq c}$ for both. In Eq.~(\ref{vg}) $m_{\phi}$
represents the soft SUSY breaking mass term, $\phi$ the radial coordinate of the complex scalar field
 ${\bf\Phi}=\phi\exp(i\theta)$ ($\in\mathbb{C}$) and the second term is the so called
 A-term which has a periodicity of $2\pi$ in 2 D along with
an extra phase $\theta_{A}$. The radiative correction slightly affects the soft term and the value
of the saddle point.

 For $n=4$ we get an extremum for the principal values of $\theta$ at $\theta=\frac{(m\pi-\theta_{A})}{4}$
 (where ${\bf m\in \mathbb{Z}}$)

\be\begin{array}{ll}\label{sd} \displaystyle \phi_{0}=\sqrt{\frac{M_{p}}{4\lambda_{4}(3+D_{3})}}\left[A\left(1+\frac{D_{2}}{2}\right)
\pm \sqrt{A^{2}\left(1+\frac{D_{2}}{2}\right)^{2}-8m^{2}_{\phi}(1+D_{1})(3+D_{3})}\right]^{\frac{1}{2}},
\end{array}
\ee

which appears from the constraint $V^{'}(\phi_{0})=0$ as a
necessary condition for {\it saddle point}. However, this condition alone will not lead to
{\it saddle point}. Rather, we have to make the potential sufficiently flat which can be achieved by vanishing 
higher derivatives of the potential. In this article, we consider non-vanishing fourth derivative of the potential 
resulting in saddle point. This will imply more fine-tuning but increased precision level in the information obtained 
from RG flow. 

 As discussed, $V^{''''}(\phi_{0})<0$ will give us secondary local minimum.
This leads to constraint relations:

\bea
\label{Ast}\displaystyle A &=&\sqrt{2(3+D_{3})G_{1}G_{2}G_{3}}m_{\phi}(\phi_{0}),
\eea

\be\begin{array}{llll} \label{con2}\displaystyle  D_{3}=\frac{M_{p}A_{0}}{4\lambda_{4,0}\phi^{2}_{0}\left(37+60\log\left(\frac{\phi_{0}}{\mu_{0}}\right)\right)}
\left\{D_{2}\left(13+12\log\left(\frac{\phi_{0}}{\mu_{0}}\right)\right)
-\frac{2m^{2}_{\phi}(\phi_{0})D_{1}M_{p}}{\lambda_{4,0}A_{0}\phi^{2}_{0}}
+6\left(1-\frac{20\lambda_{4,0}\phi^{2}_{0}}{M_{p}A_{0}}\right)\right\}\end{array},\ee
one each for $V^{''}(\phi_{0})=0$ and $V^{'''}(\phi_{0})=0$. In this context:
 \begin{eqnarray}G_{1}&=&\left[\frac{(1+D_{1})}{(3+D_{3})}(15+11D_{3})-(1+3D_{1})\right]^{2},\\
G_{2}&=&\left[(1+D_{1})\left(3+\frac{7}{2}D_{3}\right)-(1+3D_{1})\left(1+\frac{D_{2}}{2}\right)\right]^{-1},\\
G_{3}&=&\left[\frac{\left(1+\frac{D_{2}}{2}\right)}{(3+D_{3})}(15+11D_{3})-\left(3+\frac{7}{2}D_{3}\right)\right]^{-1}.\end{eqnarray}
For the limit $|D_{1}|\ll 1$,$|D_{2}|\ll 1$ and $|D_{3}|\ll 1$ which gives:
\begin{eqnarray}\phi_{0}&=&\phi^{tree}_{0}\left[1+\frac{D_{1}}{2}-\frac{D_{3}}{6}\right]^{\frac{1}{2}},\\
A&\simeq& A_{tree}\left[1+\frac{D_{1}}{2}-\frac{D_{3}}{6}\right],\end{eqnarray}
where \begin{eqnarray}\phi^{tree}_{0}&=&\sqrt{\frac{m_{\phi}(\phi_{0})M_{p}}{\lambda_{4}\sqrt{6}}},\\ A_{tree}&=&2\sqrt{6}m_{\phi}(\phi_{0}),\end{eqnarray}
represents tree level expressions. This means, during RG flow mentioning two parameters only ($D_{1}$ and $D_{2}$) will
suffice instead of the usual three parameters in earlier MSSM models. This results in more precise information in RG flow. 
One may get tempted to vanish further higher derivatives of the potential in order to evaluate other
 unknown parameters ($D_{1}$ and $D_{2}$) without going into RG flow but this will make the effective inflaton potential
 in the vicinity of {\it saddle point} non-renormalizable. So, this is the highest level of precision 
constraint one can impose on RG flow parameters.

Consequently, around the {\it saddle point} $\phi_{0}$, the inflaton potential can be expanded in a Taylor series as,

\be\label{hgkl} V(\phi)=\tilde{C}_{0}+\tilde{C}_{4}(\phi-\phi_{0})^{4},\ee
where 
\be\begin{array}{lll}\label{co}\displaystyle \tilde{C}_{0}=V(\phi_{0})\\\displaystyle ~~~~=\frac{m^{3}_{\phi}(\phi_{0})M_{p}}{6\sqrt{6}\lambda_{4}}
\left\{3\left(1+\frac{D_{1}}{2}-\frac{D_{3}}{6}\right)
\left[1+D_{1}\log\left(\frac{\phi^{2}_{0}}{\mu^{2}_{0}}\right)\right]
-3\left(1+\frac{D_{1}}{2}-\frac{D_{3}}{6}\right)^{2}\left[1+D_{2}\log\left(\frac{\phi^{2}_{0}}{\mu^{2}_{0}}\right)\right]\right.
\\  \left.~~~~~~~~~~~~~~~~~~~~~~~~~~~~~~~~~~~~~~~~~~~~~~~~~~~~~~~~~~~~~~~~~~~~~~~~~~~\displaystyle +\left(1+\frac{D_{1}}{2}-\frac{D_{3}}{6}\right)^{2}\left[1+D_{2}\log\left(\frac{\phi^{2}_{0}}{\mu^{2}_{0}}\right)\right]
\right\}\end{array}\ee
and
\be\begin{array}{lll}\label{co1}\displaystyle \tilde{C}_{4}=\frac{1}{{4!}}V^{''''}(\phi_{0})\\
\displaystyle~~~~=\frac{m^{2}_{\phi}(\phi_{0})}{24\sqrt{6}\phi^{2}_{0}}\left(1+\frac{D_{1}}{2}-\frac{D_{3}}{6}\right)
\left\{\left\{\left[\left(\frac{360}{\sqrt{6}}-12\sqrt{6}\right)+(684D_{3}-50\sqrt{6}D_{2})\right]
\left(1+\frac{D_{1}}{2}-\frac{D_{3}}{6}\right)\right.\right.
\\  \left.\left.~~~~~~~~~~~~~~~~~~~~~~~~~~~~~~~~~~~\displaystyle -\frac{2\sqrt{6}D_{1}}{\left(1+\frac{D_{1}}{2}-\frac{D_{3}}{6}\right)}\right\}
  +\left(1+\frac{D_{1}}{2}-\frac{D_{3}}{6}\right)\left(\frac{360D_{3}}{\sqrt{6}}
-12\sqrt{6}D_{2}\right)
\log\left(\frac{\phi^{2}_{0}}{\mu^{2}_{0}}\right)\right\}.\end{array}\ee

In what follows we shall model MSSM inflation with the above potential.

\subsection{Modeling inflation \& parameter estimation}

 For the best fit value of the model parameters: \begin{eqnarray}\tilde{C}_{0}&=&2.867\times 10^{-36}M^{4}_{p},~~~~~
\tilde{C}_{4}=-1.685\times 10^{-13}\end{eqnarray} for the no. of e-foldings ${\cal N}=70$ the cosmological parameters obtained from our model is:
\begin{eqnarray}
P_{s}=\Del^{2}_{s}&=&2.498\times 10^{-9},
~~~~~ n_{s}=0.957,\nonumber\\
r&=&1.240\times 10^{-29},~~~~~ \alpha_{s}=-0.612\times 10^{-3}~~~ \kappa_{s}=1.749\times 10^{-5}\nonumber.\end{eqnarray}

\begin{figure}[htb]
{\centerline{\includegraphics[width=15cm, height=7cm]{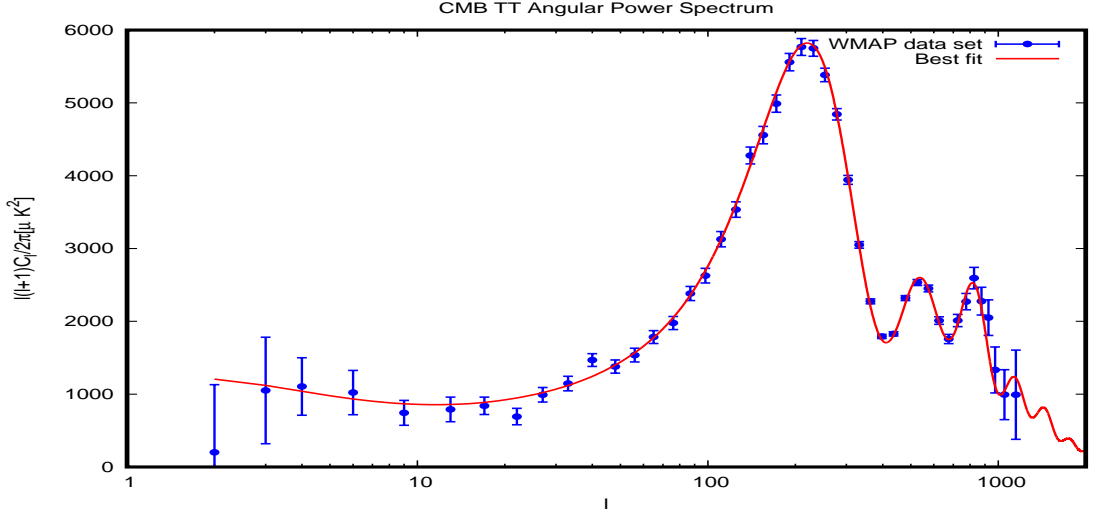}}}
\caption{Variation of CMB angular power spectrum $~$ ($l(l+1)C^{TT}_{l}/2\pi$) for
best fit proposed model and WMAP seven years data with the multipoles l for scalar mode \cite{Choudhury:2011jt}.} \label{figVr1181}
\end{figure}

Further, we use the publicly available code CAMB \cite{CAMB} to verify our results directly with observation.
To operate CAMB at the pivot scale $k_0=0.002~{\rm Mpc}^{-1}$ the values of the initial parameter space
are taken for $\tilde{C}_{0}=2.867\times10^{-36}M^{4}_{p}$ and ${\cal N}=70$.
 Additionally WMAP dataset \cite{Spergel:2006hy,WMAP7} for
$\Lambda$CDM background has been used in CAMB to obtain CMB angular power spectrum.
In Table~\ref{tab2} we have given all the input parameters for CAMB.
Table~\ref{tab3} shows the CAMB output, which is in good agreement with WMAP seven years data.
In Fig.~\ref{figVr1181}
we have plotted CAMB output of CMB TT angular power spectrum
$C_l^{TT}$ for best fit with WMAP seven years data for scalar mode, which explicitly show
the agreement of our model with WMAP dataset.

\begin{table}[htb]
\begin{center}
\begin{tabular}{|c|c|c|c|c|c|c|c|c|c|}
\hline $H_0$ & $\tau_{Reion}$ &$\Omega_b h^2$& $\Omega_c h^2
$& $T_{CMB}$
 \\
km/sec/MPc& & && K\\
 \hline
71.0&0.09&0.0226&0.1120&2.725\\
\hline
\end{tabular}
\caption{Input parameters \cite{Choudhury:2011jt}.}\label{tab2}
\end{center}
\begin{center}
\begin{tabular}{|c|c|c|c|c|c|c|c|c|c|}
\hline $t_0$ & $z_{Reion}$ &$\Omega_m$&$\Omega_{\Lambda}$&$\Omega_k$&$\eta_{Rec}$& $\eta_0$
 \\
Gyr& & && &Mpc & Mpc\\
 \hline
13.707&10.704&0.2670&0.7330&0.0&285.10&14345.1\\
\hline
\end{tabular}
\caption{Output obtained from CAMB \cite{Choudhury:2011jt}.}\label{tab3}
\end{center}
\end{table}

\subsection{Analyzing primordial Black hole formation}
Now in the context of any running mass model one can expand the
spectral index with the following parameterization \cite{Kosowsky:1995aa}:
\be \label{ntwer}
 n({\cal R}) = n_z (k_0) - \frac{\alpha_z(k_0)}{2!}\ln\left( k_0
 {\cal R} \right) + \frac{\kappa_z(k_0)}{3!}\ln^2\left( k_0 {\cal R} \right)+.....
\ee
with ${\cal R} \ll 1/k_0$, i.e. $\ln(k_0 {\cal R}) < 0$. This is identified to be the significant
 contribution to the Primordial Black Hole (PBH) formation \cite{Hidalgo}. Here the
parameterization index $ z:\left[s(scalar),t(tensor)\right]$ and the explicit form of
 the first term in the above expansion is given by
\be\label{runfunc}  n_z (k_0)=
\left\{
	\begin{array}{ll}
		 n_s (k_0)-1 & \mbox{if } z=s \\
		 n_t (k_0) & \mbox{if } z=t.
	\end{array}
\right.
\ee

Existence of the running and running of the running is
 the key feature in the formation of PBH in the radiation dominated era just after inflation \cite{Drees:2011hb}
which could form CDM in the Universe. The
initial PBHs mass $\cal M_{\text{PBH}}$ is related to the particle horizon mass
${\cal M}$ by:
\be { M_{\text{PBH}}}={M }\gamma= \frac{4\pi}{3}\gamma \rho  {\cal H}^{-3}\,\ee
at horizon entry, ${ R}=(a{\cal H})^{-1}$. This is formed when the density
fluctuation exceeds the threshold for PBH formation given as in {\it Press--Schechter theory} by~\footnote{In this chapter we fix $\gamma
\simeq 0.2$ during the radiation dominated era \cite{Carr:1975qj} for numerical estimations.}
\begin{equation} \label{fofm}
  f( \geq{\cal M}) =\gamma~\text{erf}\left[\frac{5\pi\left(1+\frac{3}{5}w\right)\varTheta_{\rm th}}
{8 (1+w)\sqrt{\frac{2\tilde{C}_{4}}{\kappa_{s}}\Gamma[\frac{(n_S({ R})+3)}{2}]}}\right].
\end{equation}
Here $\varTheta_{\rm th}$ be the threshold value of the
linearized density field $\varTheta$ smoothed on a comoving scale $\cal R$
with $n_s({ R})>3$.
  
 In general
 the mass of PBHs is expected to depend
on the amplitude, size and shape of the perturbations.
As a consequence the PBH mass is given by:
\be { M_{\text{PBH}}} = \gamma { M_{\text{eq}}} (k_{\text{eq}}{ R})^2 \left( \frac
{g_{\ast,\text{eq}}} {g_{\ast}} \right)^{\frac{1}{3}}\, ,\ee
where the subscript ``eq'' refers to the matter--radiation
equality. Here we use
$g_{\ast}=228.75$ (all degrees of freedom
in MSSM), while $g_{\ast,\text{eq}} = 3.36$
and $k_{eq} = 0.07 \Omega_{\text{m}} h^{2}\ \text{Mpc}^{-1}$~\footnote{Here we use $\Omega_{\text{m}}
h^{2} = 0.2670$ from the CAMB output.}. Consequently
 the relation between comoving scale and the PBH mass in the context of MSSM is given by:

\begin{equation} \label{R}
\frac { R} {1\ \text{Mpc}} = 1.250 \times 10^{-23}\left(
\frac { M_{\text{PBH}}} {1\ \text{g}} \right)^{\frac{1}{2}}\, .
\end{equation}

\begin{figure}[htb]
{\centerline{\includegraphics[width=13cm, height=7.5cm]{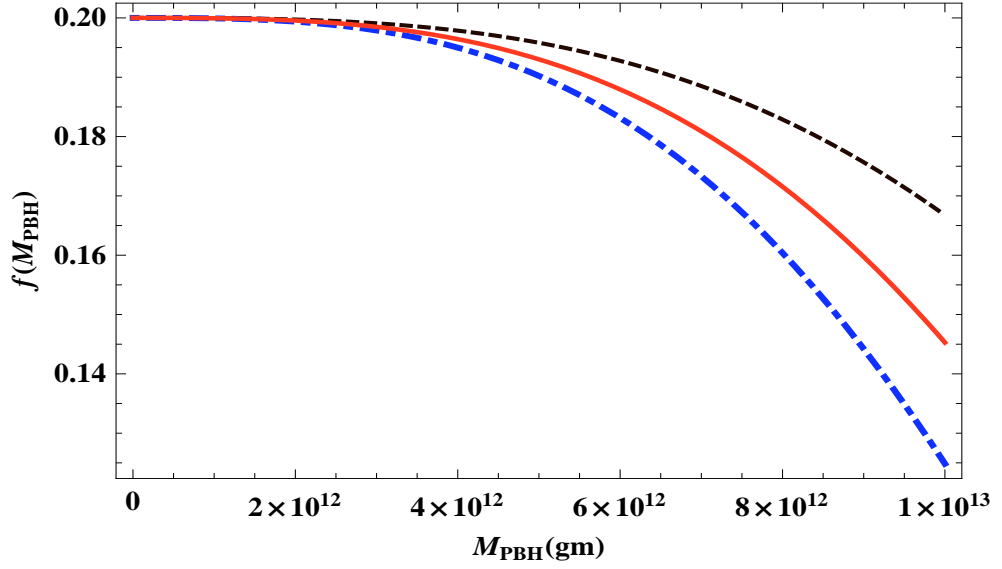}}}
\caption{$~$ Variation of the$~$
 fraction of the energy density of the universe collapsing into PBHs as a function
of the PBH mass, for three different values of the threshold
$\varTheta_{th} = 0.3(dashed), 0.5(solid), 0.7(dotdashed)$ \cite{Choudhury:2011jt}.
} \label{figVr8167}
\end{figure}

 Fig(\ref{figVr8167}) shows the behavior of {\it Press--Schechter mass function}
with respect to PBH mass. With the values of the parameters as obtained earlier, we have 
${\cal M}_{PBH}\simeq 10^{13}gm$ and
the corresponding fractional energy density $f=0.170$.

\subsection{One loop Renormalization Group flow}

 For the flat direction \textbf{QQQL,QuQd,QuLe,uude} the soft SUSY breaking masses can be expressed as:
\begin{eqnarray}\label{jkhjg}
  \displaystyle (m^{2}_{\phi})_{\textbf{QQQL}}&=&\frac{1}{4}(m^{2}_{{\bf\tilde{Q}_{a}}}
+m^{2}_{{\bf\tilde{Q}_{b}}}+m^{2}_{{\bf\tilde{Q}_{c}}}+m^{2}_{{\bf\tilde{L}_{3}}}),\nonumber
\\ \displaystyle (m^{2}_{\phi})_{\textbf{QuQd}}&=&\frac{1}{4}(m^{2}_{{\bf\tilde{Q}_{a}}}
+m^{2}_{{\bf\tilde{Q}_{b}}}+m^{2}_{{\bf\tilde{u}_{c}}}+m^{2}_{{\bf\tilde{d}_{3}}}), \nonumber
 \\  \displaystyle (m^{2}_{\phi})_{\textbf{QuLe}}&=&\frac{1}{4}(m^{2}_{{\bf\tilde{Q}_{a}}}
+m^{2}_{{\bf\tilde{u}_{b}}}+m^{2}_{{\bf\tilde{L}_{c}}}+m^{2}_{{\bf\tilde{e_{3}}}}),\nonumber
\\ \displaystyle (m^{2}_{\phi})_{\textbf{uude}}&=&\frac{1}{4}(m^{2}_{{\bf\tilde{u}_{a}}}+m^{2}_{{\bf\tilde{u}_{b}}}
+m^{2}_{{\bf\tilde{d}_{c}}}+m^{2}_{{\bf\tilde{e}_{3}}}),
\end{eqnarray}

where ${\bf1\leq a,b,c\leq 3}$ and ${\bf a\neq b\neq c}$. After neglecting the contribution from the all Yukawa
 couplings except from the top we can express the one-loop beta function as:
\cite{Nilles:1983ge}
\be\begin{array}{lll}\displaystyle \beta_{m^{2}_{a}}=\dot{\mu}m^{2}_{a}
=\frac{1}{8\pi^2}\left(m^2_{a}+|A^{33}_{U}|^2\right)\left(\lambda^{33}_{U}\right)^{2}-\frac{1}{2\pi^{2}}\sum^{3}_{\alpha=1}
g^{2}_{\alpha}|\tilde{m}_{\alpha}|^{2}\textbf{X}_{\alpha a}\end{array}\ee
where $\textbf{X}_{\alpha a}$ are the quadratic Casimir Group Invariants for the superfield $\Phi$, defined in terms
of Lie Algebra generators $T^{a}$ by
\be (T^{\alpha}T^{\alpha})^{a}_{b}=\textbf{X}_{\alpha a}\delta^{a}_{b}\ee
and $\dot{\mu}=\mu\frac{\partial}{\partial\mu}$. Here $\mu$ is the Renormalization Group (RG) scale of the effective theory.


\begin{figure}[htb]
{\centerline{\includegraphics[width=13cm, height=7cm]{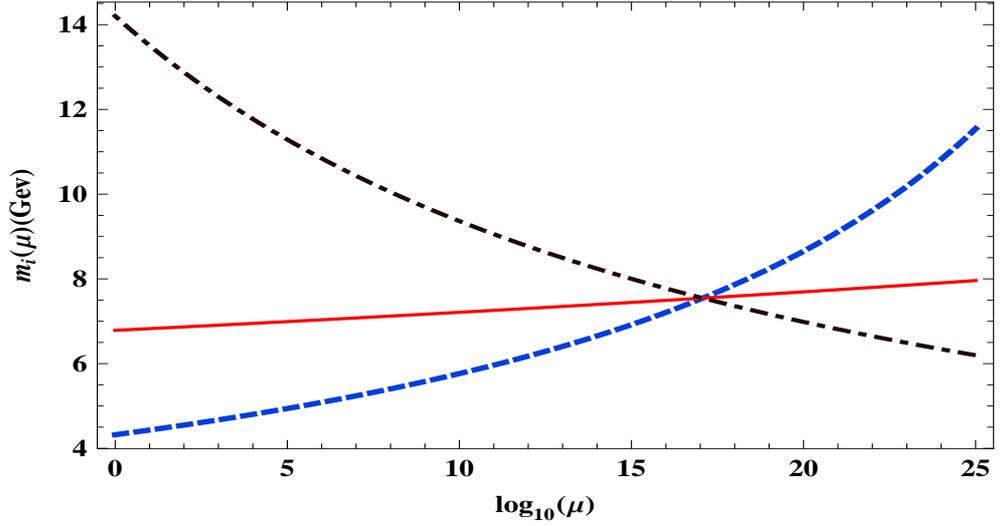}}}
\caption{Running of gaugino mass ($m_{i}(\mu)$)
 in one loop RGE for MSSM with the logarithmic scale $\log_{10}\left(\mu\right)$ \cite{Choudhury:2011jt}. Here we have used
$\mu_{0}=2.6\times 10^{7} GeV$ , $m_{i}(\mu_{0})=7.546\times10^{-3}TeV$, $\zeta=1$ $\forall$ i.} \label{figVr18}
\end{figure}

\begin{figure}[htb]
{\centerline{\includegraphics[width=13cm, height=7cm]{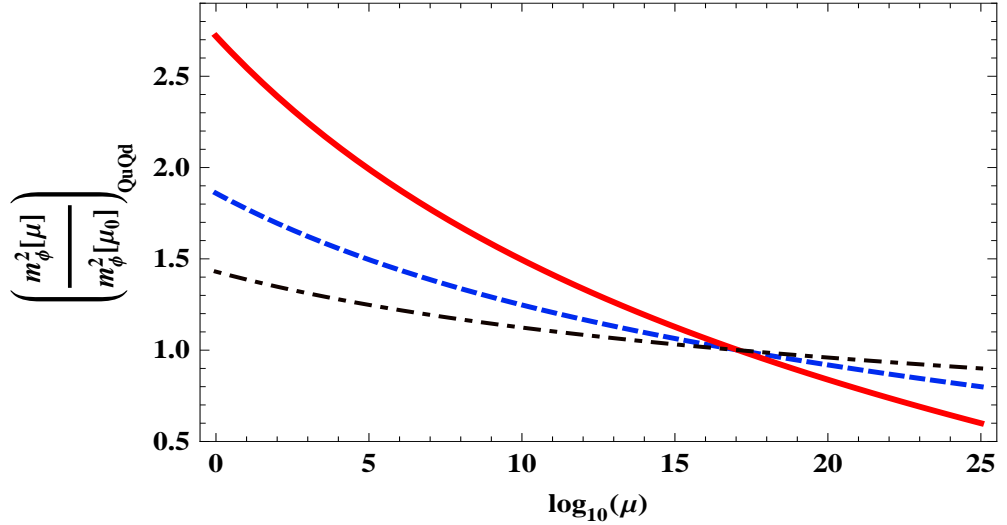}}}
\caption{Running of soft mass squared ratio $\left(\frac{m^{2}_{\phi}(\mu)}{m^{2}_{\phi}(\mu_{0})}\right)$ in one loop RGE for MSSM 
with the logarithmic scale $\log_{10}\left(\mu\right)$ where
$\mu_{0}=2.6\times 10^{7} GeV$ for $\zeta=0.5,1,2$ for $n=4$ level $~${\bf QuQd} $\forall$ i. Similar plots can be obtained for
{\bf QuLe}, {\bf QQQL} and
{\bf uude} flat directions also $\forall$ i \cite{Choudhury:2011jt}.} \label{figVr1568}
\end{figure}


\begin{figure*}[htb]
\centering
\subfigure[]{
    \includegraphics[width=5cm,height=5.5cm] {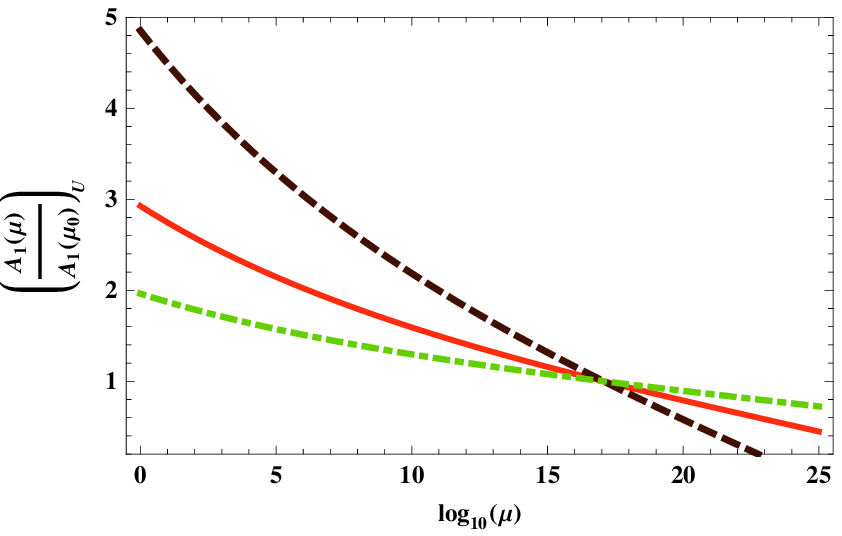}
    \label{figVr1}
}
\subfigure[]{
    \includegraphics[width=5cm,height=5.5cm] {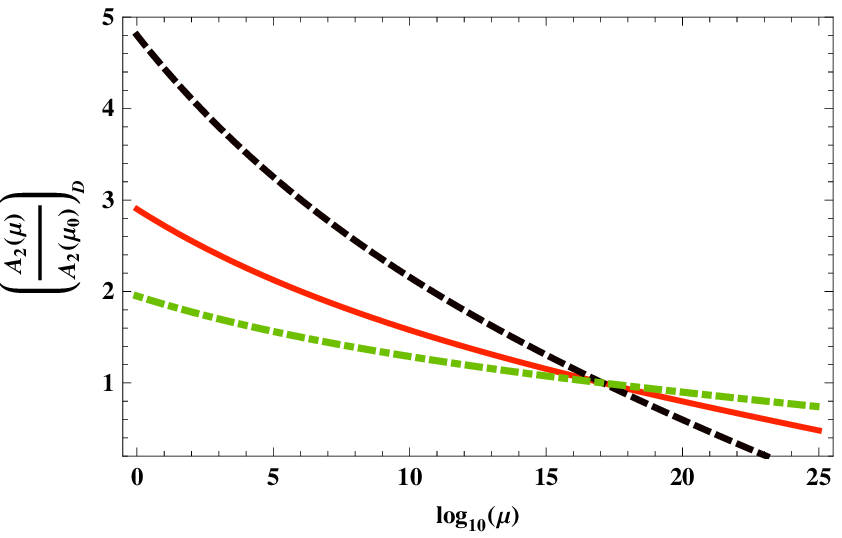}
    \label{figVr2}
}
\subfigure[]{
    \includegraphics[width=5cm,height=5.5cm] {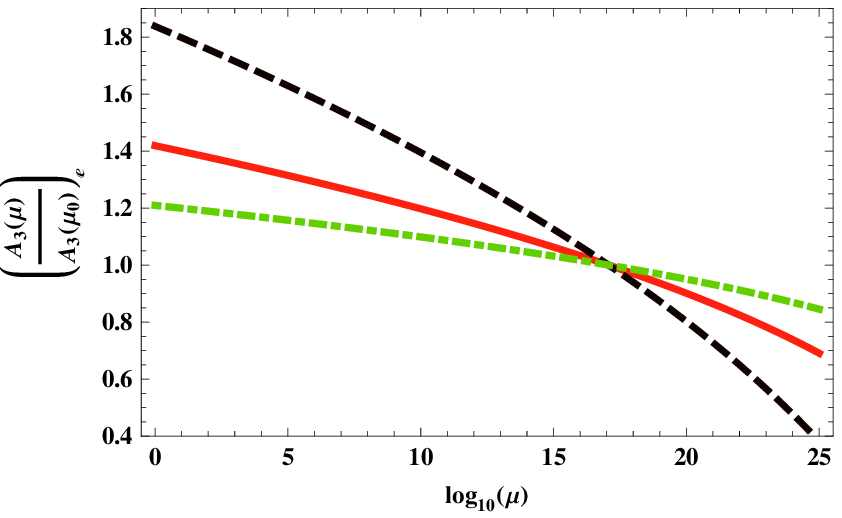}
    \label{figVr3}
}
\caption[Optional caption for list of figures]{Running of trilinear A -term ratio $\left(\frac{A_{\beta}(\mu)}{A_{\beta}(\mu_{0})}\right)$ in
 one loop RGE for MSSM with the logarithmic scale $\log_{10}\left(\mu\right)$ where
$\mu_{0}=2.6\times 10^{7} GeV$, $\zeta=0.5({\bf dotdashed}),
1({\bf solid}),2({\bf dashed})$ and $\beta=1(U),2(D),3(E)$ for $n=4$ level $\forall$ i \cite{Choudhury:2011jt}.
}
\label{figVr118}
\end{figure*}

\begin{figure}[htb]
{\centerline{\includegraphics[width=13cm, height=7cm]{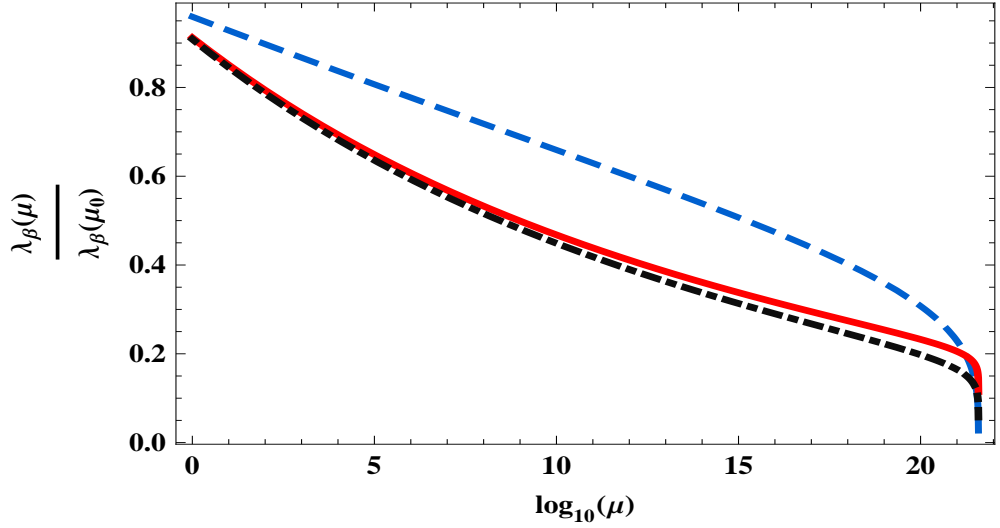}}}
\caption{Running of the ratio of the Yukawa coupling $\left(\frac{\lambda_{\beta}(\mu)}{\lambda_{\beta}(\mu_{0})}\right)$
 in one loop RGE for MSSM with the logarithmic scale $\log_{10}\left(\mu\right)$. Here we have used
$\mu_{0}=2.6\times 10^{7} GeV$ and $\beta=1(U),2(D),3(E)$ $\forall$ i \cite{Choudhury:2011jt}.} \label{figVr178}
\end{figure}

In the context of MSSM: \begin{eqnarray} \textbf{X}_{1a}&=&\frac{3{\bf Y}^{2}_{a}}{5}~({\rm for ~each~} {\bf\Phi}_{a} {\rm ~with~ weak~ hyper~ charge~} {\bf Y}_{a}),\\
\textbf{X}_{2a}&=&\frac{3}{4} ~{\rm (for~ {\bf\Phi}_{a}={\bf Q,L,H_{u},H_{d}})},
\\  ~~~~~~&=& 0  ~{\rm (for~{\bf\Phi}_{a}={\bf \bar{u},\bar{d},\bar{e}})},
\\ \textbf{X}_{3a}&=&\frac{4}{3} ~{\rm (for~{\bf\Phi}_{a}={\bf Q,\bar{u},\bar{d}})},
\\  ~~~~~~&=& 0  ~{\rm (for~{\bf\Phi}_{a}={\bf L,\bar{e},H_{u},H_{d}})},\end{eqnarray}
 where $\textbf{X}_{1a}$, $\textbf{X}_{2a}$ and $\textbf{X}_{3a}$ are applicable for ${\bf U(1)_{Y}}$,${\bf SU(2)_{L}}$ and
${\bf SU(3)_{C}}$ respectively.
 So for the flat direction content \textbf{QQQL,~QuQd,~QuLe,~uude} we have the following beta functions:\\  \\
(a) \underline{\textbf{For Soft mass}}:
\begin{eqnarray}\label{ad1}
    \dot{\mu}(m^{2}_{\phi})_{\textbf{QQQL}}&=& \frac{1}{8\pi^2}\left(3m^{2}_{Q_{3}}+m^{2}_{U_3}+|A^{33}_{U}|^2\right)\left(\lambda^{33}_{U}\right)^{2}
-\frac{1}{8\pi^{2}}\left(3g^{2}_{2}|\tilde{m}_{2}|^{2}+4g^{2}_{3}|\tilde{m}_{3}|^{2}\right),
\nonumber \\
\dot{\mu}(m^{2}_{\phi})_{\textbf{QuQd}}&=&\frac{1}{2\pi^2}\left(m^{2}_{Q_{3}}+m^{2}_{U_3}\right)\left(\lambda^{33}_{U}\right)^{2}
-\frac{1}{8\pi^{2}}\left(\frac{3}{2}g^{2}_{2}|\tilde{m}_{2}|^{2}+\frac{16}{3}g^{2}_{3}|\tilde{m}_{3}|^{2}\right),
\nonumber\\
\dot{\mu}(m^{2}_{\phi})_{\textbf{QuLe}}&=&\frac{3}{8\pi^2}\left(m^{2}_{Q_3}+m^{2}_{U_3}\right)\left(\lambda^{33}_{U}\right)^{2}
-\frac{1}{8\pi^{2}}\left(\frac{3}{2}g^{2}_{2}|\tilde{m}_{2}|^{2}+\frac{8}{3}g^{2}_{3}|\tilde{m}_{3}|^{2}\right),
\nonumber\\
\dot{\mu}(m^{2}_{\phi})_{\textbf{uude}}&=&\frac{1}{2\pi^2}\left(m^{2}_{Q_{3}}+m^{2}_{U_3}\right)\left(\lambda^{33}_{U}\right)^{2}
-\frac{1}{2\pi^{2}}g^{2}_{3}|\tilde{m}_{3}|^{2},\nonumber\\
\end{eqnarray}

(b) \underline{\textbf{For Trilinear A- term}}:
\begin{eqnarray}\label{hoj}
 \dot{\mu}A^{aa}_{D}&=&\delta_{b3}\left(\lambda^{33}_{U}\right)^{2}\frac{A^{33}_{U}}{8\pi^{2}}
-\frac{1}{4\pi^{2}}\left(\frac{7}{18}g^{2}_{1}|\tilde{m}_{1}|^{2}+
\frac{3}{2}g^{2}_{2}|\tilde{m}_{2}|^{2}+\frac{8}{3}g^{2}_{3}|\tilde{m}_{3}|^{2}\right),
\nonumber\\
\dot{\mu}A^{ab}_{U}&=&\frac{3(1+\delta_{a3})A^{33}_{U}}{8\pi^{2}}\left(\lambda^{33}_{U}\right)^{2}
-\frac{1}{4\pi^{2}}\left(\frac{13}{18}g^{2}_{1}|\tilde{m}_{1}|^{2}+\frac{3}{2}g^{2}_{2}|
\tilde{m}_{2}|^{2}+\frac{8}{3}g^{2}_{3}|\tilde{m}_{3}|^{2}\right),
\nonumber\\
\dot{\mu}A^{aa}_{E}&=&
-\frac{1}{4\pi^{2}}\left(\frac{3}{2}g^{2}_{1}|\tilde{m}_{1}|^{2}+\frac{3}{2}g^{2}_{2}|\tilde{m}_{2}|^{2}\right),\nonumber\\
\end{eqnarray}

(c) \underline{\textbf{For Fourth level Yukawa coupling}}:
\begin{eqnarray}\label{coj}
\dot{\mu}\lambda^{aa}_{U}&=&\frac{3(1+\delta_{a3})}{8\pi^{2}}\left(\lambda^{33}_{U}\right)^{3}
-\frac{\lambda^{aa}_{U}}{4\pi^{2}}\left(\frac{13}{18}g^{2}_{1}+\frac{3}{2}g^{2}_{2}+\frac{8}{3}g^{2}_{3}\right),
\nonumber\\
\dot{\mu}\lambda^{ab}_{D}&=&\delta_{b3}\left(\lambda^{33}_{U}\right)^{2}\frac{\lambda^{ab}_{D}}{8\pi^{2}}
-\frac{\lambda^{ab}_{D}}{4\pi^{2}}\left(\frac{7}{18}g^{2}_{1}+\frac{3}{2}g^{2}_{2}+\frac{8}{3}g^{2}_{3}\right),
\nonumber\\
\dot{\mu}\lambda^{aa}_{E}&=&
-\frac{\lambda^{aa}_{E}}{4\pi^{2}}\left(\frac{3}{2}g^{2}_{2}+\frac{3}{2}g^{2}_{3}\right)\nonumber
\end{eqnarray}
where all the superscript $a$ and $b$ represent generation or family indices run from 1 to 3 physically representing the
first, second and third generation respectively. For the one-loop renormalization of gauge couplings and gaugino masses, one has in general
\begin{eqnarray}\label{poi}\beta_{g_{\alpha}}&=&\dot{\mu}g_{\alpha}=\frac{g^{3}_{\alpha}}{16\pi^{2}}\left[\Sigma_{a}\textbf{I}_{\alpha a}-3\textbf{X}_{\alpha G}\right],\\
  \beta_{m_{\alpha}}&=&\dot{\mu}m_{\alpha}=\frac{g^{2}_{\alpha}m_{\alpha}}{8\pi^{2}}\left[\Sigma_{a}\textbf{I}_{\alpha a}-3\textbf{X}_{\alpha G}\right]  
   \end{eqnarray}
where $\textbf{X}_{\alpha G}$ quadratic Casimir invariant of the group $[$ 0 for $\bf{U(1)}$ and $\bf{N}$ for $\bf{SU(N)}$ $]$, $\textbf{I}_{\alpha a}$ is the Dynkin 
index of the chiral supermultiplet $\bf{\Phi}_{a}$ $[$ normalized to $\frac{1}{2}$ for each fundamental representation of $\bf{SU(N)}$ and to 
$3\bf{Y}^{2}_{a}/5$ for $\bf{U(1)}_{Y}$ $]$. For the above mentioned flat direction
the running of gauge couplings ($g_{i}(\mu)$) and gaugino masses ($m_{i}(\mu)$) obey,
\begin{eqnarray}\dot{\mu}g_{i}&=&
\frac{d_{i}}{2}g^{3}_{i},\\
\dot{\mu}\left(\frac{m_{i}}{g^{2}_{i}}\right)&=&
0~\forall~~ i\end{eqnarray}
where for $i=1({\bf U(1)_{Y}}),2({\bf SU(2)_{L}}),3({\bf SU(3)_{C}})$ here $d_{1}=\frac{11}{8\pi^{2}}$,$d_{2}=\frac{1}{8\pi^{2}}$,$d_{3}=-\frac{3}{8\pi^{2}}$ 
which is the simpler version of the equation(\ref{poi}).
Now to show explicitly that the contributions from the top Yukawa coupling ($\lambda^{33}_{U}$) are very small
for an induced electroweak group $\textbf{G}_{EW}$=$\textbf{SU(2)}_{L}$ $\otimes$ $\textbf{U(1)}_{Y}$ breakdown,
let us start with the Higgs potential \cite{Nilles:1983ge}
\be
\begin{array}{lllll}\label{yuki}\displaystyle V_{ Higgs}({\bf H,\bar{H}})=m^2_{1}|{\bf H}|^2+m^2_{2}|{\bf\bar{H}}|^2+m^{2}_{3}
\left({\bf H{\bar{H}}}+{\bf H^{\dagger}\bar{H^{\dagger}}}\right)
+\frac{1}{8}\left(g^2_{1}+g^2_{2}\right)\left[|{\bf H}|^2-|{\bf\bar{ H}}|^2\right]^2,\end{array}
\ee
where  ${\bf H}=H_{u}$ and ${\bf \bar{H}}=H_{d}$ represent the Higgs superfields and
 the relative vev of the two
Higgses are given by 
\be\label{desy}
\begin{array}{lllllll}
 v=\sqrt{{\langle {\bf H}\rangle}^2+{\langle {\bf {\bar H}}\rangle}^2}
\displaystyle=\sqrt{\frac{2\left[m^2_{1}-m^2_{2}-\left(m^2_{1}+m^2_{2}\right)cos(2{\bf\theta})\right]}{\left(g^2_{1}+g^2_{2}\right)cos(2{\bf\theta})}}
\end{array}
\ee
with \be tan(\theta)=\frac{{\langle {\bf {\bar H}}\rangle}}{{\langle {\bf H}\rangle}}.\ee Here $\theta$ represents 
an angular parameter which parameterizes MSSM. For the sake of convenience 
let us now write $cos(2\theta)$ appearing in equation(\ref{desy}) introducing new parameterization as \cite{Nilles:1983ge}:
\be cos(2\theta)=\frac{w^2-1}{w^2+1}\ee where \be w=\frac{\frac{\langle {\bf H}\rangle}{v}}{\sqrt{1-\left(\frac{\langle {\bf H}\rangle}{v}\right)^2}}.\ee  
Consequently the top Yukawa coupling can be expressed as: \be\lambda^{33}_{U}=\frac{m_{U}}{v sin(\theta)}\ee where $0\leq\theta<\frac{\pi}{2}$
and the top mass:~$43 {\rm GeV}\leq m_{U}\leq 170{\rm  GeV}\ll\mu_{GUT}$ 
comes from the RG flow \cite{Nilles:1983ge}. It is evident from the above parameterization that
as $w\rightarrow 1$, $\theta\rightarrow \frac{\pi}{4}$ which implies ${\langle {\bf H}\rangle}$ and ${\langle {\bf\bar{ H}}\rangle}$ is very large and have the same order of magnitude. 
As a result the relative vev $v$ is also large and the top Yukawa coupling is very very small
for which one can easily neglect it from the RG flow at the energy scale of MSSM inflation as mentioned earlier.
The consequence of the large vev of Higgs field can be taken care of by introducing strongly interacting gauge group  ${\bf G_{NEW}= G_{S}\otimes SU(3)_{C}}$
and its superconformal version ${\bf G_{SCONF}=SU(3)_{SC}\otimes SU(3)_{C}}$.


 In table(\ref{tab8}) we have tabulated the numerical values of vev of ${\bf H}$ and ${\bf \bar{H}}$, the angular parameter $\theta$,
tan($\theta$), $w$, the top mass $m_{U}$ and 
the top Yukawa coupling $\lambda^{33}_{U}$ contributing to the parameter space of MSSM for the $n=4$ level flat directions 
 \textbf{QQQL,~QuQd,~QuLe} and \textbf{~uude}. It should be noted that appearance of large VEV of Higgses as mentioned in table(\ref{tab8}) can easily be interpreted
when Einstein Hilbert term appears in the total action of the theory at lowest order approximation which
 is our present consideration. Consequently the contributions from the hard cutoff is sub-leading due to the soft conformal symmetry
breaking. This leads to small top Yukawa coupling in the restricted parametric space
 of MSSM characterized by the phenomenological bound: \begin{eqnarray} 43 {\rm GeV}\leq m_{U} \leq 170 {\rm GeV},~~~  1.006\leq tan(\beta)\leq  1.025\end{eqnarray}
for the n=4 flat directions.
 
 Neglecting all the sub-leading contributions arising from the top Yukawa coupling in the restricted parameter space of the MSSM,
 the solutions of these RGE for n=4 level flat directions can be written as:

\begin{eqnarray}\label{s1}
g_{i}(\mu)&=&
\frac{g_{i}(\mu_{0})}{\sqrt{1-d_{i}g^{2}_{i}(\mu_{0})\ln\left(\frac{\mu}{\mu_{0}}\right)}},
\\
\label{s1a}m_{i}(\mu)&=&m_{i}(\mu_{0})\left(\frac{g_{i}(\mu)}{g_{i}(\mu_{0})}\right)^{2},
\\
\label{s1aa}\Delta m^{2}_{\phi}&=&
\sum^{3}_{i=1}f_{F}^{i}\Delta m^{2}_{i},
\\
\label{s1aaa}\Delta A^{ab}_{\beta}&=&
\frac{1}{2}\sum^{3}_{i=1}(C_{\beta}^{i})^{ab}\Delta m_{i},
\\
\label{s1aaaa}\lambda^{ab}_{\beta}(\mu)&=&
\lambda^{ab}_{\beta}(\mu_{0})\prod^{3}_{i=1}\left(\frac{g_{i}(\mu_{0})}{g_{i}(\mu)}\right)^{(C_{\beta}^{i})^{ab}},\end{eqnarray}
Here $g_{i}(\mu_{0})$, $m_{i}(\mu_{0})$, $A_{\beta}(\mu_{0})$,  $m_{\phi}(\mu_{0})$ and  $\lambda_{\beta}(\mu_{0})$ represent the value of the gauge couplings, gaugino masses,
 trilinear couplings, soft SUSY braking masses and Yukawa couplings at the characteristic scale $\mu_{0}$. In equation(\ref{s1}-\ref{s1aaaa}) we
have used the following shorthand notation:
$\Delta W=W(\mu)-W(\mu_{0}),$ where $W=\{A_{\beta},m^{2}_{\phi},m^{2}_{i},m_{i}\}$ and the $\beta$ indices 1,2,3 represent U, D, E
respectively. In equation (\ref{s1}-\ref{s1aaaa}) $f_{F}^{i}$ and $(C_{\beta}^{i})^{ab}$ are $(4\times 3)$ and $(3\times 3)$ matrices whose entries 
are tabulated in Table(\ref{tab5}) and Table(\ref{tab6}) respectively. It is obvious from the RGE that $\beta=1,2$ implies $a=b$ and $\beta=3$ implies $a\neq b$.

\begin{table}[htb]
\small\begin{center}
\begin{tabular}{|c|c|c|c|c|}
\hline ${\bf f_{F}^{i}}$ & ${\bf i=1(U(1)_{Y})}$ &${\bf i=2(SU(2)_{L})}$&${\bf i=3(SU(3)_{C})}$\\
 \hline
F=1({\bf QQQL})&0&$\frac{3}{2}$&-$\frac{2}{3}$\\
\hline
F=2({\bf QuQd})&0&$\frac{3}{4}$&-$\frac{8}{9}$\\
\hline
F=3({\bf QuLe})&0&$\frac{3}{4}$&-$\frac{4}{9}$\\
\hline
F=4({\bf uude})&0&0&-$\frac{2}{3}$\\
\hline
\end{tabular}
\caption{Entries of $f_{F}^{i}$ matrix obtained from the solution of RGE \cite{Choudhury:2011jt}.}\label{tab5}
\end{center}
\end{table}

\begin{table}[htb]
\small\begin{center}
\begin{tabular}{|c|c|c|c|c|}
\hline ${\bf (C_{\beta}^{i})^{ab}}$ & ${\bf i=1(U(1)_{Y})}$ &${\bf i=2(SU(2)_{L})}$&${\bf i=3(SU(3)_{C})}$\\
 \hline
$\beta$=1({\bf U}),a=b&$\frac{26}{99}$&6&-$\frac{32}{9}$\\
\hline
$\beta$=2({\bf D}),a=b&$\frac{14}{99}$&6&-$\frac{32}{9}$\\
\hline
$\beta$=3({\bf E}),$a\neq b$&$\frac{6}{11}$&6&0\\
\hline
\end{tabular}
\caption{Entries of $(C_{\beta}^{i})^{ab}$ matrix obtained from the solution of RGE \cite{Choudhury:2011jt}.}\label{tab6}
\end{center}
\end{table}

Using the solutions of RGE along with the approximation that the running of the gaugino masses and gauge couplings
is very very small we get:

\begin{eqnarray}\label{cg1} D_{1}&=&
-\frac{1}{8\pi^{2}}\sum^{3}_{i=1}J_{i}\left(\frac{m_{i}}{m_{\phi_{0}}}\right)^{2}g^{2}_{i}(\mu_{0}),
\\
 D^{\beta}_{2}&=&
-\frac{1}{4\pi^{2}}\sum^{3}_{i=1}K^{\beta i}\left(\frac{m_{i}}{A_{0}}\right)g^{2}_{i}(\mu_{0}),\end{eqnarray}

where we have $J_{1}=0$,$J_{2}=3$ and $J_{3}=4$ for $i=1, 2, 3$ and all the entries of  $K^{\beta i}$ $(3\times 3)$ matrix are tabulated in
table(\ref{tab7}).

\begin{table}[htb]
\small\begin{center}
\begin{tabular}{|c|c|c|c|c|}
\hline ${\bf K^{\beta i}}$ & ${\bf i=1(U(1)_{Y})}$ &${\bf i=2(SU(2)_{L})}$&${\bf i=3(SU(3)_{C})}$\\
 \hline
$\beta$=1({\bf U})&$\frac{13}{18}$&$\frac{3}{2}$&$\frac{8}{3}$\\
\hline
$\beta$=2({\bf D})&$\frac{7}{18}$&$\frac{3}{2}$&$\frac{8}{3}$\\
\hline
$\beta$=3({\bf E})&$\frac{3}{2}$&$\frac{3}{2}$&0\\
\hline
\end{tabular}
\caption{Entries of $K^{\beta i}$ matrix \cite{Choudhury:2011jt}.}\label{tab7}
\end{center}
\end{table}
In this context the subscript `0' represents the values of parameters at the high scale $\mu_{0}$. As discussed in section III, constraining only $D_{1}$ and $D^{\beta}_{2}$ is sufficient here. Eqn(\ref{con2}) provides an 
extra constraint relation which restricts the parameters further leading to more precise information in RG flow.
For universal boundary conditions, the high scale is identified to be
the GUT scale:\begin{eqnarray} \mu_{GUT} &\approx& 3 \times
10^{16}{\rm~GeV}, \\ {\tilde m_{1}}(\mu_{GUT}) &=& {\tilde m_{2}}(\mu_{GUT}) ={\tilde m_{3}}(\mu_{GUT})= {\tilde m},\\
A_{E}(\mu_{GUT})&=&A_{U}(\mu_{GUT})=A_{D}(\mu_{GUT})=A_{0}\end{eqnarray}
with 
$g_{1}\approx 0.56, ~~ g_{2}\approx
0.72, ~~ g_{3}\approx 0.85$. Now depending upon the different phenomenological situations the $n=4$ level flat directions 
are divided into two classes. The first class deals with ${\bf QuQd, QuLe}$
which is lifted completely at $n=4$ level. The other class which is lifted by higher dimensional operators
deals with ${\bf uude, QQQL}$. Most importantly ${\bf uude, QQQL}$ take part in the proton
 decay (${\it p\rightarrow\pi^{0}e^{+}}$, ${\it p\rightarrow\pi^{+}\nu_{e}}$ etc.) \cite{Choudhury:2011jt} which introduces a stringent constraint
 on the Yukawa coupling $\lambda_{0}$ at $n=4$ level. Additionally the neutrino-antineutrino oscillation data
restricts $\lambda_{0}$ again.
Then we just use RG equations along with these restrictions to run the coupling constants and masses to the scales as mentioned in 
table(\ref{tab8}).

Considering all these values we obtain effectively:
\begin{eqnarray}D_{1} &\approx&  -0.056 \zeta^2, \\
D^{1}_{2} &\approx&  -0.074 \zeta,
~~~D^{2}_{2} \approx  -0.071 \zeta,
~~~D^{3}_{2} \approx  -0.031 \zeta,\\
D^{1}_{3}&=& D^{2}_{3}=D^{3}_{3}\approx  -0.048-0.168\zeta^{2},\end{eqnarray}
where $\zeta = m/ m_{\phi}$ is calculated at the GUT scale. Typically the running
 based on gaugino loops alone results in negative values of
 $D_{i}\forall i$. Positive values can be obtained when one includes the Yukawa couplings,
practically the top Yukawa, but the order of magnitude remains the same. 
The choice of fine tuned initial conditions directly shows
 more fine tuning is required compared to other models. It is a straightforward exercise to verify that
even if one considers all the flat directions at $n=4$ level one will arrive at the potential eqn.(\ref{hgkl})
with same $\tilde{C}_{0}$ and $\tilde{C}_{4}$. This is precisely what we have done in this paper.

The results of RG flow have been demonstrated in figs(\ref{figVr18})-(\ref{figVr178}).
In fig(\ref{figVr18}) and fig(\ref{figVr1568})
`{\it dashed}', `{\it solid}' and `{\it dotdashed}' line represents ${\bf U(1)_{Y}}$, ${\bf SU(2)_{L}}$ and ${\bf SU(3)_{C}}$
gauge group content respectively. Fig(\ref{figVr18})-fig(\ref{figVr178}) explicitly showing
 the behavior of the RGE flow of gaugino masses, soft SUSY
breaking mass, trilinear couplings and Yukawa couplings respectively.
 Additionally fig(\ref{figVr18})-fig(\ref{figVr118}) give consistent GUT scale unification.
\begin{center}
\begin{table*}
\renewcommand{\tabcolsep}{0.02pc}
{\small
\hfill{}
\begin{tabular}{|c|c|c|c|c|c|c|c|c|c|c|}
\hline ${\bf Flat~}$ & ${\bf\mu_{0}=\phi_{0}}$&$A_{0,tree} $&${\bf m_{\phi_{0}}}$&${\bf\langle {\bf H}\rangle }$& ${\bf\langle {\bf \bar{H}}\rangle }$
&$\tan\theta$&$v$&$m_{U}$ & ${\bf \lambda^{33}_{U}=\lambda_{0}}$\\
 ${\bf direction}$ & ${ GeV}$&${ GeV}$ &${ GeV}$&${ GeV}$&${ GeV}$ & &${ GeV}$&${ GeV}$&${ GeV}$\\
 \hline
${\bf QuLe}$&$2.6 \times
10^{7} $&$36.967$&$7.546$&$0.200\times 10^{16}$  & $0.458\times 10^{16}$& 1.006  &$0.500\times 10^{16} $& 43&$1.212\times 10^{-14}  $\\
\hline
${\bf QuQd}$&$2.6 \times
10^{7} $&$36.967$&$7.546$&$ 0.450\times 10^{16} $ & $0.423\times 10^{16}$ & 1.013  & $0.601\times 10^{16}$ &170 &$7.106\times 10^{-14}  $\\
\hline
${\bf QQQL}$&$1.344\times
10^{14}$&$892\times 10^{3}$&$182\times 10^{3}$&$0.188\times 10^{8}$&$0.124\times 10^{8}$& 1.025 &$0.226\times 10^{8}$& 80 &$4.945\times 10^{-6}$\\
\hline
${\bf uude}$&$2.896\times
10^{13}$&$4.142\times10^{6}$&$845\times 10^{3}$&$0.174\times 10^{6}$ &$0.157\times 10^{6}$ &  1.019 &$0.235\times 10^{6}$ &135 & $8.047\times 10^{-4}$\\
\hline
\end{tabular}}
\hfill{}
\caption{MSSM parameter values obtained from RG flow for n=4 level flat directions \cite{Choudhury:2011jt}.}\label{tab8}
\end{table*}
\end{center}

\section{MSSM inflation using inflection point technique}
\label{c3}
In the last section we have explored the possibility of low scale MSSM inflation using saddle point technique originating from D-flat directions. 
But effectively one can uplift the scale of inflation by adding an extra vacuum energy dominated SUGRA correction term to the low scale inflationary potential.
In this section instead of saddle point technique we use inflection point technique to construct the effective potential within MSSM and we explore both the cases-low scale as
well as high scale inflationary paradigm. We will
 keep $V_0$ in Eq~(\ref{hc1}) for this case, and do the computation for $n=6$ flat directions, {\bf udd} and {\bf LLe}.
\subsection{Flat potential around the inflection point}
Applying inflection-point technique as discussed in section \ref{tech}, inflationary potential $V(\phi)$ can be expanded in Taylor series as~\cite{Enqvist:2010vd}:
\begin{equation}\label{rt1}
V(\phi)=\alpha+\beta(\phi-\phi_{0})+\gamma(\phi-\phi_{0})^{3}+\kappa(\phi-\phi_{0})^{4}+\cdots\,,
   \end{equation}
where any generic potential, $V(\phi)$, has been expanded around the inflection-point, $\phi_0$, where $\alpha$ denotes the cosmological 
constant which will further determine the scale of inflation, and coefficients $\beta,~\gamma,~\kappa$ determine the shape of the potential in terms of the model parameters. Typically,
$\alpha$ can be set to zero by fine tuning, but here we wish to keep this term for generality as we are interested here for $V_0 \neq 0$.
 Note that not all of the coefficients are 
independent once we prescribe inflaton within MSSM.




In the present context let us consider two
 $D$-flat directions- $\widetilde{u}\widetilde{d}\widetilde{d}$ and $\widetilde{L}\widetilde{L}\widetilde{e}$ as the inflaton candidates \cite{Allahverdi:2006iq,Allahverdi:2006we}. Both $\widetilde{u}\widetilde{d}\widetilde{d}$, where
 $\widetilde u,~\widetilde d$ correspond to the right handed squarks, and $\widetilde{L}\widetilde{L}\widetilde{e}$, 
where $\widetilde L$ is the left handed slepton, and $\widetilde e$ is the right handed (charged) leptons,
 flat directions are lifted by higher order superpotential terms of the following simple form: 
\begin{equation} \label{supot}
W (\Phi)= {\lambda \over 6}{\Phi^6 \over M^3_{p}}\, ,
\end{equation}
 where $\lambda \sim {\cal O}(1)$ coefficient. The scalar component of $\Phi$ superfield, denoted by $\phi$, is given by~
\begin{equation} \label{infl}
\phi = {\widetilde{u} + \widetilde{d} + \widetilde{d} \over \sqrt{3}} ~ ~ ~ , ~ ~ ~ \phi = {\widetilde{L} + \widetilde{L} + 
\widetilde{e} \over \sqrt{3}},
\end{equation}
and 
the masses are given by:
$$m^2_{\phi}=\frac{m^2_{\widetilde L}+m^2_{\widetilde L}+m^2_{\widetilde e}}{3},~~~
m^2_{\phi}=\frac{m^2_{\widetilde u}+m^2_{\widetilde d}+m^2_{\widetilde d}}{3}$$ 
for the $\widetilde{u}\widetilde{d}\widetilde{d}$ and $\widetilde{L}\widetilde{L}\widetilde{e}$ flat directions, respectively.


\begin{figure}[t]
{\centerline{\includegraphics[width=14cm, height=7cm] {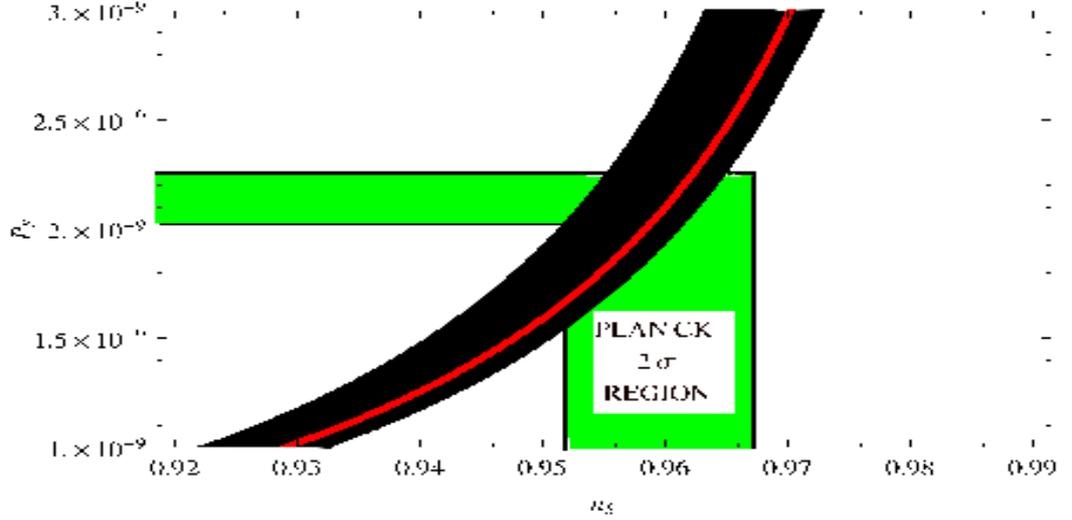}}}
\caption{ For large scale inflation, $H\gg m_\phi$, we have shown the variation of $P_{S}$~vs~$n_{S}$ \cite{Choudhury:2013jya}. The {\it red} curve shows the
model parameters, $\delta\sim 10^{-4},~\lambda =1,~c_H=2,~a_H=2.108,~\phi_0=1.129\times 10^{16}~{\rm GeV}$, for the pivot scale 
$k_{\star}=0.002~Mpc^{-1}$. The ${\it green}$ shaded region shows the  $2\sigma$ CL. range 
in $n_s$ allowed by the Planck data~\cite{Ade:2013uln}. Instead of getting a single solid {\it red} curve we get a {\it black} shaded region if 
we consider the full parameter space for high scale ($H\gg m_{\phi}$) inflation given by Eq.~(\ref{P-space}). } \label{fig2}
\end{figure}




 
 
\subsection{Low scale inflation}\label{01}
 
Since $c_H, a_H~\sim {\cal O}(1)$ the Hubble-induced terms do not play any crucial role in this case, and the scale of inflation remains very low. 
As a result the tensor-to scalar ratio, $r$, become too small to be ever detectable. This was the scenario studied in Refs.~\cite{Allahverdi:2006iq,Allahverdi:2006we}. 
In the low scale scenario, the value of $V_0\leq m_{\phi}^2\phi_0^2$, is negligible and does not contribute to the dynamics.
In this case we can set its value to $V_0=0$ from the beginning by tuning the gravitino mass~\cite{Nilles:1983ge}.
The potential can be minimized along the $\theta$ direction, which reduces to~\cite{Allahverdi:2006iq,Allahverdi:2006we}:
\begin{equation}
V(\phi) =\frac{m^2_\phi}{2}|\phi|^{2}-a_\lambda  m_{\phi}\frac{\lambda\phi^{6}} {6M^{3}_{p}}
+  \frac{\lambda^{2}|\phi|^{10}}{M^{6}_{p}}
\end{equation}
For,
 \begin{equation} \label{dev}
{a_\lambda^2 \over 40} \equiv 1 - 4 \delta^2\, ,
\end{equation}
and $\delta^2 \ll 1$, there exists a point of inflection ($\phi_0$) in $V(\phi)$, where
\begin{eqnarray}\label{vev}
&&\phi_0 = \left({m_\phi M^{3}_{p}\over \lambda \sqrt{10}}\right)^{1/4} + {\cal O}(\delta^2) \, , \label{infvev} \\
&&\, \nonumber \\
&&V^{\prime \prime}(\phi_0) = 0 \, , \label{2nd}
\end{eqnarray}
at which
\begin{eqnarray}
\label{pot}
\alpha &=&V(\phi_0) = \frac{4}{15}m_{\phi}^2\phi_0^2 + {\cal O}(\delta^2) \, , \\
\label{1st}
\beta&=&V'(\phi_0) = 4 \alpha^2 m^2_{\phi} \phi_0 \, + {\cal O}(\delta^4) \, , \\
\label{3rd}
\gamma&=&V^{\prime \prime \prime}(\phi_0) = 32\frac{m_{\phi}^2}{\phi_0} + {\cal O}(\delta^2) \, .
\end{eqnarray}
The potential is specified completely by $m_\phi$ and $\lambda$. However $m_\phi$ is determined by the soft-SUSY breaking
mass parameter,  which is well constrained by the current ATLAS~\cite{ATLAS} and CMS~\cite{CMS} data,  and we shall take $m_\phi=1$~TeV. 
For $m_\phi\sim 1$~TeV, $H\ast\sim 0.1$~GeV, and our assumption of neglecting $H$ in such a case is well justified.
We will always consider $\lambda =1$ in our analysis.


 \subsection{High scale inflation}\label{02}
 
  The SUGRA corrections become important, the Hubble-induced terms dominate the potential. This can happen
 quite naturally if there exists a previous source of effective cosmological constant term described in Ref.~\cite{Mazumdar:2011ih}. 
 In this case one can safely ignore the soft SUSY breaking mass term, and since $a_\lambda\sim {\cal O}(1)$, one can 
 safely consider only the Hubble-induced non-renormalizable term~\footnote{Within the setup of effective field theory without loosing the generality one can consider 
the non-renormalizable operators. In such a case the co-efficients of the non-renormalizable operators are suppressed by UV cut-scale $\Lambda_{UV}\sim M_{p}$ of the effective theory.}. One advantage of considering such a potential is to 
 obtain large tensor-to-scalar ratio, $r$,  which can be within the range of Planck and other future CMB B-mode polarization experiments. We will
 keep $V_0$ in this case, and  the potential simplifies to~\cite{Mazumdar:2011ih}:
\begin{equation}\label{h1a}
 V(\phi)=V_{0}+\frac{c_{H}H^{2}}{2}|\phi|^{2}-\frac{a_{H}H \phi^{6}}{6M^{3}_{p}}+\frac{|\phi|^{10}}{M^{6}_{p}}.
\end{equation}
where we have taken $\lambda=1$.
The potential admits inflection point for  $a_H^2\approx40 c_H^2$. We characterize the required fine-tuning by the 
quantity, $\delta$, defined as~\cite{Allahverdi:2006we}
\begin{equation}
\label{newbeta}
\frac{a_H^2}{40c_H^2} = 1-4\delta^2\,.
\end{equation}
When $\vert\delta\vert$ is small, a point of inflection $\phi_0$ exists such that $V^{\prime\prime}\left(\phi_0\right) =0$, with
\begin{equation}
\label{phi0}
\phi_0 = \left(\sqrt{\frac{c_H}{10}} H M_{p}^{3}\right)^{{1}/{4}}\,.
\end{equation}
For $\delta <1$, we can Taylor-expand the inflaton potential around the inflection point $\phi=\phi_{0}$ similar to 
 Eq.~(\ref{rt1}), where the coefficients are now given by:
\begin{eqnarray}\label{p1}
     \alpha&=&V(\phi_{0})=V_{0}+\left(\frac{4}{15}+\frac{4}{3}\delta^{2}\right)c_{H}H^{2}\phi^{2}_{0}+{\cal O}(\delta^{4}),\\
 \beta&=&V^{'}(\phi_{0})=4\delta^{2}c_{H}H^{2}\phi_{0}+{\cal O}(\delta^{4}),\\
 \gamma&=&\frac{V^{'''}(\phi_{0})}{3!}=\frac{c_{H}H^{2}}{\phi_{0}}\left(32-80\delta^{2}\right)+{\cal O}(\delta^{4}),\\
 \kappa&=&\frac{V^{''''}(\phi_{0})}{4!}=\frac{c_{H}H^{2}}{\phi^{2}_{0}}\left(384-1260\delta^{2}\right)+{\cal O}(\delta^{4}).
   \end{eqnarray}
Note that once we specify $c_H$ and $H$, all the terms in the potential can be determined. In this regard the potential 
indeed simplifies a lot to study the cosmological observables. 

One must also ensure that the vacuum energy density which generated the
 large cosmological constant in the first place vanishes by the end of slow-roll inflation.
 This typically happens in the case of hybrid inflation~\cite{Linde:1993cn}, and
 as discussed in~\cite{Enqvist:2010vd}. In the string
 landscape, or in the case of MSSM, this
 can happen through bubble nucleation, provided the rate of nucleation is such
 that $\Gamma_{nucl}\gg H$. In the latter case all the bubbles will belong to the
 MSSM vacuum---similar to the first order phase transition in the electroweak symmetry
 breaking scenario. However, one has to make sure that the cosmological
 constant disappears in the MSSM vacuum right at the end of inflation.


\subsection{Parameter estimation and CMB observables}

In this section our primary focus is to study the cosmological observables to match the CMB data for an  inflection-point inflation
whose potential is given by Eq.~(\ref{h1a}). 
Here the consistency relations are modified at 
the second order due to the presence of running. 
Cosmological parameter estimation can be done more precisely once we allow the higher order radiative corrections to the slow-roll parameters
 \cite{Easther:2006tv}, which we have listed in Appendix A (see Eqs.~(\ref{para 21a})-(\ref{para 21i})). In our case we have obtained the predicted power spectrum from the higher order radiative
 corrections to the slow-roll parameters. In order to illustrate this, let us consider the case when $H\gg m_\phi$. 
In this case there is a possibility of detecting large tensor-scalar ratio, $r$.


\begin{figure}[t]
\centering
\subfigure[~r vs $n_{S}$. By varying $H_{\star}$ we can probe a wide range of tensor-to-scalar ratio: $10^{-29}< r_{\star} \leq 0.12$.
The vertical line on the left corresponds to $N=50$, while the right line corresponds to $N=70$.]{
    \includegraphics[width=7cm,height=7cm] {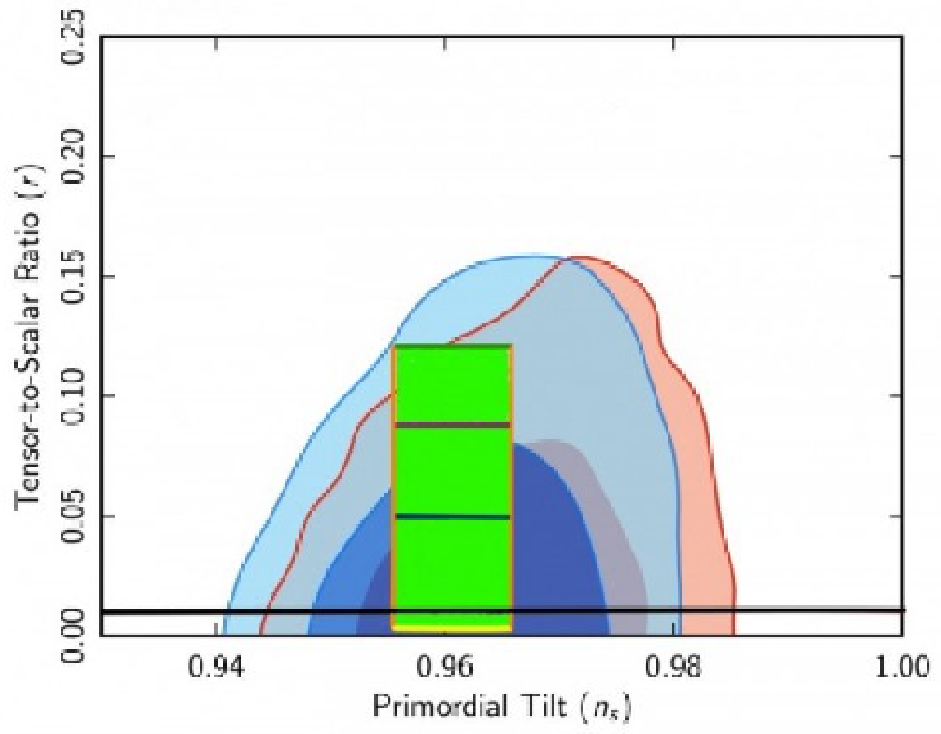}
    \label{x2}
}
\subfigure[~r vs $n_{S}$. By varying $H_{\star}$ we can probe a wide range of tensor-to-scalar ratio:
 $10^{-29}< r_{\star} \leq 0.12$. The vertical line on the left corresponds to $N=50$, while the right line corresponds to $N=70$.]{
    \includegraphics[width=7cm,height=7cm] {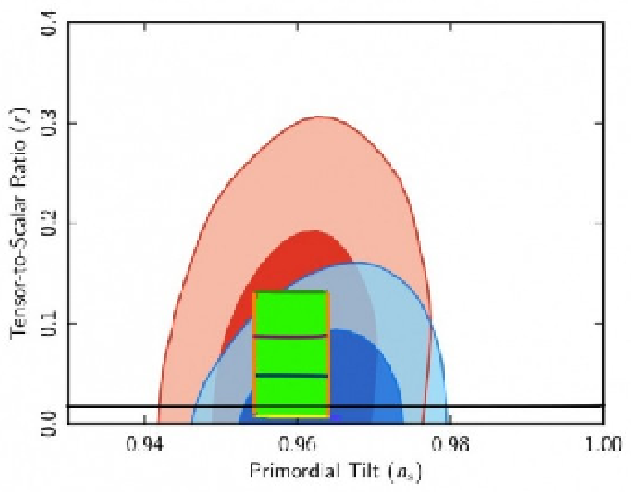}
    \label{x3}
}
\caption[Optional caption for list of figures]{ We show the joint $1\sigma$ and $2\sigma$ CL. contours in $r-n_{S}$ plane using
 \subref{x2}~Planck+WMAP-9 data with~$\Lambda$CDM+r(${\it Blue~region}$), and $\Lambda$CDM+$r+\alpha_{S}$(${\it Red~region}$),
 \subref{x3}~Planck+WMAP-9+BAO data with~$\Lambda$CDM+r(${\it Blue~region}$) and $\Lambda$CDM+$r+\alpha_{S}$(${\it Red~region}$) \cite{Choudhury:2013jya}.
The straight lines parallel to $n_{S}$ axis are drawn by varying the Hubble parameter $H_{\star}$ within
the range $10^{-1}~{\rm GeV}< H_{\star} \leq 9.241\times 10^{13}~{\rm GeV}$.
 The {\it deep green} line and the {\it yellow } line correspond to the upper and lower bound of $H_{\star}$ respectively. The {\it green}
 shaded region bounded by {\it orange} lines represent the allowed region obtained from the model.
 Additionally, the {\it black} thick line divides the low scale ($m_{\phi}\gg H$) and the
high scale ($H\gg m_{\phi}$) regions of inflation. 
}
\label{fig3}
\end{figure}


\begin{figure}[t]
\centering
\subfigure[~$\alpha_{S}$ vs $n_{S}$]{
    \includegraphics[width=7cm,height=7cm] {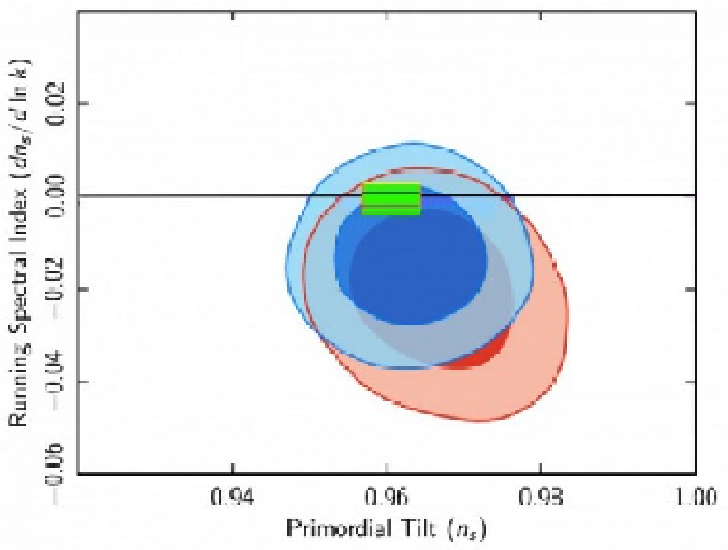}
    \label{z1}}
\subfigure[~$\kappa_{S}$ vs $\alpha_{S}$]{
    \includegraphics[width=7cm,height=7cm] {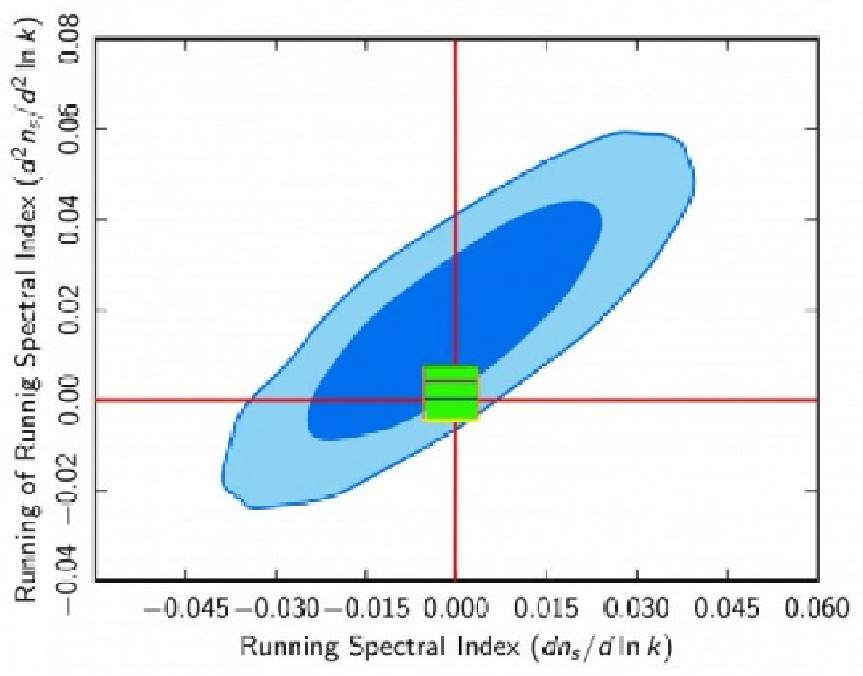}
    \label{z2}
}
\caption[Optional caption for list of figures]{ We show the joint $1\sigma$ and $2\sigma$ CL. contours in $\alpha_{S}-n_{S}$ and $\kappa_{S}-\alpha_{S}$ plane
using {\it Planck}+WMAP-9+BAO with \subref{z1}~$\Lambda$CDM+$\alpha_{S}$(${\it Blue~ region}$) and 
$\Lambda$CDM+$\alpha_{S}+r$(${\it Red~ region}$),
\subref{z2}~$\Lambda$CDM+$\alpha_{S}+\kappa_{S}$(${\it Blue~ region}$) background \cite{Choudhury:2013jya}. 
The straight lines parallel to $n_{S}$ axis are drawn by varying the Hubble parameter $H_{\star}$ within
the range $10^{-1}~{\rm GeV}< H_{\star} \leq 9.241\times 10^{13}~{\rm GeV}$.
 The {\it deep green} line and the {\it yellow } line correspond to the upper and lower bound of $H_{\star}$ respectively. The {\it green}
 shaded region bounded by {\it orange} lines represent the allowed region obtained from the model.
}
\label{fig4}
\end{figure}



We need to compute the pivot scale, $k_\ast$, when the relevant perturbations had left the Hubble patch during 
inflation. We can compute by expressing the number of e-foldings during inflation,
which is given by~\cite{Ade:2013uln}:
\be\begin{array}{llll}\label{efold}
\displaystyle N_{\star} &\approx & 71.21 - \ln \left(\frac{k_{\star}}{k_{0}}\right)  
+  \frac{1}{4}\ln{\left( \frac{V_{\star}}{M^4_{P}}\right) }
+\frac{1}{4}\ln{\left( \frac{V_{\star}}{\rho_{end}}\right) }  
+ \frac{1-3w_{int}}{12(1+w_{int})} 
\ln{\left(\frac{\rho_{rh}}{\rho_{end}} \right)},
\end{array}\ee
where $\rho_{end}$ is the energy density at the end of inflation, 
$\rho_{rh}$ is an energy scale during reheating, 
$k_{0}=a_0 H_0$ is the present Hubble scale, 
$V_{\star}$ corresponds to the potential energy when the relevant modes left the Hubble patch 
during inflation and $w_{int}$ characterizes the effective equation of state 
parameter between the end of inflation and the energy scale during reheating. For our model we have $w_{int}=1/3$ exactly for
which the contribution from the last term in Eq.~(\ref{efold}) vanishes. 
The resultant  {\it upper-bound} on the
reheat temperature at which all the MSSM {\it degrees of freedom} are in thermal equilibrium (kinetic and chemical equilibrium) is given by~\cite{Allahverdi:2011aj}
\begin{equation}\label{reh}
     T_{rh}=\left(\frac{30}{\pi^{2}g_{\star}}\right)^{\frac{1}{4}}\sqrt[4]{V_{\star}}\leq 6.654\times 10^{15}\sqrt[4]{\frac{r_{\star}}{0.12}}~{\rm GeV}.
\end{equation}
where we have used $g_{\star}=228.75$ (all the degrees of freedom in
MSSM). Since the temperature of the universe is so high, the lightest supersymmetric particle (LSP) relic density is then given by the 
standard (thermal) {\it freeze-out} mechanism~\cite{Jungman:1995df}. In particular, if
the neutralino is the LSP, then its relic density is determined
by its annihilation and coannihilation rates.
The advantage of realizing inflation in the visible sector is that it is possible to nail down the thermal 
history of the universe precisely~\footnote{At temperatures below $100$ GeV there will be no extra degrees of 
freedom in the thermal bath except that of the SM, therefore
BBN can proceed without any trouble.}.

For low scale models of inflation, i.e. $m_\phi \gg H$, the tensor modes are utterly negligible. For $m_\phi\sim 1$TeV,
and $\phi_0\sim 3\times 10^{14}$~GeV, the value of $H_{\ast} \sim 10^{-1}$~GeV, see Eq.(\ref{vev}). 
The estimation of the reheat temperature is given by the equality of the above 
Eq.~(\ref{reh}). The reheat temperature is typically $3\times 10^{8}$~GeV for the above parameters. Note that for $m_\phi\gg H$, 
the tensor to scalar ratio $r$ does not scale with the reheat temperature.

Note that saturating the upper-bound on $r\sim 0.12$ would yield a large reheating temperature of the universe. It is sufficiently large 
to create gravitino from a thermal bath.
The gravitino production from the direct decay of the inflaton will be suppressed.
In this case, the gravitino abundance is compatible with the BBN bounds, provided  the gravitino mass,
$m_{3/2}\geq {\cal O}(10)$~TeV, see~\cite{Moroi:1995fs} .The bound holds only for a decaying gravitino,  
for which the graviton will decay before the BBN. 
If gravitino happens to be the LSP,  then such a high scale model of 
inflation with large Hubble-induced corrections will be ruled out, unless there is some late entropy injection or 
there are some kinematical reasons for which the gravitino production is highly suppressed.

The Planck constraint implies that the tensor-to-scalar ratio, $r$, at the pivot scale $k=k_{\star}$, corresponds to an upper bound
on the energy scale of the Hubble induced inflection point inflation~\footnote{Here in eqn~(\ref{reh}) and eqn~(\ref{scale}) the equalities hold good in high scale inflationary regime ($H\gg m_{\phi}$). 
The inequalities are more significant once we enter low scale inflationary ($m_{\phi}\gg H$) region}:
\begin{equation}\label{scale}
     V_{\star}\leq (1.96\times 10^{16}{\rm GeV})^{4}\frac{r_{\star}}{0.12}\Rightarrow H_{\star}\leq 9.241\times 10^{13}\sqrt{\frac{r_{\star}}{0.12}}~{\rm GeV}\, .
   \end{equation}
  %
   %

Using this input we scan the model parameters for obtaining large tensor to scalar ratio, $r$, for the following values:
\begin{eqnarray}\label{P-space}
c_{H} \sim {\cal O}(1-10)\,, ~
a_{H}  \sim {\cal O}(10-100)\,,~
\lambda \sim  {\cal O}(1)\,, ~
\phi_{0} \sim  {\cal O}((1-3)\times 10^{16}GeV).
\end{eqnarray}
Now including the higher order corrections to the slow-roll parameters, 
the inflationary observables are estimated from our model as following:
\begin{eqnarray}
2.092<10^{9}P_{S}<2.297\,, ~~~
0.958<n_{S}<0.963\,, ~~~
r<0.12\,, \nonumber \\
-0.0098<\alpha_{S}<0.0003\,, ~~~
-0.0007<\kappa_{S}<0.006\,
\end{eqnarray}

which confronts the {\it Planck}+WMAP-9+BAO data set, well within $2\sigma$ CL.
Furthermore, we consider the following values of the model parameters which match the TT-spectrum of the CMB data for high scale model of 
inflation, i.e. $H\gg m_\phi$,
 \begin{eqnarray}\label{paraspace}
&&\delta\sim 10^{-4},~~~\lambda =1,~~~c_H=2,~~~a_H=12.650,
~~~ \phi_0=1.129\times 10^{16}~{\rm GeV}.
 \end{eqnarray}

In principle, we can vary $H_\ast$ from high scales to low scales. Since in our case the advantage is that the thermal history is well established, 
 we can trace the relevant number of e-foldings, 
 by varying $10^{-1}~{\rm GeV}< H_{\star} \leq 9.241\times 10^{13}~{\rm GeV}$ and consequently we can probe tensor-to-scalar ratio for a wide range: $10^{-29}< r_{\star} \leq 0.12$.

Using Eq.~(\ref{paraspace}), in Fig.~(\ref{fig2}) we have shown the behavior of the amplitude of the the power spectrum, $P_{S}$ with respect to spectral tilt, $n_{S}$
 at the pivot scale $k_{\star}=0.002~{\rm Mpc}^{-1}$ by a {\it red} curve. If we take care of the full parameter
space, see Eq.~(\ref{P-space}), there are solutions which have been shown in a black shaded region.
Furthermore, by using {\it Planck}+WMAP-9 and {\it Planck}+WMAP-9+BAO datasets
 with $\Lambda$CDM background
 along with different  combined constraints, we have shown the status of inflection point inflationary model in the 
 marginalized $1\sigma$ and $2\sigma$ CL.
 contours  in Fig.~(\ref{fig3}). The allowed region from the model is explicitly shown by the {\it green} shaded
 region bounded by {\it orange} vertical lines parallel to $r$-axis. Along the vertical lines 
the number of e-foldings varies within $50<N_{\star}<70$ (from left to right) for various ranges of $H_\ast$ as mentioned earlier.
 We have also shown a thick {\it black} line parallel to $n_{S}$ axis in Fig.~(\ref{fig3}) which 
discriminates between the low scale $(m_{\phi}\geq  H)$ and high scale $(H\gg m_{\phi})$ inflationary scenarios. 
Additionally, we have depicted various straight 
lines for the intermediate values of $H_{\star}$ within the allowed region. The model also provides very mild running, $\alpha_{s}$, 
 and running of running, $\kappa_{S}$, which is also shown in the marginalized $1\sigma$ and $2\sigma$ CL. contours in Fig.~(\ref{fig4}), 
 where we have  used {\it Planck}+WMAP-9 and {\it Planck}+WMAP-9+BAO datasets with $\Lambda$CDM background along with different combined observational constraints.




\begin{figure}[t]
\centering
\includegraphics[width=15cm,height=7cm]{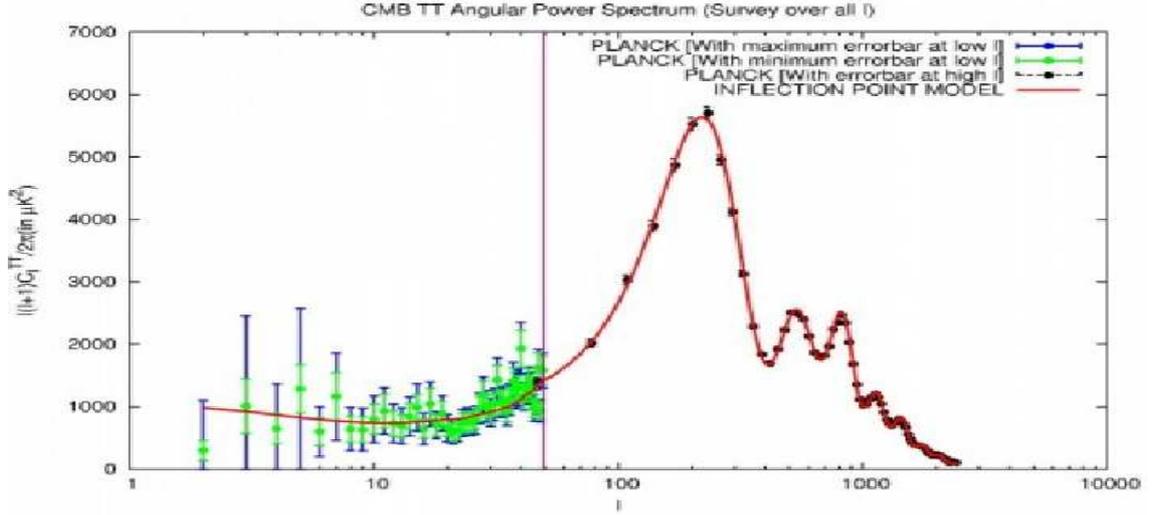}
\caption{TT-power spectrum for $\ell$~($2<l<2500$) \cite{Choudhury:2013jya}. The vertical line is drawn at $l=50$ which separates the
 low-$l$ ($2<l<50$) and high-$l$ ($50<l<2500$) region. Here the TT-power spectrum is drawn for the parameter values mentioned
in Eq.~(\ref{paraspace}) in the context of high scale ($H\gg m_{\phi}$) Hubble induced inflationary framework.}
\label{fig6}
\end{figure}

In this section we study the TT-angular power spectrum for the CMB anisotropy. For our present setup 
at low $\ell$ region $(2<l<49)$ the contributions from the running ($\alpha_{S},\alpha_{T}$), and running of running ($\kappa_{S},\kappa_{T}$) are very small.
Consequently their additional contribution to the power spectrum for scalar and tensor modes becomes unity $(\sim {\cal O}(1))$ (this is consistent with 
the initial condition at the pivot scale $k=k_{\star}$) and the original
 power spectrum becomes unchanged. As a result the proposed model will be well fitted with the {\it Planck } low-$l$ data within 
 high cosmic variance except for a few outliers. On the other hand, when we move
 towards high $\ell$ regime ($50<l<2500$) the contribution of running and running of running become stronger and
 this will enhance the power spectrum to a permissible value such that it will accurately fit {\it Planck} high-$l$ 
data within very small cosmic variance. In this way one can easily survey over all the multipoles starting from low-$l$ to high-$l$ using the
same parameterizations as mentioned in Eqs.~(\ref{paraspace}). From Figs.~(\ref{fig6}), we see that the Sachs-Wolfe plateau obtained
from our model is non flat, confirming the appearance of running, and running of the running in the spectrum 
observed for low $l$ region ($l<50$).
 For larger value of the multipole ($50<l<2500$),
CMB anisotropy spectrum is dominated by the Baryon
Acoustic Oscillations (BAO), giving rise to several ups and
downs in the spectrum, see~\cite{Hu:1996qs}.   Note  that high $l$ regions of our model are well fitted within the 
small cosmic variance observed by {\it Planck}. In the low $l$ region due to the presence of very 
large cosmic variance there may be other pre-inflationary scenarios which might be able to fit the TT-power spectrum better. 


\section{Chapter summary}
\label{c4}

In this chapter by implementing the saddle point and inflection point mechanism we have proposed two different models of inflation 
in the framework of MSSM constructed from various D-flat directions. The major outcomes of our study are:
\begin{itemize}
 \item We have demonstrated how we can construct the
 effective inflationary potential in the vicinity of the {\it saddle point} and {\it inflection point} starting
 from $n=4$ and $n=6$ level superpotential for the D-flat direction content
 \textbf{QQQL,~QuQd,~QuLe,~uude} and \textbf{udd,~LLe} respectively.

\item The effective inflaton potential
 around saddle point and inflection point, has then been utilized in estimating 
 the observable parameters and confronting them with
WMAP7 and Planck dataset using the publicly available code CAMB \cite{CAMB}, which reveals consistency of our model with latest observations.

\item  We have then explored the possibility of Primordial Black Hole formation from the 
running-mass model by estimating the mass of PBH from the model derived from $n=4$ flat directions.

\item Subsequently, we have engaged ourselves in finding out the effective parameter space
and the coupling constants appearing in the {\it saddle point} analysis for the MSSM inflation
 by exactly solving the one loop RGE in this context.

\item In case of inflection-point inflation 
we have taken potentials from two situations where the SUGRA corrections are important and as well as negligible. In the former case,
we yield significantly large tensor-to-scalar ratio, $r\leq 0.12$, for $H_\star\sim 9.24\times 10^{13}$~GeV and 
the VEV $\phi_0 =1.12\times 10^{16}$~GeV. The model tends to predict a perfect match for the spectral tilt
even for a small tensor-to-scalar ratio, $r$ for the situation where SUGRA corrections are negligible. 

\item The {\it inflection point} model fits the amplitude of the power spectrum and the spectral tilt. The model predicts mild running and running of the running 
of the spectral tilt well within the $2\sigma$ uncertainties.

\item In particularly, the high scale inflection-point inflation model fits the high 
$l$ multipoles of the {\it Planck} data quite well with the $\Lambda$CDM parameters.
Inflection point technique plays another crucial role in fitting Planck data 
for low $l$ in CMB. In CMB TT spectra the low $l$ multipoles have high uncertainties and they are
within the cosmic variance. The forthcoming polarization data from
Planck will hopefully further constrain the inflection-point model of inflation.

\item The perturbations created from the slow roll evolution of the inflaton are Gaussian and adiabatic. The amplitude and the spectral tilt  match 
very well with the {\it Planck} data. One of the advantages of the proposed model is that it is embedded fully within MSSM, and therefore, it predicts 
the right thermal history of the universe with {\it no extra relativistic degrees of freedom} other than that of the  Standard Model.

\end{itemize}



\chapter{Inflation from background supergravity}
\label{ch:FRWxc1}
\section{Introduction}
\label{b1}

Investigations for the crucial role of Supergravity (SUGRA) in explaining
cosmological inflation date back to early eighties of the last
century. Inflation can be caused by the potential energy of a scalar field. Such a potential must be relatively flat in order
to guarantee long duration of inflation and small deviation of scale invariance of primordial density fluctuations.
However, the flatness of the scalar potential can be easily destroyed by radiative corrections. 
Supersymmetry (SUSY) is one of the leading theoretical proposals to protect a scalar field from radiative corrections, which also gives an attractive
solution to the hierarchy problem of the standard model (SM) of particle physics as well as the unification
of the strong and electro-weak gauge couplings. In particular, its local version, SUGRA, would govern the dynamics of the early
Universe. However, in fact, it is a non-trivial task to incorporate inflation in SUGRA. 
This is primarily due to the well-known $\eta$-problem of SUGRA inflation, which appears in
the F-term inflation due to the fact that the energy scale of
F-term inflation is induced by all the couplings via vacuum energy
density.  Precisely, in the expression of F-term inflationary
potential a factor $\exp\left(K/M_{p}\right)$ appears, leading to
the second slow roll parameter $\eta \gg 1$, thereby violating an
essential condition for slow roll inflation. The usual wayout is
to impose additional symmetry to the framework. One such symmetry
is Nambu-Goldstone shift symmetry \cite{Kawasaki:2000yn,Yamaguchi:2000vm} under which
K$\ddot{a}$hler metric becomes diagonal which serves the purpose
of canonical normalization and stabilization of the volume of the
compactified space. Consequently, the imaginary part of the scalar
field gives a flat direction leading to a successful model of
inflation. An alternative approach is to apply non-compact
Heisenberg group transformations of two or more complex scalar
fields where one can exploit Heisenberg symmetry \cite{Binetruy:1987xj}
to solve $\eta$-problem. In addition,
it prevents a scalar field from acquiring a value larger than $M_p$. This fact implies that it is almost impossible to
realize large field inflation like chaotic inflation in SUGRA.

Of late the idea of braneworlds came forward \cite{Randall:1999ee,Randall:1999vf} as elaborately discussed in chapter \ref{ch:FRW}. From
cosmological point of view the most appealing feature of brane
cosmology is that the 4 dimensional Friedmann equations are to
some extent different from the standard ones due to the
non-trivial embedding in the $S^{1}/Z_{2}$ orbifold
\cite{Maartens:2010ar}. This opens up new perspectives to look at the
nature in general and cosmology in specific. Brane inflation in the above
framework has also been studied to some extent in Refs.~\cite{Bridgman:2001mc,Maartens:2000fg}~\footnote{There are some other approaches as well 
which are more appealing in dealing with
fundamental aspects such as possible realization in string theory
can be found in Refs.~\cite{Jones:2002cv,Nilles:2001my}. For
example, an apparent conflict between self-tuning mechanism and 
volume stabilization has been shown in \cite{Forste:2000ps}, subsequently, this problem
has been resolved in \cite{Forste:2000ft} where
the credentials of the dilatonic field in providing a natural explanation for dark energy by an effective scalar field
on the brane has been demonstrated using self-tuning mechanism in six dimensional bulk.}
In the same
vein, we construct the  brane inflaton potential of our
consideration starting from  5D SUGRA. 
In brane inflation the modified Friedman equations lead to a
modified version of the slow roll parameters \cite{Maartens:2010ar}.
So, by construction, $\eta$-problem is smoothened to some extent 
by modification of Friedmann equations on the brane \cite{Dvali:1994ms}. In a sense, this is a parallel
approach to the usual string inflationary framework where $\eta$-problem is resolved by
fine-tuning  \cite{Kachru:2003sx}. As it will appear, there is still some fine-tuning required in brane inflation, which
arises via a new avatar of five-dimensional Planck mass  
but it is softened to some extent due to the modified Friedman equations.

On the other hand, in higher-dimensional setups as in the
 case of {\it DGP model} \cite{Dvali:2000hr} where self-acceleration
is sourced by a scalar field, Infra-Red (IR) modification of gravity \cite{ArkaniHamed:2003uy} play 
a crucial role. Despite its profound success it has got some serious limitations
 \cite{DeFelice:2010aj}, which are resolved by introducing a dynamical field, {\it aka}, Galileon \cite{DeFelice:2010pv}~\footnote{The cosmological consequences of the Galileon models have 
been studied in Ref.~\cite{Silva:2009km,Chow:2009fm}.}
 arising on the brane from the bulk in the DGP setup. Very recently, a natural extension to the scenario has been brought forward by tagging Galileon with the 
good old DBI model \cite{Cederwall:1996uu,Copeland:2010jt}, resulting in ``DBI Galileon'' \cite{deRham:2010eu},
 that has reflected a rich structure from four dimensional cosmological point
of view. However, in most of the physical situations, this type of {\it effective}
gravity theories are plagued with additional degrees of freedom which often results in unwanted debris like ghosts, Laplacian instabilities etc \cite{DeFelice:2011bh}.   
In the second portion of this chapter we introduce a single scalar field model described by the D3 DBI Galileon originated
 from D4-$\bar{\rm D4}$ brane anti-brane setup in the background of 5D SUGRA. 
This prevents the framework from having extra degrees of freedom as
well as {\it Ostrogradski instabilities} \cite{Woodard:2006nt}.
Nevertheless, a consistent field theoretic derivation of the effective potential commonly used in the context of DBI Galileon cosmology has not been brought forth so far.
On top of that, it is imperative to point out that the SUGRA origin of D3 DBI Galileon is yet to be addressed. 
In this chapter we plan to address both of these issues explicitly by deriving the inflaton potential from our proposed framework of DBI Galileon.
 Moreover, 
in general appearance of non-vanishing frame functions 
in the 4D action expedites breakdown of shift symmetry. Without shift symmetry, it may happen that the theory is unstable
against large renormalization.
The background action chosen in our model preserves shift symmetry of the single scalar field which gives it a firm footing from phenomenological point of view as well.

The plan of this chapter is as follows.
First we propose a fairly general framework in the background of bulk ${\cal N}$=2, ${\cal D}$=5 SUGRA 
by taking the Randall Sundrum braneworld scenario including Einstein's Hilbert term in the gravity sector and the full DBI action 
 in D4 brane including the quadratic modification
in Einstein's Hilbert action via Gauss-Bonnet correction in the bulk.
 Hence, using dimensional reduction technique, we derive the effective action for brane inflation and DBI Galileon in D3 brane induced 
by the quadratic correction in the geometry sector
in the background of ${\cal N}$=1, ${\cal D}$=4 SUGRA
and study cosmological inflationary scenario therefrom.
We next engage ourselves to calculate the primordial power spectrum of the scalar and tensor modes, their running and other observable parameters for both the frameworks.
 We further confront our proposed models with observations by using the publicly available code CAMB \cite{CAMB}, and 
find them to fit well with observational data from WMAP7 \cite{WMAP7}.

\section{Brane inflation}
\label{b2}

\subsection{The background model in ${\cal N}=2,{\cal D}=5$ supergravity}
\label{b2a}
For systematic development of the formalism, let us  demonstrate
briefly how one can construct the effective 4D inflationary
potential of our consideration starting from ${\cal N}=2, {\cal D}=5$ SUGRA in
the bulk which leads to an effective ${\cal N}=1, {\cal D}=4$ SUGRA in the brane. As
mentioned, we consider the bulk to be five dimensional where the
fifth dimension is compactified on the orbifold $S^{1}/Z_{2}$ of
comoving radius R. The system is described by the following action \cite{Riotto:2001hu}:
 \be\label{as1} S=\frac{1}{2}\int d^{4}x\int^{+\pi
R}_{-\pi
R}dy\sqrt{-g_{5}}\left[M^{3}_{5}\left(R_{(5)}-2\Lambda_{5}\right)
+{\cal L}_{bulk}+\sum_{i}\delta(y-y_{i}){\cal L}_{4i}\right]\ee
where $M_{5}$ be the 5D quantum gravity scale, $\Lambda_{5}$ be the 5D bulk cosmological constant and ${\cal L}_{bulk}$ contains bulk field contents.
 Also the sum includes the walls at the orbifold points
$y_{i}=(0,\pi R)$ and 5-dimensional coordinates
$x^{m}=(x^{\alpha},y)$, where $y$ parameterizes the extra
dimension compactified on the closed interval $[-\pi R,+\pi R]$
and $Z_{2}$ symmetry is imposed.
The metric in ${\cal D}=5$
in conformal form is given by,
\be\label{metrix}ds^{2}_{5}=g_{mn}dx^{m}dx^{n}=e^{2A(y)}\left(ds^{2}_{4}+R^{2}\beta^{2}dy^{2}\right),\ee
where the ${\cal D}=4$ metric $ds^{2}_{4}=g^{\alpha\beta}dx_{\alpha}dx_{\beta}$ is the well known FLRW metric and $g_{5}=det(g_{mn})$.
 The numerical constant $\beta$ has been introduced just for convenience and physically determines the slope of the 
warp factor $e^{2A(y)} $. Also the product $R\beta$ stands for compact dimension which will stabilize the modulus fields appearing in the present context. 
Further solving ${\cal D}=5$ Einstein Equations the warp factor can be expressed as:
\be\label{warp}e^{2A(y)}=\frac{b^{2}_{0}}{R^{2}\left(e^{\beta y}+\frac{\Lambda_{5}b^{4}_{0}}{24R^{2}}e^{-\beta y}\right)},\ee
 where $b_{0}$ is a constant having dimension of length.

For ${\cal N}=2, {\cal D}=5$
SUGRA in the bulk Eq (\ref{as1}) can be written as \cite{Choudhury:2011sq}:
\be\label{su1}
 S=\frac{1}{2}\int d^{4}x\int^{+\pi
R}_{-\pi R}dy\sqrt{-g_{5}}\left[M^{3}_{5}\left(R_{(5)}-2\Lambda_{5}\right)+{\cal L}^{(5)}_{SUGRA}+\sum_{i}
\delta(y-y_{i}){\cal L}_{4i}\right],\ee which is a generalization of the
scenario described in \cite{Riotto:2001hu}. Written explicitly, the
contribution from bulk SUGRA in the action is given by:
\be\label{sug2}e^{-1}_{(5)}{\cal L}^{(5)}_{SUGRA}=-\frac{M^{3}_{5}R^{(5)}}{2}+\frac{i}{2}\bar{\Psi}_{i\tilde{m}}
\Gamma^{\tilde{m}\tilde{n}\tilde{q}}\nabla_{\tilde{n}}
\Psi^{i}_{\tilde{q}}-{S}_{IJ}F^{I}_{\tilde{m}\tilde{n}}F^{I\tilde{m}\tilde{n}}-\frac{g_{\mu\nu}}{2}
(D_{\tilde{m}}\phi^{\mu})(D^{\tilde{m}}\phi^{\nu})
$$$$ ~~~~~~~~~~~~~~~~~~~~~~~~~~~~~~~~~~+ {\rm Fermionic} + {\rm Chern-Simons},\ee Including the
contribution from the radion fields:
\be\begin{array}{lll}\label{we1}\displaystyle \chi = -\psi_{5}^{2},~~~T
= \frac{1}{\sqrt{2}} \left(e_{5}^{\dot 5} - i \sqrt{\frac{2}{3}}
A_{5}^{0} \right)\end{array}\ee the effective brane SUGRA counterpart turns out
to be~\footnote{In this context $\Delta(y)=e^{5}_{\dot{5}}\delta(y)$ is the modified Dirac
delta function which satisfies the normalization
conditions: \be\begin{array}{lll}\label{we2}\displaystyle   \int^{+\pi R}_{-\pi R}dy
~e^{5}_{\dot{5}}\Delta(y)=1, ~~ \int^{+\pi R}_{-\pi
R}dy~e^{5}_{\dot{5}}= {\cal V}_{5},\end{array}\nonumber\ee where ${\cal V}_{5}$ is the 5
dimensional volume.}\be\label{lsug4}
\delta(y)L_{4}=-e_{(5)}\Delta(y)\left[(\partial_{\alpha}\phi)^{\dagger}
(\partial^{\alpha}\phi)+i\bar{\chi}\bar{\sigma}^{\alpha}D_{\alpha}\chi\right].\ee
 The Chern-Simons terms can be gauged away
assuming cubic constraints and $Z_2$
symmetry.  It is useful to define the five
dimensional generalized $K\ddot{a}hler$ function ($G$) in this context as:
\be\begin{array}{lll}\displaystyle
G=-3\ln\left(\frac{T+T^{\dagger}}{\sqrt{2}}\right)+\delta(y)\frac{\sqrt{2}}{T+T^{\dagger}}K(\phi,\phi^{\dagger}),\end{array}\ee 
 which precisely represents  interaction of the radion with gauge
fields. Including the kinetic term of the five dimensional field
$\phi$ the singular terms  measured from the modified Dirac delta
function can be rearranged into a perfect square thereby leading
to the following expression for the action \be\label{modsug}
S\supset\frac{1}{2}\int d^{4}x\int^{+\pi R}_{-\pi
R}dy\sqrt{-g_{5}}e_{(4)}e^{5}_{\dot{5}}\left[g^{\alpha\beta}G_{m}^{n}(\partial_{\alpha}\phi^{m})^{\dagger}(\partial_{\beta}\phi_{n})
+\frac{1}{g_{55}}\left(\partial_{5}\phi-\sqrt{H(G)}\Delta(y)\right)^{2}\right],\ee
where the bulk F-term potential in terms of generating function  can be written as:
\be H(G)=\exp\left(\frac{G}{M^{2}_{p}}\right)\left[\left(\frac{\partial
W}{\partial \phi_{m}}+\frac{\partial G}{\partial
\phi_{m}}\frac{W}{M^{2}_{p}}\right)^{\dagger}(G_{m}^{n})^{-1}\left(\frac{\partial
W}{\partial \phi^{n}}+\frac{\partial G}{\partial
\phi^{n}}\frac{W}{M^{2}_{p}}\right)-3\frac{|W|^{2}}{M^{2}_{p}}\right].\ee
In this context we introduce the 4D Planck scale ($M_p$), which can be expressed in terms of the 5D scale ($M_5$) as:
\be\begin{array}{llll}\displaystyle M_{p}=\sqrt{\frac{e_{4}}{b_{0}}}=\sqrt{\frac{6e_{(5)}}{\lambda}}
=\sqrt{\frac{3}{4\pi \lambda}}M^{3}_{5}\end{array}\ee
where $\lambda=\frac{\Lambda_{5}b^{4}_{0}}{24R^{2}}$.
Further, imposing $Z_{2}$ symmetry to $\phi$ via
$\phi(0)=\phi(\pi
R)=0$ and compactifying around a circle $(S^1)$ by imposing the constraint condition,
$\partial_{5}\phi=\sqrt{H(G)}\left(\Delta(y)-\frac{1}{2\pi
R}\right)$ we get,

\be\label{tout}
S=\frac{1}{2}\int d^{4}x\int^{+\pi R}_{-\pi
R}dy\sqrt{-g_{5}}\left[M^{3}_{5}\left(R_{(5)}-2\Lambda_{5}\right)
+e_{(4)}e^{5}_{\dot{5}}\left\{g^{\alpha\beta}G_{m}^{n}(\partial_{\alpha}\phi^{m})^{\dagger}(\partial_{\beta}\phi_{n})
-g^{55}\frac{H(G)}{4\pi^{2}R^{2}}\right\}\right].\ee
 Now to trace out all the significant contribution from the fifth dimension using dimensional reduction technique
here we use method of separation of variable $\phi^{m}=\phi(x^{\mu},y)=\phi(x^{\mu})\chi(y)$ which leads to,

\be\begin{array}{lllll}\label{ast8}\displaystyle S=
\frac{M^{2}_{p}}{2}\int d^{4}x \sqrt{-g_{4}}\left[R_{(4)}-P\int^{+\pi R}_{-\pi R}dy \frac{4(3e^{2\beta y}+3\lambda^{2}e^{-2\beta y}-2\lambda)}{R^{2}(e^{\beta y}+\lambda
 e^{-\beta y})^{5}}\right.\\ \left.~~~~~~~~~~~~~~~~~~~~~~~~~~~~~~~~~~~~~~~~\displaystyle ~~~~~~~~~~~~~~+\left(\frac{\partial^{2}
K}{\partial\phi^{\dagger}_{\mu}\partial\phi^{\nu}}\right)(\partial_{\alpha}\phi^{\mu})^{\dag}(\partial^{\alpha}\phi_{\nu})-QV_{F}
\right].\end{array}\ee
where we define: 
\be\begin{array}{llll}\displaystyle 
P=\frac{2M^{3}_{5}\beta b^{6}_{0}}{M^{2}_{p}R^{5}},~~~~~~~~~~~~~~Q=\frac{C(T,T^{\dag})}{4\pi^{2}R^{2}}.
\end{array}\ee  Here
$C(T,T^{\dagger})$  represents an arbitrary
function of stabilized  modulus $T$ and $T^{\dagger}$. 
Eqn(\ref{ast8}) explicitly shows that the theory is reduced to an effective ${\cal N}=1, {\cal D}=4$ SUGRA theory.
For a general physical situation of ${\cal N}=1, {\cal D}=4$
SUGRA in the brane where the F-term potential on the brane
defined earlier is modified as
 \cite{Nilles:1983ge}:
\be\label{vf}
V_{F}=\exp\left(\frac{K(\phi,\phi^{\dagger})}{M^{2}_{p}}\right)\left[\left(\frac{\partial W}{\partial
\Psi_{\alpha}}+\frac{\partial K}{\partial
\Psi_{\alpha}}\frac{W}{M^{2}_{p}}\right)^{\dag}\left(\frac{\partial^{2}K}{\partial\Psi^{\alpha}\partial\Psi^{\dagger}_{\beta}}\right)^{-1}\left(\frac{\partial W}{\partial
\Psi^{\beta}}+\frac{\partial K}{\partial
\Psi^{\beta}}\frac{W}{M^{2}_{p}}\right)-3\frac{|W|^{2}}{M^{2}_{p}}\right].\ee
  Here $\Psi^{\alpha}$ is the chiral superfield and $\phi^{\alpha}$
be the complex scalar field. From now on the inflaton field $\phi$ appears to be
4-dimensional as demonstrated earlier.
 Finally in the canonical basis~\footnote{In this context we assume that
the K$\ddot{a}$hler potential is dominated by the leading order
term in canonical basis of the series representation 
i.e. $K= \sum_{\alpha}\phi^{\dagger}_{\alpha}\phi^{\alpha}$.
The superpotential in Eq~(\ref{vf}) is given by 
$W=\sum^{\infty}_{n=0}D_{n}W_{n}(\phi^{\alpha})$
 with the constraint $D_{0}=1$. Here $W_{n}(\phi^{\alpha})$ is
a holomorphic function of $\phi^{\alpha}$ in the complex plane.} Eq.~(\ref{ast8}) takes the following form~\footnote{In Eq~(\ref{astre}) the second term plays the role of
effective cosmological constant in 4D. For further discussion we absorb this contribution to the scale of the effective potential $\Delta$.}:
\be\label{astre}S=\frac{M^{2}_{p}}{2}\int d^{4}x \sqrt{-g_{4}}\left[
R_{(4)}-P\int^{+\pi R}_{-\pi R}dy \frac{4(3e^{2\beta y}+3\lambda^{2}e^{-2\beta y}-2\lambda)}{R^{2}(e^{\beta y}+\lambda
 e^{-\beta y})^{5}}+(\partial_{\alpha}\phi^{\mu})^{\dag}(\partial^{\alpha}\phi_{\mu})-QV_{F}
\right],\ee
where
the F-term potential can be recast as~\footnote{  
In this analysis we assume that the $U(1)$ gauge interaction is absent, which implies $V_{D} = 0$.} \cite{Bento:2002kp}: \be\label{totalpot}
V=V_{F}=\exp\left[\frac{1}{M^{2}_{p}}\sum_{\alpha}\phi^{\dagger}_{\alpha}\phi^{\alpha}\right]\left[\sum_{\beta}\left|\frac{\partial
W}{\partial \phi_{\beta}}\right|^{2}-3\frac{|W|^{2}}{M^{2}_{p}}\right].\ee
Next we expand the slowly varying inflaton potential derived from
F-term around the value of the inflaton field where the quantum
fluctuation is governed by, $\phi \rightarrow \tilde{\phi}+\phi$,
($\tilde{\phi}$ being the value of the inflaton field where
structure formation occurs). Further we impose $Z_{2}$ to remove all odd order terms responsible for
gravitational instabilities. Finally, the required renormalizable inflaton potential turns
out to be: \be V(\phi)=\Delta^{4}\sum^{2}_{m=0}C_{2m}\left(\frac{\phi}{M_{p}}\right)^{2m},\ee
with an additional constraint on the tree-level constant $C_{0}=1$. The mass term decides the steepness of the potential.
 Absence of this term indicates that process is slow which is compensated by brane
 tension in the braneworld scenario. For the phenomenological purpose this specific choice 
is completely viable. Now translating
the momentum integral within a specified UV cut-off ($\Lambda$) the effective potential turns out to be:
\be\label{litl}
V(\phi)=\Delta^{4}+\frac{g}{4!}\phi^{4}+\frac{g^{2}\phi^{4}}{(16\pi)^{2}}
\left[\ln\left(\frac{\phi^{2}}{\Lambda^{2}}\right)-\frac{25}{6}\right]+{\cal O}(\lambda^{3}),\ee
where the coupling constant $g=\frac{24\Delta^{4}C_{4}}{M^{4}}$ which 
is, in general at the scale $M$,  defined as:
\be\begin{array}{llll}\displaystyle
g(M)=\left[\frac{d^{4}V(\phi)}{d\phi^{4}}\right]_{\phi=M}=g+\frac{6g^{2}}{(8\pi)^{2}}\ln\left(\frac{M^{2}}{\Lambda^{2}}\right)+{\cal O}(g^{3}),\end{array}\ee
 so that the general expression
 for the effective potential  in terms of all finite physical parameters is given by:
 \be\label{hhui}
V(\phi)=\Delta^{4}+\frac{g(M)}{4!}\phi^{4}+\frac{g^{2}(M)\phi^{4}}{(16\pi)^{2}}\left[\ln\left(\frac{\phi^{2}}{M^{2}}\right)-\frac{25}{6}\right]+{\cal O}(g^3(M)).\ee
which is the Coleman Weinberg potential \cite{Coleman:1973jx}. 
After substituting the expression for $g$ in terms of $C_{4}$ the one loop corrected potential can be expressed at the mass scale $M\sim M_{p}$ as: 
\be\label{post}V(\phi)=\Delta^{4}\left[1+\left\{D_{4}+K_{4}\ln\left(\frac{\phi}{M}\right)\right\}
\left(\frac{\phi}{M}\right)^{4}\right],\ee where the one-loop co-efficients are given by:
$K_{4}=\frac{9\Delta^{4}C^{2}_{4}}{2\pi^{2}M^{4}} , ~~
D_{4}=C_{4}-\frac{25K_{4}}{12}.$
 


\begin{figure}[t]
\centering
\subfigure[~$V(\phi)$ vs $\phi$]{
    \includegraphics[width=7.5cm,height=5cm] {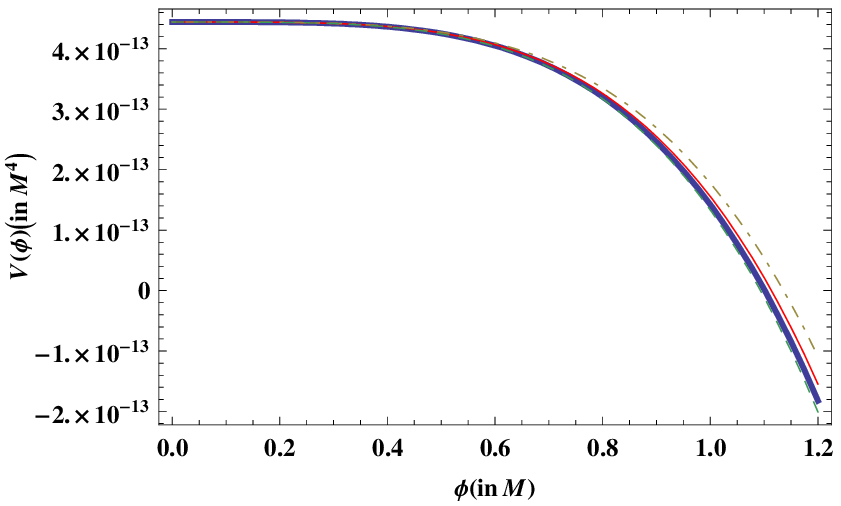}
    \label{figVr845}}
\subfigure[~1-$|\eta_{V}|$ vs inflaton field $\phi$]{
    \includegraphics[width=7.5cm,height=5cm] {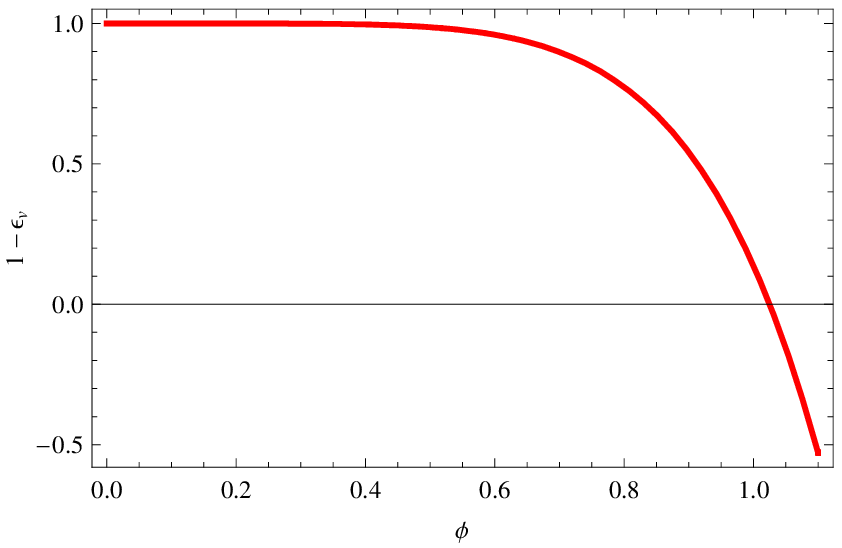}
    \label{figVr9}
}
\subfigure[~$N$ vs $\phi$]{
    \includegraphics[width=9.5cm,height=5cm] {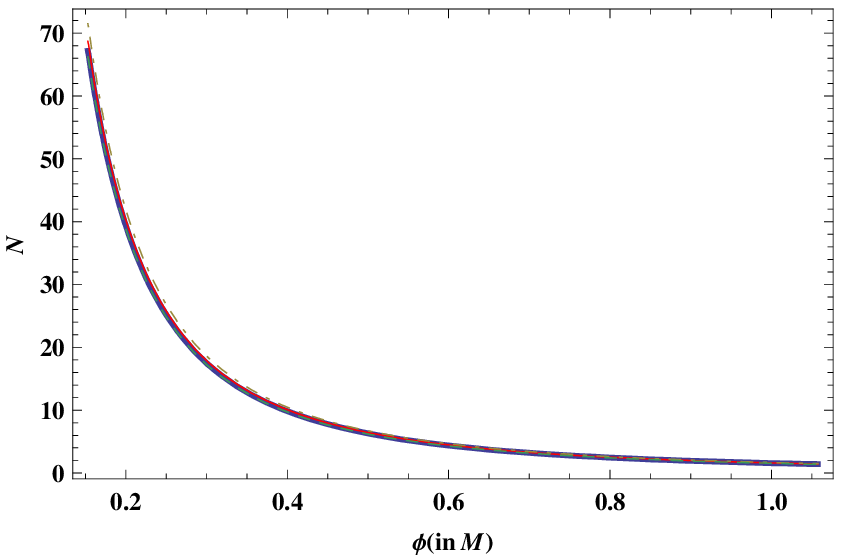}
    \label{figVr241}
}
\caption[Optional caption for list of figures]{ Variation of \subref{figVr845} one loop corrected potential ($V(\phi)$) vs inflaton field ($\phi$) \cite{Choudhury:2011sq},
  \subref{figVr9} 1-$|\eta_{V}|$ vs inflaton field $\phi$ for $C_{4}=-0.68$ \cite{Choudhury:2011sq} and \subref{figVr241} number of e-foldings ($N$) vs inflation field ($\phi$) within the
 range $-0.70<D_{4}<-0.60$ \cite{Choudhury:2011sq}.
}
\label{vfcv1}
\end{figure}


Fig.~(\ref{figVr845}) represents the inflaton potential for
different values of $C_{4}$, $D_{4}$ and $K_{4}$. From the observational constraints
the best fit model is given by the range $-0.70<D_{4}<-0.60$  so
that while doing numerical analysis we shall restrict ourselves to this
range of $D_{4}$. In what follows our primary intention will be to
engage ourselves in modeling brane inflation and to search for its
pros and cons with the derived effective potential. 

\subsection{Modeling brane inflation in ${\cal N}=1,{\cal D}=4$ supergravity}
\label{b3a}

The most appealing feature of brane
cosmology is that the 4D Friedmann equations are to
some extent different from the standard ones due to the
non-trivial embedding in the $S^{1}/Z_{2}$ manifold
\cite{Maartens:2010ar}. At high energy
regime one can neglect the contribution from Weyl term and consequently, the brane Friedmann
equations within slow-roll regime are given by: \begin{eqnarray} H^{2}&=&\frac{
V(\phi)}{3M^{2}}\left(1+\frac{V(\phi)}{2\lambda}\right),\\
\dot{H}+H^2 &=&\frac{
V(\phi)}{3M^{2}}\left(1+\frac{V(\phi)}{2\lambda}\right).
\end{eqnarray}
In the high energy regime the contribution of energy density of the scalar field is significant compared to the brane tension i.e.  
$\rho\sim V(\phi)>>\lambda$ within slow-roll. Consequently the Friedmann Eqn can be modified as, $H=V(\phi)/\sqrt{6\lambda}M$. On the other hand,
in the GR limit the energy density of the inflaton field is suppressed compared to the brane tension i.e. $\rho\sim V(\phi)<<\lambda$ and 
we get the usual Friedmann Eqn, $H=\sqrt{V(\phi)}/\sqrt{3}M$. In the braneworld scenario the modified
Freidmann equations, along with the Klein Gordon equation, lead to
new slow roll conditions  and new
expressions for observable parameters as well.
 Incorporating the potential of our consideration from
Eq (\ref{post}) the
 slow roll parameters turn out to be~\footnote{For convenience throughout the analysis we define the following functions:
\be\begin{array}{lllll}\label{h341} L(\phi)=\left[1+\frac{\alpha}{2}S(\phi)\right],~~~~
 T(\phi)=\left[1+\alpha S(\phi)\right],~~
 S(\phi)=\left[1+\{D_{4}+K_{4}\ln\left(\frac{\phi}{M}\right)\}
\left(\frac{\phi}{M}\right)^{4}\right],\\
U(\phi)=\left[(K_{4}+4D_{4})+4K_{4}\ln\left(\frac{\phi}{M}\right)\right],~~ 
E(\phi)=\left[(7K_{4}+12D_{4})+12K_{4}\ln\left(\frac{\phi}{M}\right)\right],\\
F(\phi)=\left[(26K_{4}+24D_{4})+24K_{4}\ln\left(\frac{\phi}{M}\right)\right],~~
J(\phi)=\left[(50K_{4}+24D_{4})+24K_{4}\ln\left(\frac{\phi}{M}\right)\right],\\
P(\phi)=\sqrt{\left[1+2\alpha S(\phi)L(\phi)\right]}
-2\alpha S(\phi)L(\phi)\sinh^{-1}\left[2\alpha S(\phi)L(\phi)\right]^{-1/2}
\end{array}\ee with $\alpha=\Delta^{4}/\lambda$. }:
\bea \label{first} \epsilon_{V}
&=&\frac{M^{2}_{p}}{2}\left(\frac{V^{'}}{V}\right)^{2}\frac{1+\frac{V}{\lambda}}{(1+\frac{V}{2\lambda})^{2}}
=
\frac{U^{2}(\phi)T(\phi)}{2S^{2}(\phi)L^{2}(\phi)}\left(\frac{\phi}{M}\right)^{6},
\\
\label{second}\eta_{V} &=&
M^{2}_{p}\left(\frac{V^{''}}{V}\right)\frac{1}{(1+\frac{V}{2\lambda})}
=\frac{E(\phi)}
{S(\phi)L(\phi)}\left(\frac{\phi}{M}\right)^{2},
\\
\label{third} \xi_{V} &=&
M^{4}_{p}\left(\frac{V^{'}V^{'''}}{V^{2}}\right)\frac{1}{(1+\frac{V}{2\lambda})^{2}}
=\frac{U(\phi)F(\phi)}
{S^{2}(\phi)
L^{2}(\phi)}\left(\frac{\phi}{M}\right)^{4},
\\
\label{fourth} \sigma_{V} &=&
M^{6}_{p}\frac{(V^{'})^{2}V^{''''}}{V^{3}}\frac{1}{(1+\frac{V}{2\lambda})^{3}}
=
\frac{U^{2}(\phi)J(\phi)}
{S^{3}(\phi)L^{3}(\phi)}\left(\frac{\phi}{M}\right)^{6}.\eea


Fig.~(\ref{figVr9}) depict how the slow roll parameter $\eta_{V}$ vary with the inflaton
field for the allowed range of $D_{4}$ and they give us a clear
picture of the starting point as well as the end of the cosmic
inflation. Nevertheless, it further reveals
that the $\eta$-problem is smoothened to some extent in brane cosmology~\footnote{However, we are yet to figure out if there is
any underlying dynamics that may lead to the solution of this
generic feature of SUGRA.}.

The number of e-foldings are  defined in brane cosmology 
\cite{Maartens:2010ar} for our model as:
 \be\begin{array}{lll}\label{noe}
  \displaystyle N\simeq\frac{1}{M^{2}_{p}}\int^{\phi_{i}}_{\phi_{f}}\left(\frac{V}{V^{'}}\right)\left(1+\frac{V}{2\lambda}\right)d\phi
 \simeq\frac{M^{2}}{U}\left[\frac{1}{2}\left(1+\frac{\alpha}{2}\right)\left(\frac{1}{\phi^{2}_{f}}
-\frac{1}{\phi^{2}_{i}}\right)\right.\\ \left.\displaystyle~~~~~~~~~~~~~~~~~~~~~~~~~~~~~~~~~~~~~~~~~~~~~~+\frac{D_{4}}{2M^{4}}(1+\alpha)(\phi^{2}_{i}-\phi^{2}_{f})+
\frac{\alpha
D^{2}_{4}}{12M^{8}}(\phi^{6}_{i}-\phi^{6}_{f})\right].
\end{array} \ee
Here $\phi_{i}$ and $\phi_{f}$ are the corresponding values of
the inflaton field at the start and end of inflation. Fig.~(\ref{figVr241}) represents a graphical behavior of number
of e-folding versus the inflaton field in the high energy limit
for different values of $D_{4}$ and the most satisfactory point in
this context is the number of e-folding lies within the
observational window  $56<N<70$. The end of the inflation leads
to the constraint, $\alpha=\frac{2}{|U|}|E|^{3/2}$, which is required 
for numerical estimations.

\begin{figure}[t]
\centering
\subfigure[~$l(l+1)C_l^{TT}/2\pi$ vs $l$]{
    \includegraphics[width=14cm,height=5cm] {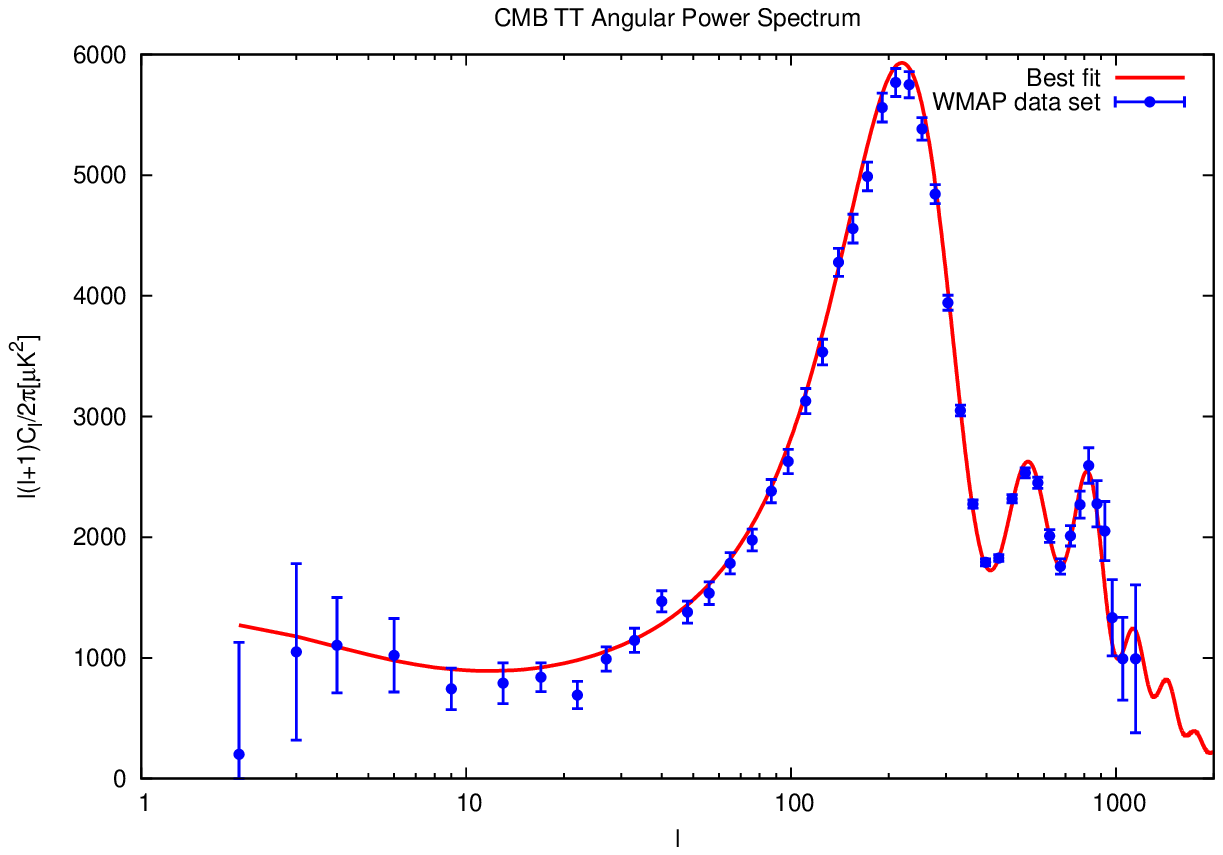}
    \label{zc1}}
\subfigure[~$l(l+1)C_l^{TE}/2\pi$ vs $l$]{
    \includegraphics[width=7cm,height=5cm] {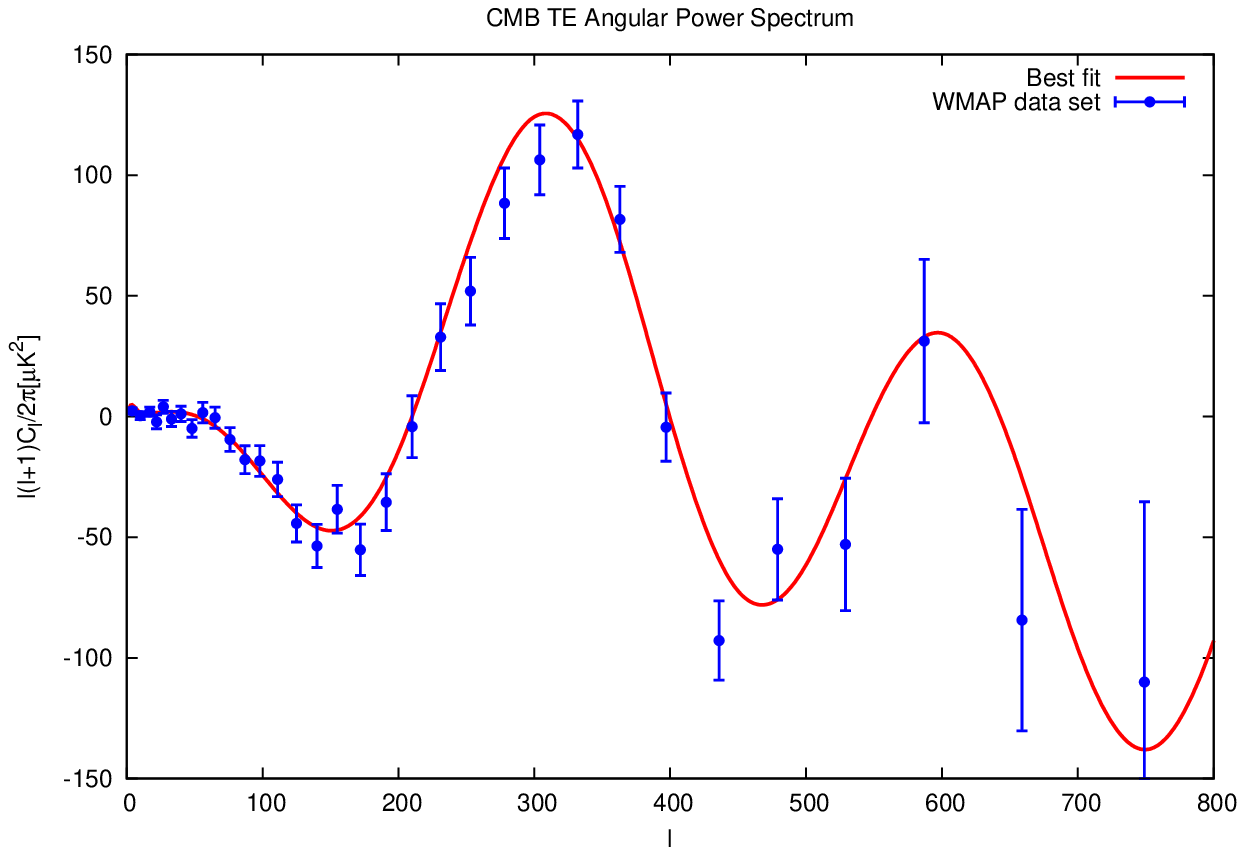}
    \label{zc2}
}
\subfigure[~$l(l+1)C_l^{EE}/2\pi$ vs $l$]{
    \includegraphics[width=7cm,height=5cm] {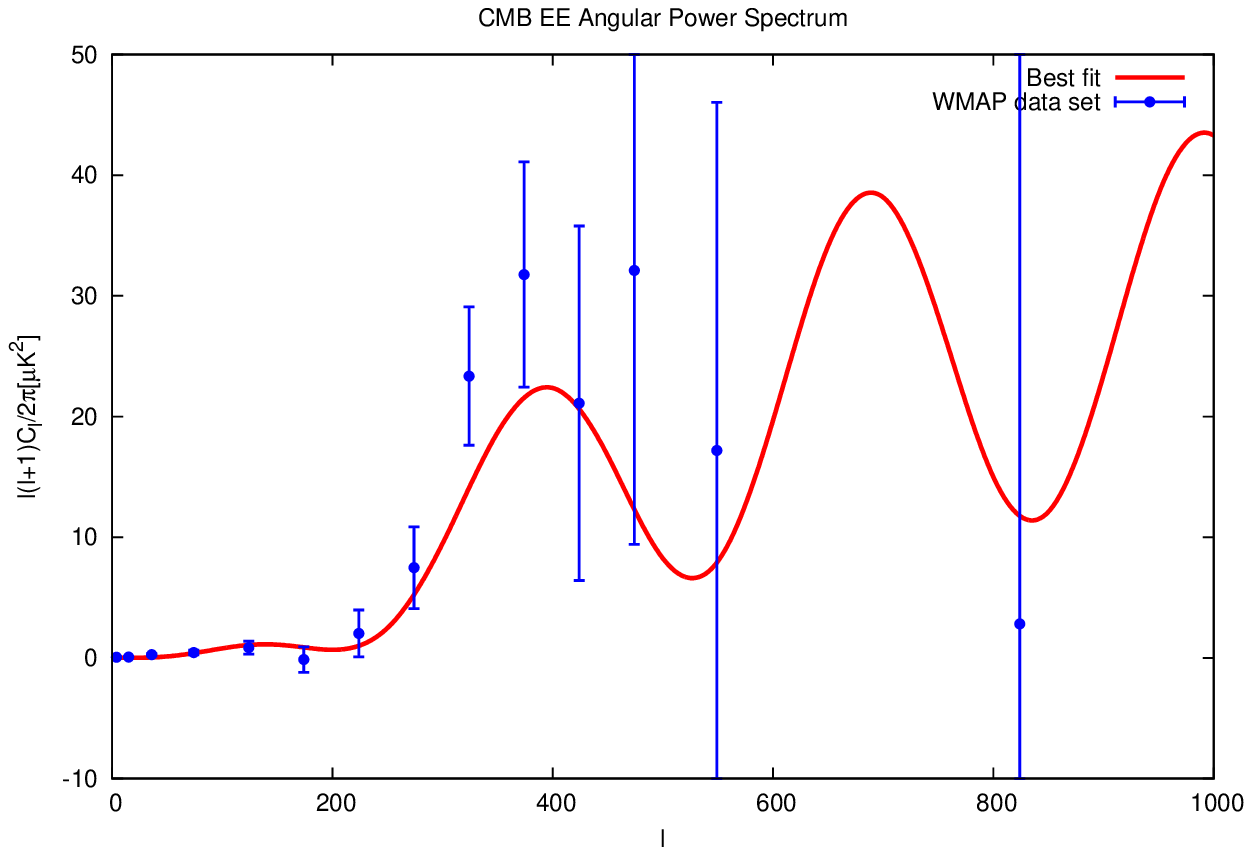}
    \label{zc3}
}
\caption[Optional caption for list of figures]{ We show the variation of CMB angular power spectrum for \subref{zc1}~TT,
\subref{zc2}~TE and \subref{zc3}~EE correlation with respect to multipole $l$ for the best fit model parameters \cite{Choudhury:2011sq}.
}
\label{tt}
\end{figure}

Let us now engage ourselves in analyzing quantum fluctuation in
our model and its observational imprints via primordial spectra
generated from cosmological perturbation. In  brane inflation the expressions for
amplitude of the scalar perturbation, tensor perturbation and
tensor to scalar ratio are given by~\footnote{
 In this context $\phi_{\star}$
represents the value of the inflaton field at the horizon
crossing represented by $k=aH$.}~\footnote{In Appeandix B we have also mentioned a set of consistency relations applicable for brane inflation.}:
 \begin{eqnarray}\label{scalar}
\Delta^{2}_{s} &\simeq&
\frac{512\pi}{75M^{6}_{p}}\left[\frac{V^{3}}
{(V^{'})^{2}}\left[1+\frac{V}{2\lambda}\right]^{3}\right]_{k=aH}
=\frac{M^{2}\alpha\lambda S^{3}(\phi_{\star})L^{3}(\phi_{\star})}{75\pi^{2}U^{2}(\phi_{\star})\phi^{6}_{\star}},\\
\label{tensor} \Delta^{2}_{t}  &\simeq&
\frac{32}{75M^{4}_{p}}\left[\frac{V\left[1+\frac{V}{2\lambda}\right]}{\left[\sqrt{1+\frac{2V}{\lambda}\left(1+\frac{V}{2\lambda}
\right)}-\frac{2V}{\lambda}\left(1+\frac{V}{2\lambda}\right)\sinh^{-1}\left[\frac{1}{\sqrt{\frac{2V}{\lambda}\left(1+\frac{V}{2\lambda}\right)}}
\right]\right]}\right]_{k=aH}\nonumber\\
&=&\frac{\lambda\alpha}{150\pi^{2}M^{4}}\frac{S(\phi_{\star})L(\phi_{\star})}{P(\phi_{\star})},\\
\label{ratio} r  &=&
16\frac{\Delta^{2}_{t}}{\Delta^{2}_{s}}\simeq
\frac{8(\phi_{\star})^{6}U^{2}(\phi_{\star})}{M^{6}
 S^{2}(\phi_{\star})L^{2}(\phi_{\star})P(\phi_{\star})}.
\end{eqnarray} 

Finally, to estimate five dimensional Planck mass from the observational parameters we use the
relation $M\sim M_{p}=\frac{M^{3}_{5}}{\sqrt{\lambda}}\sqrt{\frac{3}{4\pi}}$
and consequently from Eq (\ref{scalar})
we get an analytical expression for 5D cut-off scale in terms of model parameters: \be\label{fiv}
M_{5}=\sqrt[6]{\frac{800\pi^{4}\Delta^{2}_{s}U^{2}(\phi_{\star})}
{\alpha S^{3}(\phi_{\star})L^{3}(\phi_{\star})}}\phi_{\star}.\ee

\subsection{Parameter estimation}
\label{b4a}

\subsubsection*{Direct nuerical anlysis}

   \begin{table}[htb]
   \begin{center}\small\begin{tabular}{|c|c|c|c|c|c|c|c|c|c|c|c|c|}
   \hline $C_{4}$ & $\alpha$  & $\lambda$& $\phi_{f}$ & $\phi_{i}$ & $N$ & $\phi_{\star}$ & $\Delta^{2}_{s}$ &  $n_{s}$ &  $r$ & $\alpha_{s}$ & $M_{5}$\\
    $\simeq D_{4}$ & & $\times10^{-14}M^{4}$& $M$ & $M$ &  & $M$ & $\times10^{-9}$&  & $\times10^{-5}$ & $\times10^{-3}$ & $\times10^{-3}M$\\
   \hline
    &  &  &  &0.147 & 70 & 0.158 & 3.126 & 0.951 & 2.176 &-0.798 & \\
   -0.70  &  17.389 & 2.553 &1.017 &0.158& 60 & 0.173 & 1.835 & 0.941 &  3.706 &-1.142 &  11.792  \\
        &  &  &  &0.164 & 56 & 0.180 & 1.440 & 0.936 &  4.723 &-1.345 &  \\
   \hline
       &  &  &  &0.150 & 70 & 0.161 &2.902 &  0.951 & 2.176 & -0.798 & \\
   -0.65 &  16.757 & 2.632 & 1.036 & 0.161& 60 & 0.176 &1.704&  0.941 & 3.706 & -1.142 &  11.865 \\
     &  &  &  &0.167  & 56 & 0.184 &1.327 & 0.936 &  4.723 &  -1.345 &  \\
   \hline
       &  &  &  &0.153 & 70 & 0.165 & 2.679 & 0.951 &  2.176 & -0.798 &\\
   -0.60 &  16.099 & 2.758& 1.057 & 0.165 &60 & 0.180 & 1.573 & 0.941 &  3.706 &-1.142 & 11.944 \\
      &  &  &  &0.170 & 56 & 0.187 & 1.234 & 0.936 & 4.723&-1.345 &   \\
   \hline
  \end{tabular}
  \caption{Different observational  parameters related to the cosmological perturbation
  for our  proposed model in Eq~(\ref{post}) \cite{Choudhury:2011sq}.}
  \label{tabv1}
\end{center}
  \end{table}

Table \ref{tabv1} 
represent numerical estimation for different observational
parameters related to the cosmological perturbation as estimated
 from our model. It is worthwhile to point
out to the salient  features of estimated inflationary parameters
 obtained from our proposed model:
\begin{itemize}

\item The observable parameters help us have an
    estimation for the brane tension to be $\lambda\gg 1~
    {\rm MeV}^{4}$ provided energy scale of the inflation is in the
    vicinity of GUT scale. Also the 5D cut-off scale turns out to be $M_{5}\sim (11.792-11.944)\times 10^{-3}M$.
\item The amplitude of scalar power spectrum corresponding to different
    best fit values of $D_{4}$ is of the order
    of  $5\times 10^{5}$ and it perfectly matches with WMAP7 \cite{WMAP7}.
    The scalar spectral index  for lower values of
    $N\sim 55$ are pretty close to observational window
    $0.948<n_{s}<1$ \cite{WMAP7} whereas for higher values of
    $N\sim 70$  this lies well within the window.
    Also for our model running of the scalar spectral index
    $\alpha_{s} \sim -10^{-3}$.
\item Though the tensor to scalar ratio as estimated from our
    model is well within its upper bound fixed by WMAP7
    \cite{WMAP7} and Planck \cite{Ade:2013zuv,Ade:2013uln}, thereby facing no contradiction
    with observations, its value is even small to be detected.
      
\end{itemize}

\subsubsection*{Data analysis with CAMB}

In this context we shall make use of the cosmological code CAMB \cite{CAMB} in order
 to confront our results directly with observation.
To operate CAMB, the values of the initial parameters
associated with inflation are taken from the Table~\ref{tabv1} for $D_{4}=-0.60$. Additionally WMAP7 dataset in 
$\Lambda$CDM background has been used in CAMB to obtain CMB angular power spectrum at the pivot scale $k_*=0.002~{\rm Mpc}^{-1}$.
Table~\ref{tabv2} and table~\ref{tabv3} shows input from the WMAP7 dataset and 
the output obtained from CAMB respectively. 

\begin{table}[htb]
\begin{center}
\begin{tabular}{|c|c|c|c|c|c|c|c|c|c|}
\hline $H_0$ & $\tau_{Reion}$ &$\Omega_b h^2$& $\Omega_c h^2
$& $T_{CMB}$
 \\
km/sec/MPc& & && K\\
 \hline
71.0&0.09&0.0226&0.1119&2.725\\
\hline
\end{tabular}
\caption{Input in CAMB \cite{Choudhury:2011sq}.}\label{tabv2}
\end{center}
\end{table}

\begin{table}[htb]
\begin{center}
\begin{tabular}{|c|c|c|c|c|c|c|c|c|c|}
\hline $t_0$ & $z_{Reion}$ &$\Omega_m$&$\Omega_{\Lambda}$&$\Omega_k$&$\eta_{Rec}$& $\eta_0$
 \\
Gyr& & && &Mpc & Mpc\\
 \hline
13.708&10.692&0.2669&0.7331&0.0&285.15&14347.5\\
\hline
\end{tabular}
\caption{Output from CAMB \cite{Choudhury:2011sq}.}\label{tabv3}
\end{center}
\end{table}

The curvature perturbation is generated due to the fluctuations in
the {\it inflaton} and at the end of inflation it makes horizon
re-entry creating matter density fluctuations, which is the origin of the structure formation in
Universe. In Fig.~\ref{zc1}, Fig.~\ref{zc2} and Fig.~\ref{zc3} we confront CAMB output of CMB angular power spectrum $C_l^{TT}$, $C_l^{TE}$ and $C_l^{EE}$
 for best fit with WMAP seven years data for the scalar mode.
 From Fig.~\ref{zc1} we see that the Sachs-Wolfe plateau obtained from
  our model is almost flat confirming a nearly scale invariant spectrum. For larger value of the multipole $l$,
 CMB anisotropy spectrum is dominated by the
Baryon Acoustic Oscillations (BAO) giving rise to
several ups and downs in the spectrum. Also the peak positions
are sensitive on the dark energy and other forms of
the matter. Also fig.~\ref{zc1} is in good
 agreement with WMAP7 data for $\Lambda$CDM
background apart from the two
outliers at $l\sim 21$ and $l\sim 42$.

\section{DBI Galileon inflation}
\label{d1}

\subsection{The background model in D4 brane}
\label{d2a}
As discussed in introduction, now we will describe the features of DBI Galileon inflation in this section.
Let us demonstrate briefly the construction of DBI Galileon starting from 
${\cal N}$=2,${\cal D}$=5 SUGRA along with Gauss Bonnet correction in D4 brane set up. The full five dimensional model is described by \cite{Choudhury:2012yh}:
\be\label{totac}S^{(5)}_{Total}=S^{(5)}_{EH}+S^{(5)}_{GB}+S^{(5)}_{D4~brane}+S^{(5)}_{Bulk Sugra}\ee
where
\begin{eqnarray}\label{5eh}S^{(5)}_{EH}&=&\frac{1}{2\kappa^{2}_{5}}\int d^{5}x\sqrt{-g^{(5)}}\left[R_{(5)}-2\Lambda_{5}\right],\\
\label{5gb}S^{(5)}_{GB}&=&\frac{\alpha_{(5)}}{2\kappa^{2}_{5}}\int d^{5}x\sqrt{-g^{(5)}}
\left[R^{ABCD(5)}R^{(5)}_{ABCD}-4R^{AB(5)}R^{(5)}_{AB}+R^{2}_{(5)}\right]\end{eqnarray}
where $\alpha_{(5)}$ and $\kappa_{(5)}$ represent Gauss-Bonnet coupling and 5D gravitational coupling strength respectively. Additionally, 
$\Lambda_{(5)}$ and $g^{(5)}$ represent the 5D cosmological constant and the determinant of the 5D metric explicitly mentioned in equation(\ref{metric}).
The D4 brane action decomposed into two parts as \cite{Choudhury:2012yh}:
\be\label{d4}S^{(5)}_{D4~ brane}=S^{(5)}_{DBI}+S^{(5)}_{WZ},\ee
where the {\it DBI} action and the {\it Wess-Zumino} action are given by respectively \cite{Choudhury:2012yh}:
\begin{eqnarray}\label{dbi}S^{(5)}_{DBI}&=&-\frac{T_{(4)}}{2}\int d^{5}x \exp(-\Phi)\sqrt{-\left(\gamma^{(5)}+B^{(5)}+2\pi\alpha^{'}F^{(5)}\right)},\\
\label{WZ}\displaystyle S^{(5)}_{WZ}&=&-\frac{T_{(4)}}{2}
\int \sum_{n=0,2,4} \hat{C_{n}}\wedge \exp\left(\hat{B}_{2}+2\pi\alpha^{'}F_{2}\right)|_{4~form}\\
\displaystyle&=&\frac{1}{2}\int d^{5}x\sqrt{-g^{(5)}}\left\{\epsilon^{ABCD}\left[\partial_{A}\Phi^{I}\partial_{B}\Phi^{J}
\left(\frac{C_{IJ}B_{KL}}{4T_{(4)}}\partial_{C}\Phi^{K}\partial_{D}\Phi^{L}+\frac{\pi\alpha^{'}C_{IJ}F_{CD}}{2}\nonumber\right.
\right.\right.\\ &&\left.\left.\left.
\displaystyle+\frac{C_{0}}{8T_{(4)}}B_{IJ}B_{KL}\partial_{C}\Phi^{K}\partial_{D}\Phi^{L}+
\frac{\pi\alpha^{'}C_{0}}{2}B_{IJ}F_{CD}\right)+2\pi^{2}\alpha^{'2}T_{(4)}C_{0}F_{AB}F_{CD}-T_{(4)}\nu(\Phi)\right]\right\}\nonumber\end{eqnarray}
where $T_{(4)}$ is the D4 brane tension, $\alpha^{'}$ is the Regge Slope, $\exp(-\Phi)$
is the closed string dilaton and $C_{0}$ is the Axion. Here and through out the article hat denotes a pull-back onto the D4 brane so that $\gamma_{AB}$ 
is the 5D induced metric on the D4 brane explicitly defined in equation(\ref{ind}). Here $\gamma^{(5)}$, $B^{(5)}$ and $F^{(5)}$ represent the determinant
of the 5D induced metric ($\gamma_{AB}$) and the gauge fields ($B_{AB},F_{AB}$) respectively.
The gauge invariant combination of rank 2 field strength tensor, appearing in D4 brane, is ${\cal F}_{AB}
=B_{AB}+2\pi\alpha^{'}F_{AB}$ and $\{F_{2},B_{2}\}$ represents 2-form $U(1)$ gauge fields which have the only non-trivial components along compact direction. 
On the other hand
$C_{4}$ has components only along the non-compact space-time dimensions.
In a general flux compactification all fluxes may be turned on as the Ramond-Ramond (RR) forms
$F_{n+1}=dC_{n}$ (along with their duals) with $n=0,2,4$ and the Neveu Schwarz-Neveu Schwarz (NS-NS) flux $H_{3}=dB_{2}$.
Additionally the D4 brane frame function is defined as \cite{Choudhury:2012yh}:
 \be\label{frrr1}\nu(\Phi)=\left(\nu_{0}+\frac{\nu_{4}}{\Phi^{4}}\right)\ee which is originated 
from interaction between D4-$\bar{D4}$ brane in string theory. Here $\nu_{0}$ and $\nu_{4}$ represent the constants characterizing the
interaction strength between  D4-$\bar{D4}$ brane.

In Eq~(\ref{totac}) ${\cal N}=2, {\cal D}=5$ bulk SUGRA action is exactly similar as described in Eq~(\ref{sug2}) in \ref{b2a}.
In this context the 5-dimensional coordinates
$X^{A}=(x^{\alpha},y)$, where $y$ parameterizes the extra
dimension compactified on the closed interval $[-\pi R,+\pi R]$
and $Z_{2}$ symmetry is imposed. For computational purpose it is useful to define the five
dimensional generating function($G$) of SUGRA in this setup as \cite{Choudhury:2012yh}:
\be\label{gen}G=-3\ln\left(\frac{T+T^{\dagger}}{\sqrt{2}}\right)+K(\Phi,\Phi^{\dagger}), \ee
 where the SUGRA K$\ddot{a}$hler moduli fields are given by $T=\frac{(e^{\dot{5}}_{5}-i\sqrt{\frac{2}{3}}A^{0}_{5})}{\sqrt{2}}$ which 
is assumed to be stabilized and 
$K(\Phi,\Phi^{\dagger})$ represents generalized K$\ddot{a}$hler function.

 Including the kinetic term of the five dimensional field
$\Phi$ and rearranging into a perfect square, the 5D bulk SUGRA action can be expressed as
\be\label{modsug}
S\supset\frac{1}{2}\int d^{4}x\int^{+\pi R}_{-\pi
R}dy\sqrt{-g_{5}}e_{(4)}e^{5}_{\dot{5}}\left[g^{\alpha\beta}G_{M}^{N}(\partial_{\alpha}\Phi^{M})^{\dagger}(\partial_{\beta}\Phi_{N})
+\frac{1}{g_{55}}\left(\partial_{5}\Phi-\sqrt{V^{(5)}_{bulk}(G)})\right)^{2}\right],\ee
where the 5D potential described by
\be\label{fpotw}V^{(5)}_{bulk}(G)=\exp\left(\frac{G}{M^{2}}\right)\left[\left(\frac{\partial
W}{\partial \Phi_{M}}+\frac{\partial G}{\partial
\Phi_{M}}\frac{W}{M^{2}}\right)^{\dagger}(G_{M}^{N})^{-1}\left(\frac{\partial
W}{\partial \Phi^{N}}+\frac{\partial G}{\partial
\Phi^{N}}\frac{W}{M^{2}}\right)-3\frac{|W|^{2}}{M^{2}}\right]\ee
where $W$ physically represents the superpotential
in the context of ${\cal N}=2,{\cal D}=5$ SUGRA theory and expressed in terms of the holomorphic combination of the fields $\Phi,\Phi^{\dagger},T$ and $T^{\dagger}$.
The field equations in presence of Gauss-Bonnet term can be expressed as \cite{Choudhury:2012yh}:
\be\label{eeq}G^{(5)}_{AB}+\alpha_{(5)}H^{(5)}_{AB}=8\pi G_{(5)}T^{(5)}_{AB}-\Lambda_{(5)}g^{(5)}_{AB},\ee
where the $H^{(5)}_{AB}$ covariantly conserved Gauss-Bonnet tensor as defined in Eq~(\ref{gbp}). It is 
useful to introduce the 5D metric in conformal form:
\be\label{metric}ds^{2}_{4+1}=g_{AB}dX^{A}dX^{B}
=\frac{1}{\sqrt{h(y)}}ds^{2}_{4}+\sqrt{h(y)}\tilde{G}(y)dy^{2},\ee
with metric function:
\be\label{tgvd}\frac{1}{\sqrt{h(y)}}=\frac{R^2}{b^2_{0}\beta^{2}}\tilde{G}(y)=\frac{b^{2}_{0}}{R^{2}\left(\exp(\beta y)
+\frac{\Lambda_{(5)}b^{4}_{0}}{24R^{2}}\exp(-\beta y)\right)}
\ee and $ds^{2}_{4}=g_{\alpha\beta}dx^{\alpha}dx^{\beta}$ is FLRW counterpart. In order to write down
explicitly the expression for D4 brane action, the induced metric can be shown as
\be\label{ind}\gamma_{CD}=\frac{1}{\sqrt{h(y)}}
\left(g_{AB}+h(y)G_{AB}\partial_{C}\Phi^{A}\partial_{D}\Phi^{B}\right).\ee

Now using the scaling relations 
\be\begin{array}{llll}\displaystyle {\Phi}^{A}=\sqrt{T_{(4)}}\tilde{\Phi}^{A},~~~~ G_{AB}=\exp(-\Phi)g_{AB},~~~~  
b_{AB}=\frac{\sqrt{h(y)}}{T_{(4)}}B_{AB}\end{array}\ee the 5D action for D4 brane can be expressed in more convenient form as
\be\label{conv}S^{(5)}_{D4~brane}=\int d^{5}x \sqrt{-g^{(5)}}\left[K(\Phi,X)-G(\Phi,X)\Box^{(5)}\Phi\right],\ee
where 
\be\label{kl}K(\Phi,X)=-\frac{1}{2f(\Phi)}\left(\sqrt{\cal D}
-1\right)-\frac{V^{(5)}_{brane}(\Phi)}{2}\ee where the determinant can be expressed as 
\be \label{detdef}{\cal D}\simeq 1-2f(\Phi)G_{AB}X^{AB}+4f^{2}(\Phi)X_{A}^{[A}X_{B}^{B]}
-8f^{3}(\Phi)X_{A}^{[A}X^{B}_{B}X^{C]}_{C}+16 f^{4}(\Phi)X_{A}^{[A}X^{B}_{B}X^{C}_{C}X^{D]}_{D}\ee
which is expressed in terms of the kinetic term 
$X^{B}_{D}=-\frac{1}{2}G_{DA}\partial^{C}\Phi^{A}\partial_{C}\Phi^{B}$. In this context the 5D D'Alembertian Operator is defined as, 
$\Box^{(5)}=\frac{1}{\sqrt{-g^{(5)}}}\partial_{A}\left(\sqrt{-g^{(5)}}g^{AC}\partial_{C}\right)$.
Here we use the fact that no spatial direction along which the scalar fields are only time dependent
lead to $B^{\mu}_{\nu}=0$ and $F_{\mu\nu}=0$ in the background. Consequently {\it Maxwell's field equations} are 
unaffected in 4D after dimensional reduction. In this context the D4 brane potential is given by:
 \be\label{D4pot}V^{(5)}_{brane}(\Phi)=T_{(4)}\nu(\Phi)+\frac{1}{f(\Phi)},\ee
where 5D warped geometry motivated $Z_{2}$ symmetric frame function 
\be\label{ffn} f(\Phi)=\frac{\exp(\Phi)h(y)}{T_{(4)}}
\simeq\frac{1}{(f_{0}+f_{2}\Phi^{2}+f_{4}\Phi^{4})}\ee  is originated from higher dimensional 
field theory and the implicit D4 brane function defined as:
\be\label{implbr}G(\Phi,X)=\frac{g(\Phi)}{2(1-2f(\Phi)X)}\ee
with $g(\Phi)=g_{0}+g_{2}\Phi^{2}$. Here $g_{0}$ and $g_{2}$ are model dependent constants
characterizes the effects of possible interactions on the D4 brane.

\subsection{Modeling DBI Galileon inflation in D3 brane}
\label{d3a}

The technical details of the dimensional reduction technique are elaborately discussed in the Appendix C which can generate an effective D3 DBI Galileon theory in 4D.
Summing up all the contributions from Eq~(\ref{ghu},\ref{gbe4},\ref{br1},\ref{ast5}), the model for {\it D3 DBI Galileon} is described by the 
following effective action \cite{Choudhury:2012yh}:
\be\begin{array}{lllllll}\label{model1}
  \displaystyle S= \int d^{4}x \sqrt{-g^{(4)}}
\left[\hat{\tilde{K}}(\phi,X)
-\tilde{G}(\phi,X)\Box^{(4)}\phi+\tilde{l}_{1}R_{(4)}\right.\\ \left.
\displaystyle~~~~~~~~~~~~~~~~~~~~~~~~~~~~~~~~~~~~~~~~~~~~~+\tilde{l}_{4}\left({\cal C}(1)R^{\alpha\beta\gamma\delta(4)}R^{(4)}_{\alpha\beta\gamma\delta}
-4{\cal I}(2)R^{\alpha\beta(4)}R^{(4)}_{\alpha\beta}+{\cal A}(6)R^{2}_{(4)}\right)+\tilde{l}_{3}\right],\end{array}\ee
where
\be\begin{array}{lllll}\label{effcons} \hat{\tilde{K}}(\phi,X)=-\frac{\tilde{D}}{\tilde{f}(\phi)}\left[\sqrt{1-2QX\tilde{f}}-Q_{1}\right]
-\tilde{C}_{5}\tilde{G}(\phi,X)-2X\tilde{M}(T,T^{\dag})-V(\phi),\\ 
\tilde{M}(T,T^{\dag})=\frac{M(T,T^{\dag})}{2\kappa^{2}_{(4)}},~M(T,T^{\dag})=\frac{\sqrt{2}\beta R^{2}}{(T+T^{\dag})},~\tilde{D}=\frac{D}{2\kappa^{2}_{(4)}},\\
\tilde{G}(\phi,X)=\left(\frac{\tilde{g}(\phi)k_{1}\tilde{C}_{4}}{2(1-2\tilde{f}(\phi)Xk_{2}))}\right),~
\tilde{g}(\phi)=\tilde{g}_{0}+\tilde{g}_{2}\phi^{2},~\tilde{f}(\phi)\simeq\frac{1}{(\tilde{f}_{0}+\tilde{f}_{2}\phi^{2}+\tilde{f}_{4}\phi^{4})}\\
\tilde{l}_{1}=\left\{\frac{1}{2\kappa^{2}_{(4)}}
\left[1+\frac{\alpha_{(4)}}{R^{2}\beta^{2}}\left(24{\cal I}(2)-24{\cal A}(9)-16{\cal A}(10)\right)
\right]-\frac{\alpha_{(4)}{\cal C}(2)}{\kappa^{2}_{(4)}R^{2}\beta^{2}}\right\},\tilde{l}_{4}=\frac{\alpha_{(4)}}{2\kappa^{2}_{(4)}}, \\
\tilde{l}_{3}=\frac{1}{2\kappa^{2}_{(4)}}\left[\frac{\alpha_{(4)}}{R^{4}\beta^{4}}
\left(24{\cal C}(4)-144{\cal I}(4)-64{\cal A}(5)+144{\cal A}(7)+64{\cal A}(8)+192{\cal A}(11)\right)-\frac{3M^{3}_{5}\beta b^{6}_{0}}
{2\kappa^{2}_{(4)}M^{2}_{p}R^{5}}{\cal I}(1)\right]\end{array}.\ee
where $\alpha_{(4)},\tilde{l}_{1},\tilde{l}_{3},\tilde{l}_{4}$ are effective 4D couplings and $\kappa_{(4)}$ be the gravitational coupling strength. Here $X$ represents
the 4D kinetic term after dimensional reduction given by $X:=-\frac{1}{2}g_{\mu\nu}\partial^{\mu}\phi\partial^{\nu}\phi$.
 In this context $(T,T^{\dagger})$ are the four dimensional background SUGRA moduli fields
 which are constant after dimensional reduction. The collective effect of Eq~(\ref{scaled}) and Eq~(\ref{pot1}) gives the total {\it D3
DBI Galileon} potential as \cite{Choudhury:2012yh}:
\be\begin{array}{lllllllll}\label{modelpot} 
\displaystyle V^{}(\phi)=\bar{Q}_{2}\tilde{D}V^{(4)}_{brane}+
\tilde{Z}(T,T^{\dag})V^{(4)}_{bulk}(\phi)=\sum^{2}_{m=-2,m\neq-1}C_{2m}\phi^{2m},\end{array}\ee
where 
\be\begin{array}{llll}\label{vccvc}\displaystyle C_{0}=\left(T^{}_{3}\tilde{\nu}_{0}+\beta R {\cal I}(2)\tilde{f}_{0}+\tilde{Z}(T,T^{\dag}){\cal A}(13)v^4\right),\\
C_{-4}=T^{}_{3}\tilde{\nu}_{4},\\ C_{2}=\left(\beta R {\cal I}(2)\tilde{f}_{2}-gv^{2}\tilde{Z}(T,T^{\dag}){\cal A}(13)\right),\\
C_{4}=\left(\beta R {\cal I}(2)\tilde{f}_{4}+\frac{\tilde{Z}(T,T^{\dag}){\cal A}(13)g^{2}}{4}\right)\end{array}\ee are tree level 
constants. Now we want to see the effect of one-loop radiative correction to the derived potential. After doing proper analysis
throughout it comes out that the one-loop correction does not effect the superpotential due to the cancellation of all tadpole terms appearing in the theory.
 On the other hand one-loop radiative correction in the K$\ddot{a}$hler potential results in \cite{Choudhury:2012yh}:
\be\begin{array}{llll}\label{bvcxv}
  \displaystyle  \delta {\cal K}^{1-loop}(\phi,\phi^{\dagger})
=\int^{\Lambda_{UV}}_{p}\frac{d^{4}p}{(2\pi)^{4}p^{2}}\left[\frac{1}{2}Tr\ln \hat{\cal K}(\phi,\phi^{\dagger})
\right.\\ \left.~~~~~~~~~~~~~~~~~~~~~~~~~~~~~~~~~~~~~~~\displaystyle +\frac{1}{2}Tr\ln\left(\hat{\cal K}(\phi,\phi^{\dagger})p^{2}-\hat{\cal W}^{\dagger}(\phi,\phi^{\dagger})\left(\hat{\cal K}(\phi,\phi^{\dagger})^{-1}\right)^{\dagger}
\hat{\cal W}(\phi,\phi^{\dagger})\right)\right]\\
~~~~~~~~~~~~~~~~~~~~\displaystyle =\frac{\Lambda^{2}_{UV}}{16\pi^{2}}\ln\left(det\left[\hat{\cal K}(\phi,\phi^{\dagger})\right]\right)
-\frac{1}{32\pi^{2}}Tr\left({\cal M}^{2}_{\phi}\left[\frac{{\cal M}^{2}_{\phi}}{\Lambda^{2}_{UV}}-1\right]\right)
   \end{array}\ee
where $\Lambda_{UV}=M\sim M_p$ is used as a UV cut-off of the theory appearing in the context of cut-off regularization. In this connection
the chiral mass matrix is given by \be\label{massmatrix}
{\cal M}^{2}_{\phi}=\hat{\cal K}^{-\frac{1}{2}}(\phi,\phi^{\dagger})\hat{\cal W}^{\dagger}(\phi,\phi^{\dagger})
\left(\hat{\cal K}(\phi,\phi^{\dagger})^{-1}\right)^{\dagger}\hat{\cal W}(\phi,\phi^{\dagger})\hat{\cal K}^{-\frac{1}{2}}(\phi,\phi^{\dagger}).\ee
Now including the contribution 
from one-loop radiative correction both from brane and bulk SUGRA, the renormalizable Coleman Weinberg potential is as under~\footnote{To compute the trace part here
 we use the supertrace identity,
$STr \left( {\cal M}^{\alpha}\right) \equiv \sum_{i}\left( -1\right) ^{2j_{i}}\left(
2j_{i}+1\right) m_{i}^{\alpha}$.} \cite{Choudhury:2012yh}:
\be\begin{array}{lll}\label{loop}
V(\phi)=V_{tree}(\phi)+\del V_{1-loop}(\phi)\\~~~~~~~
\displaystyle=\underbrace{\sum^{2}_{m=-2,m\neq-1}C_{2m}\phi^{2m}}_{Tree-level~ contribution}
+
\underbrace{\lim_{\epsilon\rightarrow 0}\sum^{0}_{n=-2,n\neq-1}B_{2m}\left(\int^{\Lambda_{UV}=M}_{p}\frac{d^{4}p}
{(2\pi)^{4}}\frac{1}{\left(p^{2}-2C_{2}+i\epsilon\right)^{2}}\right)\phi^{2n}}_{One-loop~ correction~ in~ D3~ brane}
\\~~~\displaystyle +\underbrace{\sum^{2}_{q=0}\frac{\phi^{2q}}{64\pi ^{2}}\left[ \Lambda^{4}_{UV} STr\left(
{\cal M}^{0}\right) \ln \left( \frac{\Lambda^{2}_{UV}}{\phi^{2}}\right)
+2\Lambda^{2}_{UV} STr\left( {\cal M}^{2}\right) + STr \left( {\cal M}^{4}\ln \left(
\frac{{\cal M}^{2}}{\Lambda^{2}_{UV}}\right) \right) \right]}_{One-loop~ correction~ in~ the~ bulk~ {\cal N}=1,~ {\cal D}=4 ~SUGRA}\\
\\~~~~~~~\displaystyle=\sum^{2}_{m=-2,m\neq-1}\left[1+D_{2m}\ln\left(\frac{\phi}{M}\right)\right]\phi^{2m},\end{array}\ee
where $D_{0}=0$, $D_{2m}=\frac{\bar{B}_{2m}+A_{2m}}{C_{2m}}$. Here the 4D effective potential respect the 
{\it Galilean symmetry}: $\phi\rightarrow \phi+b_{\mu}x^{\mu}+c$ which taske care both shift and spacetime translational symmetry.


\begin{figure}[t]
\centering
\subfigure[~$V(\phi)$ vs $\phi$]{
    \includegraphics[width=12.5cm,height=6cm] {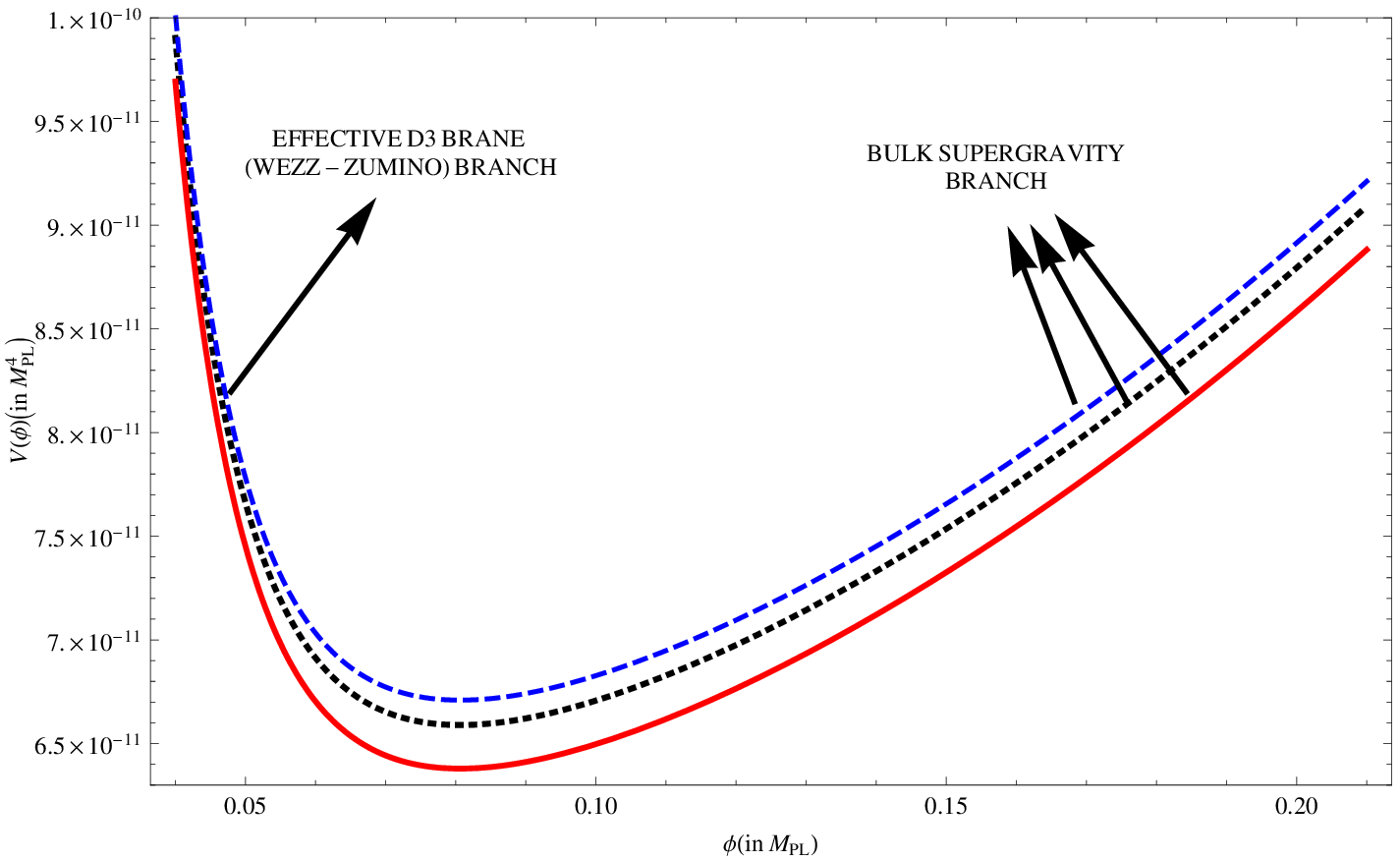}
    \label{figgv}}
\subfigure[~$N$ vs $\phi$]{
    \includegraphics[width=12.5cm,height=6cm] {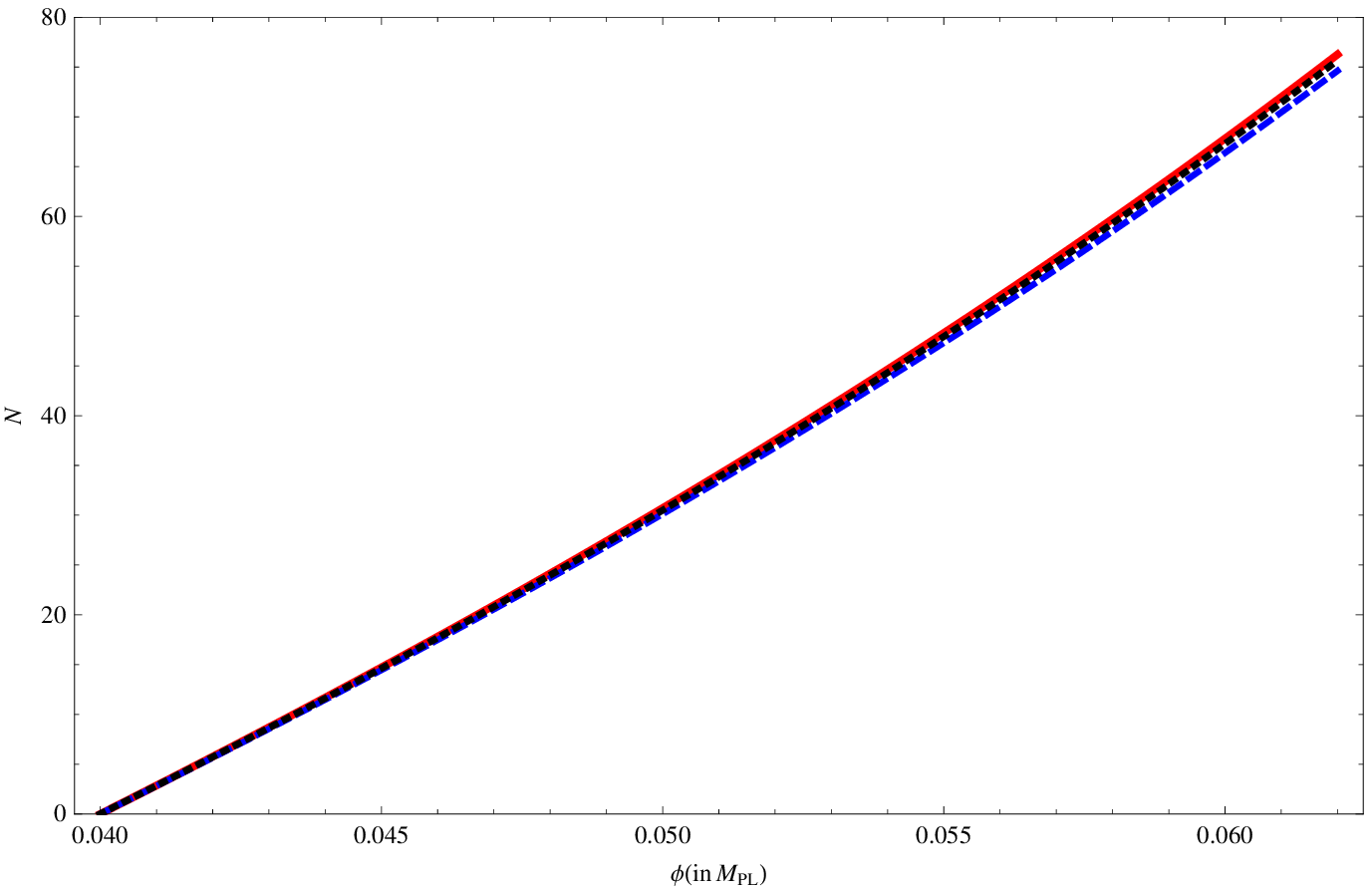}
    \label{fig:subfiggh1}
}
\caption[Optional caption for list of figures]{ Variation of \subref{figgv} one loop corrected potential ($V(\phi)$) vs inflaton field ($\phi$) \cite{Choudhury:2012yh}
 and \subref{fig:subfiggh1} number of e-foldings ($N$) vs inflation field ($\phi$) for best fit values of $C_{i}\forall i$ \cite{Choudhury:2012yh}.
}
\label{vfcv}
\end{figure}

Fig.~(\ref{figgv}) represents the inflaton potential for
different values of $C_{2m}$ and $D_{2m}$. From the observational constraints
the best fit model is given by the range: $5.67\times 10^{-11}M^{4}_{p}<C_{0}<6\times 10^{-11}M^{4}_{p}$,
 $1.01\times 10^{-16}M^{8}_{p}<C_{-4}<2\times 10^{-16}M^{8}_{p}$, $7.27\times 10^{-10}M^{2}_{p}<C_{2}< 7.31\times 10^{-10}M^{2}_{p}$,
$2.01\times 10^{-14}<C_{4}<2.45\times 10^{-14}$, $0.014 <D_{-4}<0.021$, $0.002<D_{2}< 0.012$  and $0.011<D_{4}<0.019$ so
that while doing numerical analysis, we shall restrict ourselves to this
range. 

Hence using Eq~(\ref{model1}) the modified {\it Friedmann } and
{\it Klein-Gordon} equations can be expressed as:
\bea\label{fr1}
H^{4}&=&\frac{\Lambda^{}_{(4)}+8\pi G_{(4)}V(\phi)}{\tilde{g}_{1}},\\
\label{fr2}
\dot{\phi}^{2}\left(e_{2}(\phi)+9e_{3}(\phi)H^{2}\right)&=&\left\{V^{'}(\phi)+\tilde{C}_{5}\tilde{g}^{'}(\phi)k_{1}
-\frac{\tilde{D}\tilde{f}^{'}(\phi)}{\tilde{f}(\phi)}\left(1-Q_{1}\right)\right\},\eea
where $e_{2}(\phi)=\tilde{M}(T,T^{\dag})J_{\phi}+2\tilde{g}(\phi)
\tilde{f}^{'}(\phi)k_{1}k_{2}
+8\tilde{f}(\phi)\tilde{f}^{'}(\phi)\tilde{g}(\phi)k_{1}k^{2}_{2}
+2\tilde{g}^{'}(\phi)\tilde{f}(\phi)k_{1}k_{2}-g^{''}(\phi)k_{1}$ and 
$e_{3}(\phi)=2\tilde{C}_{4}\tilde{f}(\phi)\tilde{g}(\phi)k_{1}k_{2}$
provided $|e_{3}(\phi)|\gg|e_{2}(\phi)|$ in the slow-roll regime.
Here we have fixed the signature of $\dot{\phi}$ so that the scalar
field rolls down the potential. Additionally ghost instabilities are 
avoided provided the coefficient of $\dot{\phi}^{2}>0$. Consequently 
the potential dependent slow-roll parameters can be expressed as:
\bea \label{first} \epsilon_{V}
:&=&\frac{M^{2}_{p}}{2}\left(\frac{V^{'}}{V}\right)^{2}\frac{1}{\sqrt{{\cal G}(\phi)V^{'}(\phi)}},
\\
\label{second}\eta_{V} :&=&
M^{2}_{p}\left(\frac{V^{''}}{V}\right)\frac{1}{\sqrt{{\cal G}(\phi)V^{'}(\phi)}},
\\
\label{third} \xi_{V} :&=&
M^{4}_{p}\left(\frac{V^{'}V^{'''}}{V^{2}}\right)\frac{1}{{\cal G}(\phi)V^{'}(\phi)},
\\
\label{fourth} \sigma_{V} :&=&
M^{6}_{p}\left(\frac{(V^{'})^{2}V^{''''}}{V^{3}}\right)\frac{1}{\left({\cal G}(\phi)V^{'}(\phi)\right)^{\frac{3}{2}}},
\eea
where ${\cal G}(\phi)=\frac{16e_{3}(\phi)M^{2}_{p}}{\tilde{g}_{1}V(\phi)}$. In this connection Galileon terms effectively flatten the potential
due to the presence of the {\it flattening factor} $\frac{1}{\sqrt{{\cal G}(\phi)V^{'}(\phi)}}\ll 1$. This implies that in the presence of Galileon
like derivative interaction slow-roll inflation can take place even if the potential is rather steep.

The number of e-foldings for {\it D3 DBI Galileon} can be expressed as 
\be\label{iop}
{\cal N}=\frac{1}{8M_{p}}\int^{\phi_{f}}_{\phi_{i}}\frac{\sqrt[4]{{\cal G}(\phi)V^{'}(\phi)}}{\sqrt{\epsilon_{V}}}d\phi,\ee
where $\phi_{i}$ and $\phi_{f}$ are the corresponding values of
the inflaton field at the beginning and end of inflation.


Fig.~(\ref{fig:subfiggh1}) represents a graphical behavior of number
of e-folding versus the inflaton field 
for different values of $C_{i}\forall i$~\footnote{Here the end of the inflation leads
to the extra constraint $V^{'}(\phi_{f})=\sqrt{V^{''}(\phi_{f})V(\phi_{f})}$.}.
\subsection{Quantum fluctuations and CMB observables}
\label{d4a}

Let us now engage ourselves in analyzing quantum fluctuation in
our model and its observational imprints via primordial spectra
generated from cosmological perturbation. To serve this purpose we use {\it ADM formalism} \cite{kifer}~\footnote{
In {\it ADM 
formalism} the line element can be written as:$$ds^{2}=-N^{2}dt^{2}+h_{ij}(N^{i}dt+dx^{i})
(N^{j}dt+dx^{j})\,,$$
where $N$ and $N^i$ ($i=1, 2, 3$) are the lapse and shift functions, respectively.
In this context we consider scalar metric perturbations 
about the flat FLRW background. 
Here we expand the lapse $N$ and the shift vector
$N^{i}$, as $N=1+\alpha$ and $N_{i}=\partial_{i}\psi+\tilde{N}_{i}$, respectively. Here $\partial_{i}\psi$ is the irrotational part and 
$\tilde{N}_{i}$ be the incompressible vector part ($\tilde{N}_{i,i}=\partial_{i}\tilde{N}_{i}=0$).
These are actually non-dynamical Lagrange multipliers in the action, so that it is sufficient to know $N$ and $N^i$
up to first order. This implies their equation of motion is purely algebraic. 
To fix the time and spatial reparameterization we choose the uniform-field gauge 
with $\delta\phi=0$, which fixes the temporal component
of a gauge-transformation vector $\xi^{\mu}$. After that by fixing
the spatial part of $\xi^{\mu}$ we gauge away a field
$\varepsilon$ that appears as a form $\varepsilon_{,ij}$ inside $h_{ij}$. Consequently the metric on three dimensional constant time slice
can be expressed as $h_{ij}=a^{2}(t)e^{2{\zeta}}\delta_{ij}$. Finally
at linear level of the perturbation theory one can write: $$ds^{2}=-(1+2\alpha)\, dt^{2}+2(\partial_{i}\psi+\tilde{N}_{i})\, dt\, dx^{i}
+a^2(t)\, (1+2\zeta)\,\delta_{ij}dx^{i}dx^{j}\,.$$} to expand the 4D effective action up to second order as:
\begin{equation}
\label{eqxz}
S^{\zeta}_2=\int dt\,d^3x\,a^3
\left[ -3t_1 \dot{\zeta}^2 
+\frac{2w_1}{a^2} \dot{\zeta} \partial^2\psi
-\frac{t_{2}}{a^2} \alpha \partial^2\psi
-\frac{2 t_{1}}{a^2} \alpha \partial^2{\zeta}
+3 t_{2}\,\alpha\,\dot{\zeta}
+\frac13 t_{3} \alpha^2
+\frac{t_{4}}{a^2}\,\partial_i{\zeta}\,\partial_i{\zeta}\right]\,,
\end{equation}
where the effect of effective Gauss-Bonnet coupling and the DBI Galileon features 
are explicitly appearing in the co-efficients of the second order perturbative action as:
\begin{eqnarray}
t_{1} &=&t_{4} \approx  \tilde{l}_{1}\,,\nonumber\\
t_{2} & \approx & \left(2H\tilde{l}_{1}-2\dot{\phi}X\tilde{G}_{X}\right)\,,\\
t_{3} &\approx& 
-9\tilde{l}_{1}H^{2}+3\left(X\hat{\tilde{K}}_{X}+2X^{2}\hat{\tilde{K}}_{XX}\right)
+18H\dot{\phi}
\left(2X\tilde{G}_{X}+X^{2}\tilde{G}_{XX}\right)
-6(X G_{,\phi}+X^2 G_{,\phi X})\,.\nonumber
\end{eqnarray}

It is important to mention here that, in the action (\ref{eqxz}), both the coefficients of the terms 
$\alpha {\zeta}$ and ${\zeta}^2$ vanish by using the background equations of motion. 
Furthermore, in (\ref{eqxz}), the term quadratic in $\psi$ vanishes by making use of integrations by parts.
The equations of motion for $\psi$ and $\alpha$, derived from (\ref{eqxz}),
lead to the following two-fold constraint relations~\footnote{Here we introduce new set of parameters:
$$s^{S}_{V}=\frac{\dot{c_{s}}}{Hc_{s}}=\frac{4\sqrt{V^{'}(\phi)}M^{2}_{p}}{\sqrt{{\cal G}(\phi)}}\frac{d}{d\phi}\left(\ln c_{s}\right),
~~s^{T}_{V}=
\frac{\dot{c_{T}}}{Hc_{T}}=\frac{M_{p}\sqrt{\tilde{g}_{1}V^{'}(\phi)}}{2\sqrt{e_{3}(\phi)V(\phi)}}\frac{d}{d\phi}\left(\ln\left[1+{\cal O}(
\epsilon^2_{V})\right]\right),~~\delta_{GX}=\frac{\dot{\phi}X \tilde{G}_{X}}{\tilde{l}_{1}},$$
$$\eta_{s}=
\frac{\dot{\epsilon_{s}}}{H\epsilon_{s}}=\frac{4\sqrt{V^{'}(\phi)}M^{2}_{p}}{\sqrt{{\cal G}(\phi)}}\frac{\left[\epsilon^{'}_{V}\left(1+{\cal O}(\epsilon_{V})\right)+\delta_{GX\phi}
\right]}{\left[\epsilon_{V}+\delta_{GX}+{\cal O}(\epsilon^2_{V})\right]},
~~\delta_{V}=
\frac{\dot{Y_s}}{HY_{s}}=\frac{4\sqrt{V^{'}(\phi)}M^{2}_{p}}{\sqrt{{\cal G}(\phi)}}\frac{d}{d\phi}\left(\ln Y_s\right).$$}: 
\begin{eqnarray}\label{eqxc1}
\alpha &=& {\cal J} \dot{{\zeta}}\,,\\
\label{eqxc2}\frac{1}{a^2}\,\partial^2\psi &=& \frac{2t_3}{3t_2}\,\alpha
+3\dot{\zeta}-\frac{2t_1}{t_2}\frac1{a^2}
\partial^2{\zeta}\,,
\end{eqnarray}
where 
\begin{equation}
{\cal J} \equiv \frac{2t_1}{t_2}=\frac{2\tilde{l}_{1}}
{\left(2H\tilde{l}_{1}-2\dot{\phi}X\tilde{G}_{X}\right)}\, =\frac{1}{H} \left[1+
\delta_{GX}+{\cal O} (\epsilon^2_{V}) \right]\,.
\label{L1expansion}
\end{equation}

Then substituting Eq~(\ref{eqxc1}) and Eq~(\ref{eqxc2})
into Eq~(\ref{eqxz}) and integrating the term 
$\dot{\zeta}\partial^2{\zeta}$ by parts the second order  
action stated in Eq~(\ref{eqxz}) can be re-expressed as:
\be\label{ac2}
S^{\zeta}_{2}=\int dt d^{3}xa^{3}Y_{S}\left[\dot{\zeta}^{2}-\frac{c^{2}_{s}}{a^{2}}(\partial\zeta)^{2}\right],\ee
where 
\begin{eqnarray}\label{contr} Y_{S}&=&\frac{t_{1}\left(4t_{1}t_{3}+9t^{2}_{2}\right)}{3t^{2}_{2}}, ~~~
\label{contr1}c^{2}_{s}=\frac{3\left(2Ht_{2}t^{2}_{1}-t_{4}t^{2}_{2}-2t^{2}_{1}\dot{t}_{2}\right)}{t_{1}\left(4t_{1}t_{3}+9t^{2}_{2}\right)}.
\end{eqnarray}
 It is important to mention here that {\it ghosts} and {\it Laplacian} instabilities
can be avoided iff $c^{2}_{s}>0, Y_{S}>0$. Now using Eq~(\ref{eqxc1}) and Eq~(\ref{contr}) in Eq~(\ref{eqxc2}) we 
get:
\begin{eqnarray}\label{contr3} \psi &=& -{\cal J}\zeta+\partial^{-2}\left(\frac{a^{2}Y_{S}\dot{\zeta}}{t_{1}}\right).
\end{eqnarray}

 For future convenience, we have introduced a new parameter defined as:
\be\begin{array}{lll}\label{contr4}\displaystyle \epsilon_{s}=\frac{Y_{S}c^{2}_{s}}{\tilde{l}_{1}}=
\frac{\left(2Ht_{2}t^{2}_{1}-t_{4}t^{2}_{2}-2t^{2}_{1}\dot{t}_{2}\right)}{t^{2}_{2}\tilde{l}_{1}}=\epsilon_{V}+\delta_{GX}+{\cal O}(\epsilon^{2}_{V}).
\end{array}\ee
Now varying the action stated in Eq~(\ref{ac2}) and expressing the solution at the linear level in terms of Fourier modes,
we arrive at the {\it Mukhanov Sasaki Equation} for Galileon scalar mode. 
\be\label{ghy}
v^{''}_{\vec{k}}+\left(c^{2}_{s}k^{2}-\frac{z^{''}}{z}\right)v_{\vec{k}}=0,
\ee
where $c^2_{s}$ takes into account the nontrivial modification due to Galileon.
Similarly
for tensor modes, Eq~(\ref{ac2}) can be recast as:
\be\label{ac3}
S^{h}_{2}=\int dt d^{3}xa^{3}Y_{T}\left[\dot{h}^{2}_{ij}
-\frac{c^{2}_{s}}{a^{2}}(\partial h_{ij})^{2}\right],\ee
where 
\begin{eqnarray}\label{yui} Y_{T}&=&\frac{t_{1}}{4}=\frac{\tilde{l}_{1}}{4},~~ c^{2}_{T}=\frac{t_{4}}{t_{1}}=1+{\cal O}(\epsilon^{2}).\end{eqnarray}
For tensor modes we use the normalization condition $e^{\lambda}_{ij}e^{\lambda^{'}}_{ij}=2\delta^{\lambda\lambda^{'}}$ and traceless condition $e_{ii}=0$ 
for polarization tensor.
Following the same prescription we can establish Eq~(\ref{ghy}) for tensor modes provided $c_{s}$
is replaced by $c_{T}$. 
The {\it Bunch-Davies} mode function turns out to be (Throughout the paper we
have used {\it DS} for {\it de-Sitter} results and {\it BDS} for {\it beyond de-Sitter} results.)
\be\label{ght}
u_{\zeta}(\eta,k)=
\left\{
	\begin{array}{ll}
                    \displaystyle\frac{iH\exp(-ikc_{s}\eta)}{2\sqrt{Y_{S}}(c_{s}k)^{\frac{3}{2}}}\left(1+ikc_{s}\eta\right)& \mbox{~~~~ \it:{\cal DS}}  \\
         \displaystyle  \frac{\sqrt{-k\eta c_{s}}}{a\sqrt{2Y_{S}}}{\cal H}^{(1)}_{\nu_{s}}(-k\eta c_{s})& \mbox{~~~~~\it:{\cal BDS }} .
          \end{array}
\right.
\ee
where $\nu_{s}=\left(\frac{3-\epsilon_{V}-2s^{S}_{V}+\delta_{V}}{2(1-
\epsilon_{V}-s^{S}_{V})}\right)$ and in the {\it super-Hubble}
 limit we have:
\begin{eqnarray}{\cal H}^{(1)}_{\nu_{s}}&\rightarrow&\frac{\left(-kc_{s}\eta\right)^{-\nu_{s}}
\exp\left(i[\nu_{s}-\frac{1}{2}]\frac{\pi}{2}\right)2^{\nu_{s}
-\frac{3}{2}}}{\sqrt{2c_{s}k}}\left(\frac{\Gamma(\nu_{s})}{\Gamma(\frac{3}{2})}\right)\end{eqnarray}.

Now using eqn(\ref{ght}) the {\it two-point} correlation function for scalar modes can be expressed as:
\be\label{tpt}
\langle0|\zeta(\vec{k})\zeta(\vec{k^{'}})|0\rangle=\frac{2\pi^{2}}{k^{3}}(2\pi)^{3}{\cal P}_{\zeta}(k)\delta^{3}(\vec{k}+\vec{k^{'}})
=(2\pi)^{3}|u_{\zeta}(\eta,k)|^{2}\delta^{3}(\vec{k}+\vec{k^{'}}),\ee
where the {\it dimensionless Power spectrum} for scalar modes ${\cal P}_{\zeta}(k)$ at the horizon crossing turns out to be:
\be\label{ght1}
{\cal P}_{\zeta}(k_{\star})=\frac{k^{3}_{\star}}{2\pi^{2}}|u_{\zeta}(k_{\star})|^{2}=
\left\{
	\begin{array}{ll}
                    \displaystyle\left(\frac{\sqrt{V(\phi)}}{8\pi^{2}c_{s}\epsilon_{s}\tilde{l}_{1}\sqrt{\tilde{g}_{1}}M_{p}}\right)_{\star}
& \mbox{~ \it:{\cal DS}}  \\
         \displaystyle\left(2^{2\nu_{s}-3}\left|\frac{\Gamma(\nu_{s})}{\Gamma(\frac{3}{2})}\right|^{2}\frac{\left(1-\epsilon_{V}-s^{S}_{V}\right)^{2}\sqrt{V(\phi)}}{8\pi^{2}Y_{S}
c^{3}_{s}\sqrt{\tilde{g}_{1}}M_{p}}\right)_{\star}& \mbox{~~\it:{\cal BDS }} .
          \end{array}
\right.
\ee
 $\star$ corresponds to the horizon crossing.
Similarly using the tensor version of eqn(\ref{ght}) the {\it two-point} correlation function for tensor modes can be expressed as:
\be\label{tptz}
\langle0|h_{ij}(\vec{k})h_{ij}(\vec{k^{'}})|0\rangle=\frac{2\pi^{2}}{k^{3}}(2\pi)^{3}{\cal P}_{T}(k)\delta^{3}(\vec{k}+\vec{k^{'}})
=(2\pi)^{3}|u_{\zeta}(\eta,k)|^{2}\delta^{3}(\vec{k}+\vec{k^{'}}),\ee
where ${\cal P}_{T}(k)=[{\cal P}_{T}(k)]_{ij;ij}$ and the corresponding {\it dimensionless Power spectrum} for tensor modes reads:
\be\label{ght2}
{\cal P}_{T}(k_{\star})=\frac{k^{3}_{\star}}{2\pi^{2}}|u_{h}(k_{\star})|^{2}\left(\sum_{\lambda = +,\times}e^{\lambda}_{ij}e^{\lambda}_{ij}\right)=
\left\{
	\begin{array}{ll}
                    \displaystyle\left(\frac{\sqrt{V(\phi)}}{2\pi^{2}c_{T}\epsilon_{T}\tilde{l}_{1}\sqrt{\tilde{g}_{1}}M_{p}}\right)_{\star}
& \mbox{~ \it:{\cal DS}}  \\
         \displaystyle \left(2^{2\nu_{T}-3}\left|\frac{\Gamma(\nu_{T})}{\Gamma(\frac{3}{2})}\right|^{2}\frac{\left(1-\epsilon_{V}-s^{T}_{V}\right)^{2}\sqrt{V(\phi)}}{2\pi^{2}Y_{T}
c^{3}_{T}\sqrt{\tilde{g}_{1}}M_{p}}\right)_{\star}& \mbox{~~\it:{\cal BDS }} .
          \end{array}
\right.
\ee
Consequently the ratio of tensor to scalar power spectrum can be expressed as:
\be\label{ght3}
{r}(k_{\star})=\frac{{\cal P}_{T}(k_{\star})}{{\cal P}_{\zeta}(k_{\star})}=
\left\{
	\begin{array}{ll}
                    \displaystyle \left(16\epsilon_{s}c_{s}\left[1-\frac{3}{2}{\cal O}(\epsilon^{2}_{T})\right]\right)_{\star}
& \mbox{~ \it:{\cal DS}}  \\
         \displaystyle \left(16. 2^{2(\nu_{T}-\nu_{s})}\left|\frac{\Gamma(\nu_{T})}{\Gamma(\nu_{s})}\right|^{2}\left(\frac{1-\epsilon_{V}-s^{T}_{V}}
{1-\epsilon_{V}-s^{S}_{V}}\right)^{2}c_{s}\epsilon_{s}\left[1-\frac{3}{2}{\cal O}(\epsilon^{2}_{T})\right]\right)_{\star}& \mbox{~~\it:{\cal BDS }} .
          \end{array}
\right.
\ee
Further, the scale dependence of the perturbations, described by
the scalar and tensor spectral indices, as follows:
\be\label{ght3}
{n}_{\zeta}-1=\left(\frac{d\ln{\cal P}_{\zeta}}{d\ln k}\right)_{\star}=
\left\{
	\begin{array}{ll}
                    \displaystyle \left(-2\epsilon_{V}-\eta_{s}-s^{S}_{V}\right)_{\star}
=\left(-2\epsilon_{s}-\eta_{s}-s^{S}_{V}+2\delta_{GX}+2{\cal O}(\epsilon^{2}_{s})\right)_{\star}& \mbox{ \it:{\cal DS}}  \\
         \displaystyle  \left(3-2\nu_{s}\right)=-\left(\frac{2\epsilon_{V}+s^{S}_{V}+\delta_{V}}{1-\epsilon_{V}-s^{S}_{V}}\right)_{\star}& \mbox{\it:{\cal BDS }} .
          \end{array}
\right.
\ee
\be\label{ght4}
{n}_{T}=\left(\frac{d\ln{\cal P}_{T}}{d\ln k}\right)_{\star}=
\left\{
	\begin{array}{ll}
                    \displaystyle -2\epsilon_{V}|_{\star}
=\left(-2\epsilon_{s}+2\delta_{GX}+2{\cal O}(\epsilon^{2}_{s})\right)_{\star}& \mbox{ \it:{\cal DS}}  \\
         \displaystyle  \left(3-2\nu_{T}\right)=-\left(\frac{s^{T}_{V}}{1-\epsilon_{V}-s^{T}_{V}}\right)_{\star}& \mbox{\it:{\cal BDS }} .
          \end{array}
\right.
\ee
Consistently, the consistency relation is also modified to:
\be\label{ght5}
{r}=
\left\{
	\begin{array}{ll}
                    \displaystyle-\left(8c_{s}\left(n_{T}-2\delta_{GX}
-2{\cal O}\left(\frac{n^{2}_{T}}{4}\right)\right)\left[1-\frac{3}{2}{\cal O}(\epsilon^{2}_{T})\right]\right)_{\star}
& \mbox{~ \it:{\cal DS}}  \\
         \displaystyle
\displaystyle \left(8. 2^{(n_{\zeta}-n_{T})}\left|\frac{\Gamma(\frac{3-n_{T}}{2})}{\Gamma(\frac{3-n_{\zeta}}{2})}\right|^{2}\left(\frac{\frac{s^{T}_{V}}{n_{T}}}
{s^{T}_{V}\left(1-\frac{1}{n_{T}}\right)-s^{S}_{V}}\right)^{2}c_{s}\left(2\left[1+s^{T}_{V}\left(\frac{1}{n_{T}}-1\right)\right]
\right.\right.\\ \left.\left.\displaystyle ~~~~~~~~~~~~~+2\delta_{GX}+2{\cal O}(\epsilon^{2}_{V})\right)
\left[1-\frac{3}{2}{\cal O}(\epsilon^{2}_{T})\right]\right)_{\star} & \mbox{~~\it:{\cal BDS }} .
          \end{array}
\right.
\ee

The expressions for the
running of the scalar and tensor spectral index in this specific 
model with respect to the logarithmic pivot scale at the horizon crossing are given by:

\be\label{ght6}
{\alpha}_{\zeta}=\left(\frac{d{n}_{\zeta}}{d\ln k}\right)_{\star}=
\left\{
	\begin{array}{ll}
                    \displaystyle \left\{\frac{4\sqrt{V^{'}(\phi)}M^{2}_{p}}{\sqrt{{\cal G}(\phi)}}\left(-2\epsilon^{~'}_{V}-\eta^{'}_{s}-s^{S~'}_{V}\right)\right\}_{\star}& \mbox{ \it:{\cal DS}}  \\
         \displaystyle  \frac{4\sqrt{V^{'}(\phi)}M^{2}_{p}}{\sqrt{{\cal G}(\phi)}\left(1-\epsilon_{V}-s^{S}_{V}\right)^{2}}
\left[\underbrace{s^{S~'}_{V}\epsilon^{~'}_{V}}+\underbrace{\delta^{~'}_{V}\epsilon^{~'}_{V}}
+\underbrace{s^{S~'}_{V}\delta^{~'}_{V}}+\left(2\epsilon^{~'}_{V}+s^{S~'}_{V}+\delta^{~'}_{V}\right)\right]& \mbox{\it:{\cal BDS }} .
          \end{array}
\right.
\ee
\be\label{ght7}
{\alpha}_{T}=\left(\frac{d{n}_{T}}{d\ln k}\right)_{\star}=
\left\{
	\begin{array}{ll}
                    \displaystyle -2\left\{\frac{4\sqrt{V^{'}(\phi)}M^{2}_{p}}{\sqrt{{\cal G}(\phi)}}\epsilon^{~'}_{V}\right\}_{\star}& \mbox{ \it:{\cal DS}}  \\
         \displaystyle -\left[\frac{4\sqrt{V^{'}(\phi)}M^{2}_{p}}{\sqrt{{\cal G}(\phi)}}
\left(\frac{s^{T~'}_{V}}{1-\epsilon_{V}-s^{T}_{V}}\right)+\frac{s_{T}\left(\epsilon^{~'}_{V}+s^{T~'}_{V}\right)}{\left(1-\epsilon_{V}-s^{T}_{V}\right)^{2}}\right]_{\star}& \mbox{\it:{\cal BDS }} .
          \end{array}
\right.
\ee

Here we have used a shorthand notation $\underbrace{ab}=a^{'}b-ab^{'}$ where $'=\frac{d}{d\phi}$.
We also use the operator identity $\frac{d}{d\ln k}:=\frac{4\sqrt{V^{'}(\phi)}M^{2}_{p}}{\sqrt{{\cal G}(\phi)}}\frac{d}{d\phi}$ to compute all the inflationary observables. 

\begin{figure}[t]
\centering
\subfigure[~$l(l+1)C_l^{TT}/2\pi$ vs $l$]{
    \includegraphics[width=7cm,height=4.5cm] {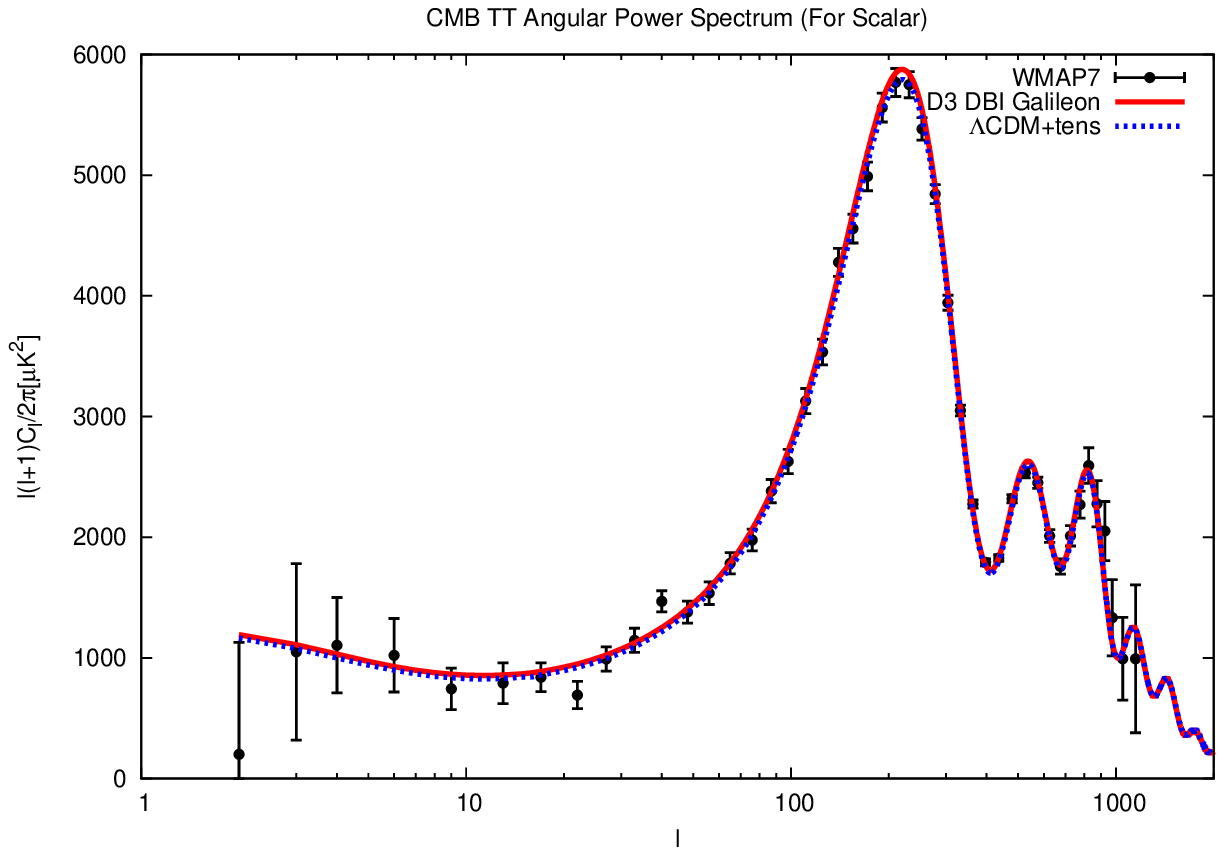}
    \label{zc1dbi}}
\subfigure[~$l(l+1)C_l^{TE}/2\pi$ vs $l$]{
    \includegraphics[width=7cm,height=4.5cm] {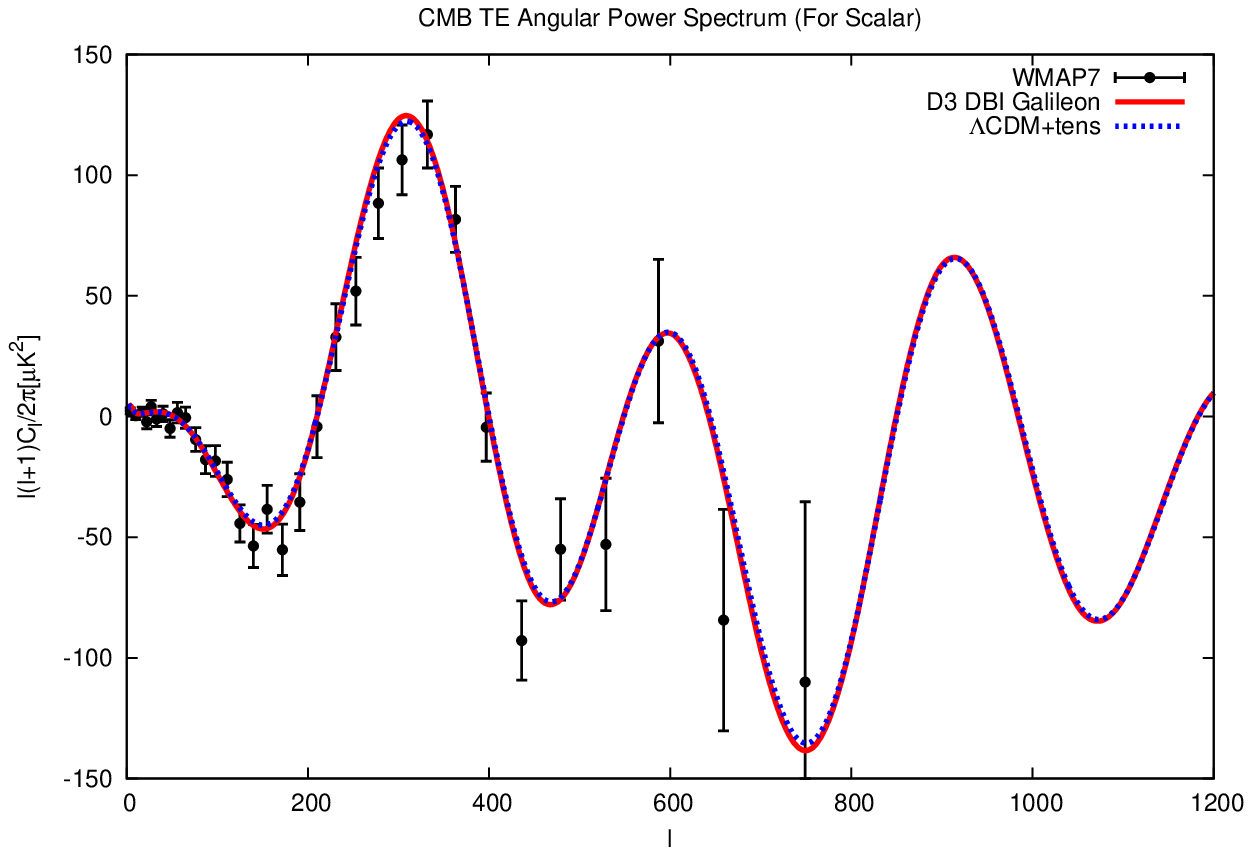}
    \label{zc2dbi}
}
\subfigure[~$l(l+1)C_l^{EE}/2\pi$ vs $l$]{
    \includegraphics[width=7cm,height=4.5cm] {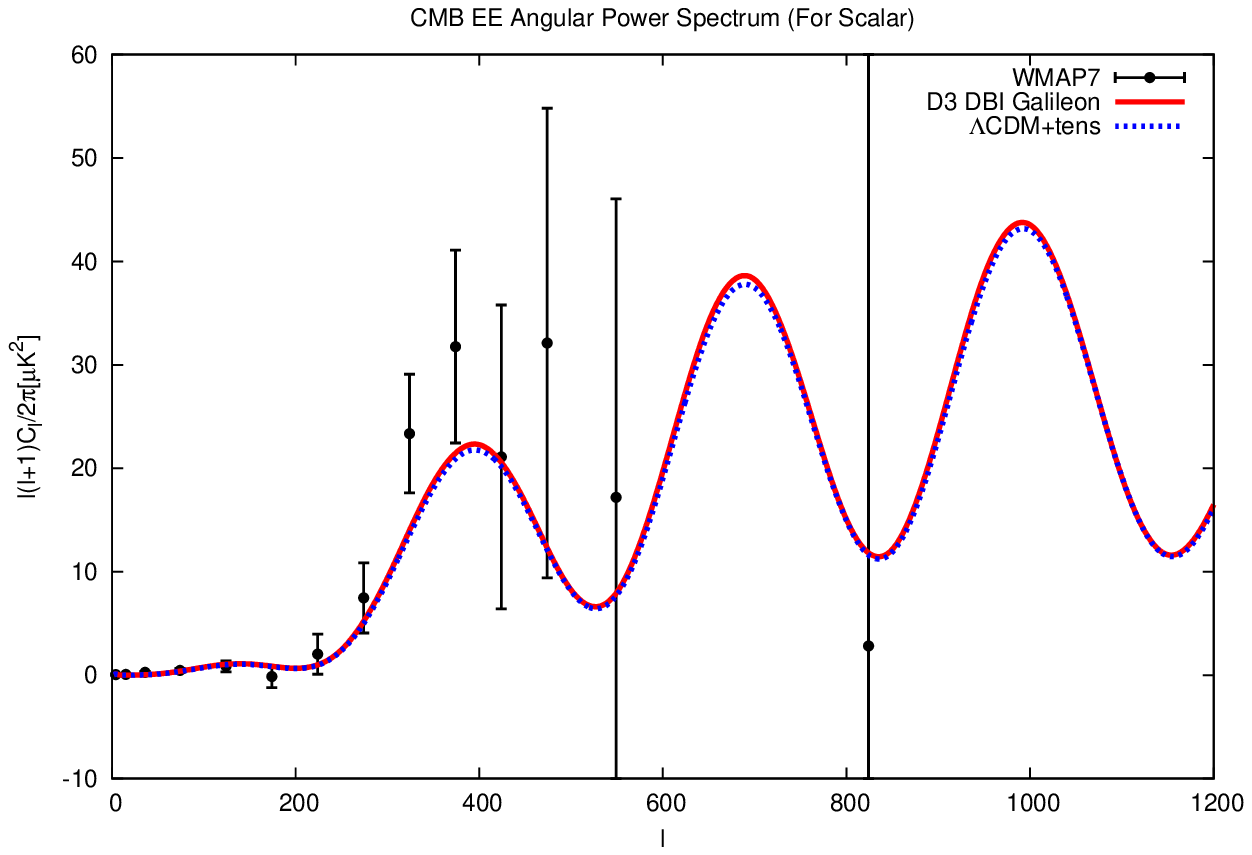}
    \label{zc3dbi}
}
\subfigure[~$P_{MAT}(k)$ vs $k$]{
    \includegraphics[width=7cm,height=4.5cm] {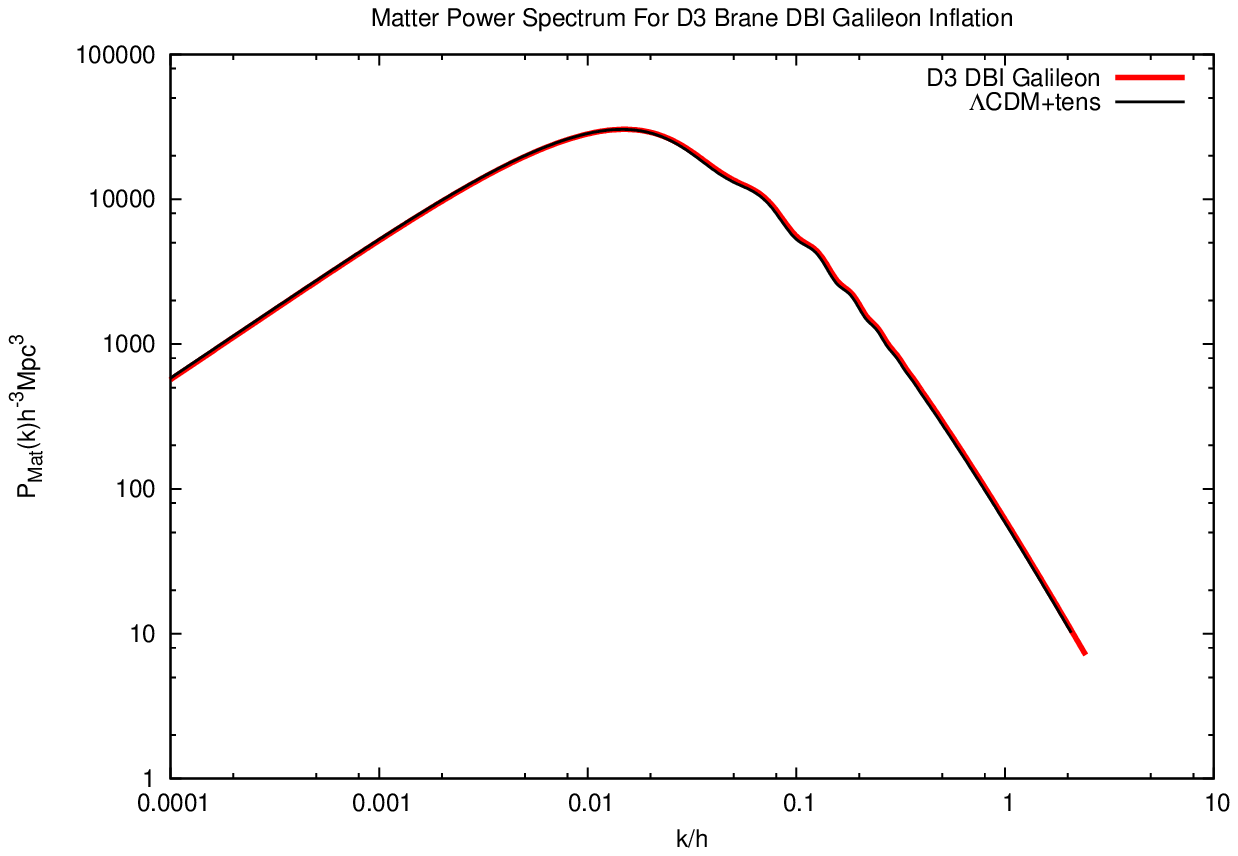}
    \label{zc4dbi}
}
\caption[Optional caption for list of figures]{ Variation of CMB angular power spectrum for \subref{zc1dbi}~TT,
\subref{zc2dbi}~TE and \subref{zc3dbi}~EE correlation with respect to multipole $l$ for the best fit model parameters.
Also in \subref{zc4dbi} we show the variation of matter power spectrum with respect to the momentum scale \cite{Choudhury:2012yh}.
}
\label{figdbi}
\end{figure}

\subsection{Parameter estimation using CAMB}
\label{d5a}
Using the parameter space for the model parameters ($C_{i},D_{i}$) 
we have estimated the window of the cosmological parameters from our model which confronts
 observational data well in $56<N<70$. In Table(\ref{tabg21}) we have tabulated the relevant observational parameters 
estimated from our model for both {\it DS} and {\it BDS} limit.

\begin{table}[h]
\begin{center}
\begin{tabular}{|c|c|c|c|c|c|c|c|c|c|}
\hline Scheme & ${\cal P}_{\zeta}$ & $r$ &$n_{\zeta}$& $\alpha_{\zeta}$
 \\  &$\times 10^{-9}$
 & & & $\times \left(-10^{-3}\right)$\\
 \hline
DS&~2.401~-~2.601~&~0.215~-~0.242~&~0.964~-~0.966~&~2.240~-~2.249~\\
\hline
BDS&~ 2.471~-~2.561~&~0.232~-~0.250~&~0.962~-~0.964~&~4.008~-~4.012~\\
\hline
\end{tabular}
\caption{Model Dependent Observational Parameters \cite{Choudhury:2012yh}.}\label{tabg21}
\end{center}
\end{table}


\begin{table}[h]
\begin{center}
\begin{tabular}{|c|c|c|c|c|c|c|c|c|c|}
\hline $H_0$ & $\tau_{Reion}$ &$\Omega_b h^2$& $\Omega_c h^2
$& $T_{CMB}$
 \\
km/sec/MPc& & && K\\
 \hline
71.0&0.09&0.0226&0.1119&2.725\\
\hline
\end{tabular}
\caption{Input parameters in CAMB \cite{Choudhury:2012yh}.}\label{tabg2}
\end{center}
\end{table}

\begin{table}[h]
\begin{center}
\begin{tabular}{|c|c|c|c|c|c|c|c|c|c|}
\hline $t_0$ & $z_{Reion}$ &$\Omega_m$&$\Omega_{\Lambda}$&$\Omega_k$&$\eta_{Rec}$& $\eta_0$
 \\
Gyr& & && &Mpc & Mpc\\
 \hline
13.707&10.704&0.2670&0.7329&0.0&285.10&14345.1\\
\hline
\end{tabular}
\caption{Output parameters from CAMB \cite{Choudhury:2012yh}.}\label{tabg3}
\end{center}
\end{table}
Further, we use the publicly available code CAMB \cite{CAMB}
to verify our results directly with observation. To
operate CAMB at the pivot scale $k_{0} = 0.002 Mpc^{-1}$
the values of the initial parameter space are taken for
 ̃lower bound of $C_{i}^{'}$s and $N = 70$. Additionally
WMAP7 years dataset for $\Lambda CDM$ background has
been used in CAMB to obtain CMB angular power spectrum. In Table(\ref{tabg2}) we have given all the input parameters
for CAMB. Table(\ref{tabg3}) shows the CAMB output, which is
in good fit with WMAP7 \cite{WMAP7} data. In
fig.~(\ref{figdbi})(a)-figure(\ref{figdbi})(c) we have plotted CAMB output of CMB TT, TE and EE
 angular power spectrum $C^{TT}_{l}, C^{TE}_{l}, C^{EE}_{l}$ for the best fit with WMAP7
 data for scalar mode, which explicitly show the
agreement of our model with WMAP7 dataset. The small scale modes have no
impact in the CMB anisotropy spectrum only the large scale modes
have little contribution. Hence in fig.~(\ref{zc4dbi}) we
 have plotted the variation of matter power spectrum
with respect to the momentum scale which is in concordance with observational results.

\section{Chapter summary}
\label{bd}

In this chapter we have studied single field inflation in the context of Randall-Sundrum brane and DBI Galileon
induced D3 brane respectively. The major outcomes of our study are:
\begin{itemize}
 \item We have demonstrated the technical details of construction mechanism of an one-loop
4D inflationary potential via dimensional reduction starting from ${\cal N}=2, {\cal D}=5$ SUGRA in
the bulk which leads to an effective field theoretic picture within ${\cal N}=1, {\cal D}=4$ SUGRA embedded in the brane for both the cases.

\item Hence we have studied inflation using the one loop effective potential
by estimating the observable parameters originated from primordial quantum fluctuation for scalar and tensor modes.

\item We have further confronted our results with WMAP7 \cite{WMAP7} dataset by
using the cosmological code CAMB \cite{CAMB}.

\item Hence we have generated the theoretical CMB angular power spectra from 
TT, TE and EE correlation for scalar modes from both the proposed inflationary frameworks and fit with the observed CMB angular power spectra obtained from WMAP7 data.

\item On the top of that in this chapter we have proposed new sets of inflationary consistency relations in the case of braneworld and DBI Galileon framework which is 
quite different from the results obtained from usual General Relativistic framework.
\end{itemize}

\chapter{Reheating \& Leptogenesis in brane inflation}
\label{ch:FRWxc2}

\section{Introduction}
\label{g1}

It is now  well accepted  fact that the post big bang universe \cite{Hawking:1969sw}
 passed through different phases having cosmo-phenomenological
significance. One of the significant phases, namely, reheating \cite{Mazumdar:2010sa,Kofman:1994rk}
 plays the pivotal role in explaining
 production of different particles from inflaton and vacuum energy. As we look back
in time reheating was completed within approximately the first second
after the big bang. At that time nucleosynthesis \cite{Burles:2000ju},
or the formation of light nuclei occurred. On the other hand
the mysterious force that drives the inflationary phase is conventionally described
 by a scalar field, named  inflaton which oscillates
near the minimum of its effective potential and produces
elementary particles \cite{Palma:2000md}. These particles interact with each
other and eventually they come to a state of thermal
equilibrium at some arbitrary temperature. This process completes
when all the energy of the classical
scalar field transfer to the thermal energy of elementary
particles. Since long theoretical physicists have been investigating reheating as a perturbative phase \cite{Allahverdi:2010xz}, or one
 in which single inflaton quanta decayed individually into ordinary matter and
 radiation~\footnote{The recent theoretical studies have shown that in many cases the decay
 occurs through a non-perturbative process \cite{Traschen:1990sw}, in which the particles behave in an
 ordered manner. Non-perturbative processes involved at reheating are extremely
 more efficient than the perturbative ones \cite{Allahverdi:2007zz} and often more difficult to investigate
in practice.}. In short there is
 no existence of a complete theory which explains non-perturbative effects during reheating
 for the total time scale.

     Besides production of gravitinos during perturbative reheating \cite{Copeland:2005qe}  its
 decay plays a significant role in the context of leptogenesis  \cite{Davidson:2008bu}. More precisely
 two types of gravitinos are produced in this epoch - stable \cite{Buchmuller:2006tt} and unstable \cite{Kohri:2005wn}.
 Stable ones and decay products of
unstable ones directly or indirectly stimulate the light element
abundances during big bang nucleosynthesis. Most importantly the
 unstable one has important cosmological consequences out of which the major one directly affects the expansion
rate of the universe. In order to explain cosmological consequences at a time by a single physical entity, it
is customary to explain everything in terms of gravitino energy density which is directly
 proportional to the gravitino number density or gravitino
abundance. This gravitino abundance is obtained by
considering gravitino production in the radiation dominated
era following reheating \cite{Choi:1999xm}. Gravitinos are originated through thermal scattering \cite{Pradler:2006qh} in the
early universe and are usually related to the reheating temperature.
 Particle physics phenomenology usually requires that under instantaneous decay
 approximation \cite{kolb} reheating temperature is maximum during
  reheating.

 In this chapter we have studied extensively reheating and leptogenesis
in a typical brane inflation model in the framework
of ${\cal N}=1,{\cal D}=4$ SUGRA in braneworld
 which was proposed in the chapter \ref{ch:FRWxc1} in section \ref{b2}. The standard results of reheating and
 leptogenesis are obtained considering four dimensional Einsteinian gravity~\cite{kolb,Rangarajan:2008zb}~\footnote{For standard results of reheating mechanism and leptogenesis
in Einsteinian GR see also Appendix D.}. 
Here we show that the results are dramatically different if one considers brane inflation starting from higher dimension resulting in effective non-Einsteinian gravity in 
four dimension. This has serious implication for the production of the heavy
   Majorana neutrinos needed for leptogenesis \cite{Frere:1999rh}.
We further estimate different  parameters related to reheating and leptogenesis at the epoch of
 phase transition. Finally, we have given an estimate of CP violation which is the indirect evidence of
the baryon asymmetry in the present context.

\section{Background model}
\label{g2}

 From the knowledge of particle physics it is known that during the epoch of reheating inflatons
decay into different particle constituents \cite{Kofman:1994rk} are directly related to the
 trilinear coupling of the inflaton field. There might be a possibility of collision originated
 through quartic coupling and driven by background scalar field. For example here the contribution from the heavy Majorana neutrino comes
from the seesaw Lagrangian \be {\cal L}_{Majo}=-h_{ij}\bar{l}_{L,i}{\cal H}\psi_{j}-\frac{1}{2}\sum_{i}{\cal M}_{i}\bar{\psi}_{i}\psi_{i}+h.c.,\ee where
where $i, j = 1, 2, 3$ denote the generation indices, $h$ is the Yukawa coupling,
 $l_{L}$ and ${\cal H}$ are
the lepton and the Higgs doublets, respectively, and ${\cal M}_{i}$ is the lepton-number-violating mass
term of the right-handed neutrino. In this chapter, we assume the hierarchical mass spectrum for the
heavy neutrinos, ${\cal M}_{1}\ll {\cal M}_{2}, {\cal M}_{3}$, for simplicity.
Now using the another assumption, inflaton mass $ m_{\phi}\gg m_{\sigma}$,$m_{\phi}\gg
m_{\psi}$~\footnote{Here $m_{\sigma}$ and $m_{\psi}$ be the background scalar field mass and fermion mass respectively.} the {\it total inflaton decay width} for the positively and negatively charged
$\phi(\phi^{+},\phi^{-})$ scalar fields as well as the fermionic field $\psi$~\footnote{For the heavy Majorana neutrinos the decay
process $\psi\rightarrow l_{L}{\cal H}$, $\psi\rightarrow \bar{l}_{L}{\cal H}$ predominates.}
 is given by: \be\Gamma_{total}\simeq \frac{C^2}{16\pi
m_{\phi}}+\frac{h^{2}m_{\psi}}{4\pi}\sim
\frac{1}{(2\pi)^{3}}\left(\frac{\Delta^{6}}{M^{5}}\right)\ee where
the coupling strength
$C\sim m_{\phi}\left(\frac{\Delta^{2}}{M^{2}}\right)$ and $h\sim\left(\frac{\Delta^{2}}{M^{2}}\right)$ and the background
scalar field is $\sigma$.

Now to construct the thermodynamical observable the effective number of particles incorporating relativistic degrees of freedom is defined \cite{kolb} as
~\footnote{For the phenomenological
estimation \cite{Linde:2005ht} $N^{*}\sim 10^{2}-10^{4}$ and for realistic
models $N^{*}\sim 10^{2}-10^{3}$.}:
 \be N^{*}=N^{*}_{B}+\frac{7}{8}N^{*}_{F},\ee  where $N^{*}_{B}=\sum_{i}N^{*}_{Bi}$ and $N^{*}_{F}=\sum_{j}N^{*}_{Fj}$. Here $N^{*}_{B}$ represents the number of
bosonic degrees of freedom with mass $m_{\phi}\ll T$ and
$N^{*}_{F}$ represents number of fermionic degrees of freedom with
mass $m_{\psi}\ll T$. Here `i' and `j' stand for different
bosonic and fermionic species respectively.  For convenience let us express reheating
temperature on the brane as:
\be\label{ga}\Gamma_{total}=3H(T^{br})=\sqrt{\frac{3\rho(t_{reh})}{M^{2}}
\left[1+\frac{\rho(t_{reh})}{2\lambda}\right]},\ee
where $\lambda$, $H(T^{br})$ and $\rho(t_{reh})$ be the brane tension, Hubble parameter and energy density during reheating respectively.
In Eq~(\ref{ga}) the correction term in the Hubble parameter is significant in the high energy limit where the energy density
 is large compared to the brane tension i.e. $\rho>>\lambda$. On the other hand, in the limit $\rho<<\lambda$ we get back the standard result in Einsteinian gravity. 
It is worth mentioning that the brane reheating temperature does not depend on the initial value
of the inflaton field and is solely determined by the elementary particle theory of the early universe.

\section{Phase transition in brane inflation}
\label{g3}

Phase transition in braneworld scenario is weakly
first order in nature \cite{Narlikar:1990bu}. So it is convenient to write the
brane reheating temperature in terms of the critical parameters. To serve this purpose
the critical density and the critical temperature or transition temperature can be written as :
\be\label{roz}\rho(t_{c})=2\lambda=\frac{3}{16\pi^{2}}\frac{M^{6}_{5}}{M^{2}},~~~T_{c}=\sqrt{\frac{3}{\pi}\sqrt{\frac{5}{\pi
N^{*}}}\frac{M^{3}_{5}}{M}}\ee
which makes a bridge between the phenomenology and observation. In the high energy limit 5D Planck mass ($M_{5}$)
can be expressed in terms of our model parameters as~\footnote{Here $\Delta^{2}_{s}$ represents the amplitude of the scalar perturbation defined in Eq~(\ref{scalar})
 in chapter \ref{ch:FRWxc1}. Most importantly here the subscript $\star$
represents here the epoch of horizon crossing ($k=aH$) and $\alpha$ represents a dimensionless model parameter defined as, $\alpha=\frac{\Delta^{4}}{\lambda}$, where $\Delta$
represents the energy scale of brane inflation.}:
 \be M_{5}=\sqrt[6]{\frac{6400\pi^{4}\Delta^{2}_{s}(K_{4}+4D_{4})^{2}}{\alpha^{4}}}\phi_{\star}.\ee

The major thermodynamic quantities -- critical density ($\rho_{c}$), critical pressure ($P_{c}$), critical entropy ($S_{c}$)
 -- and the Hubble parameter at the critical temperature ($H_{c}$) related to the phase
transition designated by the following fashion for our model:
\begin{eqnarray}
 \label{comp}\rho_c &=& 1200 \phi^{4}_{\star}A(\phi_{\star}) ~\forall \gamma\in J,\\
\label{compv1} P_c &=& 400 \phi^{4}_{\star}A(\phi_{\star})= \frac{\rho_c}{3} ~\forall \gamma\in J,\\
\label{compv2} S_c &=& \frac{1600\phi^{4}_{\star}A(\phi_{\star})}{T_{c}}=\frac{4 \rho_c}{3 T_c}~ \forall \gamma\in J,\\
\label{compv3} H_c &=& \frac{20\sqrt{A(\phi_{\star})}\phi^{2}_{\star}}{M}=\frac{\sqrt{\rho_c}}{\sqrt{3} M}~\forall \gamma\in J
\end{eqnarray}

where we have defined a dimensionless characteristic quantity: \bea A(\phi_{\star})&=&
\frac{\pi^{2}(K_{4}+4D_{4})^{2}\Delta^{2}_{s}\phi^{2}_{\star}}{\alpha^{4}M^{2}}\eea
 at the horizon crossing in this context.
 The above mentioned physical
 quantities are function of the critical or transition temperature which is defined as
\be\begin{array}{ll}\label{ubi}\displaystyle T_{c}:=\sqrt[4]{\left\{C_{\gamma}
\frac{A(\phi_{\star})\phi^{4}_{\star}}{\pi^{2}N^{*}_{\gamma}}\right\}}
~~ {\rm with} ~~ C_{\gamma}=\left(36000,\frac{288000}{7},19200\right)\forall \gamma\in J\end{array}\ee
 with gauge group $J:=SU(2)_{L}\otimes U(1)_{Y}$ and the species index $\gamma=1(B\Rightarrow Boson),2(F\Rightarrow Fermion),
3(M\Rightarrow Mixture)$.

\section{Reheating temperature}
\label{g4}

 In the present context the reheating temperature can be written \cite{Felipe:2004wh} as:
 \bea\label{reh1}
 T^{br}&=&\frac{T_{c}}{\sqrt[4]{2}}\sqrt[4]{\left[\sqrt{1+\frac{5}{\pi^{3}N^{*}}
\left(\frac{\Gamma_{total}M_{PL}}{T^{2}_{c}}\right)^{2}}-1\right]}.
\eea
In the high energy regime the reheating temperature can berecast as:
\bea
T^{brh}&=&\sqrt[4]{\left\{\sqrt{\frac{10}{N^{*}}}\frac{2\sqrt{2}M\Gamma_{total} T^{2}_{c}}{3\pi}\right\}}.
\eea
But in this context we are confining ourselves into
the Standard Model regime. So to construct a fruitful model of reheating in the context of Standard Model gauge group, we rewrite all general
 principal components in terms of physical degrees of freedom in a compact fashion. We consider a one to one high energy mapping such that:
\be\begin{array}{ll}\label{f1}\displaystyle 
T^{br}_{\gamma}=\frac{T_{c\gamma}}{\sqrt[4]{2}}\sqrt[4]{\left[\sqrt{1+\frac{Z_{\gamma}}{\pi^{3}N^{*}_{\gamma}}\left(\frac{\Gamma_{total}M_{PL}}
{T^{2}_{c\gamma}}\right)^{2}}-1\right]}\Rightarrow 
T^{brh}_{\gamma}=\sqrt[4]{\left\{\frac{W_{\gamma}(K_{4}+4D_{4})\Delta_{s}\phi^{3}_{\star}\Gamma_{total}}{\pi
N^{*}_{\gamma}\alpha^{2}}\right\}}\forall \gamma \in J\end{array}\ee it maps the actual brane reheating temperature ($T^{br}_{\gamma}$) to its high energy value ($T^{brh}_{\gamma}$) in the Standard Model gauge group $J:=SU(2)_{L}\otimes U(1)_{Y}$
with $Z_{\gamma}=\left(5,\frac{40}{7},\frac{8}{3}\right)$,$W_{\gamma}=\left(600,\frac{4800}{7},320\right)$ and $\gamma=1(B),2(F),3(M)$. 
Most importantly the superscript `br' and `brh' stands for parameters before and after high energy mapping respectively.
 Here it should be mentioned that the
brane reheating temperature incorporates all the effects of heavy Majorana neutrinos as well as the other fermions and bosons
through the total decay width $\Gamma_{total}$.

The reheating temperature for different species can readily be calculated from our model as proposed in in chapter \ref{ch:FRWxc1}.
 For a typical value of model parameters, 
$C_{4}\simeq D_{4}=-0.7$, we have:\\
\begin{enumerate}
 \item For boson:~~ $T^{brh}_{B}\simeq7.6\times 10^{10}~{\rm GeV}$,
\item For fermion:~~ $T^{brh}_{F}\simeq7.8\times 10^{10}~{\rm GeV}$,
\item For mixture of boson and fermion:~~ $T^{brh}_{M}\simeq6.5\times 10^{10}~{\rm GeV}$.
\end{enumerate}
  These results are significantly different from GR value $T^{reh}\leq10^{9}~{\rm GeV}$ and is a characteristic feature of brane inflation with non-Einstenian framework.

\section{Gravitino production via leptogenesis}
\label{g5}

Let us now move on to studying how the self interacting term of our model is directly related to the
leptogenesis through the production of thermal gravitinos which is a special ingredient for the heavy  Majorana neutrinos
in the leptogenesis. Let us start with a physical situation where the inflaton field starts
oscillating when the inflationary epoch ends at a cosmic time
$t=t_{osc}\simeq t_{f}$. Throughout the analysis we have assumed that the
universe is reheated through the perturbative decay of the
inflaton field for which the reheating phenomenology in brane is described
by the Boltzmann equation \cite{kolb}:
\be\label{rfg} \dot{\rho_{r}}+4H\rho_{r}=\Gamma_{\phi}\rho_{\phi},\ee
where in braneworld
\be\begin{array}{ll}\label{rgg1} \displaystyle H^{2}=\frac{8\pi}{3M^{2}_{PL}}\left(\rho_{r}+\rho_{\phi}\right)\left[1+\frac{(\rho_{r}+\rho_{\phi})}{2\lambda}\right]
=H^{2}_{osc}\left(\frac{a_{osc}}{a}\right)^{4}\left[1+\frac{\alpha}{2}\left(\frac{a_{osc}}{a}\right)^{4}\right].\end{array}\ee
Here $\rho_{r}$ and $\rho_{\phi}$ represent the energy density of radiation and inflaton respectively and
$\Gamma_{\phi}$ is the rate of dissipation of the inflaton field energy
density. At $t=t_{osc}$ epoch the Hubble parameter is designated by \cite{kolb}:\be\label{ddg}
H_{osc}=\sqrt{\frac{8\pi}{3}}\frac{\Delta^{2}}{M_{PL}}=\frac{\Delta^{2}}{\sqrt{3}M}.\ee
 Assuming $\Gamma_{\phi}\gg H$ from we get \be\label{gh1}
\rho_{\phi}=\Delta^{4}\left(\frac{a_{osc}}{a}\right)^{4}\exp\left[-\Gamma_{\phi}(t-t_{osc})\right].\ee
It is worthwhile to mention here that the inflaton field $\phi$ follows
 an equation of state similar to radiation rather than matter i.e. $\omega_{\phi}=\frac{P_{\phi}}{\rho_{\phi}}$~\footnote{
Here pressure can be expressed as, $P_{\phi}=\rho_{\phi}-2V(\phi)$, where $V(\phi)$ is the potential derived in Eq~(\ref{post}) of chapter \ref{ch:FRWxc1}.}
Now solving Friedmann equation the dynamical character of the scale factor can be expressed as \be\label{dt1}
a(t)=a_{osc}\sqrt[4]{\left[\left[\sqrt{1+\frac{\alpha}{2}}+2H_{osc}(t-t_{osc})\right]^{2}-\frac{\alpha}{2}\right]},\ee
where we use a specific notation $a(t_{osc})=a_{osc}$.

  Plugging Eq~(\ref{dt1}) and Eq~(\ref{gh1}) in Eq~(\ref{rfg})
we get: \be\begin{array}{ll}\label{opp}
\displaystyle\dot{\rho_{r}}+\frac{2H_{osc}}{\left[\left[\sqrt{1+\frac{\alpha}{2}}+2H_{osc}(t-t_{osc})\right]^{2}-\frac{\alpha}{2}\right]}\rho_{r}
=\frac{\Gamma_{\phi}\Delta^{4}\exp\left[-\Gamma_{\phi}(t-t_{osc})\right]
}{\left[\left[\sqrt{1+\frac{\alpha}{2}}+2H_{osc}(t-t_{osc})\right]^{2}-\frac{\alpha}{2}\right]}.\end{array}\ee
 As a whole phenomenological construction of gravitino abundance is governed by the above equation. But Eq~(\ref{opp})
is not exactly analytically solvable. So we are confining our attention to the high energy limit where
the Friedmann equation (\ref{rgg1}) can be approximated as:
\be\label{ty} H^{2}=\frac{8\pi}{6\lambda
       M^{2}_{PL}}\left(\rho_{r}+\rho_{\phi}\right)^{2}=\frac{\alpha}{2}H^{2}_{osc}\left(\frac{a_{osc}}{a}\right)^{8},\ee
     whose solution is given by: \be\label{olpiuy}a(t)=a_{osc}\sqrt[4]{\left[1+2\sqrt{2\alpha}H_{osc}(t-t_{osc})\right]}.\ee

     Now using an physically viable assumption $t\leq \Gamma^{-1}_{\phi}$ the exact solution of the Eq~(\ref{opp}) in the high energy limit can be written as: \bea\label{sd1}
        \rho_{r} &\simeq& \frac{3M^{2}H^{2}_{osc}\Gamma_{\phi}(t-t_{osc})}{\left[1+2\sqrt{2\alpha}H_{osc}(t-t_{osc})\right]}
            =\frac{3M^{2}H_{osc}\Gamma_{\phi}}{2\sqrt{2\alpha}}\left(\frac{a_{osc}}{a}\right)^{4}\left[\left(\frac{a}{a_{osc}}\right)^{4}-1\right].\eea

Our intention is to find out the extremum temperature during
reheating epoch which is one of the prime components for the determination of gravitino abundance. In the braneworld scenario this extremum temperature
is given by: \be\label{cv1}
 T^{bh}_{ex}=\sqrt[4]{\left[\frac{13\sqrt{3}\Delta^{2}M\Gamma_{\phi}}{N^{*}\pi^{2}}\sqrt{\frac{1}{2\alpha}}\right]}
=\sqrt[4]{\left\{\frac{45\Gamma_{\phi}M^{3}_{5}}{8N^{*}\pi^{3}}\right\}}\ee
and it is less than the reheating temperature in brane ($T^{brh}$). This phenomenon is different from standard GR results \cite{Rangarajan:2008zb} where we see that the reheating temperature shoots up to a maximum value and it gives the upper
bound of the reheating temperature. But in the present context of brane inflation this situation is completely different i.e. at first temperature falls down to a minimum which fixes the lower bound of the reheating temperature and rises to a maximum at the end of reheating
 epoch. Using Eq~(\ref{opp}), Eq~(\ref{cv1}) and the thermodynamic background of energy density
 of radiation we can express the scale factor in terms of temperature as \be\label{xc1}
 a(T)=
\left\{
	\begin{array}{ll}
                    \frac{a_{osc}}{\sqrt[4]{\left[1-32\left(\frac{T}{T^{bh}_{ex}}\right)^{4}\right]}}& \mbox{if } t=t_{osc}\simeq t_{f} \\
           \frac{a_{osc}}{\sqrt[4]{\left[32\left(\frac{T}{T^{bh}_{ex}}\right)^{4}-1\right]}}& \mbox{if } t_{osc}(\simeq t_{f})<t\leq t_{reh}.
          \end{array}
\right.
\ee
 It is worth mentioning that if we break the time scale into two parts $t_{osc}<t<t_{ex}$ and $t_{ex}<t<t_{reh}$,
as done in GR the scale
factor and hence the remaining results have same expressions in these two different zones. This is in sharp contrast with standard GR
results except at $t=t_{f}$, where they have different values in the two different regimes.


\begin{figure}[htb]
\centering
{\includegraphics[width=10cm, height=6.5cm] {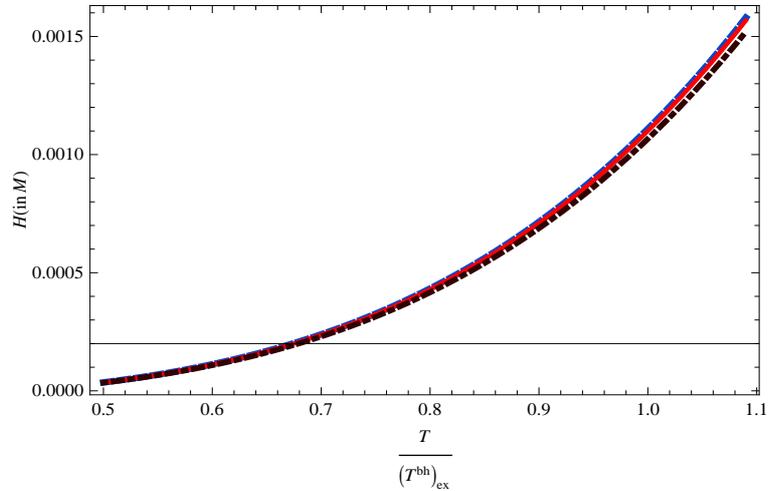}}
\caption{In the above figure we have plotted the variation of the Hubble parameter
  with respect to dimensionless parameter $\frac{T}{T^{bh}_{ex}}$ in the domain $-0.70<D_{4}<-0.60 $,
which shows the smooth behavior of Hubble parameter except $x\leq 0$ i.e. at $\frac{T}{T^{bh}_{ex}}\leq\frac{1}{\sqrt[4]{32}}$ \cite{Choudhury:2011rz}.
Most importantly here the equality corresponds to the end of reheating epoch and the beginning of radiation dominated era which is the
direct outcome of the
first expression at $t\simeq t_{f}$ for the scale factor ($a(T)$) stated in Eq~(\ref{xc1}). The rest of the the part
follows the second expression given in Eq~(\ref{xc1}) in the interval $t_{f}<t<t_{reh}$  plotted in the above figure.
Additionally the vertical scale corresponds to $M=2.43\times 10^{18}~{\rm GeV}$.} \label{figVr400}
\end{figure}


 Let us now use this phenomenological background to derive the expression of the gravitino production during two thermal epochs -
  reheating and radiation dominated era. It is well known that gravitinos are produced by the scattering of the inflaton decay products \cite{Ferrantelli:2010as}.
 The master equation of gravitino as obtained from `Boltzmann equation.' is given by \cite{Pradler:2006qh}
~\footnote{In this context, $n=\frac{\zeta(3)T^{3}}{\pi^{2}}$ is the number density of scatterers bosons in thermal bath with
$\zeta(3)$=1.20206.... Here $\Sigma_{total}$ is the total scattering cross section for thermal gravitino production, $v$ is the relative velocity of the incoming particles with $\langle v\rangle=1$ where $\langle...\rangle$ represents the thermal average. The factor $\frac{m_{\frac{3}{2}}}{\langle E_{\frac{3}{2}}\rangle}$ represents the averaged Lorentz factor which comes from the decay of gravitinos can be neglected due to weak interaction.
 For the gauge group $E:=SU(3)_{C}\bigotimes SU(2)_{L}\bigotimes U(1)_{Y}$ the thermal gravitino production rate is given by, 
$$\displaystyle\langle\Sigma_{total}|v|\rangle=\frac{\tilde{\alpha}}{M^{2}}=\frac{3\pi}{16\zeta(3)M^{2}}\sum^{3}_{i=1}\left[1+\frac{M^{2}_{i}}{3m^{2}_{\tilde{G}}}\right]
C_{i}g^{2}_{i}\ln\left(\frac{K_{i}}{g_{i}}\right),$$
where $i=1,2,3$ stands for the three gauge groups $U(1)_{Y}$,$SU(2)_{L}$ and $SU(3)_{C}$ respectively. Here $M_{i}$ represent gaugino mass parameters and $g_{i}(T)$
 represents gaugino coupling constant at finite temperature from MSSM RGE
$$ g_{i}(T)\simeq\frac{1}{\sqrt{\frac{1}{g^{2}_{i}(M_{Z})}-\frac{b_{i}}{8\pi^{2}}\ln\left(\frac{T}{M_{Z}}\right)}}$$
with $b_{1}=11,b_{2}=1,b_{3}=-3$. Here $C_{i}$ and $K_{i}$ represents the constant associated with the gauge groups with
 $C_{1}=11,C_{2}=27,C_{3}=72$ and $K_{1}=1.266,K_{2}=1.312,K_{3}=1.271$.}:
  \be\label{bn1}
\frac{dn_{\tilde{G}}}{dt}+3Hn_{\tilde{G}}=
\langle\Sigma_{total}|v|\rangle n^{2}-\frac{m_{\frac{3}{2}}n_{\tilde{G}}}{\langle E_{\frac{3}{2}}\rangle\tau_{\frac{3}{2}}},\ee

For convenience let us recast Eq~(\ref{bn1}) as
 \be\label{uiu}\dot{T}\frac{dn_{\tilde{G}}}{dT}+3Hn_{\tilde{G}}=\langle\Sigma_{total}|v|\rangle
n^{2},\ee where a boundary condition $T=T^{bh}_{ex}$ ,$\dot{T}=0$ is introduced.
In terms of a dimensionless variable \be\label{az1} x=32\left(\frac{T}{T^{bh}_{ex}}\right)^{4}-1\ee
 Eq~(\ref{uiu}) can be expressed as \be\label{iop}\frac{dn_{\tilde{G}}}{dx}+\frac{d_{1}}{x}n_{\tilde{G}}=-\frac{d_{3}(x+1)^{\frac{3}{2}}}{x^{2}}\ee
where \be  d_{1}=-\frac{3}{4},~~~d_{3}=\frac{(T^{bh}_{ex})^{6}}{32}
\frac{\tilde{\alpha}}{M^{2}}\left(\frac{\zeta(3)}{\pi^{2}}\right)^{2}\frac{\sqrt{\lambda}}{4\sqrt{3}H^{2}_{osc}M}.\ee
The exact solution of the Eq~(\ref{iop}) is given by~\footnote{Using the properties of Gaussian hypergeometric function for $x>>1$ Eq~(\ref{d1})
 reduces to the following simpler form \cite{Choudhury:2011rz}:
\be\begin{array}{ll}
    \label{zac} \displaystyle n_{\tilde{G}}(x)\simeq2d_{3}x^{\frac{1}{4}}\sqrt{1+x}\frac{\Gamma(\frac{3}{2})\Gamma(\frac{1}{2})}{\Gamma(1)}
   \left\{\frac{1}{\Gamma(\frac{3}{4})}+\frac{1}{\Gamma(-\frac{5}{4})}-\frac{2}{\Gamma(-\frac{1}{4})}\right\}.\end{array}
\ee
Using the boundary condition at $T=T^{bh}_{ex}$ in Eq~(\ref{zac}) the numerical value of gravitino abundance turns out to be
$n_{\tilde{G}}(x_{ex})=62.023d_{3}$.} \cite{Choudhury:2011rz}:
\be
\begin{array}{ll}
\label{d1}\displaystyle n_{\tilde{G}}(x)=\frac{2d_{3}}{x^{d_{1}}}\sqrt{x+1}\left(-2\,_2F_1\left[\frac{1}{2};1-d_{1};\frac{3}{2};x+1\right]
+_2F_1\left[\frac{1}{2};2-d_{1};\frac{3}{2};x+1\right] \right.
 \\ \left.~~~~~~~~~~~~~~~~~~~~~~~~~~~~~~~~~~~~~~~~~~~~~~~~~~~~~~~~~~~~~~~~~~~~~~~\displaystyle +_2F_1\left[\frac{1}{2};-d_{1};\frac{3}{2};x+1\right]\right).
\end{array}
\ee

where $_2F_1\left[a,b,c,d\right]$ be the Hypergeomeometric function defined for $|d|<1$ by the power series as:
\bea
_2F_1\left[a,b,c,d\right]&=& \sum^{\infty}_{n=0}\frac{(a)_{n}(b)_{n}d_{n}}{(c)_{n}n!}
\eea
It is undefined (or infinite) if $c$ equals a non-positive integer. Here $(a)_n$ is the (rising) Pochhammer symbol, which is defined by:
\be\label{gcgv}  (a)_n=
\left\{
	\begin{array}{ll}
		 1 & \mbox{if } n=0 \\
		 a(a+1)....(a+n-1) & \mbox{if } n>0.
	\end{array}
\right.
\ee
\begin{figure}[htb]
\centering
{\includegraphics[width=10cm, height=6.5cm] {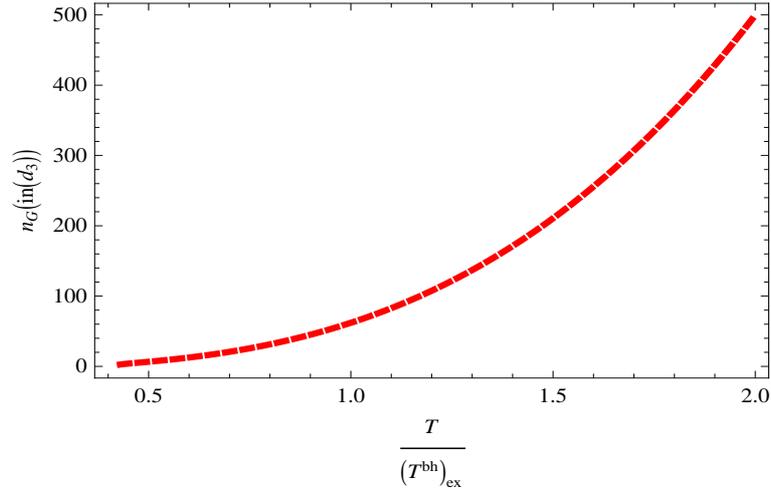}}
\caption{In the above diagram we have plotted variation of gravitino number density in a physical volume  vs scaled temperature in braneworld scenario \cite{Choudhury:2011rz}.
Here we have used the fundamental scale $d_{3}=4.596\times10^{-44}\tilde{\alpha}M^{3}$
, where $\tilde{\alpha}$ is a dimensionless number depends on the species of the MSSM gauge group $E$. For an
example $n=4$ level flat direction content  ${\bf QQQL, QuQd, QuLe, uude }$ of MSSM gives $\tilde{\alpha}\simeq 15.694$ in the absence
of top Yukawa coupling. Most importantly
4D effective Planck mass $M=2.43\times
10^{18}~{\rm GeV}$. From the plot it is obvious that the gravitino number density is monotonically increasing function of the dimensionless
variable $\frac{T}{T^{bh}_{ex}}$ except at $x\leq 0 $ which implies $\frac{T}{T^{bh}_{ex}}\leq\frac{1}{\sqrt[4]{32}}$. } \label{figVr900}
\end{figure}

Let us now find out the exact analytical expression for the gravitino abundance at reheating
 temperature $T^{brh}$ in the high energy limit. To serve this purpose substituting $T=T^{brh}$ in Eq~(\ref{d1}) we get \cite{Choudhury:2011rz}:
\be\begin{array}{ll}\label{bhu}
  \displaystyle n_{\tilde{G}}(T^{brh})=8\sqrt{2}d_{3} \left(32\left(\frac{T^{brh}}{T^{bh}_{ex}}\right)^{4}-1\right)^{\frac{3}{4}}
 \left(\frac{T^{brh}}{T^{bh}_{ex}}\right)^{2}G\left(\frac{T^{brh}}{T^{bh}_{ex}}\right)
 \end{array}\ee
where \be\begin{array}{ll}\label{sctt}  \displaystyle G\left(\frac{T^{brh}}{T^{bh}_{ex}}\right)=\left(-2\,_2F_1\left[\frac{1}{2};\frac{7}{4};\frac{3}{2};32\left(\frac{T^{brh}}{T^{bh}_{ex}}\right)^{2}\right]
   +_2F_1\left[\frac{1}{2};\frac{11}{4};\frac{3}{2};32\left(\frac{T^{brh}}{T^{bh}_{ex}}\right)^{2}\right]
   \right.\\ \left. \displaystyle ~~~~~~~~~~~~~~~~~~~~~ ~~~~~~~~~~~~~~~~~~~~~~~~~~~~~~~~~~~~~~~~ +_2F_1\left[\frac{1}{2};\frac{3}{4};\frac{3}{2};32\left(\frac{T^{brh}}{T^{bh}_{ex}}\right)^{2}\right]\right)
\end{array}\ee
along with an extra constraint \cite{Choudhury:2011rz}: \be G\left(\frac{T^{brh}}{T^{bh}_{ex}}>>\frac{1}{\sqrt[4]{32}}\right)
=\frac{\pi}{2}\left(32\left(\frac{T^{brh}}{T^{bh}_{ex}}\right)^{4}-1\right)^{-\frac{1}{2}}
 \left\{\frac{1}{\Gamma(\frac{3}{4})}+\frac{1}{\Gamma(-\frac{5}{4})}-\frac{2}{\Gamma(-\frac{1}{4})}\right\}.\ee
 It is convenient to express the abundance of any species `$\sigma$'  as \cite{Rangarajan:2008zb},  $ Y_{b}=\frac{n_{b}}{s}$ where $n_{b}$
 is the number density of the species `$b$' in a physical volume and `s' is the entropy density given by $s=\frac{2\pi^{2}}{45}N^{*}T^{3}$.
 Here the master equation. for gravitino can be expressed as \cite{Choudhury:2011rz}:\be\label{zx1}
 \dot{T}\frac{dY^{br}_{\tilde{G}}}{dT}=\langle\Sigma_{total}|v|\rangle Y^{br}_{\tilde{G}}n\ee

Using Eq~(\ref{dt1}) and Eq~(\ref{xc1}) the time-temperature relation can be found as \cite{Choudhury:2011rz}:
\be\label{bnmj}T=\frac{T^{br}}{\sqrt[4]{\left[\left[\sqrt{1+\frac{\alpha}{2}}+2H_{reh}
(t-t_{reh})\right]^{2}-\frac{\alpha}{2}\right]}}.\ee

Eliminating $\dot{T}$ we get the solution of the master Eq~(\ref{zx1}) in the radiation dominated era as \cite{Choudhury:2011rz}:\be\label{ft}
 Y^{br}_{\tilde{G}}(T_{f})=Y^{br}_{\tilde{G}}(T^{br})+Y^{br-rad}_{\tilde{G}}(T_{f})\ee
 where\be\begin{array}{ll}\label{dct2}\displaystyle Y^{br-rad}_{\tilde{G}}(T_{f})=\sqrt{\frac{90}{\pi^{2}N^{*}}}\left(\frac{45\sqrt{2}}{2\pi^{2}N^{*}\sqrt{\alpha}}\right)\left(\frac{\tilde{\alpha}}{M}\right)\left(\frac{\zeta(3)}{\pi^{2}}\right)^{2}
 \\ ~~~~~~~~~~~~~~~~~~~~~~~~~~~~~~~\displaystyle\times\frac{T^{br}}{T_{f}\sqrt{1+\frac{\pi^{2}}{60\lambda}N^{*}(T^{br})^{4}}}
\left(T^{br}\,_2F_1\left[\frac{1}{4};\frac{1}{2};\frac{5}{4}; -\frac{2(T^{br})^{4}}{\alpha T^{4}_{f}}\right]\right.
\\ \left.~~~~~~~~~~~~~~~~~~~~~~~~~~~~~~~~~~~~~~~~~~~~~~~~~~~~~~~~~~~~~~~~~~~~~~~\displaystyle-T_{f}\,_2F_1\left[\frac{1}{4};\frac{1}{2};\frac{5}{4};-\frac{2}{\alpha}\right]\right)\end{array}\ee
 But in Eq~(\ref{ft}) the first term on the right-hand side is not
 exactly computable. As mentioned earlier to find out exact expression we have used here the high energy mapping.

In the radiation dominated era the dynamical behavior of temperature can be mapped as
 \be\begin{array}{l}\label{hj2}\displaystyle T=
\left(\frac{T^{br}}{\sqrt[4]{\left[\left[\sqrt{1+\frac{\alpha}{2}}+2H_{reh}
(t-t_{reh})\right]^{2}-\frac{\alpha}{2}\right]}}\displaystyle\Longrightarrow \frac{T^{brh}}{
\left[1+2\sqrt{2\alpha}H_{reh}(t-t_{reh})\right]^{\frac{1}{4}}}\right)\end{array}\ee

Using this map we finally have \cite{Choudhury:2011rz}:
\be\begin{array}{ll}\label{fgh4}
     \displaystyle Y^{brh}_{\tilde{G}}(T_{f})=\left(\frac{\tilde{\alpha}}{M}\right)
     \left(\frac{\zeta(3)}{\pi^{2}}\right)^{2}\left(\frac{45\sqrt{3\lambda}}{2\pi^{3}\Delta^{2}N^{*}}\right) \left[\left(\frac{60\sqrt{\lambda}}{\pi N^{*}T_{f}}\right)
     \left(1-\frac{T_{f}}{T^{brh}}\right)\right.\\ \left.\displaystyle ~~~~~~~~~~~~~~~~~~~~~~~~~~~~~~~~~~~~~~~+\left(\frac{(T^{bh}_{ex})^{4}}{16\Delta^{2}T^{brh}}\right)
\left(32\left(\frac{T^{brh}}{T^{bh}_{ex}}\right)^{4}-1\right)^{\frac{3}{4}}G\left(\frac{T^{brh}}{T^{bh}_{ex}}\right)\right].\end{array}\ee

where
 \be\begin{array}{l}\label{yh2}
    \displaystyle Y^{b-rad}_{\tilde{G}}(T_{f})=\left(\frac{6\tilde{\alpha}}{M}\right)\left(\frac{\zeta(3)}{\pi^{2}}\right)^{2}
     \sqrt{\frac{3\lambda}{\alpha}}\left(\frac{15}{\pi^{2}N^{*}}\right)^{2}\left(\frac{1}{T_{f}}-\frac{1}{T^{brh}}\right),\end{array}\ee
     and \be\begin{array}{l}\label{rd6} \displaystyle Y^{brh}_{\tilde{G}}\simeq Y^{b}_{\tilde{G}}=\frac{n_{\tilde{G}}}{s}=\left(\frac{360\sqrt{2}d_{3}}{2\pi^{2}N^{*}(T^{brh})^{3}}
\right)\left(32\left(\frac{T^{brh}}{T^{bh}_{ex}}\right)^{4}-1\right)^{\frac{3}{4}}
         \left(\frac{T^{brh}}{T^{bh}_{ex}}\right)^{2}G\left(\frac{T^{brh}}{T^{bh}_{ex}}\right)\end{array}.\ee
The gravitino dark matter abundance and
the baryon asymmetry is connected through \cite{Choudhury:2011rz}:\be Y^{brh}_{\tilde{G}}\simeq\frac{\Theta_{CP}{\cal D}}{N^{*}}\ee
 where ${\cal D}(\leq 1)$ is the
dilution factor and the leading contribution is given
by the interference between the tree level and the one-loop level decay amplitudes. 
In the case of the hierarchical mass spectrum for the heavy neutrinos, the lepton asymmetry
 in the universe is generated dominantly by CP-violating out-of-equilibrium decay of
the lightest heavy neutrino. In the present context the
CP-violating parameter is described as \cite{Okada:2005kv}:
\bea\Theta_{CP}&=&
\frac{\Gamma(\psi\rightarrow\bar{l}_{L}{\cal H})
-\Gamma(\psi\rightarrow l_{L}{\cal H^{\star}})}{\Gamma(\psi\rightarrow\bar{l}_{L}{\cal H})
+\Gamma(\psi\rightarrow l_{L}{\cal H^{\star}})}= \frac{3{\cal M}_{1} m_{\nu}}{16\pi v^2}\sin\delta_{CP},\eea
where $m_{\nu}$ is the heaviest light neutrino mass, $v = 174 ~{\rm GeV}$ is the VEV of Higgs and $\delta_{CP}$ is an effective CP phase
 which parameterize each entries of the CKM matrix. Particularly $\delta_{CP}$ acts as a probe of
 flavor structure in SUGRA theories.
The complete wash out situation corresponds to ${\cal D}=1$.

\begin{figure}[htb]
\centering
{\includegraphics[width=10cm, height=6.5cm] {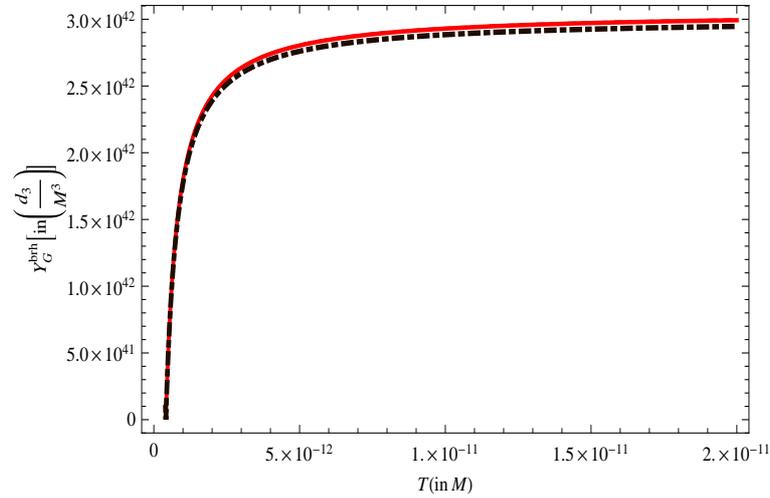}}
\caption{
Here we have plotted the variation of total gravitino abundance vs temperature in the domain $-0.70<D_{4}<-0.60$,
which clearly shows that  gravitino abundance  at zero temperature shoots up initially to maximum and then becomes
 constant with respect to temperature during radiation dominated era in braneworld scenario \cite{Choudhury:2011rz}.
As mentioned earlier we have used the fundamental scale $\frac{d_{3}}{M^{3}}=4.596\times10^{-44}\tilde{\alpha}$
, where $\tilde{\alpha}$ is a dimensionless number which depends on the species of the MSSM and $M=2.43\times
10^{18}~{\rm GeV}$. } \label{figVr600}
\end{figure}

\section{Numerical analysis for MSSM flat direction}
\label{g6}

Embedding MSSM from String theory via braneworld is a long standing unresolved problem in the context of theoretical physics. 
If such embedding is really possible then one can study various particle cosmological aspects from MSSM flat direction in braneworld scenario. 
The procedure of embedding needs to be checked throughly, but there are some positive efforts towards this field of research is already been discussed in refs.~\cite{Nilles:2008gq}.
In this chapter we assume that such embedding is true and just as an example we have further studied the numerical estimations for MSSM D-flat
 directions proposed in chapter \ref{ch:FRWxc} section \ref{c2}.   
 Through out all the numerical estimation we have taken decay width
 $\Gamma_{\phi}\simeq2.9\times 10^{-3}~{\rm GeV}$,  mass of the
inflaton $m_{\phi}\simeq10^{13}~{\rm GeV}$, final temperature and time at the end of reheating $T_{f}\simeq10^{6}~{\rm GeV}$ and
 $t_{f}\simeq1.4\times 10^{31}~{\rm GeV}$ respectively.  Let us discuss our results in following:
\begin{itemize}
 \item For a typical value of $C_{4}\simeq D_{4}=-0.7$
 extremum (minimum) temperature during reheating can be estimated as
 $ T^{bh}_{ex}\simeq7.0\times10^{10}~{\rm GeV}$. This clearly
 shows deviation from standard GR phenomenology \cite{Rangarajan:2008zb}
 where the extremum (maximum) temperature during reheating $T_{max}\simeq 1.3\times 10^{12}~{\rm GeV}$.

\item Similarly 
for $C_{4}\simeq D_{4}=-0.7$ the critical temperature for different particle species and gravitino abundance at different temperatures 
obtained from our model are: for boson $T_{cB}\simeq3.2\times 10^{14}~{\rm GeV}$, for fermion $T_{cF}\simeq3.3\times 10^{14}~{\rm GeV}$,
 for mixture of species $T_{cM}\simeq2.8\times 10^{14}~{\rm GeV}$, at reheating temperature $Y^{b}_{\tilde{G}}(T^{brh})\simeq8.1\times 10^{-34}
~{\rm GeV}^{-3}d_{3}$ and at the end of reheating $Y^{b-rad}_{\tilde{G}}(T_{f})\simeq2.1\times 10^{-13}~{\rm GeV}^{-3}d_{3}$~\footnote{We have calculated all the abundances in the fundamental unit of
$d_{3}$ i.e. $d_{3}=6.594\tilde{\alpha}\times10^{11}~{\rm GeV}^{3}$, where $\tilde{\alpha}$ is a dimensionless
characteristic constant originated through the thermal gravitino production rate in the context of MSSM.}.

\item Most significantly for
 different flat direction contents the phenomenological parameter is different and can be calculated
from MSSM RGE flow at the one-loop level for that flat direction. To obtain a conservative
estimate of gravitino abundance we have taken here gaugino masses $M_{i}\rightarrow 0$ for all gauge subgroups within MSSM.
 For example the fourth level flat directions
${\bf QQQL, QuQd, QuLe, uude }$ give $\tilde{\alpha}=15.694$ for a specific choice of the
$U_{Y}(1)$, $SU_{L}(2)$ and $SU_{C}(3)$ gauge couplings $g_{1}=0.56$, $g_{2}=0.72$ and $g_{3}=0.85$ respectively
 obtained from the {\it universal mSUGRA boundary condition} and consistent with electroweak extrapolation of the solution of MSSM RGE
flow from the energy scale of brane inflation $\Delta=0.2\times10^{16}~{\rm GeV}$ for our model.

\item The linear dependence on $T^{brh}$ makes simple to revise
the constraints on $T^{brh}$ based on the lower limit on the
gravitino abundance - the lower bound on  $T^{brh}$ is increased by a factor of 1.074. Since $T^{bh}_{ex} \propto T^{brh}$,
 $T^{bh}_{ex}$ is
not affected much. Therefore models of leptogenesis that
invoke a small $T^{bh}_{ex}$ to create heavy Majorana neutrinos
are not significantly affected.

\item Within $55<N<70$  and  $T^{brh}\simeq6.5\times 10^{10}~{\rm GeV}$  the entropy density
changes. As a consequence the total gravitino abundance changes according to fig~(\ref{figVr600}).
It is easily seen that $P=\frac{\rho}{3}$, $S=\frac{\rho+P}{T}$ consistency relations are valid in this context.

\item It is worthwhile to mention here that in brane pressure and entropy density of the universe falls down to a minimum due to the minimum temperature during
reheating epoch. However during radiation dominated era total entropy density is almost constant for both the cases. This clearly shows
the deviation from standard GR phenomenology.

\item Throughout the analysis we have not included the effect of $\exp[-\Gamma_{\phi} (t - t_{osc})]$
in the energy density of inflaton $\rho_{\phi}$. One might be concerned that this
will lead to inaccuracies close to $t_{brh}$ when most of
the gravitinos are produced. However if one writes
$\rho_{\phi}\simeq a^{-4} \exp(-\Gamma_{\phi} t)\simeq t^{-2}\exp(-\Gamma_{\phi} t)$ for $t >> t^{bh}_{ex}$
then $\dot{\rho_{\phi}}/\rho_{\phi} = -2/t -\Gamma_{\phi}$. Therefore even till close to
 ̇$t_{brh}=\Gamma^{-1}_{\phi}\rho_{\phi}$ decreases primarily due to the expansion
of the universe. Furthermore, near $t_{brh}$ it increases as $T^{-1/2}\simeq \sqrt{a}\simeq t^{\frac{1}{8}}$ in brane
 which is again different from GR phenomenology where $T^{-1/2}\simeq \sqrt{a}\simeq t^{\frac{1}{4}}$. 

\item The thermal leptogenesis in the braneworld can take place if the lightest heavy
neutrino mass lying in the range
$T^{brh} < {\cal M}_{1} < T_{c}$. This confirms that the
upper bound of 5D Planck mass $M_{5}< 10^{16}~{\rm GeV}$ (for our model $M_{5}\simeq7.8\times 10^{15}~{\rm GeV}$ for $C_{4}\simeq D_{4}=-0.7$),
 which coincides with the
leptogenesis bound implied by the observed baryon asymmetry. 

\item It is important to mention here that in the standard cosmology,
 the thermal leptogenesis in SUGRA models is hard to be successful,
 since the reheating temperature after inflation is severely
constrained to be $T^{reh}\leq10^{9}~{\rm GeV}$ due to the gravitino problem. However, as pointed
out in \cite{Felipe:2004wh}, the constraint on the reheating temperature is replaced by
the transition temperature in the brane world cosmology. As a result the gravitino problem can be solved even if the
reheating temperature is much higher. In fact, such inflation
models are possible but limited and our model is also in that category. 

\item Here we are using a preferable
 value of the heaviest light neutrino mass from
atmospheric neutrino oscillation data
$m_{\nu} \simeq 0.05 ~{\rm eV}$ and for sufficient
 baryon asymmetry the lightest neutrino mass ${\cal M}_{1}\simeq 10^{10}~{\rm GeV}$.

\item  For complete washout situation (${\cal D}=1$) in our model the effective CP phase lying within the window
 $2.704\times10^{-9} <\delta_{CP}< 2.784\times10^{-9}$, where $\delta_{CP}$ is measured in degree.
 Most significantly it indicates that the amount
of CP violation in braneworld scenario is very small and identified with the {\it soft CP phase}. Consequently it has negligibly small
contribution to ${\cal K}$ and ${\cal B}$ physics phenomenology.
\end{itemize}

\section{Chapter summary}
\label{g7}

In this chapter we have explored the phenomenological features of reheating in brane
cosmology on the background of SUGRA. The major outcomes of our study are:
\begin{itemize}
 \item We have exhibited
 the process of construction of a fruitful theory of reheating for an effective
4D inflationary potential in ${\cal N}=1, {\cal D}=4$ SUGRA in the brane derived from  ${\cal N}=2, {\cal D}=5$ SUGRA in
the bulk.

\item We have employed the proposed
setup in reheating model building by analyzing the reheating temperature in the context of brane inflation, followed
by analytical and numerical estimation of different phenomenological parameters.

\item It is worthwhile to mention that we get a lot of new
results in the context of braneworld compared to standard GR case.
 Most importantly we get a different numerical value of reheating temperature
 as well the extremum temperature compared to the standard GR results.
Next using the extremization principle we justify that the extremum temperature
 is the minimum temperature during reheating which again shows
deviation from standard GR inspired phenomenology. All these facts are reflected
 in the numerical results of the gravitino abundance in reheating and
radiation dominated era. In the context of phase transition
 we also get different numerical results for different
parameters for standard model particle constituents.

\item We have further engaged ourselves in investigating for the effect of perturbative reheating.
 To this end we propose a theory which reflects the effect of particle production
through collision and decay thereby showing a direct connection with the thermalization phenomena.
 To show this internal link more explicitly we put forward both analytical and numerical expressions for
 the gravitino abundance in a physical volume in the reheating epoch. Next we have found out
 the gravitino abundance in the radiation dominated era.
 Last but not the least we have expressed the total
 gravitino abundance in terms of final temperature during reheating. Most significantly the precision level of all
 estimated numerical results is the outcome of the $4{\rm D}$ effective field theory which is analyzes
 with the arrival of lots of sophisticated techniques.

\item Apart from the aforesaid success in estimating phenomenological
parameters there are some added advantages of our model with reheating in brane which are worth
mentioning. One of the most significant features in the context of braneworld phenomenology is the validity
 of leptogenesis for our model which consequently shows the production of heavy Majorana neutrinos
in the brane.
\end{itemize}

\chapter{Primordial non-Gaussianity from ${\cal N}=1$ supergravity using $\delta N$ formalism }
\label{ch:FRWxc3}
\section{Introduction}
\label{h1}
The inflationary paradigm is a very rich idea to explain various aspects of the early universe, which creates
 the perturbations and the matter. The simplest prescription to explain inflation is made via a single scalar field slowly rolling down a flat potential, 
which has unique predictions for the Cosmic Microwave Background (CMB) observables. 
In this scenario, the induced cosmological perturbations are generically random Gaussian in nature with
a small tilt and running in the primordial spectrum, which can be conveniently described in terms of two-point correlation function.
 But a big issue may crop up in model discrimination and also in the removal of the degeneracy of cosmological
parameters obtained from CMB observations \cite{Ade:2013zuv,Ade:2013uln,Spergel:2006hy,Ade:2013ydc}.  
Non-Gaussianity has emerged as a powerful observational tool to discriminate between different
 classes of inflationary models. The Planck data show no significant evidence in favour of primordial non-Gaussianity, the current limits \cite{Ade:2013ydc} are
yet to achieve the high statistical accuracy expected from the single-field inflationary models and for this opportunities are galore
for the detection of large non-Gaussianity from various types of inflationary models.
To achieve this goal, apart from the huge success of cosmological linear perturbation
theory, the general focus of the theoretical physicists has now shifted towards the study of nonlinear evolution of
cosmological perturbations. Typically any types of
nonlinearities are expected to be small; but, that can be estimated via non-Gaussian n-point correlations of cosmological perturbations.

An overview of calculating non-Gaussianity from different methods have been discussed in chapter~\ref{ch:FRW} section~\ref{ng}.
Here we will employ the ``$\delta N$ formalism'' (where $N$ being the number of e-foldings)
 \cite{Sasaki:1995aw,Wands:2000dp,Lyth:2004gb,Lyth:2005fi,Mazumdar:2012jj,Sugiyama:2012tj}, which is a well accepted tool for computing non-linear evolution of cosmological
perturbations on large scales ($k<<aH$), which is derived using the ``separate universe''
 approach \cite{Sasaki:1995aw,Wands:2000dp}. Particularly, it provides a fruitful technique to
compute the expression for the curvature perturbation $\zeta$ without explicitly
solving the perturbed field equations from which the various
 local non-Gaussian parameters, $f_{NL}^{\mathrm{local}},\tau_{NL}^{\mathrm{local}},g_{NL}^{\mathrm{local}}$ and
 CMB dipolar asymmetry parameter \cite{Erickcek:2008sm,Lyth:2013vha}, $A_{CMB}$ are
 easily computable~\footnote{One can also compute all the non-Gaussian parameters, $f_{NL}^{\mathrm{local}},\tau_{NL}^{\mathrm{local}},g_{NL}^{\mathrm{local}}$ 
and CMB dipolar asymmetry parameter, $A_{CMB}$ using {\it In-In formalism} in the quantum regime. But the inflationary 
dynamics responsible for the interactions between the
modes occurs at the super-horizon scales within the effective theory setup proposed in this chapter. 
Here we use $\delta N$ formalism as-(1) it perfectly holds good at the super-horizon scales
and (2) are also independent of any kind of intrinsic non-Gaussianities generated at the scale of horizon
crossing.}.

 We will be using the following constraints on the amplitude of the power spectrum, $P_s$, spectral tilt, $n_s$, tensor-to-scalar ratio, $r$, sound speed, $c_{s}$, local type
 of non-Gaussianity, $f_{NL}^{\mathrm{local}}$ and $\tau_{NL}^{\mathrm{local}}$ and CMB dipolar asymmetry from Planck data
 throughout the paper \cite{Ade:2013zuv,Ade:2013uln,Ade:2013ydc,Ade:2013nlj}:

\begin{eqnarray}
\label{pow} \displaystyle \ln(10^{10}P_{s})=3.089^{+0.024}_{-0.027}~~(~at~2\sigma~~ CL)\,, \\
\label{ns} \displaystyle  n_{s}=0.9603\pm 0.0073~~(~at~2\sigma~~ CL)\,, \\
\label{rten} \displaystyle r\leq 0.12~~(~at~2\sigma~~ CL)\,, \\
\label{f-nl}
\displaystyle 0.02 \leq c_s \leq 1~~(~at~2\sigma~~ CL)\,, \\
\label{f-nl1}\displaystyle f_{NL}^{\mathrm{local}}=2.7\pm 5.8~~(~at~1\sigma~~ CL)\,,
\\ \label{f-nl2}\displaystyle \tau_{NL}^{\mathrm{local}}\leq2800~~(~at~2\sigma~~ CL)\,, \\
\label{f-nl3}\displaystyle A_{CMB}=0.07\pm 0.02~~(~at~2\sigma~~ CL).
\end{eqnarray}

In this chapter we will concentrate our study for Hubble induced inflection point MSSM inflation derived from various higher dimensional Planck scale suppressed 
non-minimal K\"ahler operators
in ${\cal N}=1$ supergravity (SUGRA) which satisfies the observable universe, and it is well motivated for providing an example of visible sector inflation.

In section \ref{perturb} we start our discussion with cosmological perturbation scenario for sound speed $c_s \neq 1$. Hence in section \ref{setup}, we will describe the setup with one heavy and one light superfield which are coupled via non-minimal interactions through
 K\"ahler potential. In section \ref{N-M} we discuss very briefly the role of various types of Planck suppressed non-minimal K\"ahler
 corrections to model a Hubble induced MSSM inflation for any D-flat directions. Hence in section \ref{BNV} we present a quantitative
 analysis to compute the expression for the local types of non-Gaussian parameters and CMB dipolar asymmetry
parameter which characterize the 
bispectrum and trispectrum using the $\delta N$ formalism. For the numerical estimations we analyze the results in the context of 
two D-flat direction, $\widetilde L\widetilde L\widetilde e$ and $\widetilde u\widetilde d\widetilde d$ within
 the framework of MSSM inflation \cite{Allahverdi:2006iq}.

\section{ Cosmological perturbations for $c_s\neq 1$}\label{perturb}

In this section we briefly recall some of the important formulae when $c_s\neq 1$,
 the scalar and tensor perturbations are given by~\cite{Ade:2013uln}:
\begin{eqnarray}\label{scalaras}
P_{S}(k)& = &{\cal P_{S}}\left(\frac{k}{c_{s}k_{\star}}\right)^{n_{S}-1}\,,\nonumber \\
\label{tensoras}P_{T}(k)& = &{\cal P_{T}}\left(\frac{k}{c_{s}k_{\star}}\right)^{n_{T}},
\end{eqnarray}
where the speed of sound at the Hubble patch is given by, $c_{s}k_{\star}=aH$ (where
$k_{\star}\sim 0.002~Mpc^{-1}$). The amplitude of the scalar and tensor perturbations can be recast in terms of the potential, as~\cite{Ade:2013uln}:
\begin{eqnarray}\label{scalar}
{\cal P_{S}}& = &\frac{V_{\star}}{24\pi^{2}M^{4}_{p}c_{s}\epsilon_{V}}\,, \\
\label{tensor}{\cal P_{T}}& = &\frac{2V_{\star}}{3\pi^{2}M^{4}_{p}}c^{\frac{2\epsilon_{V}}{1-\epsilon_{V}}}_{s},
\end{eqnarray}
where running of the spectral tilt for the scalar and tensor modes can be expressed 
at $c_{s}k_{\star}=aH$, as:
\begin{eqnarray}\label{scalartilt}
n_{S}-1& = &2\eta_{V}-6\epsilon_{V}-s\,, \\
\label{tensortilt}n_{T}& = &-2\epsilon_{V}.
\end{eqnarray}
where running of the sound speed is
defined by  an additional slow-roll parameter, $s$, as:
\be\begin{array}{llll}\label{cvq1}
    \displaystyle s=\frac{\dot{c_{s}}}{Hc_{s}}=\sqrt{\frac{3}{V}}\frac{\dot{c_{s}}}{c_{s}}M_{p}.
   \end{array}\ee
In all the above expressions, the standard slow-roll parameters are defined by:
\be\begin{array}{llll}\label{cvq1}
    \displaystyle \epsilon_{V}=\frac{M^{2}_{p}}{2}\left(\frac{V^{\prime}}{V}\right)^{2}, ~~~~~
     \eta_{V}={M^{2}_{p}}\left(\frac{V^{\prime\prime}}{V}\right).
   \end{array}\ee
Finally, the single field consistency relation between tensor-to-scalar ratio
 and tensor spectral tilt is modified by~\cite{Ade:2013uln}:
\be\begin{array}{llll}\label{sconfo}
    \displaystyle r_{\star}=16\epsilon_{V}c^{\frac{1+\epsilon_{V}}{1-\epsilon_{V}}}_{s}=-8n_{T}c^{\frac{1-\frac{n_{T}}{2}}{1+\frac{n_{T}}{2}}}_{s}.
   \end{array}\ee
 Using the results for $c_{s}\neq 1$ stated in Eqs.~(\ref{scalar}-\ref{sconfo}), the upper bound on the numerical value of the Hubble 
 parameter during inflation is given by:
\begin{equation}\label{hubble}
     H\leq 9.241\times 10^{13}\sqrt{\frac{r_{\star}}{0.12}}~c^{\frac{\epsilon_{V}}{\epsilon_{V}-1}}_{s}~{\rm GeV}\,
   \end{equation}
where $r_{\star}$ is the tensor-to-scalar ratio at the pivot scale of momentum $k_{\star}\sim 0.002 Mpc^{-1}$.
An equivalent statement can be made in terms of the upper bound
on the energy scale of  inflation for $c_{s}\neq 1$ as:
\begin{equation}\label{scale}
     V_{\star}\leq (1.96\times 10^{16}{\rm GeV})^{4}\frac{r_{\star}}{0.12}~c^{\frac{2\epsilon_{V}}{\epsilon_{V}-1}}_{s}.
   \end{equation}
Here in Eqs.~(\ref{hubble}) and (\ref{scale}), the equalities will hold good for a high scale model of inflation.

Furthermore, for a sub-Planckian slow-roll models of inflation, one can express the tensor-to-scalar ratio, $r_{\star}$, at the
pivot scale, $k_{\star}\sim 0.002~Mpc^{-1}$, in terms of the field displacement, $\Delta \phi$, during
the observed $\Delta{ N}\approx 17$ e-foldings of inflation, for $c_{s}\neq 1$~\cite{Choudhury:2013iaa,Choudhury:2013woa}:
\be\begin{array}{llll}\label{con10}
   \displaystyle \frac{3}{25\sqrt{c_{s}}}\sqrt{\frac{r_{\star}}{0.12}}\left|\left\{\frac{3}{400}\left(\frac{r_{\star}}{0.12}\right)-\frac{\eta_{V}(k_{\star})}{2}-\frac{1}{2}
\,\right\}\right|
\displaystyle \approx \frac{\left |\Delta\phi\right|}{M_{p}}\,.
   \end{array}\ee
 where $\Delta\phi=\phi_{cmb}-\phi_{e}<<M_{p}$, where $\phi_{cmb}$ and $\phi_{e}$
 are the values of the inflaton field at the horizon crossing and at the end of inflation.



\section{Planck suppressed non-minimal K\"ahler operators within ${\cal N}=1$ SUGRA}\label{setup}

\subsection{ The Superpotential}
In this section we concentrate on two sectors; heavy hidden sector denoted by the superfield $S$, 
and the light visible sector denoted by $\Phi$ where they interact only via gravitation. 
Specifically the inflaton superfield $\Phi$ is made up of $D$-flat 
direction within MSSM and they are usually lifted by the $F$-term \cite{Enqvist:2003gh}
of the non-renormalizable operators as appearing in the superpotential.
In the present setup for the simplest situation we start with the following simplified expression for the superpotential made up of 
the superfields $S$ and $\Phi$ as given by:
\begin{eqnarray}
 W(\Phi,S)&=&W(\Phi)+W(S)\,=\frac{\lambda \Phi^n}{n M_{p}^{n-3}}+\frac{M_s}{2}S^2\,,
\end{eqnarray}
where for MSSM D-flat directions, $n\geq 3$ (In the present context $n$ characterizes the dimension of the non-renormalizable operator) and the coupling, $\lambda\sim {\cal O}(1)$.
The scale $M_s$ characterizes the scale of heavy physics which belongs to the hidden sector of the effective theory.
 Furthermore, I will assume that the VEV, $\langle s \rangle=M_{s}\leq M_{p}$ and $\langle \phi\rangle=\phi_{0}  \leq M_{p}$, 
where both $s$ and $\phi$ are fields corresponding
to the super field $S$ and $\Phi$. We also concentrate on two MSSM flat directions, $\widetilde L\widetilde L\widetilde e$ and $\widetilde u\widetilde d\widetilde d$, 
which can drive inflation with $n=6$ via $R$-parity invariant $(\widetilde L\widetilde L\widetilde e)(\widetilde L\widetilde L\widetilde e)/M^{3}_{p}$
 and $(\widetilde u\widetilde d\widetilde d)(\widetilde u\widetilde d\widetilde d)/M^{3}_{p}$ operators in the visible sector, which are lifted by 
themselves~\cite{Dine:1995kz}, where $\widetilde u,~\widetilde d$ denote the right handed squarks, and $\widetilde L$ denotes that left handed sleptons and 
$\widetilde e$ denotes the right handed slepton. 
 

\subsection{ The K\"ahler potential}
In this paper I consider the following simplest choice of the holomorphic K\"ahler potential
which produces minimal kinetic term,  and the K\"ahler correction of the form:
\begin{eqnarray}\label{klj}
K(s,\phi,s^{\dagger},\phi^{\dagger})=\underbrace{s^{\dagger}s+\phi^{\dagger}\phi}_{minimal~part}+\underbrace{\delta K}_{non-miniml~part}\,,
\end{eqnarray}
where $\delta K$ represent
the higher order non-minimal K\"ahler corrections which are extremely hard to compute from the original string theory background. 
in a more generalized prescription such corrections allow the mixing between the hidden sector heavy fields and the soft SUSY breaking visible sector MSSM fields.  
Using Eq~(\ref{klj}) the most
general ${\cal N} = 1$ SUGRA kinetic term for $(s,\phi)$ field can be written in presence of the non-minimal K\"ahler corrections
through the K\"ahler metric as:
\be\begin{array}{llll}\label{kinf}
    \displaystyle {\cal L}_{Kin}=\left(1+\frac{\partial \delta K}{\partial\phi\partial\phi^{\dagger}}\right)(\partial_{\mu}\phi^{\dagger})(\partial^{\mu}\phi)+
                                  \left(1+\frac{\partial \delta K}{\partial s\partial s^{\dagger}}\right)(\partial_{\mu}s^{\dagger})(\partial^{\mu}s)\\
\displaystyle ~~~~~~~~~~~~~~~~~~~~~~~~~~+\left(\frac{\partial \delta K}{\partial\phi\partial s^{\dagger}}\right)(\partial_{\mu}s^{\dagger})(\partial^{\mu}\phi)+
\left(\frac{\partial \delta K}{\partial s\partial\phi^{\dagger}}\right)(\partial_{\mu}\phi^{\dagger})(\partial^{\mu}s).
   \end{array}
\ee
In this paper I consider the following gauge invariant non-minimal Planck scale suppressed K\"ahler operators within ${\cal N}=1$ SUGRA~\cite{Dine:1995kz}:
\begin{eqnarray}~\label{corrt-1}
 \delta K^{(1)} &=& \frac{a}{M_{p}^2}\phi^\dag \phi s^\dag s + h.c.+\cdots\,, \\
 \label{corrt-2}
 \delta K^{(2)}&=&\frac{b}{2M_{p}}s^\dag\phi \phi + h.c.+\cdots\,, \\
 \label{corrt-3}
 \delta K^{(3)}&=&\frac{c}{4M_{p}^2}s^\dag s^\dag\phi \phi + h.c.+\cdots\,, \\
 \label{corrt-4}
 \delta K^{(4)}&=&\frac{d}{M_{p}}s\phi^\dag \phi + h.c. +\cdots\,,
\end{eqnarray}
where $a,~b,~c,~d$ are dimensionless non-minimal coupling parameters. The $\cdots$ contain higher order non minimal terms which has been ignored in this paper. 

\section{Effective field theory potential construction from ${\cal N} = 1$
SUGRA}\label{N-M}

In this section we will consider two interesting possibilities, one which is the simplest and provides an excellent model 
for inflation with a complete decoupling of the heavy field. Inflation occurs via the slow roll of $\phi$ field within an MSSM vacuum, where inflation would end in a vacuum 
with an  {\it enhanced gauge symmetry}, where the entire electroweak symmetry will be completely restored.

\subsection{Heavy field is dynamically frozen}\label{frozen}

Let us first assume that the dynamics of the heavy field $s$ is completely frozen during the onset and the rest of the course of slow roll inflation driven by $\phi$. 
The full potential can be found in Table~\ref{tab1}. Note that
the potential for $s$ field, $V(s)$ contains  soft term and the corresponding $A$-term:
\begin{equation}
V(s)\sim M_s^2|s|^2+A'M_ss^2\,,
\end {equation}
where $A'$ is a dimensional quantity, and it is roughly proportional to $A'\sim M_s\gg {\rm TeV}$.
In this case there are two possibilities which we briefly mention below:

\begin{itemize}

\item We can imagine that the heavy field, $s$,  to have a  
global minimum at:
\begin{equation}
\langle s \rangle =0,~~~~~~~~~~~\langle \dot s \rangle =0\,,~~~~~~~~V(s)=0\,.
\end{equation}
In this particular setup, the kinetic terms for each cases, i.e. $1, 2, 3, 4$, become canonical for the $\phi$ field,  therefore the heavy field is completely decoupled from the dynamics. 
One can check them from Table-\ref{tab1}.
This is most ideal situation for a single field dominated model of inflation, where the overall potential for 
along $\phi$ direction simplifies to:
\begin{equation}
V(\phi) =m_\phi^2\lvert\phi\rvert^2+\bigg(A\frac{\lambda\phi^n}{nM_{p}^{n-3}}
 +h.c.\bigg)+\lambda^2\frac{\lvert\phi\rvert^{2(n-1)}}{M_{p}^{2(n-3)}}\,.
\end{equation}
The overall potential is solely dominated by the $\phi$ field, therefore Hubble expansion rate becomes, 
$H_{inf} \propto V(\phi)/M_{p}^2$. 

In this setup inflation can occur near a saddle point or an inflection point, where $\phi_0\ll M_p$, and $m_\phi \gg H_{inf}$, first discussed in Refs.
~\cite{Allahverdi:2006iq,Allahverdi:2006we}. During inflation the Hubble 
expansion rate is smaller than the soft SUSY breaking mass term and the $A$-term, i.e. $A\sim m_\phi \gg H(t)$ for $a_H \sim c_H\sim {\cal O}(1)$ in Eq.~(\ref{hc1}), such that the SUGRA corrections are unimportant. This scenario has been discussed extensively, and has been extremely successful with the Planck data explaining the spectral tilt right on the 
observed central value, with Gaussian perturbations with the right amplitude~\cite{Choudhury:2013jya}. 

\item On the other hand, if 
\begin{equation}
\langle s\rangle\sim M_s \ll M_p\,,~~~~~~~~~~~ \langle \dot s\rangle =0\,,~~~~~~~~~~~V(s)=M_s^4\,,
\end{equation}
then the kinetic term for $\phi$ field will be canonical for cases $K^{2}$ and $K^{3}$ by virtue $\dot s =0$, see Table-\ref{tab1}.
However for cases $K^{1}$ and $K^{4}$, the departure from canonical for the $\phi$ field will depend on $M_s$. If
$\langle M_s\rangle \ll M_p$,  and $ a,~d \sim {\cal O}(1)$, see Table-\ref{tab1}, then the kinetic term for $\phi$ will be virtually
canonical, and as a consequence $c_s\approx 1$, while the potential will see a modification:
\begin{equation}
V_{total} = M_s^4+c_{H}H^{2}|\phi|^{2}+ \bigg(a_H H\frac{\lambda\phi^n}{nM_{p}^{n-3}}
 +h.c.\bigg)+\lambda^2\frac{\lvert\phi\rvert^{2(n-1)}}{M_{p}^{2(n-3)}}\,.
\end{equation}
This large vacuum energy density, i.e. $M_s^4\gg (\rm TeV)^4$, would  yield a large Hubble expansion rate, i.e. $H_{inf}^2\sim M_s^4/M_p^2\gg m^2_\phi\sim {\cal O}({\rm TeV})^2$. 
Therefore, the Hubble induced mass and and the $A$-term would dominate the potential over the soft terms.  Inspite of large mass, $c_H$, and  $a_{H}$-term, there is {\it no} SUGRA-$\eta$
problem, provided inflation occurs near the {\it saddle point} or the {\it inflection point}~\cite{Choudhury:2013jya,Mazumdar:2011ih}. We will not discuss this case any further, we will
now focus on a slightly non-trivial scenario, where high scale physics can alter some of the key cosmological predictions. 

\end{itemize}


 \subsection{Heavy field is oscillating during the onset of inflation}\label{oscillating}

One dramatic way the heavy field can influence the dynamics of primordial perturbations is via
coherent oscillations around its minimum,  while $\phi$ still plays the role of a slow roll inflaton~\footnote{There could
be other scenarios where the influence of heavy field is felt throughout the inflationary dynamics, see for instance in 
Refs.~\cite{Assassi:2013gxa,Achucarro:2010da}. Here 
we will discuss a slightly simpler scenario where both heavy and light sectors are coupled gravitationally via the K\"ahler correction.}. Furthermore,  the heavy field would only influence the first few e-foldings of inflation, once the heavy field 
is settled down its effect would be felt only via the vacuum energy density. Inspite of this short-lived phase, the heavy field can influence the dynamics and the perturbations for
the light field as we shall discuss below. 

Let us imagine the heavy field is coherently oscillating around a VEV, $\langle s\rangle \sim M_s$, during the 
initial phase of inflation, such that
\begin{equation}
V(s)\neq 0,~~~~~~~~~~~~~~~~\langle s\rangle \neq 0,~~~~~~~~~~~~~~~~\langle \dot  s\rangle \neq 0\,.
\end{equation}
The origin of coherent oscillations of $s$ field need not be completely ad-hoc, such a scenario might arise quite naturally from the hidden sector moduli field which is  coherently oscillating before being damped away by the initial phase of inflation, see for instance~\cite{Douglas:2006es}. This is particularly plausible for high string scale moduli, where the moduli mass can be heavy and can be stabilised early on in the history of the universe. There could also be a possibility of a smooth second order phase transition from one vacuum to another
during the intermittent phases of inflation~\cite{Burgess:2005sb}. Such a possibility can arise within MSSM where there are multiple false vacua  at high energies~\cite{Allahverdi:2008bt}. Irrespective of the origin of this heavy field,
during this transient period, the heavy field with an effective mass, $M_s \gg H_{inf}$, can coherently oscillate around its vacuum. We can set its 
initial amplitude of the oscillations to be of the order $M_s$. 
\begin{equation}\label{heavyf}
 s(t)\sim M_{s}-M_{s}\sin(M_st)\,.
 \end{equation}
This also implies that at the lowest order approximation, $\langle s \rangle \sim M_{s}$ and $\langle \dot s\rangle\sim M^{2}_{s}$~\footnote{At this point one might
say why we had taken the amplitude of oscillations for the heavy field to be $M_s$. In some scenarios, it is possible to envisage the amplitude of the oscillations to be $M_p$.
This would not alter much of our discussion, therefore for the sake of simplicity we will consider the initial amplitude for the $s$ field to be displaced by $M_s$, the same as that of the VEV.}.
The contribution to the potential due to the time dependent oscillating heavy field, see Eq~(\ref{heavyf}), is averaged over a full cycle ($0<t_{osc}<H_{inf}^{-1}$) is given by:
\be\begin{array}{lll}\label{avg}
    \displaystyle \langle V(s) \rangle \approx M_s^{2} \langle s^{2}(t) \rangle \sim H_{inf}^{2} M^{2}_{p}
   \end{array}
\ee
The $s$ field provides at the lowest order corrections to the kinetic term for the $\phi$ field, 
and to the overall potential, see Table~(\ref{tab1}), for both kinetic and potential  terms. 

At this point one might worry, the coherent oscillations of the $s$ field might trigger particle creation from 
the time dependent vacuum, see Refs.~\cite{Traschen:1990sw,Kofman:2004yc}, for a review see~\cite{Allahverdi:2010xz}. 
First of all, if we assume that the heavy field is coupled to other fields gravitationally, then the particle creation may not be 
sufficient to back react into the inflationary potential. Furthermore, inflation would also dilute the quanta created during this 
transient phase.

Since the kinetic terms for the $4$ cases tabulated in Table~(\ref{tab1}) are now no longer canonical, they would contribute to 
the speed of sound, $c_s \neq 1$, which we can summarize case by case below:
%
\be\label{sigbz}
 c_s=\sqrt{\frac{\dot p}{\dot \rho}}\approx \left\{
	\begin{array}{ll}
                    \displaystyle \sqrt{\frac{{\bf X}_{1}(t)-{\bf X}_{2}(t)-\dot{\widehat{V}}}{{\bf X}_{1}(t)+{\bf X}_{3}(t)+\dot{\widehat{V}}}} & \mbox{ for $\underline{\bf Case ~I}$}  
                    \\\\
         \displaystyle \sqrt{\frac{{\bf Y}_{1}(t)-{\bf Y}_{2}(t)-\dot{\widehat{V}}}{{\bf Y}_{1}(t)+{\bf Y}_{3}(t)+\dot{\widehat{V}}}} & \mbox{ for $\underline{\bf Case ~II}$}\\\\
\displaystyle \sqrt{\frac{{\bf Z}_{1}(t)-{\bf Z}_{2}(t)-\dot{\widehat{V}}}{{\bf Z}_{1}(t)+{\bf Z}_{3}(t)+\dot{\widehat{V}}}} & \mbox{ for $\underline{\bf Case ~III}$}\\\\
\displaystyle \sqrt{\frac{{\bf W}_{1}(t)-{\bf W}_{2}(t)-\dot{\widehat{V}}}{{\bf W}_{1}(t)+{\bf W}_{3}(t)+\dot{\widehat{V}}}} & \mbox{ for $\underline{\bf Case ~IV}$} .
          \end{array}
\right. 
\ee
%
where $p$ is the effective pressure and $\rho$ is the energy density. The dot denotes derivative w.r.t. physical time, $t$.
All the symbols, i.e. $X_1,~X_2,~Y_1,~Y_2,~Z_1,~Z_2,~W_1,~W_2$, appearing in Eq~(\ref{sigbz}) are explicitly mentioned in 
the Appendix F. Additionally, here we have defined, $\widehat{V}=V(\phi)-V(s)$~\footnote{ As a side remark, our analysis will be very useful
for the Affleck-Dine (AD) baryogenesis~\cite{Dine:1995kz}, especially when the minimum of the AD field is rotating in presence of the inflaton 
oscillations. Effectively, the AD field will have non-canonical kinetic terms, this has never been taken into account in the literature and one
should take the non-canonical kinetic terms for the AD field in presence of the inflaton oscillations in order to correctly estimate the baryon asymmetry.
The role of $s$ field will be that of an inflaton and $\phi$ field will be that of an AD field.}.

\begin{figure*}[htb]
\centering
\subfigure[$\textcolor{blue}{\bf \underline{Case ~I}}$]{
    \includegraphics[width=7.2cm,height=6.5cm] {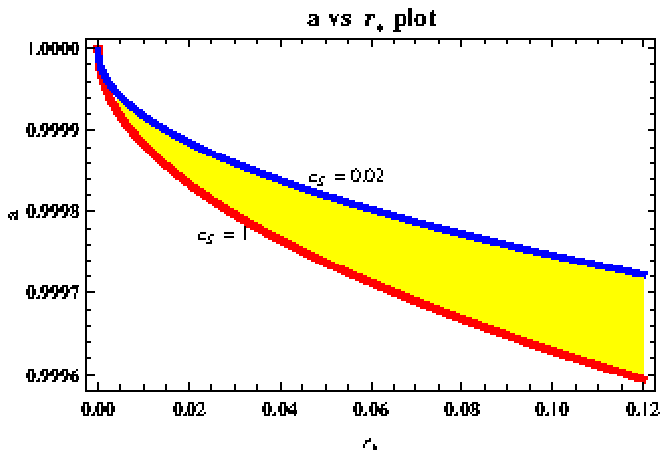}
    \label{fig:subfig1}
}
\subfigure[$\textcolor{blue}{\bf \underline{Case ~II}}$]{
    \includegraphics[width=7.2cm,height=6.5cm] {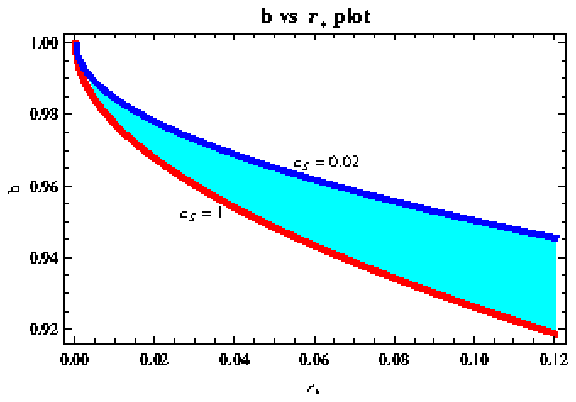}
    \label{fig:subfig2}
}
\subfigure[$\textcolor{blue}{\bf \underline{Case ~III}}$]{
    \includegraphics[width=7.2cm,height=6.5cm] {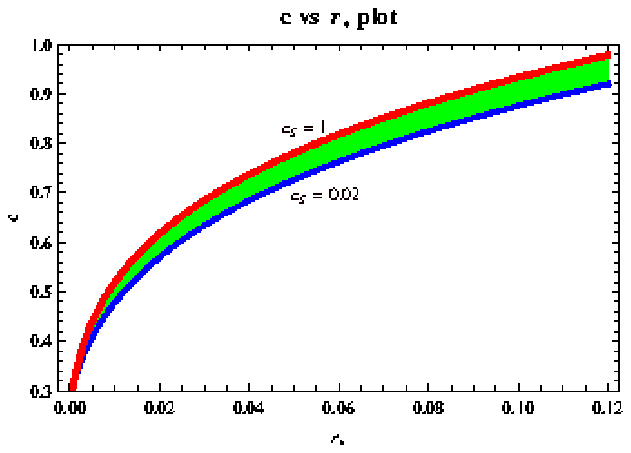}
    \label{fig:subfig3}
}
\subfigure[$\textcolor{blue}{\bf \underline{Case ~IV}}$]{
    \includegraphics[width=7.2cm,height=6.5cm] {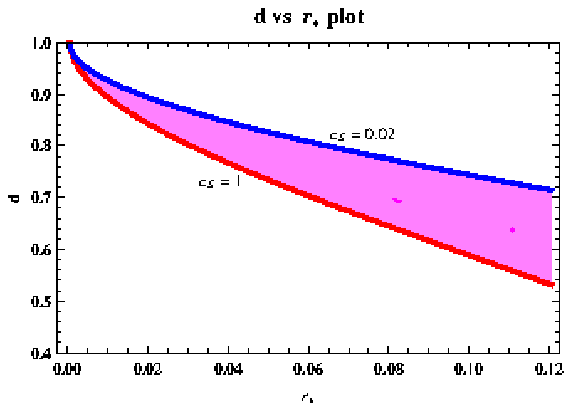}
    \label{fig:subfig4}
}
\caption[Optional caption for list of figures]{We show the constraints on 
  the non-renormalizable  K\"ahler operators, ``a'',``b'',``c'' and ``d'' 
with respect to the tensor-to-scalar ratio $r_{\star}$ at the pivot scale $k_{\star}=0.002~Mpc^{-1}$
when the heavy field $s$ is oscillating during the initial phase of inflation, especially at the time 
when the interesting perturbations are leaving the Hubble patch for $H_{inf} >>m_{\phi}\sim{\cal O}({\rm TeV})$ \cite{Choudhury:2014uxa}. 
All the shaded regions represent the allowed parameter space for the Hubble induced inflation satisfying
the Planck $2\sigma$ constraints on the amplitude of power spectrum $2.092\times 10^{-9}<P_S< 2.297\times 10^{-9}$ and spectral tilt $0.958 < n_S<0.963$.
 The dark coloured boundaries are obtained from the allowed range of the 
speed of sound $c_{s}$, within the window $0.02\leq c_{s}\leq 1$. }  
\label{fz}
\end{figure*}


 \subsection{Constraining non-renormalizable operators, i.e. $a,~b,~c,~d$, and $M_s$}\label{02}
 
For the potential under consideration, we have $V(s)=3H^2M^{2}_{p}\sim M^{2}_{s}s^2>>m_{\phi}^2|\phi|^2$, 
where $m_{\phi}\sim{\cal O}(\rm TeV)$ is the soft mass. In this case the 
contributions from the Hubble-induced terms are important compared to the soft SUSY breaking mass, $m_\phi$, and the $A$ term for 
all the four cases tabulated in Table-(\ref{tab1}).
After stabilizing the angular direction of the complex scalar field $\phi= |\phi|\exp[i\theta]$
~\cite{Allahverdi:2006iq,Allahverdi:2006we,Mazumdar:2011ih}, reduces to a simple form along the real direction, which is dominated by a 
single scale, i.e. $H\sim H_{inf}$:
\begin{equation}\label{h1a}
 V(\phi)=V(s)+c_{H}H^{2}|\phi|^{2}-
 \frac{a_{H}H\phi^n}{nM_{p}^{n-3}}+\frac{\lambda^2\lvert\phi\rvert^{2(n-1)}}{M_{p}^{2(n-3)}},
\end{equation}
where we take $\lambda=1$, and,the Hubble-induced mass parameter $c_{H}$, for $s<<M_{p}$, is defined as~\footnote{See Appendix-G and Appendix-H for details.}:
\be\label{effective}
c_{H}=\left\{
	\begin{array}{ll}
                    \displaystyle  3(1-a)\,, & \mbox{ for $\underline{\bf Case ~I}$}  \\
         \displaystyle  3(1+b^2)\,, & \mbox{ for $\underline{\bf Case ~II}$}\\
\displaystyle  3\,, & \mbox{ for $\underline{\bf Case ~III}$}\\
\displaystyle  3(1+d^2)\,, & \mbox{ for $\underline{\bf Case ~IV}$}.
          \end{array}
\right.
\ee
Note that for only third case, i.e. $K^{3}$, the Hubble induced mass term does not contain any K\"ahler correction, i.e. $\delta K$.
Similarly, we can express $a_H$, see Appendix~H for full expressions.
Note that for all $4$ cases, the kinetic terms are all non-minimal, and we have already listed in Table-(\ref{tab1}).
Fortunately for this class of potential given by Eq~(\ref{h1a}), inflection point inflation can be accommodated, when  $a_H^2\approx 8(n-1)c_H$. 
This can be characterized by a fine-tuning parameter, $\delta$, which is defined as~\cite{Allahverdi:2006we}:
\begin{equation}
\label{newbeta}
\frac{a_H^2}{8(n-1)c_H} = 1-\left(\frac{n-2}{2}\right)^2\delta^2\,.
\end{equation}
When $\vert\delta\vert$ is small~\footnote{We will consider a moderate tuning of order $\delta \sim 10^{-4}$ between $c_H$ and $a_H$.}, a point of inflection $\phi_0$ exists,
such that $V^{\prime\prime}\left(\phi_0\right) =0$, with
\begin{equation}
\label{phi0}
\phi_0 = \left(\sqrt{\frac{c_H}{(n-1)}} H M_{p}^{n-3}\right)^{{1}/{n-2}}\, +{\cal O}(\delta^2).
\end{equation}

For $\delta <1$, one can Taylor-expand the inflaton potential around an inflection 
point, $\phi=\phi_{0}$, as~\cite{Enqvist:2010vd,Choudhury:2013jya,Mazumdar:2011ih,Allahverdi:2008bt}:
\be\label{rt1a}
V(\phi)=\alpha+\beta(\phi-\phi_{0})+\gamma(\phi-\phi_{0})^{3}+\kappa(\phi-\phi_{0})^{4}+\cdots\,,
\ee 
where $\alpha$ denotes the height of the potential, and the coefficients $\beta,~\gamma,~\kappa$ determine the shape of the 
potential in terms of the model parameters~\footnote{The analytical expressions for the co-efficients appearing in the {\it inflection point} potential, 
$\alpha,\beta,\gamma$ and $\kappa$, can be expressed in terms of 
the mass parameter $c_{H}$, Hubble scale $H$ and, the VEV of the inflaton $\phi_{0}$ and tuning parameter $\delta$ are explicitly mentioned in the appendix E.}.
 Note that once the numerical values of $c_H$ and $H$ are specified, all the terms in the potential are determined. 

For $c_{s}\neq 1$, the upper bound on the numerical value of the scale of the heavy string moduli field ($M_{s}$) are expressed as:

\begin{equation}\label{hscalecon1}
     \displaystyle M_{s}\leq 1.77\times 10^{16}\times\sqrt[4]{\frac{r_{*}}{0.12}}~c^{\frac{\epsilon_{V}}{2(\epsilon_{V}-1)}}_{s}~{\rm GeV}\,.
    \end{equation}
where $r_{*}$ is the tensor-to-scalar ratio at the pivot scale of momentum $k_{*}\sim 0.002 Mpc^{-1}$.


\begin{figure*}[htb]
\centering
\subfigure[$\textcolor{blue}{\bf \underline{Case ~I}}$]{
    \includegraphics[width=7.2cm,height=5.9cm] {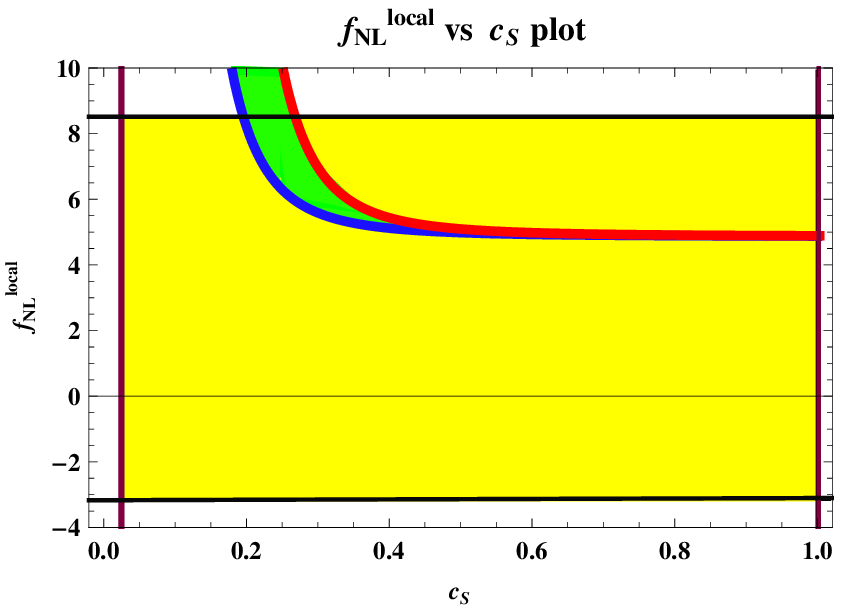}
    \label{fig:subfig1}
}
\subfigure[$\textcolor{blue}{\bf \underline{Case ~II}}$]{
    \includegraphics[width=7.2cm,height=5.9cm] {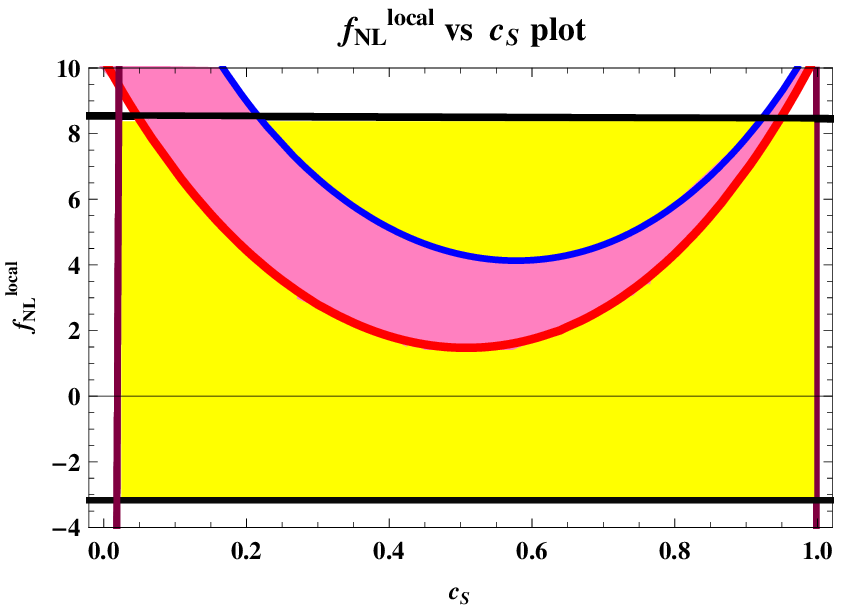}
    \label{fig:subfig2}
}
\subfigure[$\textcolor{blue}{\bf \underline{Case ~III}}$]{
    \includegraphics[width=7.2cm,height=5.9cm] {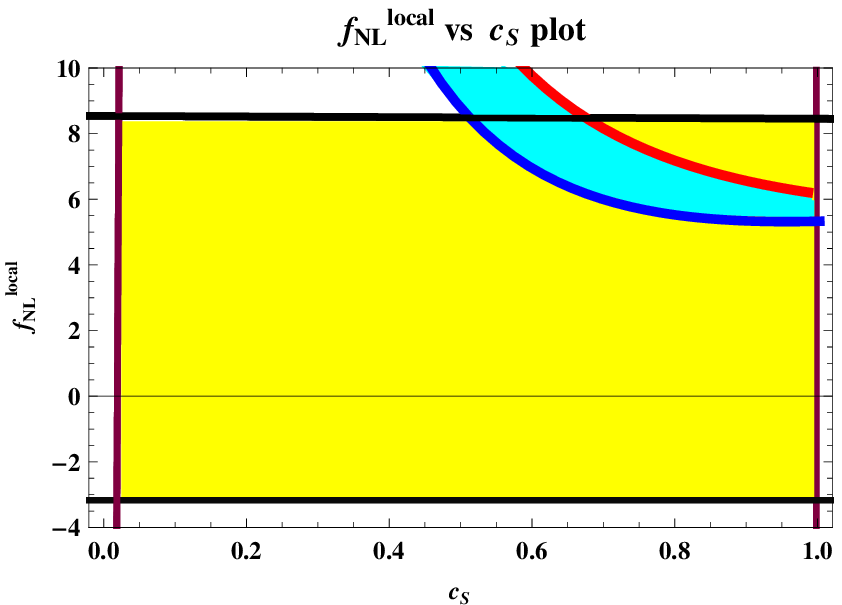}
    \label{fig:subfig3}
}
\subfigure[$\textcolor{blue}{\bf \underline{Case ~IV}}$]{
    \includegraphics[width=7.2cm,height=5.9cm] {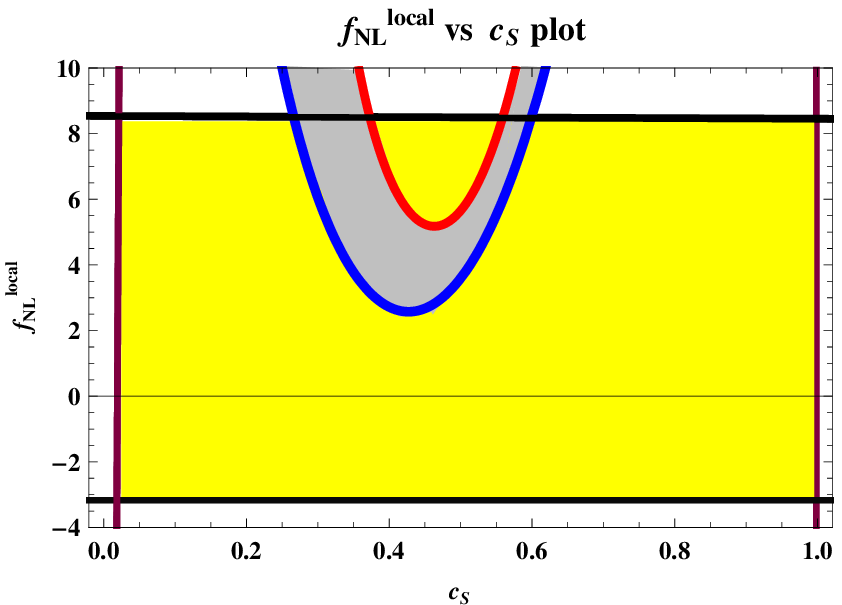}
    \label{fig:subfig4}
}
\caption[Optional caption for list of figures]{Behaviour
 of the local type of non-Gaussian parameter $f_{NL}^{\mathrm{local}}$ computed from the effective theory of ${\cal N}=1$ supergravity
with respect to the sound speed $c_{s}$ in the Hubble induced inflection point inflationary regime, represented
by $H>>m_{\phi}$ \cite{Choudhury:2014uxa}. The shaded yellow region represents the allowed parameter space for Hubble induced inflation which satisfies 
the combined Planck constraints on the $f_{NL}^{\mathrm{local}}$ (within $1\sigma$ CL) and sound speed $c_{s}$ (within $2\sigma$ CL).
 The red, blue coloured boundaries and the bounded dark coloured regions are obtained from the scanning range of the 
scale of the of heavy scalar degrees freedom $M_{s}$ corresponds to the hidden sector,
 within the window $9.50\times 10^{10}~{\rm GeV}\,\leq M_s \leq 1.77\times10^{16}~{\rm GeV}\,$. The four distinctive
 features are obtained by varying the model parameters of the effective theory of ${\cal N}=1$ SUGRA, $c_{H},a_{H},M_{s}$ and $\phi_{0}$, 
subject to the constraint as stated in Eq~(\ref{P-space}-\ref{pj0}). The overlapping region between the dark coloured and yellow region 
satisfied the combined constraints on the $f_{NL}^{\mathrm{local}}$ and $c_{s}$ within our proposed framework and the rest of the regions are 
excluded.} 
\label{fza}
\end{figure*}

\begin{figure*}[htb]
\centering
\subfigure[$\textcolor{blue}{\bf \underline{Case ~I}}$]{
    \includegraphics[width=7.2cm,height=6.5cm] {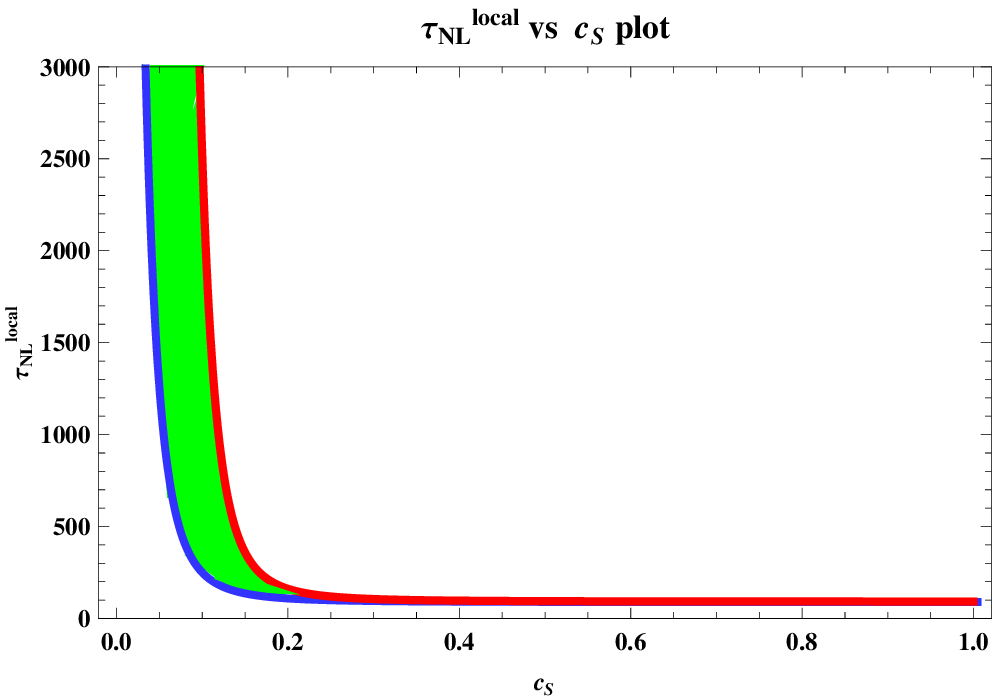}
    \label{fig:subfig1}
}
\subfigure[$\textcolor{blue}{\bf \underline{Case ~II}}$]{
    \includegraphics[width=7.2cm,height=6.5cm] {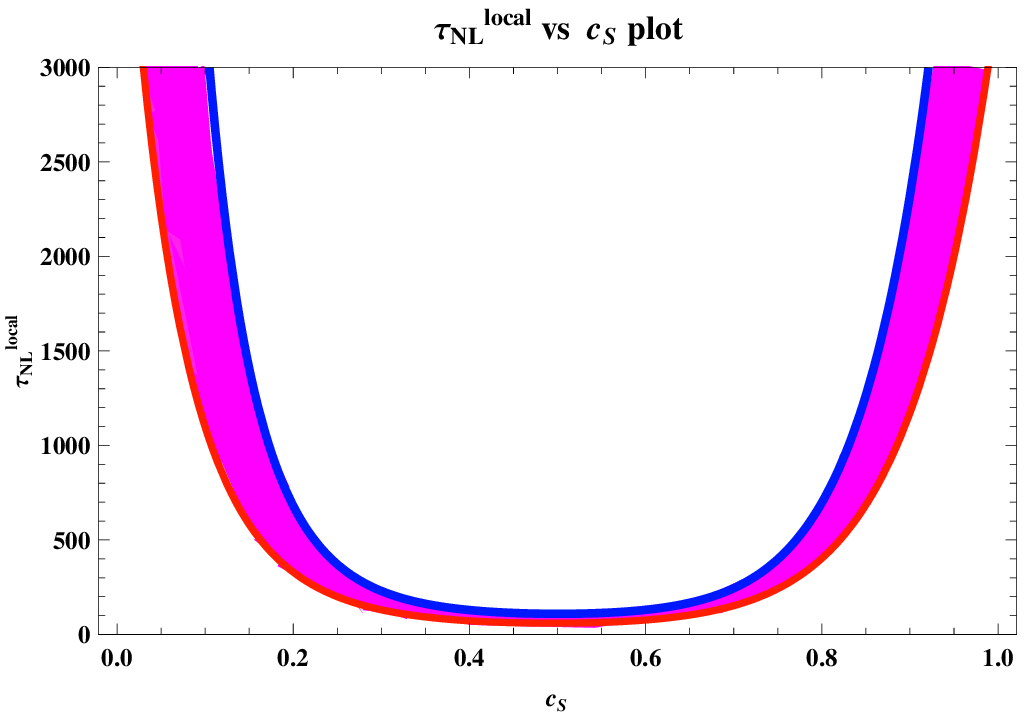}
    \label{fig:subfig2}
}
\subfigure[$\textcolor{blue}{\bf \underline{Case ~III}}$]{
    \includegraphics[width=7.2cm,height=6.5cm] {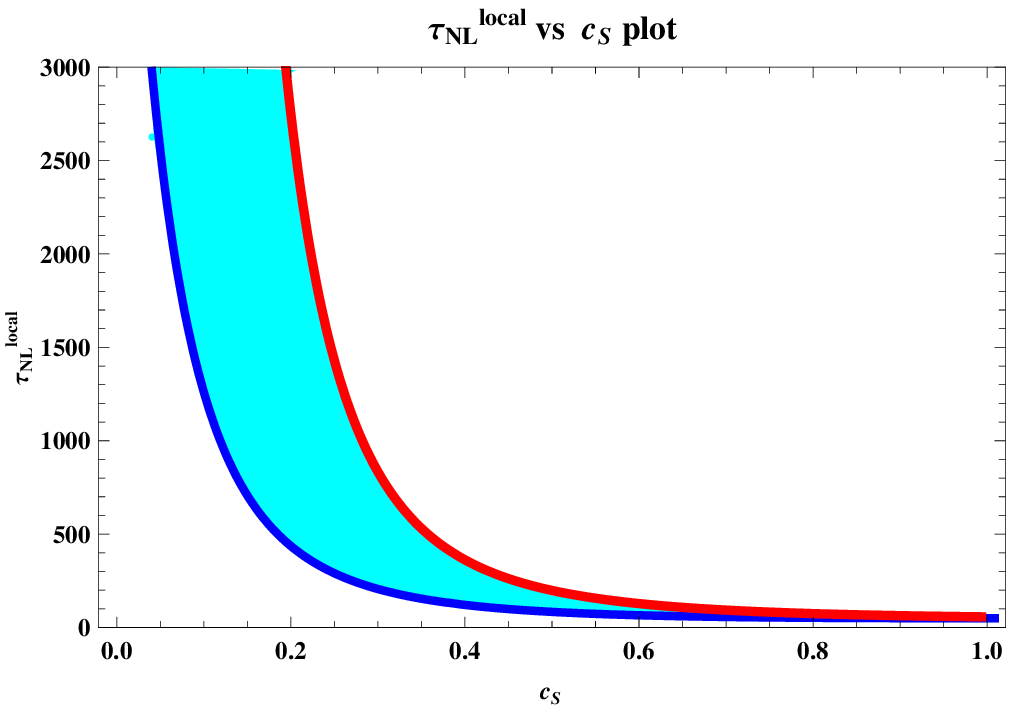}
    \label{fig:subfig3}
}
\subfigure[$\textcolor{blue}{\bf \underline{Case ~IV}}$]{
    \includegraphics[width=7.2cm,height=6.5cm] {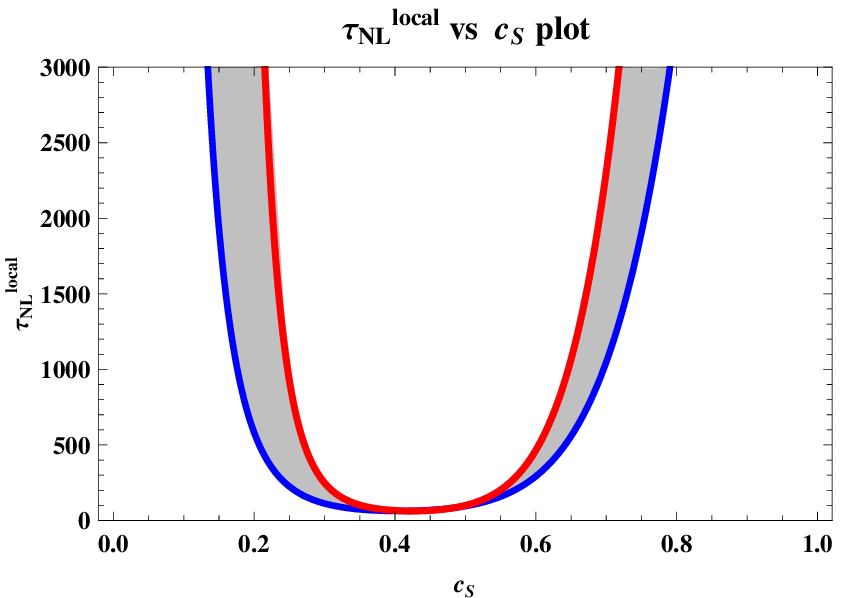}
    \label{fig:subfig4}
}
\caption[Optional caption for list of figures]{Behaviour
 of the local type of non-Gaussian parameter $\tau_{NL}^{\mathrm{local}}$ computed from the effective theory of ${\cal N}=1$ supergravity
with respect to the sound speed $c_{s}$ in the Hubble induced inflationary regime is represented
by $H>>m_{\phi}$ \cite{Choudhury:2014uxa}. The red and blue coloured boundaries are obtained from the scanning range of the 
scale of the of heavy scalar degrees freedom $M_{s}$ corresponds to the hidden sector,
 within the window $9.50\times 10^{10}~{\rm GeV}\,\leq M_s \leq 1.77\times10^{16}~{\rm GeV}\,$. The four distinctive
 features are obtained by varying the model parameters of the effective theory of ${\cal N}=1$ SUGRA, $c_{H},a_{H},M_{s}$ and $\phi_{0}$, 
subject to the constraint as stated in Eq~(\ref{P-space}-\ref{pj0}). The dark coloured region 
satisfied the combined constraints on the $f_{NL}^{\mathrm{local}}$ and $c_{s}$ within the proposed framework. As Planck puts an upper bound,
$\tau_{NL}^{\mathrm{local}}\leq2800$, the rest of the region above the $\tau_{NL}^{\mathrm{local}}=2800$ line is 
excluded.} 
\label{fzat}
\end{figure*}

\begin{figure*}[htb]
\centering
\subfigure[$\textcolor{blue}{\bf \underline{Case ~I}}$]{
    \includegraphics[width=7.2cm,height=6.5cm] {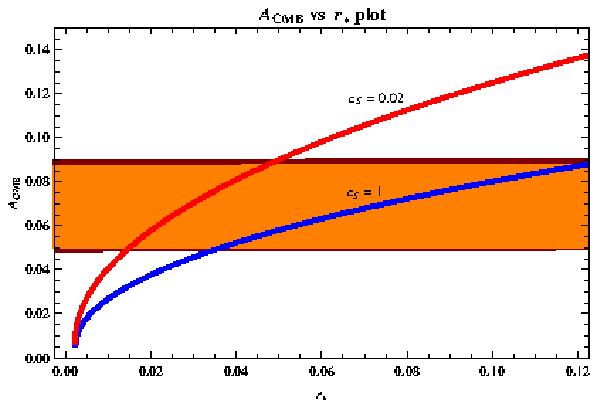}
    \label{fig:subfig1}
}
\subfigure[$\textcolor{blue}{\bf \underline{Case ~II}}$]{
    \includegraphics[width=7.2cm,height=6.5cm] {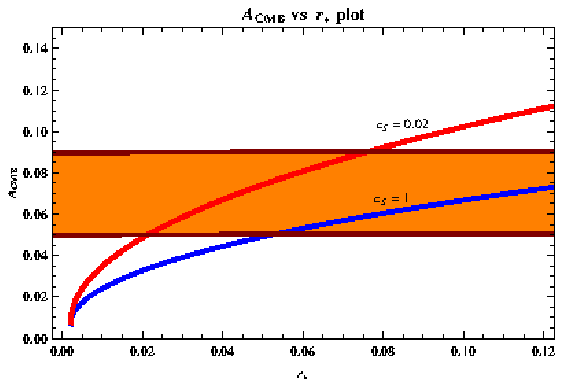}
    \label{fig:subfig2}
}
\subfigure[$\textcolor{blue}{\bf \underline{Case ~III}}$]{
    \includegraphics[width=7.2cm,height=6.5cm] {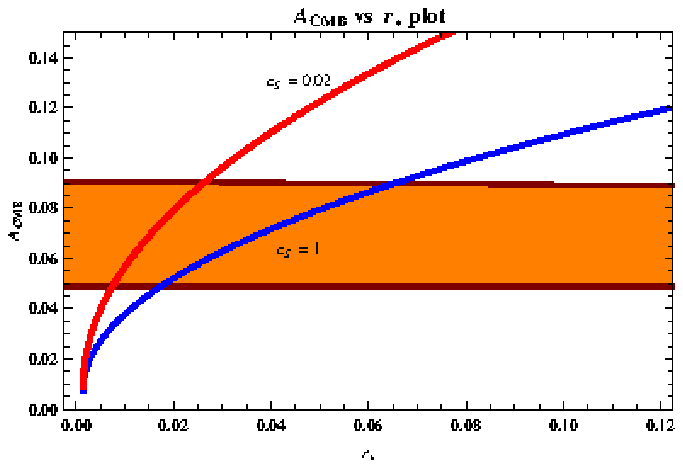}
    \label{fig:subfig3}
}
\subfigure[$\textcolor{blue}{\bf \underline{Case ~IV}}$]{
    \includegraphics[width=7.2cm,height=6.5cm] {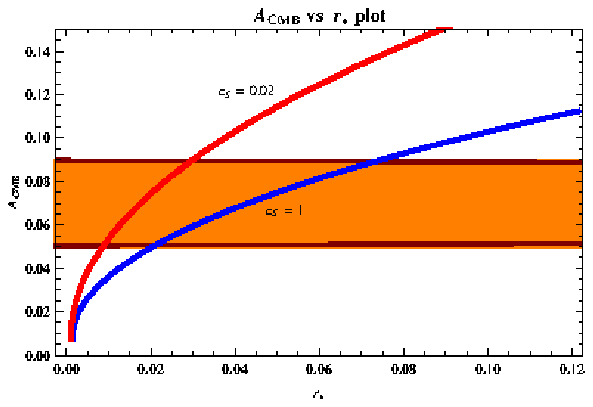}
    \label{fig:subfig4}
}
\caption[Optional caption for list of figures]{Behaviour
 of the CMB dipolar asymmetry parameter $A_{CMB}$ computed from the effective theory of ${\cal N}=1$ supergravity
with respect to the tensor-to-scalar ratio $r_{*}$ at the pivot scale, $k_{*}\sim 0.002~{\rm Mpc}^{-1}$ for the Hubble induced inflation \cite{Choudhury:2014uxa}.
 The red and blue coloured boundaries are obtained by fixing the sound speed at $c_{S}=0.02$ and $c_{S}=1$. The four distinctive
 features are obtained by varying the model parameters of the effective theory of ${\cal N}=1$ SUGRA, $c_{H},a_{H},M_{s}$ and $\phi_{0}$
subject to the constraint as stated in Eq~(\ref{P-space}-\ref{pj0}). The orange dark coloured region 
satisfied the Planck constraint on the $A_{CMB}$ within the proposed framework. Here only the region bounded by the red, blue and brown colour is the allowed one and
the rest of the region is 
excluded by the Planck data.} 
\label{fzata}
\end{figure*}
\section{ Calculation of non-Gaussianity in $\delta N$ for $c_s\neq 1$}\label{BNV}

In this section we have used the $\delta N$ formalism \cite{Sasaki:1995aw,Wands:2000dp,Lyth:2004gb,Lyth:2005fi,Mazumdar:2012jj,Sugiyama:2012tj}
to compute the local type of non-Gaussianity, $f_{NL}^{\rm local}$ from the
 prescribed setup for $c_s\neq 1$. 
In the non-attractor regime, the $\delta N$
formalism shows various non trivial features which has to be taken into account during explicit calculations.

It is important to mention that during the computation the solution
attains the attractor behaviour in the present context and consequently the significant contribution comes from only on the
cosmological perturbations of the scalar field
trajectories with respect to the inflaton field value defined at the initial
hypersurface, $\phi$. This is because the velocity of the inflaton field, $\dot\phi$, is solely 
characterized by the inflaton field $\phi$. Nonetheless, during the computation, specifically in the non-attractor regime of physical solution, 
both the information from the inflaton field value $\phi$ and also the velocity of the inflaton field $\dot \phi$ are necessarily required to determine the 
cosmological trajectory \cite{Namjoo:2012aa} relevant for present discussion.

Additionally it is important to note that, to compute the scalar-field cosmological trajectories explicitly, here we first of all we need to solve the classical equation
of motion of the scalar field and this is in general a second-order differential equation in a prescribed theoretical background. 
This can be solved by providing two initial conditions on the inflaton field $\phi$ and he velocity of the inflaton field, $\dot\phi$ on the
initial hypersurface. 
\begin{table*}
\centering
\tiny\tiny
\begin{tabular}{|c|c|c|}
\hline
\hline
\hline
\footnotesize{\bf Non--minimal} & \footnotesize{\bf Non--canonical} & \footnotesize{\bf Potential}\\
\footnotesize{\bf K\"{a}hler potential} &\footnotesize{\bf kinetic term} & \footnotesize $V(\phi)$ \\
 & \footnotesize$\mathcal{L}_{Kin}=K_{ij^*}(\partial_\mu\Phi^{i})(\partial^\mu \Phi^{j^*})$ &  \footnotesize{\bf for } $ |s|<<M_{p}$   \\
\hline\hline\hline
 \footnotesize $\displaystyle K^{(1)}= \phi^\dag \phi+s^\dag s$ & 
\footnotesize$\mathcal{L}_{Kin}=\displaystyle\bigg(1+\frac{a\lvert s\rvert^2}{M_{p}^2}\bigg)(\partial_\mu\phi)(\partial^\mu\phi^{\dagger})$
  &\footnotesize  $V(s)+\bigg(m_\phi^2+3(1-a)H^2\bigg)\lvert\phi\rvert^2\footnotesize \displaystyle-
 \frac{A\phi^n}{nM_{p}^{n-3}}$
\\
 \footnotesize $+\frac{a}{M_{p}^2}\phi^\dag \phi s^\dag s $ &\footnotesize $\displaystyle+\frac{a}{M_{p}^2}
\big\{\phi^{\dagger} s(\partial_\mu\phi)(\partial^\mu s^{\dagger})$  
&  \footnotesize $\displaystyle-\left( 1 + a \frac{|s|^2}{M_{p}^2} \right)\Big(1-\frac{3}{n}\Big)\frac{ s^2}{M^{2}_{p}}
\frac{\lambda M_{s}\phi^n}{M^{n-3}_{p}}$
\\
&\footnotesize $\displaystyle+\phi s^{\dagger}(\partial_\mu s)(\partial^\mu \phi^{\dagger})\big\}$ &  
\footnotesize$\displaystyle-\left( 1 - a \frac{|s|^2}{M_{p}^2} \right)\Big(a-\frac{1}{n}\Big) \frac{(s^\dagger)^2}{M^{2}_{p}} 
\frac{\lambda M_{s}\phi^n}{M^{n-3}_{p}}$
\\
 &\footnotesize $\displaystyle+\bigg(1+\frac{a\lvert \phi\rvert^2}{M_{p}^2}\bigg)(\partial_\mu s)(\partial^\mu s^{\dagger})$ & 
\footnotesize $\displaystyle+\lambda^2\frac{\lvert\phi\rvert^{2(n-1)}}{M_{p}^{2(n-3)}}+h.c.$\\

\hline
 \footnotesize$\displaystyle K^{(2)}=\phi^\dag\phi +s^\dag s$
&  \footnotesize $\mathcal{L}_{Kin}=\displaystyle(\partial_\mu\phi)(\partial^\mu\phi^{\dagger})+(\partial_\mu s)(\partial^\mu s^{\dagger})$
 &   \footnotesize $V(s)+\bigg(m_\phi^2+3(1+b^2)H^2\bigg)\lvert\phi\rvert^2\displaystyle-A \frac{\phi^n}{nM_{p}^{n-3}}$
\\ 
\footnotesize $+\frac{b}{2M_{p}}s^\dag\phi \phi + h.c.$ & \footnotesize $\displaystyle+\frac{b \phi }{2M_{p}}(\partial_\mu\phi)(\partial^\mu s^{\dagger})$ 
&  \footnotesize $\displaystyle-\left\{\left(1-\frac{3}{n}\right)\phi +\frac{ b  \phi^\dagger  s}{n M_{p}} \right\}\frac{\lambda \phi^{n-1}M_{s} s^2}{M_{p}^{n-1}}$ 
\\ 
 &  \footnotesize $\displaystyle+\frac{b \phi^{\dagger}}{2M_{p}}(\partial_\mu s)(\partial^\mu\phi^{\dagger})$ 
& \footnotesize $\displaystyle-\Big(\frac{s^\dagger \phi}{M_{p}}-bn\phi^\dagger \Big)\frac{2M_{s}\lambda \phi^{n-1} s^{\dagger}}{n M^{n-2}_{p}}$  
\\
 & &  \footnotesize  $\displaystyle-\frac{b M_{s} s^2}{2M_{p}^2}  
\left( \frac{2M_{s} s}{M_{p}}-\frac{M_{s} s^2 s^\dagger}{M_{p}^3} \right) \phi \phi$
 \\
 & &  \footnotesize $\displaystyle-\frac{4M^{2}_{s} b |s|^2 s^\dagger}{M^3_{p}} \phi \phi\displaystyle+\lambda^2\frac{\lvert\phi\rvert^{2(n-1)}}{M_{p}^{2(n-3)}}+h.c.$\\
\hline
  \footnotesize $\displaystyle K^{(3)}=\phi \phi^\dag+ss^\dag$ & 
 \footnotesize $\mathcal{L}_{Kin}=\displaystyle(\partial_\mu\phi)(\partial^\mu\phi^{\dagger})+(\partial_\mu s)
(\partial^\mu s^{\dagger})$ & \footnotesize $V(s)+\bigg(m_\phi^2+3H^2\bigg)\lvert\phi\rvert^2\displaystyle-
 A\frac{\phi^n}{nM_{p}^{n-3}}$\\
  \footnotesize $+\frac{c}{4M_{p}^2}s^\dag s^\dag\phi \phi + h.c. $ &  \footnotesize $\displaystyle +\frac{c s^{\dagger}
 \phi}{4M_{p}^2}(\partial_\mu \phi)(\partial^\mu s^{\dagger})$ &  
  \footnotesize $\displaystyle-\left\{\left(1-\frac{3}{n}\right)\phi  +\frac{c\phi^\dagger  ss}{2M_{p}^2}\right\} \frac{\lambda \phi^{n-1}M_{s} s^2}{M_{p}^{n-1}}$ 
\\
 &  \footnotesize $\displaystyle+\frac{c s \phi^{\dagger}}{4M_{p}^2}(\partial_\mu s)(\partial^\mu\phi^{\dagger})$ &
 \footnotesize $\displaystyle+\frac{c M^{2}_{s} s^2s^\dagger s \phi\phi  }{M^4_{p}}-\frac{M^{2}_{s} c 
|s|^2 s^\dagger s^\dagger}{M^4_{p}} \phi \phi\displaystyle+\lambda^2\frac{\lvert\phi\rvert^{2(n-1)}}{M_{p}^{2(n-3)}}$
 \\
& &  \footnotesize $\displaystyle-\Big(\frac{s^\dagger \phi}{M_{p}}-\frac{cn\phi^\dagger s}{M_{p}} \Big)\frac{2M_{s}\lambda \phi^{n-1} s^{\dagger}}{n M^{n-2}_{p}}+h.c.$
\\
\hline
  \footnotesize$\displaystyle K^{(4)}=\phi \phi^\dag+ss^\dag$ 
& 
 \footnotesize $\mathcal{L}_{Kin}=\displaystyle\bigg(\frac{d s}{M_{p}}+\frac{d s^{\dagger}}{M_{p}}+1\bigg)(\partial_\mu\phi)(\partial^\mu\phi^{\dagger})$ 
&  \footnotesize $V(s)+\bigg(m_\phi^2+3(1+d^2)H^2\bigg)\lvert\phi\rvert^2\displaystyle-A\frac{\phi^n}{nM_{p}^{n-3}}$\\
   \footnotesize $+\frac{d}{M_{p}}s\phi^\dag \phi + h.c.$ &  \footnotesize $\displaystyle+(\partial_\mu s)(\partial^\mu s^{\dagger})$ 
 &  \footnotesize $\displaystyle-\left(1-\frac{3}{n}\right) \frac{\lambda \phi^{n}M_{s} s^2}{M_{p}^{n-1}}
\displaystyle-\Big(\frac{s^\dagger}{M_{p}} -d\Big)\frac{2M_{s}\lambda \phi^{n} s^{\dagger}}{n M^{n-2}_{p}}$
 \\
 &  \footnotesize $\displaystyle+\frac{d \phi^{\dagger}}{M_{p}}(\partial_\mu \phi)(\partial^\mu s^{\dagger})
\displaystyle+\frac{d \phi }{M_{p}}(\partial_\mu s)(\partial^\mu\phi^{\dagger})$ 
&  \footnotesize $\displaystyle+\lambda^2\frac{\lvert\phi\rvert^{2(n-1)}}{M_{p}^{2(n-3)}}+h.c.$
\\
\hline
\hline
\hline
\end{tabular}
\vspace{.4cm}
\caption{\label{tab1}
 Various supergravity effective potentials and non-canonical kinetic terms for $|s|<<M_{p}$ in presence non-niminmal K\"{a}hler potential \cite{Choudhury:2014uxa}. Here both $\phi$ and $s$ are complex fields, and so are the $A$-terms.} 
\end{table*}
\subsection{General conventions}
In the present context, further we have neglected the canonical kinetic term
during the non-attractor phase for simplicity. The background equation
of motion for the four physical situations are given by
\be\label{svb}
 0=\left\{
	\begin{array}{ll}
                     \small\small \left(\ddot{\phi}+3H\dot{\phi}\right)\left[1+\frac{aM^{2}_{s}}{2M^{2}_{p}}\left(1+\sin(M_{s}t)\right)^{2}\right]
+\frac{2aM^{3}_{s}}{M^{2}_{p}}\left[\left(2\dot{\phi}
+3H\phi\right)\cos(M_{s}t)\right.\\ \left.~~~~~~~~~~~~~~~~~~~~~~~~~~~~~~~~~~~~~~~~~-\phi M_{s}\sin(M_{s}t)\right]\left(1+\sin(M_{s}t)\right)
+V^{'}(\phi) & \mbox{ for $\underline{\bf Case ~I}$}  \\ \\
   \ddot{\phi}+3H\dot{\phi}+\frac{bM^{2}_{s}}{M_{p}}\left[\left(\dot{\phi}+3H\phi\right)\cos(M_{s}t)-M_{s}\phi\sin(M_{s}t)\right]+V^{'}(\phi)& \mbox{ for $\underline{\bf Case ~II}$}  \\ \\ 
    \ddot{\phi}+3H\dot{\phi}+\frac{cM^{3}_{s}}{2M^{2}_{p}}\left[\left(\dot{\phi}+3H\phi\right)\cos(M_{s}t)\left(1+\sin(M_{s}t)\right)+M_{s}\phi\left(\cos^{2}(M_{s}t)
\right.\right.\\ \left.\left.~~~~~~~~~~~~~~~~~~~~~~~~~~~~~~~~~~~~~~~~~~~~-\sin(M_{s}t)\left(1+\sin(M_{s}t)\right)\right)\right]+V^{'}(\phi) & \mbox{ for $\underline{\bf Case ~III}$}  \\ \\
    \left(\ddot{\phi}+3H\dot{\phi}\right)\left[1+\frac{2dM_{s}}{M_{p}}\left(1+\sin(M_{s}t)\right)\right]
+\frac{2dM^{2}_{s}}{M_{p}}\left[\left(2\dot{\phi}
+3H\phi\right)\cos(M_{s}t)\right.\\ \left.~~~~~~~~~~~~~~~~~~~~~~~~~~~~~~~~~~~~~~~~~~~~~~~~~~~~~~~~~-\phi M_{s}\sin(M_{s}t)\right]
+V^{'}(\phi)      & \mbox{ for $\underline{\bf Case ~IV}$}.
          \end{array}
\right.
\ee
From the Eq~(\ref{svb}), it is obvious that the determination of a general analytical solution is
too much complicated. To simplify the task here we consider a particular solution,
\be\phi=\phi_{L}\propto e^{\vartheta Ht} ~~~(~{\rm i.e.}~~~ \phi=\phi_{L}(N)=\phi_*e^{-\vartheta N}),\ee
and further my prime objective is to obtain a more generalized solution for the background up to the
second order in perturbations around this particular solution. 
Here we also assume that the non-attractor phase ends when the inflaton field value is achieved at $\phi=\phi_*$.
Let me define a perturbative parameter,
$$\Delta\equiv\phi-\phi_{0}-\phi_{L}\,=\Delta_{1}+\Delta_{2}+\cdots , $$ which represents the difference
between the true background solution and the reference solution to solve the background Eq~(\ref{svb}) perturbatively. Here 
$\Delta_{1}$ and $\Delta_{2}$ are the general linearized and second order perturbative solution of the background field equations.
The $\cdots$ contribution comes from the higher order perturbation scenario which we will neglect for further computation.

\subsection{Linearized perturbation}

Let us consider the contribution from the linear perturbation, $\Delta_{1}$. Consequently in the leading order the background linearized perturbative equation of
motion takes the following form:
\be\label{svblin}
 0\approx\left\{
	\begin{array}{ll}
                     \small\small \left(\ddot{\Delta_{1}}+3H\dot{\Delta_{1}}+\vartheta H^{2}(3+\vartheta)\phi_{L}\right)
\left[1+\frac{aM^{2}_{s}}{4M^{2}_{p}}\right]
-\frac{aM^{3}_{s}}{M^{2}_{p}}(\Delta_{1}+\phi_{L}) M_{s}
+\beta & \mbox{ for $\underline{\bf Case ~I}$}  \\ \\
   \ddot{\Delta_{1}}+3H\dot{\Delta_{1}}+\vartheta H^{2}(3+\vartheta)\phi_{L}+\beta & \mbox{ for $\underline{\bf Case ~II}$}  \\ \\
    \ddot{\Delta_{1}}+3H\dot{\Delta_{1}}+\vartheta H^{2}(3+\vartheta)\phi_{L}+\beta & \mbox{ for $\underline{\bf Case ~III}$}  \\ \\
    \left(\ddot{\Delta_{1}}+3H\dot{\Delta_{1}}+\vartheta H^{2}(3+\vartheta)\phi_{L}\right)\left[1+\frac{2dM_{s}}{M_{p}}\right]
+\beta   & \mbox{ for $\underline{\bf Case ~IV}$}.
          \end{array}
\right.
\ee
where we have neglected the higher powers of $\Delta_{1}$ in the linearized approximation.
The general solution is given by
\be\label{svblin2}
 \Delta_{1}\approx\left\{
	\begin{array}{ll}
                     \small\small {\bf C}_{1}e^{\frac{1}{2}
\left(-3H-\sqrt{\frac{4aM^{4}_{s}}{M^{2}_{p}\left(1+\frac{aM^{2}_{s}}{4M^{2}_{p}}\right)}+9H^{2}}\right)t}
+{\bf C}_{2}e^{\frac{1}{2}
\left(-3H+\sqrt{\frac{4aM^{4}_{s}}{M^{2}_{p}\left(1+\frac{aM^{2}_{s}}{4M^{2}_{p}}\right)}+9H^{2}}\right)t}\\
~~~~~~~~~~~~~~~~~~~~~~~~~~~~~~~~~~~~~~~~~~~~~~~~~~~~~~+\phi_{*}e^{\vartheta Ht}
-\frac{\beta M^{2}_{p}}{aM^{4}_{s}} & \mbox{ for $\underline{\bf Case ~I}$}  \\ \\
  {\bf C}_{3}-\frac{{\bf C}_{4}}{3H}e^{-3Ht}-\frac{\beta t}{3H}-\phi_{*}e^{\vartheta Ht} & \mbox{ for $\underline{\bf Case ~II}$}  \\ \\
    {\bf C}_{5}-\frac{{\bf C}_{6}}{3H}e^{-3Ht}-\frac{\beta t}{3H}-\phi_{*}e^{\vartheta Ht} & \mbox{ for $\underline{\bf Case ~III}$}  \\ \\
    {\bf C}_{7}-\frac{{\bf C}_{8}}{3H}e^{-3Ht}-\frac{\beta t}{3H\left(1+\frac{2dM_{s}}{M_{p}}\right)}-\phi_{*}e^{\vartheta Ht}  & \mbox{ for $\underline{\bf Case ~IV}$}.
          \end{array}
\right.
\ee
where ${\bf C}_{i}\forall i(=1,2,....,8)$ are dimensionful arbitrary integration constants which can be fixed by imposing the boundary conditions.

\subsection{Second-order perturbation}

Next we have considered the contribution from the second-order perturbation, $\Delta_2$.
 Consequently in the leading order the background Second-order perturbative equation of
motion takes the following form:
\be\label{svbsec2}
 \Pi_{s}\approx\left\{
	\begin{array}{ll}
                     \small\small \left(\ddot{\Delta_{2}}+3H\dot{\Delta_{2}}+\vartheta H^{2}(3+\vartheta)\phi_{L}\right)
\left[1+\frac{aM^{2}_{s}}{4M^{2}_{p}}\right]
-\frac{aM^{3}_{s}}{M^{2}_{p}}(\Delta_{2}+\phi_{L}) M_{s}
+\beta & \mbox{ for $\underline{\bf Case ~I}$}  \\ \\
   \ddot{\Delta_{2}}+3H\dot{\Delta_{2}}+\vartheta H^{2}(3+\vartheta)\phi_{L}+\beta & \mbox{ for $\underline{\bf Case ~II}$}  \\ \\
    \ddot{\Delta_{2}}+3H\dot{\Delta_{2}}+\vartheta H^{2}(3+\vartheta)\phi_{L}+\beta & \mbox{ for $\underline{\bf Case ~III}$}  \\ 
    \left(\ddot{\Delta_{2}}+3H\dot{\Delta_{2}}+\vartheta H^{2}(3+\vartheta)\phi_{L}\right)\left[1+\frac{2dM_{s}}{M_{p}}\right]
+\beta   & \mbox{ for $\underline{\bf Case ~IV}$}.
          \end{array}
\right.
\ee
where the source term, $\Pi_{s}$, for the sub-Planckian Hubble induced inflection point inflation within ${\cal N}=1$ SUGRA is given by
\be
\Pi_{s}=3\gamma(\Delta_{1}+\phi_{L})^{2}\,.
\ee
Now to solve Eq~(\ref{svbsec2}) in presence of non-linear source term, let me assume that the contribution from $\phi_{L}$ is sub-dominant.
Consequently the general solution in presence of second-order perturbation is given by:
\be\label{svbper4}
 \Delta_{2}\approx\left\{
	\begin{array}{ll}
                     \small\small {\bf G}_{1}e^{\frac{1}{2}
\left(-3H-\sqrt{\frac{4aM^{4}_{s}}{M^{2}_{p}\left(1+\frac{aM^{2}_{s}}{4M^{2}_{p}}\right)}+9H^{2}}\right)t}
+{\bf G}_{2}e^{\frac{1}{2}
\left(-3H+\sqrt{\frac{4aM^{4}_{s}}{M^{2}_{p}\left(1+\frac{aM^{2}_{s}}{4M^{2}_{p}}\right)}+9H^{2}}\right)t}+\Sigma_{s}(t) & \mbox{ for $\underline{\bf Case ~I}$}  \\ 
  {\bf G}_{5}-\frac{12{\bf G}_{6}}{H}e^{-3Ht}+\Xi_{s}(t) & \mbox{ for $\underline{\bf Case ~II}$}  \\ 
     {\bf G}_{5}-\frac{12{\bf G}_{6}}{H}e^{-3Ht}+\Psi_{s}(t) & \mbox{ for $\underline{\bf Case ~III}$}  \\ 
    {\bf G}_{7}-\frac{12{\bf G}_{8}}{H}e^{-3Ht}+\Theta_{s}(t) & \mbox{ for $\underline{\bf Case ~IV}$}.
          \end{array}
\right.
\ee
where the time dependent functions $\Sigma_{s}(t),\Xi_{s}(t),\Psi_{s}(t)$ and $\Theta_{s}(t)$ are explicitly mentioned in the Appendix J .
Here ${\bf G}_{i}\forall i(=1,2,....,8)$ are dimensionful arbitrary integration constants which can be fixed by imposing the boundary conditions.
\subsection{$\delta N$ at the final hypersurface}

Here our prime objective is to compute the perturbations of the number of
$e$-folds, $\delta N$.
The truncated background solution of $\phi$ up to the
second-order perturbations around the reference trajectory, 
$\phi_L\propto e^{-\vartheta N}$ in terms of $N$ is given by,
\be
\phi =\phi_{0}+\left\{
	\begin{array}{ll}
              \frac{\phi_*}{1+\widehat{{\bf C}}_{1}+\widehat{{\bf C}}_{2}-\frac{\beta M^{2}_{p}}{a \phi_{*}M^{4}_{s}}
+\widehat{{\bf G}}_{1}+\widehat{{\bf G}}_{2}+\widehat{\Sigma}_{s}(0)}\,
\left(e^{-\vartheta N} +\widehat{\Delta}_{1}(N)
 +\widehat{\Delta}_{2}(N)\right)\,& \mbox{ for $\underline{\bf Case ~I}$}  \\ \\
   \frac{\phi_*}{1+\widehat{{\bf C}}_{3}-\frac{\widehat{{\bf C}}_{4}}{3H}+\widehat{{\bf G}}_{3}-\frac{12\widehat{{\bf G}}_{4}}{H}
+\widehat{\Xi}_{s}(0)}\,
\left(e^{-\vartheta N} +\widehat{\Delta}_{1}(N)
 +\widehat{\Delta}_{2}(N)\right)\, & \mbox{ for $\underline{\bf Case ~II}$}  \\ \\
    \frac{\phi_*}{1+\widehat{{\bf C}}_{5}-\frac{\widehat{{\bf C}}_{6}}{3H}+\widehat{{\bf G}}_{5}-\frac{12\widehat{{\bf G}}_{6}}{H}
+\widehat{\Psi}_{s}(0)}\,
\left(e^{-\vartheta N} +\widehat{\Delta}_{1}(N)
 +\widehat{\Delta}_{2}(N)\right)\, & \mbox{ for $\underline{\bf Case ~III}$}  \\ \\
    \frac{\phi_*}{1+\widehat{{\bf C}}_{7}-\frac{\widehat{{\bf C}}_{8}}{3H}+\widehat{{\bf G}}_{7}-\frac{12\widehat{{\bf G}}_{8}}{H}
+\widehat{\Theta}_{s}(0)}\,
\left(e^{-\vartheta N} +\widehat{\Delta}_{1}(N)
 +\widehat{\Delta}_{2}(N)\right)\,   & \mbox{ for $\underline{\bf Case ~IV}$}.
          \end{array}
\right.,
\label{phi-lambda2}
\ee
where the symbol $~\widehat{}~$ is introduced to rescale the integration constants as well as the perturbative solutions by $\phi_{*}$.
Here we have neglected the contribution from $e^{-\vartheta N}$ in $\widehat{\Delta}_{2}(N)$ to avoid over counting in the Eq~(\ref{phi-lambda2}).
It is important to note that in the present context of all 
 these sets of scaled
integration constants parameterizes
different trajectories, and we have set $\phi(0,\widehat{{\bf W}}_{k})=\phi_*$
for any value of $\widehat{{\bf W}}_{k}\forall k=([1,2],[3,4],[5,6],[7,8])$ in accordance with the assumption that 
the end of the non-attractor phase is determined only by the value of the
scalar field, $\phi=\phi_*$. Here $\widehat{{\bf W}}_{k}=\widehat{{\bf C}}_{k},\widehat{{\bf G}}_{k}$
represent collection of all integration constants.

Further inverting Eq~(\ref{phi-lambda2}) for a fixed set of $\widehat{{\bf W}}_{k}$, we have obtained
$N$ as a implicit function of $\phi$ and $\widehat{{\bf W}}_{k}$. Then the $\delta N$ formula can be
expressed in the present context as:
\begin{eqnarray}
\delta N=N(\phi+\delta\phi,\widehat{{\bf W}}_{k})-N(\phi,0)
=\sum_{k}\sum_{n,m}\frac{1}{n!m!}
\partial^{n}_{\phi^{n}}\partial^{m}_{\widehat{{\bf W}}^{m}_{k}}N(\phi,0)
\delta\phi^n\widehat{{\bf W}}^{m}_{k}\,.
\end{eqnarray}
Here we have introduced the shift in the inflaton field $\phi\to\phi+\delta\phi$ and the number of e-folds $N\to N+\delta N$
on both sides of Eq~(\ref{phi-lambda2}) to compute $\delta N$ iteratively.
In the present context we have obtained, perturbative
solutions of the scalar-field trajectories around the particular
reference solution, $\phi_L=\phi_*e^{\vartheta Ht}$, which are
valid only when the perturbed trajectories are not far away from the
reference solution. Additionally, since we have neglected the sub-dominant solution,
$\Delta_1\propto e^{\vartheta Ht}$, my approximation holds good only at sufficiently 
late times. These imply that here we should choose the initial 
time as close as possible to the final time for which $N\lesssim 1$. 
Then the simplest choice is
to take the initial time to be infinitesimally close to $t=t_*$.

Now perturbing the number of e-folds $N$ up to the second order at the epoch $t=t_*$, we get \cite{Chen:2013eea}:
\begin{eqnarray}\label{efolN}
\zeta=\delta N =  N_{,\phi}\delta\phi
+  \frac12  N_{,\phi\phi}\delta\phi^2
+ \frac16  N_{,\phi\phi\phi}\delta\phi^3 +\cdots .~~~~
\end{eqnarray}
where we have used the $\widehat{{\bf W}}$-independence of $N$ at $N=0$ for which, $N_{,\widehat{{\bf W}}_{k}}=0=N_{,\widehat{{\bf W}}_{p}\widehat{{\bf W}}_{q}}$.
Here $\cdots$ corresponds to the higher order contributions, which are negligibly small compared to the leading order contributions.
By taking the derivatives of both sides of Eq.~(\ref{phi-lambda2})
and setting $N=0=\widehat{{\bf W}}_{k}(=\widehat{{\bf C}}_{k},\widehat{{\bf G}}_{k})\forall k$ at the end,
my next task is to identify $\delta\phi_*$ and $\widehat{{\bf W}}_{k}(=\widehat{{\bf C}}_{k},\widehat{{\bf G}}_{k})$
which are actually generated from quantum fluctuations on flat slicing, $\delta\phi$.
To serve this purpose let me consider the evolution of $\delta\phi$ on super-horizon scales.
The shift in the inflaton field can be expressed here as:
\be\begin{array}{llll}\label{eqrt1}
    \displaystyle \delta\phi(N)=\delta\phi_{1}(N)+\delta\phi_{2}(N)=\phi_{*}\left(\widehat{\Delta}_{1}(N)+\widehat{\Delta}_{2}(N)\right)
   \end{array}\ee
where the subscript ``1'' and ``2'' represent the solution at the linear and the second order respectively.
It is important to note that both the solutions include the features of growing and decaying mode.
Now imposing the boundary condition from the end of the non-attractor
phase, where $N=0$, we get:
\be
\delta\phi_1(0)=\delta\phi_{1*}=\phi_{*}\widehat{\Delta}_{1}(0)\,=\left\{
	\begin{array}{ll}
                     \small\small \phi_{*}\left(\widehat{{\bf C}}_{1}
+\widehat{{\bf C}}_{2}
-\frac{\beta M^{2}_{p}}{a\phi_{*}M^{4}_{s}}\right) & \mbox{ for $\underline{\bf Case ~I}$}  \\ \\
  \phi_{*}\left(\widehat{{\bf C}}_{3}-\frac{\widehat{{\bf C}}_{4}}{3H}\right) & \mbox{ for $\underline{\bf Case ~II}$}  \\ \\
    \phi_{*}\left(\widehat{{\bf C}}_{5}-\frac{\widehat{{\bf C}}_{6}}{3H}\right) & \mbox{ for $\underline{\bf Case ~III}$}  \\ \\
    \phi_{*}\left(\widehat{{\bf C}}_{7}-\frac{\widehat{{\bf C}}_{8}}{3H}\right)  & \mbox{ for $\underline{\bf Case ~IV}$}.
          \end{array}
\right.
\label{dphistar}
\ee
\be
\delta\phi_2(0)=\delta\phi_{2*}=\phi_{*}\widehat{\Delta}_{2}(0)\,=\left\{
	\begin{array}{ll}
                     \small\small \phi_{*}\left(\widehat{{\bf G}}_{1}
+\widehat{{\bf G}}_{2}
+\widehat{\Sigma}_{s}(0)\right) & \mbox{ for $\underline{\bf Case ~I}$}  \\ \\
  \phi_{*}\left(\widehat{{\bf G}}_{3}-\frac{12\widehat{{\bf G}}_{4}}{H}+\widehat{\Xi}_{s}(0)\right) & \mbox{ for $\underline{\bf Case ~II}$}  \\ \\
    \phi_{*}\left(\widehat{{\bf G}}_{5}-\frac{12\widehat{{\bf G}}_{6}}{H}+\widehat{\Psi}_{s}(0)\right) & \mbox{ for $\underline{\bf Case ~III}$}  \\ \\
    \phi_{*}\left(\widehat{{\bf G}}_{7}-\frac{12\widehat{{\bf G}}_{8}}{H}+\widehat{\Theta}_{s}(0)\right)  & \mbox{ for $\underline{\bf Case ~IV}$}.
          \end{array}
\right.
\label{dphistar2}
\ee
from which we have obtained:
\begin{eqnarray}
\delta\phi_*=\delta\phi(0)=\delta\phi_{1*}+\delta\phi_{2*}=\phi_{*}\left(\widehat{\Delta}_{1}(0)+\widehat{\Delta}_{2}(0)\right)\,.
\label{fluctuations}
\end{eqnarray}

Further neglecting the mixing between the solutions corresponding to the
 linearized and second order perturbation,
the analytical expression for $\delta N$ can be expressed as: 
\be\begin{array}{lll}\label{delbn}\displaystyle
\displaystyle \zeta=\delta N = -\dfrac{(\delta\phi_{1*}+\delta\phi_{2*})}{\vartheta \phi_*}+\frac{(\delta\phi^{2}_{1*}+\delta\phi^{2}_{2*})}{2\vartheta\phi^{2}_{*}}+\cdots
\end{array}\ee

\subsection{Computation of local type of non-Gaussianity and CMB dipolar asymmetry}

Further using the results obtained for $\delta N$ as mentioned in the in chapter~\ref{ch:FRW} section~\ref{ng},
 the non-Gaussian parameter corresponding to the local type of non-Gaussianity $f_{NL}^{\rm local}$
, $g_{NL}^{\rm local}$ and $\tau_{NL}^{\rm local}$ can be computed as:
\be\begin{array}{lll}\label{d1}\displaystyle
\displaystyle f_{NL}^{\mathrm{local}} = \frac{5}{6}
\frac{N_{,\phi\phi}}{N^{2}_{,\phi}}+\cdots
=\frac{5\vartheta}{6}+\cdots\,
\end{array}\ee

\be\begin{array}{lll}\label{d11}\displaystyle
\displaystyle \tau_{NL}^{\mathrm{local}} =
\frac{N^{2}_{,\phi\phi}}{N^{4}_{,\phi}}+\cdots
= \vartheta^{2}+\cdots\,
\end{array}\ee

\be\begin{array}{lll}\label{d111}\displaystyle
\displaystyle g_{NL}^{\mathrm{local}} =
\frac{25}{54}\frac{N_{,\phi\phi\phi}}{N^{3}_{,\phi}}+\cdots
= \frac{25\vartheta^{2}}{108}+\cdots\,
\end{array}\ee
where the parameter $\vartheta$, appearing in all the physical situations, 
can be expressed in terms of the sound speed ($c_{s}$), potential dependent slow roll parameter $(\epsilon_{V},\eta_{V})$ and the model parameters $(\alpha,\beta,\gamma)$ as:
\be\label{cx1}
\vartheta\approx\left[\eta_{V}\left(1+\frac{1}{c^{2}_{s}}\right)^{2}+\epsilon_{V}\left(1-\frac{1}{c^{4}_{s}}\right)\right]\approx\left[\frac{6\gamma\phi_{*}M^{2}_{p}}{\alpha}\left(1+\frac{1}{c^{2}_{s}}\right)^{2}
+\frac{\beta^{2}M^{2}_{p}}{2\alpha^{2}}\left(1-\frac{1}{c^{4}_{s}}\right)\right].
\ee

where the sound speed $c_{s}$ can be expressed in terms of non-canonical K\"ahler corrections, $a,b,c,d$ and the scale of heavy field, $M_{s}$,
as:
%
\be\label{sigbz}
 c_s\approx \left[\sqrt{\frac{{\bf \Sigma}_{1}(t)-{\bf \Sigma}_{2}(t)-\dot{\widehat{V}}}{{\bf \Sigma}_{1}(t)+{\bf \Sigma}_{3}(t)+\dot{\widehat{V}}}}\right]_{t=t_{*}}
\ee
%
The dot denotes derivative w.r.t. physical time, $t$. Here $\widehat{V}=V(\phi)-V(S)$ and the symbol
 $\Sigma= X,Y,Z,W$, appearing for the four cases in Eq~(\ref{sigbz}) are mentioned in 
the Appendix F. 

Additionally, it is important to note that
the well-known {\it Suyama-Yamaguchi} consistency relation \cite{Suyama:2007bg,Ichikawa:2008iq} between
 the three and four point non-Gaussian parameters, $f_{NL}^{\mathrm{local}}$, $\tau_{NL}^{\mathrm{local}}$ and $g_{NL}^{\mathrm{local}}$ 
 violates \cite{Smith:2011if,Choudhury:2012kw} in the present context due to the appearance of $\cdots$ terms in Eq~(\ref{d1},\ref{d11},\ref{d111}). As the contributions
 form $\cdots$ terms are positive, the consistency relation is modified as:
\be\label{conq}
g_{NL}^{\mathrm{local}}=\frac{25}{108}\tau_{NL}^{\mathrm{local}}=\frac{9}{25}\left(f_{NL}^{\mathrm{local}}\right)^{2}+\cdots.\ee
However, it is important to note that since $\cdots$ terms are small, the amount of violation is also small.

Further using $\delta N$ formalism, the CMB dipolar asymmetry parameter for single field inflationary framework can be expressed as \cite{Kohri:2013kqa}:

\be\begin{array}{llll}\label{eqaq}
    \displaystyle A_{CMB}=\frac{1}{4}\frac{\Delta P_{s}(k)}{P_{s}(k)}\approx \frac{1}{2}\frac{\Delta (\delta N)}{\delta N}=
\frac{3}{5}f_{NL}^{\mathrm{local}}|N_{,\phi}\Delta\phi|+\frac{27}{50}g_{NL}^{\mathrm{local}}|N_{,\phi}\Delta\phi|^{2}+\cdots
   \end{array}\ee
where $|N_{,\phi}\Delta\phi|<1$ for which the perturbative expansion is valid here. In it the field excursion
 $\Delta\phi=\phi_{cmb}-\phi_{e}\approx \phi_{*}-\phi_{e}$. Hence substituting Eq~(\ref{d1},\ref{d11},\ref{d111}) in Eq~(\ref{eqaq}) we derive
the following simplified expression in terms of the tensor-to-scalar ratio and sound speed as:
\be\begin{array}{llll}\label{eqaq1}
    \displaystyle A_{CMB}=
\frac{1}{2}\frac{|\Delta\phi|}{\phi_*}+\frac{1}{8}\left(\frac{|\Delta\phi|}{\phi_*}\right)^{2}+\cdots
\\
~~~~~~~~\approx \displaystyle \frac{3M_{p}}{50\phi_*\sqrt{c_{s}}}\sqrt{\frac{r_{\star}}{0.12}}\left|\left\{\frac{3}{400}
\left(\frac{r_{\star}}{0.12}\right)-\frac{3\gamma\phi_{*}M^{2}_{p}}{\alpha}-\frac{1}{2}
\,\right\}\right|+\cdots
   \end{array}\ee
where $\frac{|\Delta\phi|}{\phi_*}<1$ in the present sub-Planckian setup.

\subsection{Constraining local type of non-Gaussianity and CMB dipolar asymmetry via multi parameter scanning}

Our present job is to scan the parameter space for $c_H,~a_H$, by fixing $\lambda ={\cal O}(1)$ and $\delta \sim 10^{-4}$. 
In order to satisfy the inflationary paradigm, the Planck observational constraints, as stated in the introduction of this chapter, we obtain the following
constraints on our parameters for $H_{inf}\geq m_{\phi}\sim {\cal O}(\rm TeV)$:
\begin{eqnarray}\label{P-space}
c_{H} &\sim &{\cal O}(10-10^{-6})\,,
\\
a_{H} & \sim &{\cal O}(30 - 10^{-3} )\,,
\\
M_s &\sim & {\cal O}(9.50\times 10^{10}-1.77\times10^{16})~{\rm GeV}\,,
\,.
\end{eqnarray}

Inflation would not occur outside the scanning region since, at least, one of the constraints would be violated.
Note that for the above ranges, the VEV of the inflaton, $\langle \phi \rangle =\phi_0$, gets automatically fixed by Eq.~(\ref{phi0}), 
in the sub-Planckian scale as:%
\begin{equation}\label{pj0}
\phi_{0} \sim {\cal O}(10^{14} -10^{17})~{\rm GeV}\,
\end{equation}
which bounds the tensor-to-scalar ratio within, $10^{-22}\leq r_{*}\leq0.12$ for the present setup.
This analysis will further constrain the non-minimal K\"ahler coupling parameters $a,b,c,d$~
\footnote{The analytical expressions for the non-minimal coupling parameters, $a,b,c,d$ can 
be expressed in terms of the scale (VEV) of the heavy field $M_{s}$ are explicitly mentioned in the Appendix I.}
appearing in the higher dimensional Planck scale suppressed 
operators within the following range:
\begin{eqnarray}\label{abcd-space}
a &\sim &{\cal O}(1-0.99)\,,
\\
b & \sim &{\cal O}(1-0.92)\,,
\\
c &\sim & {\cal O}(0.3-1)\,,
\\
d &\sim & {\cal O}(1-0.5)\,.
\end{eqnarray}

In Fig~(\ref{fza}) and Fig~(\ref{fzat}) we have shown the behaviour of the local type of non-Gaussian parameter $f_{NL}^{\mathrm{local}}$ and $\tau_{NL}^{\mathrm{local}}$
with respect to the sound speed $c_{s}$ in the Hubble induced inflationary regime ($H>>m_{\phi}$).
  In Fig~(\ref{fza}), the shaded yellow region represent the allowed parameter space for Hubble induced inflation which satisfies 
the combined Planck constraints on the $f_{NL}^{\mathrm{local}}$ and $c_{s}$. For all the four cases, the region above the $f_{NL}^{\mathrm{local}}=8.5$
line is observationally excluded by the Planck data. The four distinctive
 features are obtained by varying the model parameters of the effective theory of ${\cal N}=1$ SUGRA, $c_{H},a_{H}$ and $M_{s}$
subject to the constraint as stated in Eq~(\ref{P-space}-\ref{pj0}). As Planck puts an upper bound,
$\tau_{NL}^{\mathrm{local}}\leq2800$, the rest of the region above the $\tau_{NL}^{\mathrm{local}}=2800$ line in Fig~(\ref{fzat}) is 
excluded. In the present setup we have obtained the following stringent bound on
the $f_{NL}^{\mathrm{local}}$, $\tau_{NL}^{\mathrm{local}}$ and $g_{NL}^{\mathrm{local}}$ within the following range: 
\be\begin{array}{lll}\label{we1}
    \displaystyle 5\leq f_{NL}^{\mathrm{local}}\leq8.5,~~~100\leq\tau_{NL}^{\mathrm{local}}\leq2800,~~~23.2\leq g_{NL}^{\mathrm{local}}\leq648.2 & \mbox{ for $\underline{\bf Case ~I}$}  \\ 
  1\leq f_{NL}^{\mathrm{local}}\leq8.5,~~~150\leq\tau_{NL}^{\mathrm{local}}\leq2800,~~~34.7\leq g_{NL}^{\mathrm{local}}\leq648.2 & \mbox{ for $\underline{\bf Case ~II}$}  \\ 
    5\leq f_{NL}^{\mathrm{local}}\leq8.5,~~~~75\leq\tau_{NL}^{\mathrm{local}}\leq2800,~~~17.4\leq g_{NL}^{\mathrm{local}}\leq648.2 & \mbox{ for $\underline{\bf Case ~III}$}  \\ 
    2\leq f_{NL}^{\mathrm{local}}\leq8.5,~~~110\leq\tau_{NL}^{\mathrm{local}}\leq2800,~~~25.5\leq g_{NL}^{\mathrm{local}}\leq648.2 & \mbox{ for $\underline{\bf Case ~IV}$}.
          \end{array}
\ee 
Here the theoretical upper and lower bound on $f_{NL}^{\mathrm{local}}$~\footnote{In the prescribed setup the consistency relation between the
non-Gaussian parameter $f_{NL}^{\mathrm{local}}$ and the spectral tilt $n_{s}$ \cite{Maldacena:2002vr}, $f_{NL}^{\mathrm{local}}\sim\frac{5}{12}(1-n_{s})$, does not hold as in the present 
setup sound speed, $c_{s}\neq 1$ and for such non-minimal ${\cal N}=1$ SUGRA setup, Planck data favours lower values of the sound speed (within $0.02<c_{s}<1$).}
 satisfy both the constraints on the $f_{NL}^{\mathrm{local}}$ and $c_{s}$
observed by Planck data. Also it is important to note that, within this prescribed framework, $\tau_{NL}^{\mathrm{local}}$ is bounded by below
for all the four cases and consequently it is possible to put a stringent lower bound on $\tau_{NL}^{\mathrm{local}}$ which satisfies 
the constraints on $\tau_{NL}^{\mathrm{local}}$ and $c_{s}$ both. Till date the observational results obtained from Planck do not give any significant 
constraint on $g_{NL}^{\mathrm{local}}$. However in this paper we have provided a theoretical lower and upper bound of $g_{NL}^{\mathrm{local}}$ using the 
consistency relation between $\tau_{NL}^{\mathrm{local}}$ and $g_{NL}^{\mathrm{local}}$ as stated in Eq~(\ref{conq}).

Finally, in Fig~(\ref{fzata}) we have shown the behaviour
 of the CMB dipolar asymmetry parameter $A_{CMB}$ 
with respect to the tensor-to-scalar ratio $r_{*}$ within, $10^{-22}\leq r_{*}\leq0.12$, at the pivot scale, $k_{*}\sim 0.002~{\rm Mpc}^{-1}$ for the Hubble induced inflation.
 Here the red and blue coloured boundaries are obtained by fixing the sound speed at $c_{S}=0.02$ and $c_{S}=1$. The orange dark coloured region 
satisfied the Planck constraint
 on the $A_{CMB}$ i.e. $0.05\leq A_{CMB}\leq 0.09$~\footnote{The upper bound of the CMB dipolar asymmetry parameter ($A_{CMB}$) can be expressed in terms of the
non-Gaussian parameter $f_{NL}^{\mathrm{local}}$ through a consistency relation as \cite{Namjoo:2013fka}, $A_{CMB}\lesssim 10^{-1}f_{NL}^{\mathrm{local}}$, which perfectly holds good
in the present effective theory setup.} 
for $10^{-22}\leq r_{*}\leq0.12$ within our proposed framework. In Fig~(\ref{fzata}) 
only the region bounded by the red, blue and brown colour is the allowed one and
the rest of the region ($A_{CMB}<0.02$ and $A_{CMB}>0.09$) is 
excluded by the Planck data.

\section{Conclusion}

In this chapter we have studied primordial non-Gaussian features from ${\cal N}=1$ SUGRA inflationary framework in presence of Planck scale suppressed non-minimal
 K\"{a}hler operators using $\delta N$ formalism. The major outcomes of our study in this chapter are:
\begin{itemize}
 \item Here we have shown that in any general class of ${\cal N}=1$
SUGRA inflationary framework, the behaviour of K\"ahler potential in presence of non-minimal K\"ahler corrections in effective theory setup 
are constrained via the non-minimal couplings of the non-renormalizable gauge invariant K\"ahler higher dimensional Planck scale suppressed 
operators from the observational constraint on non-Gaussianity, sound speed and CMB dipolar asymmetry as obtained from the Planck data.

\item  In the present setup the hidden sector based heavy field is settled down in its potential via its Hubble induced vacuum energy density. 
In particular, for the numerical estimations in this paper we have used a very particular kind of inflection point inflationary model, which 
is fully embedded within MSSM, where the inflaton is made up of $\widetilde L\widetilde L\widetilde e$ and $\widetilde u\widetilde d\widetilde d$
gauge invariant D-flat directions. However the prescribed methodology holds good for other kinds of inflationary models too. 

\item Further, we have scanned the multi-parameter region characterized by the Hubble induced mass parameter, $c_{H}$, A-term, $a_{H}$ and the scale of the 
heavy field $M_{s}$, where we have satisfied the current Planck observational constraints on the, 
inflationary parameters: $P_S,~n_S,~c_s,r_*$ (within $2\sigma$~CL), non-Gaussian parameters: $f_{NL}^{local},\tau_{NL}^{local}$ 
(within $1\sigma-1.5\sigma$~CL) and CMB dipolar asymmetry parameter $A_{CMB}$ (within $2\sigma$~CL).

\item In our analysis the non-minimal K\"ahler couplings, $a,b,c,d$ are fixed within $\sim {\cal O}(1)$ in the proposed effective theory setup.

\item Finally, using this methodology, we have obtained 
the theoretical upper and lower bound on the non-Gaussian parameters within the range, ${\cal O}(1-5)\leq f_{NL}\leq8.5$, ${\cal O}(75-150)\leq\tau_{NL}<2800$ 
and ${\cal O}(17.4-34.7)\leq g_{NL}\leq 648.2$, and the
 CMB dipolar asymmetry parameter within, $0.05\leq A_{CMB}\leq0.09$, which satisfy 
the observational constraints stated in Eq~(\ref{pow}-\ref{f-nl3}), as obtained from Planck data. 
\end{itemize}








\chapter{Summary \& Conclusion}
\label{ch:FRWxc4}

The early universe provides
an arena where various ideas about quantum field theory can be tested and the initial
singularity of the Big Bang model is a prime example where a quantum field theoretic prescription is compulsory. Fluctuations of the metric during
inflation, imprinted in primordial B-mode perturbations of the CMB, are
the most vivid evidences conceivable for the reality of field theory.
Inflation defers the singularity problem, allowing us to make predictions
for the initial conditions that emerge from the aftermath of the Big Bang.
However, as we have discussed in the thesis, the inflationary paradigm retains a subtle
sensitivity to (sub) Planckian-scale interactions. This is a challenge for microscopic
 theories of inflation, as well as, an opportunity for using the early
universe as a natural laboratory to test (sub) Planckian-scale physics. To fulfill this expectation, inflationary scenarios
  must be developed to an unprecedented
level of completeness and sophistication.
The last decade of research on inflation has witnessed a
number of significant advances. The development of various field theoretic tools
 has led to vastly improved understanding of the background models, and in turn, to a
sharply improved understanding 
of the associated inflationary models. In special cases it has been possible to
characterize the Planck-scale suppressed non-renormalizable operators to the inflatonary action.
 On the other hand, many critical challenges still remain in existence, as for examples reheating phenomena, leptogenesis, the connection among the
 inflationary sector and
the Standard Model and Supersymmetric degrees of freedom are not clearly known.
 These continue to be
zeroth-order challenge for deriving inflation using various field theoretic tools and have
stymied many attempts to construct inflationary models. 
Furthermore, in most of the cases, the non-renormalizable Planck-suppressed corrections and other renormalizable loop corrections to the inflationary
action are only partially characterized. Finally, and most importantly, there
is not a single observation that gives direct evidence for a field-theoretic origin for inflation and reheating,
 although an unambiguous detection of primordial gravitational waves
produced by quantum fluctuations of the metric during inflation would directly prove
 the quantization of the gravitational field. 
A striking feature of present day observations is the extraordinary simplicity
of the primordial curvature fluctuations, which are, approximately Gaussian,
adiabatic, and, nearly scale-invariant. In contrast, the ultraviolet completions, involving many interacting fields
and a landscape of quantized parameters, are made. The simplicity of the data motivates
various theories of inflation but it does not constrain the
ultraviolet completions in the same way. It is important to
understand whether the simplicity of the data can emerge from the apparent complexity of the ultraviolet completion. To serve this purpose one should determine which
details of the short-distance physics decouple and which should leave subtle traces
in the data.
On the other hand, it is an important open problem
to determine the relative probabilities of different inflationary models in a
broader setting. More generally, deriving specific predictions from the field theoretic tools as a whole, rather than from individual models, are distant goal
that could require a new approach to measure problem.

Keeping the above points in mind, in this thesis we have studied the following aspects:-
\begin{enumerate}
 \item Inflationary model building from various field theoretic setup covering both low-scale and high-scale models.

\item  Estimation of cosmological parameters from the proposed models of inflation and confronting them with observational data.

\item  Estimation of scale of inflation from the proposed setup and generation of primordial gravitational waves. 

\item  Quantification of reheating temperature and study of non-trivial features of leptogenesis scenario in braneworld.


\item Estimation of primordial non-Gaussianity and CMB asymmetry parameter using $\delta N$ formalism.
\end{enumerate}

We briefly summarize our works discussed in the various chapters of the thesis as follows:- 
\begin{itemize}
\item  In the chapter \ref{ch:FRW} we review the various field theoretic approaches applicable to modeling early universe- especially in the context of 
cosmological inflation. Here we first start our discussion with -Supersymmetry, Supergravity and braneworld scenario and then we have discussed the particle physics and 
cosmology connection in the context of inflation, reheating, PBH formation and primordial non-Gaussianity in which all of these field theoretic tools are applied.

\item  In the chapter \ref{ch:FRWxc} we have proposed two different models of inflation 
in the framework of MSSM with various flat directions using saddle point and inflection point mechanism.
 We have demonstrated how we can construct the
 effective inflationary potential in the vicinity of the {\it saddle point} starting
 from $n=4$ and $n=6$ level superpotential for the D-flat direction content
 \textbf{QQQL,~QuQd,~QuLe,~uude} and \textbf{udd,~LLe} respectively. The effective inflaton potential
 around saddle point and inflection point have been utilized in estimating 
 the observable parameters and confronting them with
WMAP7 and Planck dataset using the publicly available code CAMB, which reveals consistency of our model with latest observations. 
 One of the advantages of the proposed model is that it is embedded fully within MSSM, and therefore, it predicts 
the right thermal history of the universe with {\it no extra relativistic degrees of freedom} other than that of the  Standard Model.

\item  In the chapter \ref{ch:FRWxc1} we have studied single field inflation in the context of Randall-Sundrum braneworld and DBI Galileon
induced D3 brane respectively. We have demonstrated the technical details of construction mechanism of an one-loop
4D inflationary potential via dimensional reduction starting from ${\cal N}=2, {\cal D}=5$ supergravity in
the bulk which leads to an effective field theoretic picture within ${\cal N}=1, {\cal D}=4$ supergravity embedded in the brane for both the cases. 
 Hence we have studied inflation using the one loop effective potential
by estimating the observable parameters originated from primordial quantum fluctuation for scalar and tensor modes.
We have further confronted our results with WMAP7 \cite{WMAP7} dataset by
using the cosmological code CAMB \cite{CAMB}. 
Additionally we have proposed new sets of inflationary consistency relations in the case of Randall-Sundrum braneworld and DBI Galileon framework which is 
different from the results obtained from usual GR framework.

\item In the chapter \ref{ch:FRWxc2} we have explored the features of reheating in brane
cosmology on the background of supergravity. We have exhibited
 the process of construction of a fruitful theory of reheating for an effective
4D inflationary potential in ${\cal N}=1, {\cal D}=4$ supergravity in the brane derived from  ${\cal N}=2, {\cal D}=5$ supergravity in
the bulk. We have employed this
setup in reheating model building by analyzing the reheating temperature in the context of brane inflation, followed
by analytical and numerical estimation of different phenomenological parameters.
 It is worthwhile to mention that we get a lot of new
results in the context of braneworld compared to standard GR case.
We further we propose a theory which reflects the effect of particle production
through collision and decay thereby showing a direct connection with the thermalization phenomena.
 To show this internal link more explicitly we put forward both analytical and numerical expressions for
 the gravitino abundance in a physical volume in the reheating epoch. 
 Also the validity
 of leptogenesis for our model shows the production of heavy Majorana neutrinos
in the brane.

\item In the chapter \ref{ch:FRWxc3} we have shown that in any general class of ${\cal N}=1$
SUGRA inflationary framework, the behaviour of K\"ahler potential in presence of non-minimal K\"ahler corrections in effective theory setup 
are constrained via the non-minimal couplings of the non-renormalizable gauge invariant K\"ahler higher dimensional Planck scale suppressed 
operators from the observational constraint on non-Gaussianity, sound speed and CMB dipolar asymmetry as obtained from the Planck data.
In the present setup the hidden sector based heavy field is settled down in its potential via its Hubble induced vacuum energy density. 
In particular, for the numerical estimations in this paper we have used a very particular kind of inflection point inflationary model, which 
is fully embedded within MSSM, where the inflaton is made up of $\widetilde L\widetilde L\widetilde e$ and $\widetilde u\widetilde d\widetilde d$
gauge invariant D-flat directions. However the prescribed methodology holds good for other kinds of inflationary models too.

\end{itemize}

The future prospects of the work studied in the thesis are:-
\begin{itemize}

\item \underline{\bf Cosmology from Effective Field Theory:}

 One of the prime goals is to study the various cosmo-phenomenological aspects
of Effective Field Theory (EFT) of (eternal) inflation in a model independent fashion by
imposing the constraints on various cosmological parameters obtained from the recent
observational probe which we have not addressed in the thesis elaborately. The beauty of the EFT technique is not to bother about the background field
theoretical origin and also to tightly constrain (and rule out) various existing models of
inflation available in literature. But this has a generic power to modify the background
dynamics of inflation and other cosmological features by incorporating the effects of
interactions in the effective action via higher order radiative corrections. Specifically in the
context of eternal inflation it resolves the Infra Red (IR) problem by taking into account all
possible higher order loop contributions. The earlier works in this area did not address this
issue properly. Our future aim is to address this serious issue in a very comprehensive
manner. Additionally we want to extend the EFT approach in the context of dark matter, dark energy and large scale structure formation.

\item \underline{\bf Primordial non-Gaussianity:}

Single field models of inflation are expected to produce small local type of non- Gaussianity
in the squeezed limit according to consistency relation proposed in \cite{Maldacena:2002vr}.
Nevertheless it is not necessary that it will always hold good in every physical prescription.
One of my future interests is to compute the various types of non-Gaussianity-local,
orthogonal and equilateral from bispectrum and trispectrum and to study the features in
squeezed limit configuration from the proposed inflationary models discussed in the thesis. Another focus of mine
is to study the violations/modifications of various inflationary consistency relations by
proposing model independent techniques from which one can constrain various proposed
model of inflation available in literature. We are also keenly interested in studying the features
of various CMB cross correlations from scalar and tensor cosmological perturbations from
the proposed models. Further using using EFT approach we want to constrain the features of
local type of primordial non-Gaussianity computed from the bispectrum and trispectrum
using -In-In formalism and $\delta N$ formalism.

\item \underline{\bf CMB Polarization:}

At present, one of the most challenging
area of research in the field of theoretical physics is to explain the unexplored features of
Polarization data which is already released by Planck. Another future goal
of mine is to constrain/ rule out various models of inflation using the results of CMB
Polarization (via TT, TE, BB, EB correlations) to be observed by Planck. Besides, our aim is
to study reconstruction technique of the primordial power spectrum and also the inflationary
potential using the polarization data in a model independent fashion by which it is possible
to rule out various models available in literature and also to break the degeneracy among the
cosmological parameters as obtained from various models.

\item \underline{\bf Gravitational Waves:}

One of the important focusing issues of the present day research, is to
search for the Primordial Gravitational Waves (PGW) through the large scale B-mode
signal. It is a common notion amongst theoretical physicists that the gravitational waves are
generated just after the Big-Bang and during inflationary era their amplitude will be
amplified to a permissible value. But the signatures of such gravity waves are yet to be
detected. This, in turn, it will provide the best information along with CMB about the era of
cosmic inflation. Even if the gravity waves are detected, we will not be able to draw positive
conclusion about its origin- primordial or stochastic. As the CMB Polarized B-mode signals
are generated from lensing of purely E-mode signals, the separation of pure primordial
gravity waves from observed data will be an essential subject matter for the present day
research. In future I would like to carry on my research in the extraction of pure (unlensed)
B-mode signal from the lensed data by introducing a completely new theoretical tool. As the
South Pole Telescope (SPT) has claimed about the detection of CMB (lensed) B-
mode polarization \cite{Hanson:2013hsb}, a lot of avenues would remain open for future work. This methodology
can also be applied to the inflationary scenario to constrain inflationary observables further
as inflationary tensor perturbations are believed to be the main source of primordial gravity
waves. After extraction of the unlensed B-mode signal, our aim is to propose a
theoretical/semi-analytical tool by which it is possible to subtract the effect of primordial
non-Gaussianity. This will clearly quantify the effect of inflationary tensor perturbations via
primordial gravity waves (or tensor-to-scalar ratio) present in the unlensed B-mode signal.

\item  \underline{\bf Particle Cosmology:}

Apart from that one of the most promising branches of research in present day theoretical
physics is ``particle cosmology''. Particle physics examines nature on the smallest scales,
while cosmology studies the universe on the largest scales. One of the focus of our future
research is to elaborately study various cosmo-particle phenomenological aspects like-
re/preheating, leptogenesis, baryogenesis and dark matter etc from Beyond the Standard
Model (BSM) physics (Example: Supersymmetry and Supergravity, MSSM, NMSSM, CMSSM
etc.) where the structure of the effective four dimensional Field equations are subsequently
modified in presence of modifications in the background geometry sector in light of recent
collider and observational constraints.

\end{itemize}

\mychapter{Appendix}
\subsubsection*{\LARGE A. Higher order slow-roll corrections within GR}
\label{ap1}
 \begin{eqnarray}\label{para 21a} \displaystyle P_{S}(k_{\star}) 
&=&\left[1-(2{\cal C}_{E}+1)\epsilon_{V}+{\cal C}_{E}\eta_{V}\right]^{2}\frac{V}{24\pi^{2}M^{4}_{p}\epsilon_{V}},\\
\label{para 21b} 
P_{T}(k_{\star})&=&\left[1-({\cal C}_{E}+1)\epsilon_{V}\right]^{2}\frac{2V}{3\pi^{2}M^{4}_{p}},
\\
\label{para 21c} \displaystyle  n_{S}(k_{\star})-1
 &\approx& (2\eta_{V}
-6\epsilon_{V})-2{\cal C}_{E}\xi^{2}_{V}+\frac{2}{3}\eta^{2}_{V}+2(8{\cal C}_{E}+3)\epsilon^{2}_{V}
+2\epsilon_{V}\eta_{V}\left(6{\cal C}_{E}+\frac{7}{3}\right)
\nonumber \\ &&~~~~~~~~~~~~~~~~~~~~~~~~~~-4{\cal C}_{E}({\cal C}_{E}+1)\xi^{2}_{V}\epsilon_{V}+2C^{2}_{E}\eta_{V}\xi^{2}_{V},\\
\label{para 21d}  \displaystyle n_{T}(k_{\star})
&\approx&-2\epsilon_{V}+2\left(2{\cal C}_{E}+\frac{5}{3}\right)\epsilon_{V}\eta_{V}-2\left(4{\cal C}_{E}+\frac{13}{3}\right)\epsilon^{2}_{V},\\
 \label{para 21e} \displaystyle  r(k_{\star})&=&16\epsilon_{V}\frac{\left[1-({\cal C}_{E}+1)\epsilon_{V}\right]^{2}}{\left[1-(2{\cal C}_{E}+1)\epsilon_{V}
+{\cal C}_{E}\eta_{V}\right]^{2}}\approx16\epsilon_{V}\left[1+2{\cal C}_{E}(\epsilon_{V}-\eta_{V})\right],\\
  \label{para 21f}  \displaystyle \alpha_{S}(k_{\star})&\approx&\left(16\eta_{V}\epsilon_{V}-24\epsilon^{2}_{V}-2\xi^{2}_{V}\right)-2{\cal C}_{E}(4\epsilon_{V}\xi^{2}_{V}
-\eta_{V}\xi^{2}_{V}-\sigma^{3}_{V})+\frac{4}{3}\eta_{V}(2\eta_{V}\epsilon_{V}-\xi^{2}_{V})
 \nonumber\\ &&~~~
\displaystyle+4(8{\cal C}_{E}+3)\epsilon_{V}(4\epsilon^{2}_{V}-2\eta_{V}\epsilon_{V})
 \nonumber \\ &&~~~
-4{\cal C}_{E}({\cal C}_{E}+1)\left[\epsilon_{V}(4\epsilon_{V}\xi^{2}_{V}-\eta_{V}\xi^{2}_{V}-\sigma^{3}_{V})+\xi^{2}_{V}(4\epsilon^{2}_{V}-2\eta_{V}\epsilon_{V})\right]
+2{\cal C}^{2}_{E}\xi^{2}_{V}(2\eta_{V}\epsilon_{V}-\xi^{2}_{V})\nonumber \\  &&
~~~~+2{\cal C}^{2}_{E}\eta_{V}(4\epsilon_{V}\xi^{2}_{V}-\eta_{V}\xi^{2}_{V}-\sigma^{3}_{V}),\\
\label{para 21g} \displaystyle\alpha_{T}(k_{\star})&\approx&(4\eta_{V}\epsilon_{V}-8\epsilon^{2}_{V})+2\left(2{\cal C}_{E}
+\frac{5}{3}\right)\left[\epsilon_{V}(2\eta_{V}\epsilon_{V}-\xi^{2}_{V})
+\eta_{V}(4\epsilon^{2}_{V}-2\eta_{V}\epsilon_{V})\right]
\nonumber\\ &&~~~~~~~~~~~~~~~~~~~~~~~~~~~~~~~-4\left(4{\cal C}_{E}+\frac{13}{3}\right)
\epsilon_{V}(4\epsilon^{2}_{V}-2\eta_{V}\epsilon_{V}),\end{eqnarray}
 \be\begin{array}{lllll}\label{para 21h} \displaystyle   \kappa_{S}(k_{\star})\approx192\epsilon^{2}_{V}\eta_{V}-192\epsilon^{3}_{V}+2\sigma^{3}_{V}
-24\epsilon_{V}\xi^{2}_{V}+2\eta_{V}\xi^{2}_{V}-32\eta^{2}_{V}\epsilon_{V}\\~~~~~~~~~~\displaystyle-8{\cal C}_{E}\left[\epsilon_{V}(4\epsilon_{V}\xi^{2}_{V}-\eta_{V}\xi^{2}_{V}-\sigma^{3}_{V})
+\xi^{2}_{V}(4\epsilon^{2}_{V}-2\eta_{V}\epsilon_{V})\right]
\\~~~~~~~~~\displaystyle +2{\cal C}_{E}\left[\eta_{V}(4\epsilon_{V}\xi^{2}_{V}-\eta_{V}\xi^{2}_{V}-\sigma^{3}_{V})
+\xi^{2}_{V}(2\eta_{V}\epsilon_{V}-\xi^{2}_{V})\right]+4{\cal C}_{E}\sigma^{3}_{V}(3\epsilon_{V}-\eta_{V})+\frac{4}{3}(2\eta_{V}\epsilon_{V}-\xi^{2}_{V})^{2}\\
~~~~~~~~~\displaystyle +\frac{4}{3}\eta_{V}\left[2\eta_{V}(4\epsilon^{2}_{V}-2\eta_{V}\epsilon_{V})+2\epsilon_{V}(2\eta_{V}\epsilon_{V}-\xi^{2}_{V})-(4\epsilon_{V}\xi^{2}_{V}-\eta_{V}\xi^{2}_{V}-\sigma^{3}_{V})\right] 
 \\ ~~~~~~~~~\displaystyle+4(8{\cal C}_{E}+3)(4\epsilon^{2}_{V}-2\eta_{V}\epsilon_{V})^{2}+16(8{\cal C}_{E}+3)\epsilon_{V}\left[(2\epsilon_{V}-\eta_{V})(4\epsilon^{2}_{V}-2\eta_{V}\epsilon_{V})
-\epsilon_{V}(2\eta_{V}\epsilon_{V}-\xi^{2}_{V})\right]
\\ \displaystyle ~~~~~~~~~+4\left(6{\cal C}_{E}+\frac{7}{3}\right)(2\eta_{V}\epsilon_{V}-\xi^{2}_{V})(4\epsilon^{2}_{V}-2\eta_{V}\epsilon_{V})
\\ \displaystyle~~~~~~~~~+2\left(6{\cal C}_{E}+\frac{7}{3}\right)\epsilon_{V}\left[2(2\eta_{V}\epsilon_{V}-\xi^{2}_{V})\epsilon_{V}+2\eta_{V}(4\epsilon^{2}_{V}-2\eta_{V}\epsilon_{V})
\right.\\ \left.\displaystyle ~~~~~~~~~-(4\epsilon_{V}\xi^{2}_{V}
-\eta_{V}\xi^{2}_{V}-\sigma^{3}_{V})\right]-4{\cal C}_{E}({\cal C}_{E}+1)(4\epsilon^{2}_{V}-2\eta_{V}\epsilon_{V})(4\epsilon_{V}\xi^{2}_{V}-\eta_{V}\xi^{2}_{V}-\sigma^{3}_{V})
\\ \displaystyle~~~~~~~~~
-4{\cal C}_{E}({\cal C}_{E}+1)\epsilon_{V}\left[(4\epsilon_{V}-\eta_{V})(4\epsilon_{V}\xi^{2}_{V}-\eta_{V}\xi^{2}_{V}-\sigma^{3}_{V})+(16\epsilon^{2}_{V}+\xi^{2}_{V}
-10\eta_{V}\epsilon_{V})\xi^{2}_{V}\right.\\ \left.~~~~~~~~~~~\displaystyle -2\sigma^{3}_{V}(3\epsilon_{V}-\eta_{V})\right]
-4{\cal C}_{E}({\cal C}_{E}+1)\left[(4\epsilon_{V}\xi^{2}_{V}-\eta_{V}\xi^{2}_{V}-\sigma^{3}_{V})(4\epsilon^{2}_{V}-2\eta_{V}\epsilon_{V})
\right.\\ \left.~~~~~~~~~\displaystyle +2\xi^{2}_{V}((4\epsilon_{V}-\eta_{V})[4\epsilon^{2}_{V}-2\eta_{V}\epsilon_{V}]-\epsilon_{V}[2\eta_{V}\epsilon_{V}-\xi^{2}_{V}])\right]
\\ \displaystyle ~~~~~~~~~+2{\cal C}^{2}_{E}[(4\epsilon_{V}\xi^{2}_{V}-\eta_{V}\xi^{2}_{V}-\sigma^{3}_{V})(2\eta_{V}\epsilon_{V}-\xi^{2}_{V})
+\xi^{2}_{V}(2\eta_{V}(4\epsilon^{2}_{V}-2\eta_{V}\epsilon_{V})+2\epsilon_{V}(2\eta_{V}\epsilon_{V}-\xi^{2}_{V})\\ \displaystyle~~~~~~~~~~
-[4\epsilon_{V}\xi^{2}_{V}-\eta_{V}\xi^{2}_{V}
-\sigma^{3}_{V}])]+2{\cal C}^{2}_{E}[(2\eta_{V}\epsilon_{V}-\xi^{2}_{V})
(4\epsilon_{V}\xi^{2}_{V}-\eta_{V}\xi^{2}_{V}-\sigma^{3}_{V})\\~~~~~~~~~+\eta_{V}([4\epsilon_{V}-\eta_{V}]
[4\epsilon_{V}\xi^{2}_{V}-\eta_{V}\xi^{2}_{V}-\sigma^{3}_{V}]+\xi^{2}_{V}[16\epsilon^{2}_{V}+\xi^{2}_{V}
-10\eta_{V}\epsilon_{V}]-2\sigma^{3}_{V}[3\epsilon_{V}-\eta_{V}])],\end{array}\ee
 \be\begin{array}{lllll}\label{para 21i}  \kappa_{T}(k_{\star})\approx 56\eta_{V}\epsilon^{2}_{V}-64\epsilon^{3}_{V}
-8\eta^{2}_{V}\epsilon_{V}-4\epsilon_{V}\xi^{2}_{V}+2\left(2{\cal C}_{E}
+\frac{5}{3}\right)\left[(2\eta_{V}\epsilon_{V}-\xi^{2}_{V})
(4\epsilon^{2}_{V}-2\eta_{V}\epsilon_{V})\right.\\ \left.~~~~~~~~~\displaystyle 
+\epsilon_{V}(2\eta_{V}[4\epsilon^{2}_{V}-2\eta_{V}\epsilon_{V}]+2\epsilon_{V}[2\eta_{V}\epsilon_{V}-\xi^{2}_{V}]-
[4\epsilon_{V}\xi^{2}_{V}-\eta_{V}\xi^{2}_{V}-\sigma^{3}_{V}])
\right.\\ \left.~~~~~~~~~+\eta_{V}(8\epsilon_{V}[4\epsilon^{2}_{V}-2\eta_{V}\epsilon_{V}]-2\eta_{V}[4\epsilon^{2}_{V}-2\eta_{V}\epsilon_{V}]
-2\epsilon_{V}[2\eta_{V}\epsilon_{V}-\xi^{2}_{V}])\right]\\ \displaystyle~~~~~~~~~~
-4\left(4{\cal C}_{E}+\frac{13}{3}\right)[(4\epsilon^{2}_{V}-2\eta_{V}\epsilon_{V})^{2}+\epsilon_{V}(
(8\epsilon_{V}-2\eta_{V})[4\epsilon^{2}_{V}-\epsilon_{V}]-2\epsilon_{V}[2\eta_{V}\epsilon_{V}-\xi^{2}_{V}])].
\end{array}\ee

\subsubsection*{\LARGE B. Consistency relations in brane inflation}\label{ap2}

In the context of RS single braneworld the spectral indices $(n_S, n_T)$, running $(\alpha_S, \alpha_T)$ and running of the running $(\kappa_T,\kappa_S)$ 
at the momentum pivot scale $k_*\approx k =aH$ 
can be expressed as \cite{Choudhury:2011sq}:
\begin{eqnarray}
 n_S(k_*) -1&=& 2\eta_{V}(\phi_*)-6\epsilon_{V}(k_*),\\
n_T(k_*) &=& -3\epsilon_{V}(k_*)=-\frac{r_{V}(k_*)}{8},\\
\alpha_S(k_*) &=& 16\eta_{V}(k_*)\epsilon_{V}(k_*)-18\epsilon^{2}_{V}(k_*)-2\xi^{2}_{V}(k_*),\\
\alpha_T(k_*) &=& 6\eta_{V}(k_*)\epsilon_{V}(k_*)-9\epsilon^{2}_{V}(k_*),\\
\kappa_S(k_*)&=&152\eta_{V}(k_*)\epsilon^{2}_{V}(k_*)-32\epsilon_{V}(k_*)\eta^{2}_{V}(k_*)-108\epsilon^{3}_{V}(k_*)
-24\xi^{2}_{V}(k_*)\epsilon_{V}(k_*)\nonumber \\ && ~~~~~~~~~~~~~~~~~~~~~~~~~~~~~~~~~~~~~~~~~~~~+2\eta_{V}(k_*)\xi^{2}_{V}(k_*)+2\sigma^{3}_{V}(k_*),\\
\kappa_T(k_*)&=&66\eta_{V}(k_*)\epsilon^{2}_{V}(k_*)-12\epsilon_{V}(k_*)\eta^{2}_{V}(k_*)
-54\epsilon^{3}_{V}(k_*)-6\epsilon_{V}(k_*)\xi^{2}(k_*).
\end{eqnarray}
from which we get the following set of consistency relations:
\begin{eqnarray}
 \label{wq1}n_T(k_*)-n_S(k_*)+1&=&\left(\frac{d\ln r(k)}{d\ln k}\right)_*=\frac{r(k_*)}{8}-2\eta_{V}(k_*),\\
\label{wq2}\alpha_T(k_*)-\alpha_S(k_*)&=&\left(\frac{d^2\ln r(k)}{d\ln k^2}\right)_*=\left(\frac{r(k_*)}{8}\right)^2-\frac{20}{3}\left(\frac{r(k_*)}{8}\right)
+2\xi^2_{V}(k_*),\\
\label{wq3}\kappa_T(k_*)-\kappa_S(k_*)&=&\left(\frac{d^3\ln r(k)}{d\ln k^3}\right)_* \nonumber\\&=&
2\left(\frac{r(k_*)}{8}\right)^3-\frac{86}{9}\left(\frac{r(k_*)}{8}\right)^2
\nonumber\\ &&+\frac{4}{3}\left(6\xi^2_{V}(k_*)+5\eta^{2}_{V}(k_*)\right)\left(\frac{r(k_*)}{8}\right)
\nonumber\\ && +2\eta(k_*)\xi^2_{V}(k_*)+2\sigma^{3}_{V}(k_*).
\end{eqnarray}
Here Eq~(\ref{wq1}-\ref{wq3})) represent the running, running of the running and running of the double running of tensor-to-scalar ratio.

\subsubsection*{\LARGE C. Dimensional reduction technique for DBI Galileon}\label{ap3}

In this section we employ dimensional reduction technique to derive 
a ${\cal N}$=1, ${\cal D}$=4 SUGRA and the inflaton potential therefrom that results in DBI Galileon on the D3 brane.
In this framework the 5D and 4D {\it Riemann tensor}, {\it Ricci tensor} and {\it Ricci scalar} are related through the following expressions:
\be\begin{array}{lllll}\label{r1}
  \displaystyle R^{(5)}_{\alpha\beta\gamma\delta}=R^{(4)}_{\alpha\beta\gamma\delta}
+\frac{\exp(2A(y))}{R^{2}\beta^{2}}\left(\frac{dA(y)}{dy}\right)^{2}\left[g^{(4)}_{\gamma\beta}g^{(4)}_{\alpha\delta}
-g^{(4)}_{\alpha\gamma}g^{(4)}_{\delta\beta}\right]
   \end{array}
\ee
\be\begin{array}{llll}\label{r2}
    \displaystyle R^{(5)}_{\alpha\beta}=R^{(4)}_{\alpha\beta}
-\frac{3g^{(4)}_{\alpha\beta}}{R^{2}\beta^{2}}\left(\frac{dA(y)}{dy}\right)^{2}
   \end{array}
\ee
\be\begin{array}{llll}\label{r5}
    \displaystyle R_{(5)}=\exp(2A(y))\left[R_{(4)}-\frac{12}{\beta^{2}R^{2}}\left(\frac{dA(y)}{dy}\right)^{2}
-\frac{8}{\beta^{2}R^{2}}\left(\frac{d^{2}A(y)}{dy^{2}}\right)-2\Lambda_{5}\exp(2A(y))\right]
   \end{array}
\ee
which is necessarily required for dimensional reduction. For convenience we deal with different contributions to the action
(\ref{totac}) separately.
\\
\underline{\bf I. The Einstein-Hilbert action}:-\\
\\
After integrating out the contribution from the five dimension, the  Einstein Hilbert action in four dimension can be written as:
\be\begin{array}{llllll}\label{ghu}\displaystyle S^{(4)}_{EH}=\displaystyle\frac{1}{2{\kappa}^{2}_{4}}\int d^{4}x\sqrt{-g_{(4)}}\left[R_{(4)}-\frac{3M^{3}_{5}\beta b^{6}_{0}}{M^{2}_{p}R^{5}}
{\cal I}(1)\right],\end{array}\ee
where the explicit expression for ${\cal I}(1)$ is mentioned in Appendix C. In this context $R_{(4)}$ is the 4D Ricci scalar. It is important to mention here that
the 5D Planck mass ($M_{5}$) and 4D Planck mass ($M_{p}$) are related through the following relation: 
\be\begin{array}{lllll}\label{mass}
  \tiny \displaystyle M^{2}_{p}=M^{3}_{(5)}\beta R\int^{+\pi R}_{-\pi R}dy \exp(3A(y))=\frac{M^{3}_{(5)}b^{3}_{0}}{3R^{2}T^{3/2}_{(4)}}\left[\exp(\beta \pi R)P_{1}-\exp(-\beta\pi R)P_{2}\right]
   \end{array}
\ee 
where $P_{1}$ and $P_{2}$ is defined as:
\be\begin{array}{llll}
P_{1}=\left\{\frac{3\sqrt{T_{(4)}}}{\sqrt{\exp(\beta\pi R)+T_{(4)}\exp(-\beta\pi R)}}
-\sqrt{\frac{\exp(2\pi\beta R)+T_{(4)}}{\exp(\beta\pi R)+T_{(4)}\exp(-\beta\pi R)}}
~_2F_1\left[\frac{1}{2};\frac{3}{7};\frac{7}{4};-\frac{\exp(2\beta\pi R)}{T_{(4)}}\right]\right\},\\
P_{2}=\left\{\frac{3\sqrt{T_{(4)}}}{\sqrt{\exp(-\beta\pi R)+T_{(4)}\exp(\beta\pi R)}}
-\sqrt{\frac{\exp(-2\beta\pi R)+T_{(4)}}{\exp(-\beta\pi R)+T_{(4)}\exp(\beta\pi R)}}
~_2F_1\left[\frac{1}{2};\frac{3}{7};\frac{7}{4};-\frac{\exp(-2\beta\pi R)}{T_{(4)}}\right]\right\}.
\end{array}\ee

\underline{\bf II. The higher curvature gravity action}:-\\

Using Eq~(\ref{r1})-Eq~(\ref{r5}) in Eq~(\ref{5gb}) we get 
\be\begin{array}{lllll}\label{gbe4}
     \displaystyle S^{(4)}_{GB}
=\frac{\alpha_{(4)}}{2\kappa^{2}_{(4)}}\int d^{4}x\sqrt{-g_{(4)}}
\left[\left({\cal C}(1)R^{\alpha\beta\gamma\delta(4)}R^{(4)}_{\alpha\beta\gamma\delta}
-4{\cal I}(2)R^{\alpha\beta(4)}R^{(4)}_{\alpha\beta}+{\cal A}(6)R^{2}_{(4)}\right)\right.\\ \left.
\displaystyle ~~~~~~~~~~~~~~~~~~~~~~~~~~~~+\frac{2{\cal C}(2)}{R^{2}\beta^{2}}R^{(4)}_{\alpha\beta\gamma\delta}\left(g^{\gamma\beta(4)}g^{\delta\alpha(4)}
-g^{\gamma\alpha(4)}g^{\delta\beta(4)}\right)
+\frac{{\cal G}(1)}{R^{4}\beta^{4}}+\frac{{\cal G}(2)}{R^{2}\beta^{2}}R_{(4)}\right]
   \end{array}
\ee
where the co-efficients after dimensional reduction are given by:
\be\begin{array}{llll}{\cal G}(1)=24{\cal C}(4)-144{\cal I}(4)-64{\cal A}(5)+144{\cal A}(7)+64{\cal A}(8)+192{\cal A}(11),\\
{\cal G}(2)=24{\cal I}(2)-24{\cal A}(9)-16{\cal A}(10).\end{array}\ee The scaling relationship between 4D and 5D 
coupling constant is given by:
\be\alpha_{(4)}=\frac{\kappa^{2}_{(4)}\beta R}{\kappa^{2}_{(5)}}\alpha_{(5)}\ee where $\kappa_{(4)}$ and $\kappa_{(5)}$ are 
gravitational couplings in 4D and 5D respectively. Explicit form of each of the constants appearing in Eq~(\ref{gbe4}) are mentioned in \cite{Choudhury:2012yh}.

\underline{\bf III. The D3 Brane Action}:-\\

To reduce the D4 brane action we employ the method of separation of variable 
$\Phi(X^{A})=\Phi(x^{\mu},y)=\phi(x^{\mu})\exp(\frac{2\pi i y}{R})$. Consequently the  D3 
brane action turns out to be 
\be\label{br1}
   S^{(4)}_{D3~Brane}= \int d^{4}x \sqrt{-g^{(4)}}
\left[\tilde{K}(\phi,\tilde{X})-\tilde{G}(\phi,\tilde{X})\Box^{(4)}\phi\right],\ee  
where 
\be\begin{array}{llll}
  \tilde{K}(\phi,X)=\left\{-\frac{\tilde{D}}{\tilde{f}(\phi)}\left[\sqrt{1-2Q\tilde{X}\tilde{f}}-Q_{1}\right]
-\tilde{C}_{5}\tilde{G}(\phi,\tilde{X})-\bar{Q}_{2}\tilde{D}V^{(4)}_{brane}\right\},  \\
\tilde{G}(\phi,\tilde{X})=\left(\frac{\tilde{g}(\phi)k_{1}\tilde{C}_{4}}{2(1-2\tilde{f}(\phi)\tilde{X}k_{2}))}\right),
~~~\tilde{g}(\phi)=\tilde{g}_{0}+\tilde{g}_{2}\phi^{2}, \\ \tilde{D}=\frac{D}{2\kappa^{2}_{(4)}},~~~\tilde{C}_{4}=\frac{C_{4}}{2\kappa^{2}_{4}},~~~
\tilde{C}_{5}=\frac{\bar{C}_{5}\beta^{2} R^{2}}{2\kappa^{2}_{4}},~~~\bar{Q}_{2}\tilde{D}=\beta R.
   \end{array}\ee
The effective {\it Klebanov Strassler}
and {\it Coulomb } frame function on the D3 brane are hereby expressed as 
$\tilde{f}(\phi)\simeq\frac{1}{(\tilde{f}_{0}+\tilde{f}_{2}\phi^{2}+\tilde{f}_{4}\phi^{4})}$ and 
$\nu^{(4)}(\phi)=\tilde{\nu}_{0}+\frac{\tilde{\nu}_{4}}{\phi^{4}}$ with $\tilde{\nu}_{0}=\nu_{0}{\cal A}(1)$
, $\tilde{\nu}_{4}=\nu_{4}{\cal A}(12)$.
The scaled D3 brane potential turns out to be:
\be\label{scaled} \tilde{V}^{(4)}_{brane}=\bar{Q}_{2}\tilde{D}V^{(4)}_{brane}=
T^{}_{(3)}\nu^{(4)}(\phi)+\frac{\beta R{\cal I}(2)}{\tilde{f}(\phi)}\ee
where the  D3 brane tension ($T_{(3)}$) can be expressed in terms of the D4 brane tension ($T_{(4)}$), compactification radius ($R$) and
the slope parameter ($\beta$) as,  $T^{}_{(3)}=\beta R T_{(4)}$.
\\
\\
 \underline{\bf IV. The ${\cal N}$=1, ${\cal D}$=4 Supergravity Action}:-\\

Further, imposing $Z_{2}$ symmetry to $\phi$ via
$\Phi(0)=\Phi(\pi
R)=0$ and compactifying around a circle $(S^1)$
$\partial_{5}\Phi=\sqrt{V^{(5)}_{bulk}(G)}\left(1-\frac{1}{2\pi
R}\right)$ we get,
\be \label{tout}
S^{(5)}_{Bulk~Sugra}=\frac{1}{2}\int d^{4}x\int^{+\pi R}_{-\pi
R}dy\sqrt{-g^{(5)}}\left[e_{(4)}e^{5}_{\dot{5}}\left\{g^{\alpha\beta}G_{m}^{n}(\partial_{\alpha}\phi^{m})^{\dagger}(\partial_{\beta}\phi_{n})
-g^{55}\frac{V^{(5)}_{bulk}(G)}{4\pi^{2}R^{2}}\right\}\right].\ee
Now using the above mentioned ansatz for method of separation of variable we get
\be\begin{array}{lllll}\label{ast8}\displaystyle S^{(4)}_{Sugra}=
\frac{1}{2\kappa^{2}_{(4)}}\int d^{4}x \sqrt{-g^{(4)}}\left[M(T,T^{\dag})
{\cal J}_{\nu}^{\mu}(\phi,\phi^{\dag})g^{\alpha\beta (4)}(\partial_{\alpha}\phi_{\mu})^{\dag}(\partial_{\beta}\phi^{\nu})
-Z(T,T^{\dag})V^{(4)}_{F}(\phi)
\right].\end{array}\ee
where we define

\be\begin{array}{lllll}\label{talx}\displaystyle {\cal J}_{\nu}^{\mu}(\phi,\phi^{\dag})=\int^{+\pi R}_{-\pi R}dy \exp(-A(y))
\left(\frac{\partial^{2}{\cal K}(\phi\exp(\frac{2\pi i y}{R}), 
\phi^{\dag}\exp(-\frac{2\pi i y}{R}))}{\partial\phi^{\dag}_{\mu}\partial\phi^{\nu}}\right), \\ \displaystyle
M(T,T^{\dag})=\frac{\sqrt{2}\beta R^{2}}{(T+T^{\dag})}, Z(T,T^{\dag})=\frac{1}{8\pi^{2}R^{2}\beta|T+T^{\dag}|^{2}}.\end{array}\ee 
Here we have used the ansatz ${\cal W}(\phi,\phi^{\dag},T,T^{\dag})=\frac{1}{4}{\cal W}_{1}(\phi,\phi^{\dag})|T+T^{\dag}|^{2}$ 
for superpotential and the factorization ansatz ${\cal K}(\phi\exp(\frac{2\pi i y}{R}),
\phi^{\dag}\exp(-\frac{2\pi i y}{R}))
={\cal K}_{1}(\phi,\phi^{\dag}){\cal K}_{2}(\exp(\frac{2\pi i y}{R}),\exp(-\frac{2\pi i y}{R}))$
with ${\cal K}_{1}(\phi,\phi^{\dag})={\cal K}_{1}^{\alpha\beta}\phi_{\alpha}\phi^{\dag}_{\beta}$
and ${\cal K}_{2}(\exp(\frac{2\pi i y}{R}),\exp(-\frac{2\pi i y}{R}))=1$ for the {\it K$\ddot{a}$hler } using which the effective 4D F-term potential can be expressed as:
\be\begin{array}{lllllll}\label{exact}
 \displaystyle V^{(4)}_{F}= {\cal A}(13)\exp\left(\frac{{\cal K}_{1}^{\alpha\beta}\phi_{\alpha}\phi^{\dag}_{\beta}}{M^{2}}\right)
\left[\left(\frac{\partial {\cal W}_{1}}{\partial
\phi_{\alpha}}+{\cal K}_{1}^{\alpha\beta}\phi^{\dag}_{\beta}\frac{{\cal W}_{1}}{M^{2}}\right)^{\dag}
 {\cal K}_{1\alpha}^{\nu}\left(\frac{\partial {\cal W}_{1}}{\partial
\phi^{\nu}}+{\cal K}_{1\nu\eta}\phi^{\eta}\frac{{\cal W}_{1}}{M^{2}}\right)-3\frac{|{\cal W}_{1}|^{2}}{M^{2}}\right]
   \end{array}\ee
with the general {\it K$\ddot{a}$hler metric}
 ${\cal K}_{1}^{\alpha\beta}=\frac{\partial^{2}{\cal K}_{1}}{\partial\phi_{\alpha}\partial\phi^{\dag}_{\beta}}$. 
In most of the simple situations, we are interested in the {\it Canonical metric} structure defined by ${\cal K}_{1}^{\alpha\beta}=\delta^{\alpha\beta}$.
Consequently the  ${\cal N}=1,{\cal D}=4$ SUGRA action turns out to be
\be\begin{array}{lllll}\label{ast5}\displaystyle S^{(4)}_{Can~Sugra}=
\frac{1}{2\kappa^{2}_{(4)}}\int d^{4}x \sqrt{-g^{(4)}}\left[M(T,T^{\dag})
g^{\alpha\beta (4)}(\partial_{\alpha}\phi_{\mu})^{\dag}(\partial_{\beta}\phi^{\mu})
-Z(T,T^{\dag})V^{(4)}_{Can}(\phi)
\right].\end{array}\ee
where the canonical F-term potential 
 can be recast as \be\label{totalpot}
V^{(4)}=V^{(4)}_{F}={\cal A}(13)
\exp\left(\frac{\phi^{\dagger}_{\alpha}\phi^{\alpha}}{M^{2}}\right)\left[\left|\frac{\partial
{\cal W}_{1}}{\partial \phi_{\beta}}\right|^{2}-3\frac{|{\cal W}_{1}|^{2}}{M^{2}}\right].\ee
To derive the expression for the specific form of the inflaton potential we start with a specific superpotential \cite{Choudhury:2012yh}
${\cal W}_{1}=v\phi-\frac{g}{n+1}\phi^{n+1}$ with $n\ge 2$. Here $g$ and $v$ is the coupling constant and 
the VEV of $\phi$ respectively. This leads to the following form of the bulk contribution to the  potential:
\be\label{yu}
  V^{(4)}_{bulk}(\phi) ={\cal A}(13) \exp\left(|\phi|^2\right) \left[\left| \left( 1+|\phi|^2 \right)v^2-\left( 1+\frac{|\phi|^2}{n+1} \right) g\phi^n \right|^2 
             - 3 |\phi|^2 \left| v^2 - \frac{g}{n+1}\phi^n \right|^2\right].\ee
Identifying $\phi\rightarrow
\sqrt{2}~{\rm Re} (\phi)$ and
imposing renormalization condition, here we restrict ourselves to $n=2$ leading to effective ${\cal N}=1, {\cal D}=4$ SUGRA
potential:
\be\label{pot1}
V^{(4)}_{bulk}(\phi) = {\cal A}(13)\left(v^4 - gv^2 \phi^2
  + \frac{g^2}{4}\phi^{4}\right). 
\ee

\subsubsection*{\LARGE \bf D. Standard results of reheating mechanism and leptogenesis
in Einsteinian gravity:}  

\begin{itemize}
 \item {\bf I. Reheating temperature:} \be
T_R \, \sim \, 0.2 \left(\frac{100}{N^*}\right)^{1/4}\left( \Gamma_{total} M_{PL} \right)^{1/2} \, .
\ee
\item {\bf II. Extremum (maximum) temperature:}
\be T_{max}\sim \frac{0.8}{N^{*1/4}} V^{1/8}_{0}\left( \Gamma_{total} M_{PL} \right)^{1/2} \,\ee 
where $V_0$ is the vacuum energy.
\item {\bf III. Temperature-time relationship:}
\be T=\frac{T_R}{\sqrt{2H_R (t-t_R) +1}} \, 
\ee
where $H_R$ is the Hubble parameter at reheating time scale $t_R$.
\item {\bf IV. Number density of Gravitino during reheating:}
\be
Y_{\tilde{G}}(T_R)=\frac{2\alpha}{M^2_{PL}}\left( \frac{\zeta(3)}
{\pi^2} \right)^2\left(\frac{45}{2\pi^2 N^*}\right)
\frac{T_{max}^4}{H_{0} T_R}\,
\label{density-treh}
\ee
where $H_0=\frac{\sqrt{V_0}}{\sqrt{3}M_{PL}}$.
\end{itemize}

\subsubsection*{\LARGE \bf E. The model parameters $\alpha,\beta,\gamma,\kappa$:}
The model parameters characterizing the potential stated in Eq~(\ref{rt1a}) can be expressed as:  
\begin{eqnarray}\label{p1}
     \alpha&=&M^{4}_{s}+\left(\frac{(n-2)^{2}}{n(n-1)}+\frac{(n-2)^2}{n}\delta^{2}\right)c_{H}H^{2}\phi^{2}_{0}+\cdots,\\
 \beta&=&2\left(\frac{n-2}{2}\right)^{2}\delta^{2}c_{H}H^{2}\phi_{0}+\cdots,\\
 \gamma&=&\frac{c_{H}H^{2}}{\phi_{0}}\left(4(n-2)^2-\frac{(n-1)(n-2)^3}{2}\delta^{2}\right)+\cdots,\\
 \small\kappa&=&\frac{c_{H}H^{2}}{\phi^{2}_{0}}\left(12(n-2)^3-\frac{(n-1)(n-2)(n-3)
(7n^2-27n+26)}{2}\delta^{2}\right)+\cdots
   \end{eqnarray}
where the higher order $\cdots$ terms are neglected due to $\delta^{2}<<1$. During numerical estimations I fix $n=6$ for
 $\widetilde L\widetilde L\widetilde e$ and $\widetilde u\widetilde d\widetilde d$  D-flat 
directions respectively.

\subsubsection*{\LARGE \bf F. The symbol $\Sigma=X,Y,Z,W$:}

The symbols appearing in the Eq~(\ref{sigbz}), in the definition of the sound speed $c_s$ for $s<< M_{p}$, after imposing the slow-roll approxiation are given by:
 \be\begin{array}{lll}\label{de23}
 \displaystyle {\bf X}_{1}(t)= \sqrt{\frac{2\epsilon_{V}(\phi)V(\phi)}{3}}
\left\{ \sqrt{\frac{2\epsilon_{V}(\phi)V(\phi)}{3}}\frac{aM^{3}_{s}}{M^{2}_{p}}\left[2\sin(2M_{s}t)+4\cos(M_{s}t)\right]\right.\\ \left.
\displaystyle~~~~~~~~~~~~~~~~~~~~~~~~~~~~~~~~~~~~~~~~~~~~
-\frac{aM^{4}_{s}}{M^{2}_{p}}\lvert\phi\rvert\cos{\bf\Theta}\left[\cos(2M_{s}t)-\sin(M_{s}t)\right]\right\},\end{array}\ee
\be\begin{array}{lll}\label{de23a}\displaystyle {\bf Y}_{1}(t)= \sqrt{\frac{2\epsilon_{V}(\phi)V(\phi)}{3}}\left\{ \sqrt{\frac{2\epsilon_{V}(\phi)V(\phi)}{3}}\frac{2bM^{2}_{s}}{M_{p}}\cos(M_{s}t)
+\frac{bM^{3}_{s}}{M_{p}}\lvert\phi\rvert\cos{\bf\Theta}\sin(M_{s}t)\right\},\end{array}\ee
\be\begin{array}{lll}\label{de23aa}\displaystyle {\bf Z}_{1}(t)= \sqrt{\frac{2\epsilon_{V}(\phi)V(\phi)}{3}}\left\{\sqrt{\frac{2\epsilon_{V}(\phi)V(\phi)}{3}}\frac{cM^{3}_{s}}{4M^{2}_{p}}\left[2\sin(2M_{s}t)+4\cos(M_{s}t)\right]
\right.\\ \left.
\displaystyle~~~~~~~~~~~~~~~~~~~~~~~~~~~~~~~~~~~~~~~~~~~
-\frac{cM^{4}_{s}}{4M^{2}_{p}}\lvert\phi\rvert\cos{\bf\Theta}\left[\cos(2M_{s}t)-\sin(M_{s}t)\right]\right\},\end{array}\ee
\be\begin{array}{lll}\label{de23aaa}\displaystyle {\bf W}_{1}(t)= \sqrt{\frac{2\epsilon_{V}(\phi)V(\phi)}{3}}\left\{ \sqrt{\frac{2\epsilon_{V}(\phi)V(\phi)}{3}}\frac{4dM^{2}_{s}}{M_{p}}\cos(M_{s}t)
+\frac{dM^{3}_{s}}{M_{p}}\lvert\phi\rvert\cos{\bf\Theta}\sin(M_{s}t)\right\},\end{array}\ee
\be\begin{array}{lll}\label{de23aaaa}\displaystyle {\bf X}_{2}(t)=\left({\bf Y}_{2}(t)+\frac{a|\phi|^{2}M^{5}_{s}}{M^{2}_{p}}\sin(2M_{s}t)\right),\\
\displaystyle {\bf Y}_{2}(t)={\bf Z}_{2}(t)={\bf W}_{2}(t)=5M^{5}_{s}\sin(2M_{s}t)+8M^{5}_{s}\cos(M_{s}t),\\
\displaystyle {\bf X}_{3}(t)=\left({\bf Y}_{3}(t)-\frac{a|\phi|^{2}M^{5}_{s}}{M^{2}_{p}}\sin(2M_{s}t)\right),\\
\displaystyle {\bf Y}_{3}(t)={\bf Z}_{3}(t)={\bf W}_{3}(t)=3M^{5}_{s}\sin(2M_{s}t)-8M^{5}_{s}\cos(M_{s}t).\\
    \end{array}\ee
Here the complex inflaton field $\phi$ is parameterized by, $\phi=\lvert\phi\rvert\exp(i{\bf\Theta})$. Here the new parameter ${\bf\Theta}$ characterizes 
the phase factor associated with the inflaton and it has a two dimensional rotational symmetry.

\subsection*{\LARGE G. Case -1,~2,~3,~4}

\begin{itemize}

\item{$ {\bf Case-1}~~~~~~~~~~K= \phi^\dag \phi+s^\dag s+\frac{a}{M_{p}^2}\phi^\dag \phi s^\dag s $\\

For the above non-minimal K\"ahler interaction with $'a'$ being a dimensionless number.
 We have also computed the correction to the Hubble-induced mass term, for $c_{H} $ for $|I|<<M_{p}$:
\be
\small c_{H}=\left\{ 3 \left[ (1-a) + (1+a)a\frac{|s|^2}{M_{p}^2}\right]+\left[ (1+3a) + (1-3a)a\frac{|s|^2}{M_{p}^2} \right] \left( \frac{e^K|F_s|^2}{V(s)} -1 \right)
\right\}\approx
                    \displaystyle 3(1-a) \,,
\ee
where we used the fact that: $V(s)=|W_s|^2=3H^2 M^2_{p}=4M^{2}_{s}|s|^2$. Next we compute the correction to the Hubble-induced A term, 
$a_{H}H\frac{\phi^n}{nM_{p}^{n-3}}$, in presence of the non-minimal K\"ahler correction:
\be\begin{array}{lll}\label{a1}
 a_{H}H\frac{\phi^n}{nM_{p}^{n-3}}\\  =\Big(\Big[1+a\frac{|s|^2}{M_{p}^2}\Big]W_\phi \, \phi -3W(\phi)
\Big)\frac{e^K W^*(I^\dagger)}{M_{p}^2} + \left[W(\phi) \frac{I^\dagger}{M_{p}}
-a  W(\phi) \frac{I^\dagger}{M_{p}} \frac{|I|^2}{M_{p}^2}
 \right.\\ \left.~~~~~~~~~~~~~~~~~~~~~~~~~~~~~~- a W_\phi \, \phi \,  \frac{I^\dagger}{M_{p}}
   \left( 1 - a \frac{|s|^2}{M_{p}^2} \right)\right] \displaystyle \frac{e^K F^*_{\bar{s}}}{M_{p}}+h.c. \\
 \approx \left\{\left( 1 + a \frac{|s|^2}{M_{p}^2} \right)\Big(1-\frac{3}{n}\Big)\frac{ s^2}{M^{2}_{p}}
+ \left( 1 - a \frac{|s|^2}{M_{p}^2} \right)\Big(a-\frac{1}{n}\Big) \frac{(s^\dagger)^2}{M^{2}_{p}} \right\}\frac{\lambda M_{s}\phi^n}{M^{n-3}_{p}}+h.c. \,,
\end{array}\ee
which explicitly shows the  Planck suppression for $|s|<<M_{p}$ in the Hubble-induced A term.}\\

\item{$ {\bf Case-2}~~~~~~~~~~~~K=\phi^\dag\phi +s^\dag s+\frac{b}{2M_{p}}s^\dag\phi \phi + h.c.$\\

Similarly, for the above non-minimal k\"ahler correction where $'b'$ is a dimensionless number
%
we can compute the correction to the Hubble-induced mass
term, $c_{H}H^2 |\phi|^2$, for $|s|\ll M_{p}$:
\be\begin{array}{lll}
\displaystyle c_{H}=\left[ 3+ b^2\frac{e^K|W_s|^2}{H^2 M_{p}^2} + \left( \frac{e^K|F_s|^2}{V(s)} -1 \right)\right]\approx  3(1+b^2)\,,
\end{array}\ee
where  $V(s)=|W_s|^2=3H^2 M^2_{p}=4M^{2}_{s}|s|^2$. And similarly the Hubble-induced A term, $a_{H}H\frac{\phi^n}{nM_{p}^{n-3}}$, in 
presence of snon-minimal K\"ahler correction read as:
\be\begin{array}{llll}\label{a2}
 a_{H}H\frac{\phi^n}{nM_{p}^{n-3}}\\
=\Big(W_\phi \,\phi \,
- 3 W(\phi) + b W_\phi \, \phi^\dagger  \frac{s}{M_{p}} \Big) \frac{e^K W^*(I^\dagger)}{M^{2}_{p}}-\frac{b}{2} \frac{e^K W^*(I^\dagger)}{M_{p}^2}  
\left( \frac{W_s}{M_{p}}- \frac{s^\dagger}{M_{p}} \frac{W(I)}{M_{p}^2} \right) \phi \phi   
\\ ~~~~~~~~~~~~~~~~~~~~~~~~~~~~~~~~~~
+\Big( W(\phi) \frac{s^\dagger}{M_{p}}-b W_\phi \, \phi^\dagger \Big)\frac{e^K F_{\bar{s}}^*}{M_{p}} +3bH^2 \frac{s^\dagger}{M_{p}} \phi \phi+ h.c.\\
 =\left\{\left(1-\frac{3}{n}\right)\phi +\frac{ b  \phi^\dagger  s}{n M_{p}} \right\} \frac{\lambda \phi^{n-1}M_{s} s^2}{M_{p}^{n-1}}  
+\Big(\frac{s^\dagger \phi}{M_{p}}-bn\phi^\dagger \Big)\frac{2M_{s}\lambda \phi^{n-1} s^{\dagger}}{n M^{n-2}_{p}} +\frac{4M^{2}_{s} b |s|^2 s^\dagger}{M^3_{p}} \phi \phi\\
 ~~~~~~~~~~~~~~~~~~~~~~~~~~~~~~~~~~~~~~~~~~~~~~~~~~~ -\frac{b M_{s} s^2}{2M_{p}^2}  
\left( \frac{2M_{s} s}{M_{p}}-\frac{M_{s} s^2 s^\dagger}{M_{p}^3} \right) \phi \phi + h.c.
\end{array}\ee}\\

\item{${\bf Case-3}~~~~~~~~~~~~~~K=\phi \phi^\dag+ss^\dag+\frac{c}{4M_{p}^2}s^\dag s^\dag\phi \phi + h.c. $\\

In a similar way we can analyse  the above non-minimal K\"ahler interaction, where $c$ is the dimensionless number. 
We have computed the correction to the Hubble-induced mass
term, $c_{H}H^2 |\phi|^2$ for $|s|\ll M_{p}$ as:
%
\be\begin{array}{lll}
\displaystyle c_{H}=\left[ 3+ \frac{3c}{2} \frac{|s|^2}{M_P^2} + \left( 1+ \frac{3c}{2}  \frac{|s|^2}{M_P^2} 
- \frac{c^2}{4} \frac{|s|^4}{M_P^4} \right) \left( \frac{e^K|F_s|^2}{V(s)} -1 \right)\right]\approx 
                    \displaystyle 3 
\end{array}\ee
where we have used  $V(s)=|W_s|^2=3H^2 M^2_{p}=4M^{2}_{s}|s|^2$. Next we compute  the Hubble-induced A term, 
$a_{H}H\frac{\phi^n}{nM_{p}^{n-3}}$:
\be\begin{array}{llll}\label{a3}
 a_{H}H\frac{\phi^n}{nM_{p}^{n-3}}\\=\Big(W_\phi \,\phi \,
- 3 W(\phi) + \frac{c}{2} W_\phi \, \phi^\dagger  \frac{ss}{M_{p}^2} - \frac{c}{2} \frac{s^\dagger}{M_{p}}\frac{W_s}{M_{p}} \phi\phi
\Big) \frac{e^K W^*(I^\dagger)}{M^{2}_{p}}  
+\Big( W(\phi) \frac{s^\dagger}{M_{p}}- c W_\phi \, \phi^\dagger  \frac{s}{M_{p}} \Big)\frac{e^K F_{\bar{s}}^*}{M_{p}} \\ 
~~~~~~~~~~~~~~~~~~~~~~~~~~~~~~~~~~~~~~~~~~~~~~~~~~~~~~~~~~~~~~~~~~~~~~~~~~~~~~~~~+\frac{3cH^2}{4}
\frac{s^\dagger s^\dagger}{M^{2}_{p}} \phi \phi
+ h.c.\\
=\left\{\left(1-\frac{3}{n}\right)\phi  +\frac{c\phi^\dagger  ss}{2M_{p}^2}\right\} \frac{\lambda \phi^{n-1}M_{s} s^2}{M_{p}^{n-1}}
  -\frac{c M^{2}_{s} I^2I^\dagger I \phi\phi  }{M^4_{p}} 
+\Big(\frac{s^\dagger \phi}{M_{p}}-\frac{cn\phi^\dagger s }{M_{p}} \Big)\frac{2M_{s}\lambda \phi^{n-1} s^{\dagger}}{n M^{n-2}_{p}}\\~~~~~~~~~~~~~~~~~~
~~~~~~~~~~~~~~~~~~~~~~~~~~~~~~~~~~~~~~~~~~~~~~~~~~~~~~~~~~~~~~ +\frac{M^{2}_{s} c 
|s|^2 s^\dagger s^\dagger}{M^4_{p}} \phi \phi + h.c.
\end{array}\ee}\\

\item{${\bf Case-4}~~~~~~~~~~~~~~K=\phi \phi^\dag+ss^\dag+\frac{d}{M_{p}}s\phi^\dag \phi + h.c.$\\

For the above non-minimal K\"ahler potential, where $d$ is the dimensionfull number,
we can compute the Hubble-induced mass
term, $c_{H}H^2 |\phi|^2$, for $|s|\ll M_{p}$:
\be\begin{array}{lll}
 c_{H}=\left[3 \left[ 1 + d \frac{s+s^\dagger}{M_P} 
+ d^2 \left( 1 + d \frac{s+s^\dagger}{M_P} \right)^{-1} \right] +\left[ 1 + d \frac{s+s^\dagger}{M_P} 
\right.\right.\\ \left.\left.~~~~~~~~~~~~~~~~~~~~~~~~~~~+ 3 d^2 \left( 1 + d \frac{s+s^\dagger}{M_P} \right)^{-1} \right] \left( \frac{e^K|F_s|^2}{V(s)} -1 \right) \right] 
\approx 
                    \displaystyle 3(1+d^2) \,,
\end{array}\ee
where we used  $V(s)=|W_s|^2=3H^2 M^2_{p}=4M^{2}_{s}|s|^2$. Next we compute the correction to the Hubble-induced A term, $a_{H}H\frac{\phi^n}{nM_{p}^{n-3}}$,
\be\begin{array}{llll}\label{a4}
 a_{H}H\frac{\phi^n}{nM_{p}^{n-3}}\\=\Big(W_\phi \,\phi \,
- 3 W(\phi) 
\Big) \frac{e^K W^*(s^\dagger)}{M^{2}_{p}}  
+\Big( W(\phi) \frac{I^\dagger}{M_{p}}- d \, W(\phi)\Big)\frac{e^K F_{\bar{s}}^*}{M_{p}} 
+ h.c.\\
=\left(1-\frac{3}{n}\right) \frac{\lambda \phi^{n}M_{s} s^2}{M_{p}^{n-1}}  
+\Big(\frac{s^\dagger}{M_{p}} -d\Big)\frac{2M_{s}\lambda \phi^{n} s^{\dagger}}{n M^{n-2}_{p}} + h.c.
\end{array}\ee}

\end{itemize}



\subsection*{\LARGE H. Expression for $a_H$}\label{expression for $a_H$}

Using these results in Hubble induced A-term, $a_{H}$ can be computed from Eqs.~(\ref{a1}), Eq~(\ref{a2}), Eq~(\ref{a3}) and Eq~(\ref{a4}) for the four physical situations, the simplified 
expressions turn out be:
\be\label{signgxxx}
\small a_{H}\sim \left\{
	\begin{array}{ll}
                      \frac{n}{2}\left(\frac{2}{3}\right)^{\frac{3}{4}}\sqrt{\frac{H_{inf}}{M_{p}}}
\left[1+a-\frac{4}{n}+\frac{35a}{4}\sqrt{\frac{2}{3}}\left(2-\frac{3}{n}\right)\frac{H_{inf}}{M_{p}}-\frac{35a^{2}}{4}\sqrt{\frac{2}{3}}\frac{H_{inf}}{M_{p}}\right]
                    \
  & \mbox{ for $\underline{\bf Case ~I}$}  \\ \\
   \frac{1}{2}\left[3\left(1-\frac{1}{n}\right)+\frac{5b}{n}\sqrt[4]{\frac{2}{3}}\sqrt{\frac{H_{inf}}{M_{p}}}\right] \left(\frac{2}{3}\right)^{\frac{3}{4}}\sqrt{\frac{H_{inf}}{M_{p}}}
+2\sqrt{\frac{2}{3}}\left( \left(\frac{3}{2}\right)^{\frac{3}{4}}\sqrt{\frac{H_{inf}}{M_{p}}}-bn\right)
\\ +10b\left(\frac{3}{2}\right)^{\frac{5}{4}}\left(\frac{M_{p}}{\phi}\right)^{n-2}\left(\frac{H_{inf}}{M_{p}}\right)^{\frac{3}{2}}
-\frac{b}{2}\left(\frac{3}{2}\right)^{\frac{5}{4}}\left(\frac{M_{p}}{\phi}\right)^{n-2}\left(\frac{H_{inf}}{M_{p}}\right)^{\frac{3}{2}}
\left(5-\frac{67}{8}\sqrt{\frac{2}{3}}\frac{H_{inf}}{M_{p}}\right)& \mbox{ for $\underline{\bf Case ~II}$}  \\ \\
   \sqrt[4]{\frac{2}{3}}n\left[\sqrt{\frac{3}{2}}\left(1-\frac{3}{n}\right)+\frac{35c}{24}\frac{H_{inf}}{M_{p}}\right]\sqrt{\frac{H_{inf}}{M_{p}}}
+(1-cn)\sqrt[4]{\frac{2}{3}}\sqrt{\frac{H_{inf}}{M_{p}}} & \mbox{ for $\underline{\bf Case ~III}$}  \\ \\
          \sqrt[4]{\frac{2}{3}}(n-3)\sqrt{\frac{H_{inf}}{M_{p}}}+2\sqrt{\frac{2}{3}}\left[\left(\frac{3}{2}\right)^{\frac{3}{4}}\sqrt{\frac{H_{inf}}{M_{p}}}-d\right] & \mbox{ for $\underline{\bf Case ~IV}$}.
          \end{array}
\right.
\ee

\subsubsection*{\LARGE \bf I. Expression for the non-minimal couplings $a,b,c,d$:}\label{abcd}
The expressions for the non-minimal supergravity coupling parameter $a,b,c,$ and $d$ for all the four physical cases within ${\cal N}=1$ SUGRA   
 with $H_{inf}>>m_{\phi}$ can be expressed in terms of the VEV of the heavy field, $\langle s \rangle= M_{s}$ as:
\be\begin{array}{lll}\label{pspace1cv}
    \displaystyle a\sim {\cal O}\left(1-1.06\times 10^{-5}\frac{n^{2}}{(n-1)}\frac{M^{2}_{s}}{M^{2}_{p}}\right)& \mbox{ for $\underline{\bf Case ~I}$},\\
\displaystyle b\sim {\cal O}\left(\sqrt{\left|\frac{\left(3-\frac{1}{n}\right)^{2}}{100(n-1)}\frac{M^{2}_{s}}{M^{2}_{p}}-1\right|}\right)& \mbox{ for $\underline{\bf Case ~II}$},\\
\displaystyle c\sim {\cal O}\left(\frac{1}{500}\left|\frac{\pm 8.16\frac{M_{s}}{M_{p}}\sqrt{n-1}-\sqrt[4]{\frac{2}{3}}\left(\sqrt{\frac{3}{2}}(n-3)+1
\right)}{1.24\frac{M_{s}}{M_{p}}-\sqrt[4]{\frac{2}{3}}n}\right|\right)& \mbox{ for $\underline{\bf Case ~III}$},\\
\displaystyle d\sim {\cal O}\left(\sqrt{\left|2.54\times 10^{-4}\frac{\left(n-1+\sqrt{6}\right)^{2}}{(n-1)}\frac{M^{2}_{s}}{M^{2}_{p}}-1\right|}\right)& \mbox{ for $\underline{\bf Case ~IV}$}.
   \end{array}\ee


\subsubsection*{\LARGE \bf J. Expression for $\Sigma_{s}(t),\Xi_{s}(t),\Psi_{s}(t),\Theta_{s}(t)$:}\label{time dep}

\be\begin{array}{lll}\label{sig1}
    \tiny  \Sigma_{s}(t)=
\frac{e^{\left(-3H-\sqrt{\frac{4aM^{4}_{s}}{M^{2}_{p}\left(1+
\frac{aM^{2}_{s}}{4M^{2}_{p}}\right)}+9H^{2}}\right)t}}{
\frac{3aM^{4}_{s}}{M^{2}_{p}\left(1+\frac{aM^{2}_{s}}{4M^{2}_{p}}\right)}\left(
\frac{aM^{4}_{s}}{M^{2}_{p}\left(1+\frac{aM^{2}_{s}}{4M^{2}_{p}}\right)}+2H^{2}\right)}\left\{
\frac{3\gamma}{\left(1+\frac{aM^{2}_{s}}{4M^{2}_{p}}\right)}\left[\left(\frac{aM^{4}_{s}}{M^{2}_{p}\left(1+
\frac{aM^{2}_{s}}{4M^{2}_{p}}\right)}+3H^{2}\right.\right.\right.\\ \left.\left.\left. 
~~~~~~~~~~~~~~~~~~~\tiny-H\sqrt{\frac{4aM^{4}_{s}}{M^{2}_{p}\left(1+\frac{aM^{2}_{s}}{4M^{2}_{p}}\right)}+9H^{2}}\right)
\left({\bf C}^{2}_{1}+{\bf C}^{2}_{2}e^{2\left(
\sqrt{\frac{4aM^{4}_{s}}{M^{2}_{p}\left(1+\frac{aM^{2}_{s}}{4M^{2}_{p}}\right)}+9H^{2}}\right)t}\right)\right.\right. \\ \left.\left.~~~~~~~~~~~~~~~~~~~~~~~~~~
~~~~~~~~~~~~~~~~~~~~~~~~~~~~~~~~~~\tiny-6{\bf C}_{1}{\bf C}_{2}\left(
\frac{aM^{4}_{s}}{M^{2}_{p}\left(1+\frac{aM^{2}_{s}}{4M^{2}_{p}}\right)}+2H^{2}\right)\right]\right\},\end{array}\ee
\be\begin{array}{lll}\label{sig2}
\tiny \Xi_{s}(t)=\gamma\left[\frac{{\bf C}^{2}_{4}}{54H^{4}}e^{-6Ht}+
\frac{t}{81H^{5}}\left\{\beta^{2}\left(2-3Ht\right)
-3H^{2}\beta\left(9Ht{\bf C}_{3}
\right.\right.\right.\\ \left.\left.\left.~~~~~~~~~~~~~~~~~~~~~~~~
-\left[\beta t^{2}+6{\bf C}_{3}\right]\right)-81H^{4}
\left(\frac{\beta}{\gamma}-{\bf C}^{2}_{3}\right)\right\}-\frac{{\bf C}_{4}}{243H^{5}}e^{-3Ht}\right.\\ \left.
~~~~~~~~~~~~~~~~~~~~~~~~~~~~~~~~~~~~~~~~~\times\left\{\beta
 \left(2 + 6 H t + 9 H^2 t^2\right) - 18 H^{2}\left(1 + 3 H t\right) {\bf C}_{3}\right\}\right],\end{array}\ee
\be\begin{array}{lll}\label{sig3}
\tiny \Psi_{s}(t)=\gamma\left[\frac{{\bf C}^{2}_{6}}{54H^{4}}e^{-6Ht}+
\frac{t}{81H^{5}}\left\{\beta^{2}\left(2-3Ht\right)
-3H^{2}\beta\left(9Ht{\bf C}_{5}
\right.\right.\right.\\ \left.\left.\left.~~~~~~~~~~~~~~~~~~~~~~~~
-\left[\beta t^{2}+6{\bf C}_{5}\right]\right)-81H^{4}
\left(\frac{\beta}{\gamma}-{\bf C}^{2}_{5}\right)\right\}-\frac{{\bf C}_{6}}{243H^{5}}e^{-3Ht}\right.\\ \left.
~~~~~~~~~~~~~~~~~~~~~~~~~~~~~~~~~~~~~~~~~\times\left\{\beta
 \left(2 + 6 H t + 9 H^2 t^2\right) - 18 H^{2}\left(1 + 3 H t\right) {\bf C}_{5}\right\}\right],\end{array}\ee
\be\begin{array}{lll}\label{sig4}
\tiny \Theta_{s}(t)=\frac{\gamma}{\left(1+\frac{2dM_{s}}{M_{p}}\right)}\left[\frac{{\bf C}^{2}_{8}}{54H^{4}}e^{-6Ht}+
\frac{t}{81H^{5}}\left\{\frac{\beta^{2}}{\left(1+\frac{2dM_{s}}{M_{p}}\right)^{2}}\left(2-3Ht\right)
-\frac{3H^{2}\beta}{\left(1+\frac{2dM_{s}}{M_{p}}\right)}\left(9Ht{\bf C}_{7}
\right.\right.\right.\\ \left.\left.\left.~~~~~~~~~~~~~~~~~~~~~~~~
-\left[\frac{\beta t^{2}}{\left(1+\frac{2dM_{s}}{M_{p}}\right)}+6{\bf C}_{7}\right]\right)-81H^{4}
\left(\frac{\beta}{\gamma}-{\bf C}^{2}_{7}\right)\right\}-\frac{{\bf C}_{8}}{243H^{5}}e^{-3Ht}\right.\\ \left.
~~~~~~~~~~~~~~~~~~~~~~~~~~~~~~~~~~~~~~~~~\times\left\{\frac{\beta}{\left(1+\frac{2dM_{s}}{M_{p}}\right)} 
\left(2 + 6 H t + 9 H^2 t^2\right) - 18 H^{2}\left(1 + 3 H t\right) {\bf C}_{7}\right\}\right].
   \end{array}\ee

\begingroup\raggedright\endgroup


\end{document}